\documentclass[fleqn,nofootinbib,aps,reprint,amsmath,amssymb,14pt,superscriptaddress,floatfix,a4paper,letterpaper,fleqn]{revtex4-1}
\pdfoutput=1

\usepackage{amssymb}
\usepackage{amsfonts}
\usepackage{amsmath}
\usepackage{makeidx}
\usepackage[bookmarksnumbered,colorlinks,bookmarks,breaklinks=true,linkbordercolor={1 1 1},linktocpage,citecolor=blue]{hyperref}
\usepackage{graphicx,color}
\usepackage{multirow}
\usepackage{dcolumn}
\usepackage{bm}
\usepackage{fancyhdr}
\usepackage{array}
\usepackage{threeparttable}
\usepackage{verbatim}
\usepackage[activate={true,nocompatibility},final,tracking=true,kerning=true,spacing=true,factor=1100,stretch=10,shrink=10]{microtype}
\usepackage[english]{babel}
\usepackage{textgreek}

\usepackage[utf8]{inputenc}
\usepackage{xcolor}
\usepackage{adjustbox}
\usepackage{dsfont}
\usepackage{float}
\PassOptionsToPackage{hyphens}{url}
\usepackage{cleveref}

\usepackage{float}
\usepackage{placeins}

\usepackage{ifthen}
\usepackage{xinttools}

\usepackage{soul}

\newcommand\T{\rule{0pt}{2.6ex}}  

\def\etals{\textit{et al.}}

\input{sublib-table-macros}

\newfloat{floatlist}{thp}{lop}
\floatname{floatlist}{List}

\crefname{floatlist}{list}{lists}
\Crefname{floatlist}{List}{Lists}

\begin{document}

\title{
\texorpdfstring{\qquad \\ \qquad \\ \qquad \\  \qquad \\  \qquad \\ \qquad \\}{}
FENDL: A library for fusion research and applications
}
\author{G.\thinspace Schnabel}
\email{g.schnabel@iaea.org}
\affiliation{NAPC--Nuclear Data Section, International Atomic Energy Agency, Vienna, Austria}

\author{D.L.\thinspace Aldama}
\affiliation{Centro de Aplicaciones Tecnol\'ogicas y Desarrollo Nuclear, Havana, Cuba}

\author{T.\thinspace Bohm}
\affiliation{University of Wisconsin-Madison, Madison, Wisconsin, United States}

\author{U.\thinspace Fischer}
\affiliation{Karlsruhe Institute of Technology (KIT), Hermann-von-Helmholtz-Platz 1, 76344 Eggenstein-Leopoldshafen, Germany}

\author{S.\thinspace Kunieda}
\affiliation{Nuclear Science and Engineering Center, Japan Atomic Energy Agency, Tokai, Ibaraki, Japan}

\author{A.\thinspace Trkov}
\affiliation{Josef Stefan Institute, Ljubljana, Slovenia}

\author{C.\thinspace Konno}
\affiliation{Nuclear Science and Engineering Center, Japan Atomic Energy Agency, Tokai, Ibaraki, Japan}

\author{R.\thinspace Capote}
\affiliation{NAPC--Nuclear Data Section, International Atomic Energy Agency, Vienna, Austria}

\author{A.J.\thinspace Koning}
\affiliation{NAPC--Nuclear Data Section, International Atomic Energy Agency, Vienna, Austria}

\author{S. Breidokaite}
\affiliation{Lithuanian Energy Institute, Laboratory of Nuclear Installation Safety, Kaunas, Lithuania}

\author{T.\thinspace Eade}
\affiliation{United Kingdom Atomic Energy Authority, Culham Science Centre, Abingdon, Oxon, OX14 3DB, UK}

\author{M.\thinspace Fabbri}
\affiliation{Fusion for Energy, Barcelona, Spain}

\author{D.\thinspace Flammini}
\affiliation{ENEA, Fusion and Technology for Nuclear Safety and Security Department, 
 C.R. Frascati, Italy}

\author{L.\thinspace Isolan}
\affiliation{Industrial Engineering Department, University of Bologna, Bologna, Italy}

\author{I.\thinspace Kodeli}
\affiliation{Josef Stefan Institute, Ljubljana, Slovenia}

\author{M.\thinspace Košťál}
\affiliation{Research Center Rez Ltd, Husinec Rez 25068 130, Czech Republic}

\author{S.\thinspace Kwon}
\affiliation{Rokkasho Fusion Institute, National Institutes for Quantum Science and Technology, Rokkasho, Aomori, Japan}

\author{D.\thinspace Laghi}
\affiliation{NIER Engineering Castel Maggiore, Italy}
\affiliation{Industrial Engineering Department, University of Bologna, Bologna, Italy}

\author{D.\thinspace Leichtle}
\affiliation{Karlsruhe Institute of Technology (KIT), Hermann-von-Helmholtz-Platz 1, 76344 Eggenstein-Leopoldshafen, Germany}

\author{S.\thinspace Nakayama}
\affiliation{Nuclear Science and Engineering Center, Japan Atomic Energy Agency, Tokai, Ibaraki, Japan}

\author{M.\thinspace Ohta}
\affiliation{Rokkasho Fusion Institute, National Institutes for Quantum Science and Technology, Rokkasho, Aomori, Japan}

\author{L.W.\thinspace Packer}
\affiliation{United Kingdom Atomic Energy Authority, Culham Science Centre, Abingdon, Oxon, OX14 3DB, UK}

\author{Y.\thinspace Qiu}
\affiliation{Karlsruhe Institute of Technology (KIT), Hermann-von-Helmholtz-Platz 1, 76344 Eggenstein-Leopoldshafen, Germany}

\author{S.\thinspace Sato}
\affiliation{Rokkasho Fusion Institute, National Institutes for Quantum Science and Technology, Rokkasho, Aomori, Japan}

\author{M.\thinspace Sawan}
\affiliation{University of Wisconsin-Madison, Madison, Wisconsin, United States}

\author{M.\thinspace Schulc}
\affiliation{Research Center Rez Ltd, Husinec Rez 25068 130, Czech Republic}

\author{G. Stankunas}
\affiliation{Lithuanian Energy Institute, Laboratory of Nuclear Installation Safety, Kaunas, Lithuania}

\author{M.\thinspace Sumini}
\affiliation{Industrial Engineering Department, University of Bologna, Bologna, Italy}

\author{A.\thinspace Valentine}
\affiliation{United Kingdom Atomic Energy Authority, Culham Science Centre, Abingdon, Oxon, OX14 3DB, UK}

\author{R.\thinspace Villari}
\affiliation{ENEA, Fusion and Technology for Nuclear Safety and Security Department, C.R. Frascati, Italy}

\author{A.\thinspace Žohar}
\affiliation{Josef Stefan Institute, Ljubljana, Slovenia}

\pacs{}

\begin{abstract}
The Fusion Evaluated Nuclear Data Library (FENDL) is a comprehensive and validated collection of nuclear cross section data coordinated by the International Atomic Energy Agency (IAEA) Nuclear Data Section (NDS).  FENDL assembles the best nuclear data for fusion applications selected from available nuclear data libraries and has been under development for decades.  FENDL contains sub-libraries for incident neutron, proton, and deuteron cross sections including general purpose and activation files used for particle transport and nuclide inventory calculations.

In this work, we describe the history, selection of evaluations for the various sub-libraries (neutron, proton, deuteron) with the focus on transport and reactor dosimetry applications, the processing of the nuclear data for application codes (e.g. MCNP), and the development of the TENDL-2017 library which is the currently recommended activation library for FENDL. We briefly describe the IAEA IRDFF library as the recommended library for dosimetry fusion applications. We also present work on validation of the neutron sub-library using a variety of fusion relevant computational and experimental benchmarks using the MCNP transport code and ACE-formatted cross section libraries.  A variety of cross section libraries are used for the validation work including FENDL-2.1, FENDL-3.1d, FENDL-3.2, ENDF/B-VIII.0, and JEFF-3.2 with the emphasis on the FENDL libraries.

The results of the validation using computational benchmarks
showed generally good agreement among the tested neutron cross section
libraries for neutron flux, nuclear heating, and primary displacement
damage (dpa).  Gas production (H/He) in structural materials showed substantial differences to the reference FENDL-2.1 library. The results of the experimental validation showed that the performance of FENDL-3.2b is at least as good and in most cases better than FENDL-2.1.  

Future work will consider improved evaluations developed by the International Nuclear Data Evaluation Network (INDEN) for materials such as O, Cu, W, Li, B, and F.  
Additionally, work will need to be done to investigate differences in gas production in structural materials.  Covariance matrices will need to be developed or updated as availability of consistent and comprehensive uncertainty information will be needed as fusion technology and facility construction matures.  Finally, additional validation work for high energy neutrons, protons and deuterons, as well as validation work for the activation library will be needed.  

\end{abstract}

\maketitle
\lhead{FENDL: A library for fusion research}
\chead{}
\rhead{G. Schnabel \etals}
\lfoot{} \rfoot{}
\renewcommand{\headrulewidth}{0.4pt} \renewcommand{\footrulewidth}{0.4pt}
\tableofcontents
\vfill

 
\section{INTRODUCTION}
\label{sec:introduction}
The International Fusion Evaluated Nuclear Data Library (FENDL) is a comprehensive and validated compilation of nuclear cross section data, developed over several decades under the auspices of the IAEA/NDS with the objective to provide a qualified nuclear data library for fusion technology applications, addressing in particular the needs of the ITER project. FENDL comprises a set of sub-libraries for neutron, proton and deuteron induced cross sections including general purpose and activation data files for particle transport simulations and nuclide inventory calculations.

Several versions of the library were produced since the FENDL project was launched at the IAEA/NDS back in 1987~\cite{gouloFusionEvaluatedNuclear1989,pashchenkoFirstResultsFENDL11990}. FENDL-1, the first library version, was
 released in 1994~\cite{ganesanReviewUncertaintyFiles1994} with data files selected in a careful evaluation procedure from the regional nuclear data libraries developed in the US, Japan, the EU, and the Russian Federation (ENDF/B-VI, JENDL, EFF and BROND). Application libraries in continuous energy (ACE) and multi-group (VITAMIN-J group structure) data format for use with Monte Carlo and discrete ordinates codes, designated as FENDL/MC-1.0 and FENDL/MG-1.0, respectively, were provided for the early ITER design calculations and qualified in an international benchmark effort~\cite{fischerBenchmarkValidationFendl11996}.  A further update based on the data evaluations from the then state-of-the-art nuclear data libraries has led to the improved FENDL-2 library~\cite{hermanExtensionImprovementFENDL1997} which further-on served as primary nuclear data source for ITER and other fusion projects.
In 2003, an IAEA Consultants’ Meeting recommended a further update of FENDL-2 with suitable up-to-date data evaluations to remove apparent deficiencies and replace obsolete evaluations~\cite{forrestSummaryReportIAEA2003}.  A new FENDL-2 sub-version, FENDL-2.1, was compiled by the IAEA/NDS, and subsequently released together with application libraries in ACE and multi-group data format~\cite{DLAFENDL21Processing}. FENDL-2.1 since then served as reference data library for ITER neutronics calculations despite some drawbacks and deficiencies revealed, among others, in the course of the experimental FENDL-2.1 validation activities~\cite{fischerDevelopmentNeedsNuclear2007,batistoniValidationFENDL2Nuclear2007}. This exercise highlighted the need for further data improvements and provided recommendations for FENDL-3, the next major library release.

FENDL-3 was developed in the frame of an IAEA-coordinated research project over the years 2008 to 2012 \cite{forrestFENDL3LibrarySummary2012}. The new library includes major extensions and updates with regard to the covered neutron energy range (up to 150 MeV to also serve the needs of the IFMIF fusion material irradiation facility), the library contents (number of isotopes, reactions considered, etc.) and the quality (improved evaluations for many isotopes and reactions, including gas production data and secondary energy-angle distributions). FENDL-3 became also more comprehensive through the use of evaluations from the TENDL project~\cite{koningTENDLCompleteNuclear2019} in the case that data were missing in the major nuclear data libraries. After extensive benchmarking~\cite{fischerBenchmarkingFENDL3Neutron2014}, FENDL-3 was recommended and formally adopted as the new reference data library for ITER. Further updates to FENDL-3.0 were subsequently produced to solve problems identified later in some evaluations and the processing of the data files. This led to the current version FENDL-3.2b released in February 2022, which is available on the IAEA website \url{https://nds.iaea.org/fendl/}.

This paper presents a comprehensive documentation of FENDL-3.2b with its sub-libraries, the rationale for the selection of the data files and evaluations, the processing of the data for the use in neutronics design calculations, and the verification and validation analyses conducted to qualify the library for fusion applications. Future development needs and plans are also addressed.
As a final remark, a list of the many abbreviations used in this paper with their explanations can be found in \cref{apx:acronyms}.

\section{EVALUATION}
\label{sec:evaluation}

\subsection{Neutron library}

\subsubsection{Historical background}

The idea of a library for fusion neutronics calculations goes back to IAEA Specialists' Meetings in Vienna in November 1987 and in May 1989~\cite{gouloFusionEvaluatedNuclear1989}. The original concept of creating an evaluated data library for fusion applications was not to develop new independent evaluations, but to select the evaluations from existing major data libraries that were deemed most suitable for fusion neutronics. Some of the criteria were:
\begin{itemize}
\item Completeness of the evaluations for fusion application purposes.
\item Range of the resolved resonance region (early versions of MCNP did not have the          probability table treatment in the unresolved resonance range).
\item Presence of gamma-production data.
\item Correlated energy-angle representation of the double-differential data.
\end{itemize}
Based on these criteria, a preliminary FENDL-1 library was assembled and prepared for data verification and validation. It was agreed that the full library should include the data files for incident neutrons and charged particles, as well as the library for neutron dosimetry and activation.

In spite of far less powerful computers than we have today, an extensive data verification effort was shared by the participants from different laboratories. Outstanding deficiencies were identified and preparations started for the next release of the library, FENDL-2. The work was presented at the IAEA Specialists' Meetings in Vienna in November 1990~\cite{pashchenkoFirstResultsFENDL11990}. In the report, a list of candidate integral benchmark experiments is identified for library validation. The list of benchmarks was further elaborated in the IAEA Specialists' Meetings in Vienna in March 1994~\cite{ganesanReviewUncertaintyFiles1994}.

Due to the needs of the foreseen IFMIF irradiation device for material testing, it was necessary to extend the energy range of the incident neutron evaluations to at least 60~MeV. Several material evaluations were replaced by data from more recent evaluated data libraries. Based on the feedback from the ITER community, a number of additional materials were added to the library. Altogether, 34 documents related to the development and improvements to the FENDL library are available on the IAEA web site, \url{https://nds.iaea.org/publications/group\_list.php?group=INDC-NDS}, which led to the current version of the library designated FENDL-3.2b.

In this latest release we still selected FENDL evaluations from available evaluations, but in addition to major libraries, also evaluations from TENDL and INDEN projects were considered as potential candidates.

\subsubsection{Evaluations prepared by JAEA}
\label{subsubsec:jaea-evaluations}

The neutron sublibrary of FENDL-2.1 has been created by selecting evaluations from the world libraries ENDF/B~\cite{brownENDFBVIII8th2018}, JEFF~\cite{plompenJointEvaluatedFission2020a}, JENDL~\cite{JENDL4.0} and RUSFOND~\cite{zabrodskayaRUSFONDRussianNational2007}.
For the creation of FENDL-3.0, which was eventually carried over by FENDL-3.2b, the evaluations of FENDL-2.1 were adopted as a starting point in terms of the source of library for each nucleus.
New evaluations that became available in the meantime in the world library projects were used to replace the previous version in FENDL.
These evaluations were then extended to at least 60\,MeV incident energy.
The evaluated data for $^{13}$C, $^{17}$O, Ne, $^{31}$P, S, K, Sm, Lu, $^{180m}$Ta, Re, Pt, $^{234}$U, $^{138}$La were fully adopted from TENDL~\cite{koningTENDLCompleteNuclear2019}. Consult \cref{apx:add-tech-info} for a table showing the specific TENDL version adopted for each isotope.

\textbf{Extension to high energies}. National neutron data libraries such as ENDF/B, JEFF, JENDL and RUSFOND do not have high-energy data in general,
except for some isotopes relevant for accelerator applications.
The usual maximum energy limit is either 20~MeV or 30~MeV for most data files.
However, the reliable assessment of neutron irradiation damage to structural components of currently constructed fusion research facilities also depends on nuclear data at higher energies.
The IFMIF accelerator facility, currently under construction, to study the properties of materials exposed to high-intensity neutrons is an example.
This facility will produce high energy ($<$100~MeV) deuteron or proton beams, which
will subsequently produce high-intensity neutrons above several tens of MeV.

To meet demands of material engineering, the data has therefore been extended to higher energies if the available energy range was not sufficient in the original data file.
When the IAEA-coordinated research project~\cite{sawanNuclearDataLibraries2012} was close to completion in December 2011, only the JENDL/HE-2007~\cite{Watanabe2011} and TENDL-2010~\cite{koningTENDLCompleteNuclear2019} libraries contained high-energy data. Preference was given to the evaluated data in JENDL/HE-2007 when available.
Even though JENDL/HE-2007 provides data for energies up to 3\,GeV, only the data up to 150\,MeV was included in FENDL-3.0 as this energy range was deemed sufficient for fusion research at present and the foreseeable future. Pion production data in JENDL/HE-2007 was not included in FENDL.

The following list enumerates the isotopes in FENDL-3.0 whose energy range has been extended by including data from JENDL-HE/2007. The data for the low energy part up to 20 or 30 MeV, depending on the isotope, has been carried over from new evaluations at that time. 
\begin{itemize}
  \item \fbox{ENDF/B-VII.0 + JENDL/HE-2007} \mbox{}\\--- $^{1}$H, $^{19}$F, $^{35,37}$Cl, $^{39,41}$K, $^{59}$Co, $^{197}$Au, $^{235,238}$U
  \item \fbox{JENDL-4.0 + JENDL/HE-2007} \mbox{}\\--- $^{12}$C, $^{14}$N, $^{23}$Na, $^{24,25,26}$Mg, $^{40,42,43,44,46,48}$Ca, $^{46-50}$Ti, $^{51}$V, $^{69,71}$Ga, $^{90,91,92,94,96}$Zr, $^{93}$Nb, $^{92,94-98,100}$Mo, $^{181}$Ta, $^{36,38,40}$Ar, $^{64,66,67,68,70}$Zn
\end{itemize}
Because JENDL/HE-2007 provided data for each isotope relevant to accelerator applications,
the library did not cover all the isotopes relevant for fusion-related applications.
Therefore, TENDL-2010 was adopted for a number of isotopes for which data were missing in JENDL/HE-2007. The data in TENDL-2010 cover energies up to 200\,MeV and were also included up to that maximum energy in FENDL-3.0.

The isotopes for which data above 20 or 30\,MeV were taken from TENDL-2010 are given in the following list. Data for energies below 20 or 30\,MeV were adopted from new evaluations at that time.
\begin{itemize}
  \item \fbox{ENDF/B-VII.0 + TENDL-2010} \mbox{}\\--- $^{6,7}$Li, $^{9}$Be, $^{10,11}$B, $^{32,33,34,36}$S, $^{40}$K, $^{89}$Y, $^{107,109}$Ag, $^{106,108,110-114,116}$Cd, $^{121,123}$Sb, $^{136,138,140,142}$Ce, $^{130,132,134-138}$Ba, $^{162,164,166,167,168,170}$Er
  \item \fbox{JENDL-4.0 + TENDL-2010} \mbox{}\\--- $^{50}$V, $^{112,114-120,122,124}$Sn, $^{79,81}$Br, $^{133}$Cs, $^{152,154-158,160}$Gd, $^{174,176-180}$Hf
  \item \fbox{JEFF-3.1.1 + TENDL-2010} \mbox{}\\--- $^{103}$Rh, $^{127}$I
  \item \fbox{RUSFOND-2010 + TENDL-2010} \mbox{}\\--- $^{15}$N
\end{itemize}

In the merging process, the data of the two libraries have been kept as they are, without doing any normalization. The discontinuities at the energy boundary introduced by this procedure were left untreated. A desirable future improvement will be therefore to remove discontinuities at the energy boundary.

\textbf{Revision of $(n,x\alpha)$ cross section}. The reliable estimation of the $\alpha$-particle production is very important to evaluate irradiation damage to structural materials.
However, all the world's nuclear libraries poorly estimate the measured cross sections in the high-energy region (above 20~MeV)
because the nuclear model codes employed for the data evaluation were based on the phenomenological models for the pre-equilibrium process at that time. To overcome this issue, one of the authors, Satoshi Kunieda, incorporated the Iwamoto-Harada clustering pre-equilibrium model~\cite{Kunieda2012}, which predicted the $\alpha$-particle production cross section better,
into the GNASH code. The new estimation of cross sections and double-differential cross sections were finally
adopted in ENDF/B-VII.1 for some nuclei relevant for structural materials.
In FENDL-3, we adopted a new evaluation of the $(n,x\alpha)$ data based on this approach, which is useful data for fusion-related research.

\textbf{Revision of energy/angle distributions to correct KERMA factors and DPA cross sections}.

Inconsistencies in the energy/angle distributions led to non-physical KERMA factors to determine heating numbers and DPA (=displacement per atom) cross sections. These inconsistencies have been resolved in FENDL-3.2 with comparatively minor changes. The yields of the outgoing particles provided in MF6 were reviewed and corrected if necessary to improve heating and damage calculations.
Whether an ENDF files was corrected because of this issue is indicated in a designated column \textit{H} in~\cref{tbl:neutron-sublib1,tbl:neutron-sublib2} in the appendix.

\subsubsection{Evaluations from INDEN}
The International Nuclear Data Evaluation Network (INDEN) coordinated by the IAEA \cite{INDEN} produced relevant evaluations for light nuclei, resonant absorbers and structural materials. Evaluations of the structural materials $^{50,52,53,54}$Cr and $^{54,56,57}$Fe were adopted from INDEN as well as $^{10,11}$B, $^{16,18}$O and $^{139}$La (see \cite{INDEN}, recommended evaluations).
The current INDEN evaluation of $^{55}$Mn has not yet been incorporated in the current version FENDL-3.2b. Instead, FENDL still contains an earlier IAEA evaluation that has been adopted by ENDF/B-VII.1~\cite[p.~2911-2914]{chadwickENDFBVIINuclear2011}.

\textbf{Chromium isotopes}.
New evaluation of chromium isotopes were produced within the INDEN collaboration as is fully documented in~\cite{Nobre:2021}. Improvements in fusion benchmarking were reported in that publication.

\textbf{Iron isotopes}.
The OECD/NEA Data Bank organized the CIELO Pilot Project (WPEC SG-40) with the aim to improve the evaluated nuclear data of the most important nuclides for fission reactors, including the iron isotopes, of which $^{56}$Fe is the most important.
Since no better resonance evaluation was available, the old Froehner evaluation was adopted as found in the JENDL-4 library, with only minor corrections for typing errors and one spurious resonance. However, a strong dip was observed in the capture cross section below the first s-wave resonance below 28~keV. This dip was responsible for the over-prediction of reactivity in the ICSBEP benchmarks ZPR-9/34 (ICSBEP labels HEU-MET-INTER-001). $^{56}$Fe is a near-magic nucleus and it is likely to have a significant direct capture contribution, as noted in the Atlas of Resonances~\cite{ATLAS:2018}. An ad-hoc patch was applied to fill the hole in the capture cross section to make the shape closer to $1/v$ in between the resonances. With this patch the problem of the ZPR-9/34 benchmark was solved.
The CIELO $^{56}$Fe evaluation was adopted in the ENDF/B-VIII.0 evaluated nuclear data library and it showed very good performance in criticality benchmarks. Unfortunately, a problem was discovered when modelling the neutron leakage spectra from thick iron shells with a $^{252}$Cf neutron source. The problem was reported in the ENDF/B-VIII.0 library documentation because it was not possible to fix it before the library release.

The cause of the problem was that the evaluation followed closely the new experimental data measured at JRC in Geel of the inelastic cross sections~\cite{Negret:2014}. The data seem to be in contradiction with the elastic cross sections measured at Geel~\cite{Pirovano:2019}, since the sum of the elastic and the inelastic cross sections exceeds the total, which we trust. A reduction of the inelastic cross section was introduced, carefully balancing the elastic and the inelastic cross sections within their experimental uncertainties, while preserving the total cross section. The performance of the modified file was confirmed to be in agreement with the latest neutron leakage experiment through a 50~cm stainless steel cube with a $^{252}$Cf neutron source measured at Rez~\cite{Schulc:2022}.

Jansky was reporting excessive neutron flux at energies near the resonance ``windows'', particularly near 300~keV~\cite{Jansky:2018}. The original idea was that the resonance fit of the elastic cross section in the minima is too low. An ad-hoc patch was made to fill the minima and the performance of the $^{56}$Fe evaluation improved, but unfortunately a new thick transmission experiment at nELBE~\cite{Beyer:2022} proved the patch to be invalid. A further investigation revealed that the resonance data in $^{57}$Fe extend only up to 100~keV, but there exist high-resolution measurements of the $^{57}$Fe total cross section by Pandey, going up to 800~keV. Therefore, the Pandey data were used to follow the measured fluctuating total cross section and the elastic cross section was adjusted to preserve unitarity. The patch to the $^{57}$Fe evaluation solves most of the problems with the performance of the iron evaluations in simulating deep penetration problems without the need to modify the resonance region of the $^{56}$Fe evaluation.

\textbf{Boron isotopes}.

The evaluations of boron isotopes were prepared for FENDL-3.2b taking into account data from the IRDFF-II, ENDF/B-VIII.0 and FENDL-3.1d library to match IRDFF-II dosimetry data. The low energy part was generally taken from ENDF/B-VIII.0 and the high energy part from FENDL-3.1d. The alpha and triton production cross section were taken from IRDFF-II. Finally, the lumped cross section in MF3/MT5 was rescaled to reproduce the total cross section.

\textbf{Oxygen isotopes}.
The objective of the latest evaluation of $^{16}$O in FENDL-3.2b was to keep the good performance of ENDF/B-VIII.0 in the energy range below 16 MeV and to better match higher-energy TIARA leakage experiments above this energy range. 
Therefore, the evaluation of $^{16}$O was adopted from the ENDF/B-VIII.0 library.
The scattering cross section between 16.2 and 110 MeV was taken from JENDL-4/HE.
To establish consistency between the channels, all cross sections in the other reaction channels were
multiplied by the factor
\begin{equation} \nonumber
f =\frac{(\sigma_{\textrm{tot}}^{\textrm{\tiny ENDF/B-VIII}} -
          \sigma_{\textrm{scat}}^{\textrm{\tiny JENDL-4/HE}})}
        {(\sigma_{\textrm{tot}}^{\textrm{\tiny ENDF/B-VIII}} 
         - \sigma_{\textrm{scat}}^{\textrm{\tiny ENDF/B-VIII}})}
\end{equation}

Regarding $^{18}$O, the evaluation of JENDL-4.0/HE has been adopted as a basis.
The capture cross section (MF3/MT102) was taken from EAF-2010.
The gamma production yields (MF12) and associated angular distributions (MF14) were carried over from FENDL-3.1d, which is given by the TENDL-2014 evaluation.
The angle/energy distributions (MF6) were prepared from JENDL-4.0/HE and FENDL-3.1d.

\textbf{Lanthanum-139}.
FENDL-3.0 adopted the TENDL-2010 evaluation of Lanthanum-139. It was later replaced by TENDL-2014 in FENDL-3.1d. 
For the current version, the resonance parameters of ENDF/B-VIII.0 were adopted.
Consequently the total, elastic and non-elastic cross sections in the resonance region below 100\,keV were updated.

\subsection{Incident proton data}
Evaluated cross section data were compiled for 178 nuclei, where data files were taken from those in JENDL/HE-2007,
ENDF/B-VII.0 or TENDL-2011.
Firstly, data were taken from JENDL/HE-2007, which was the most recent evaluation at that time, as we recognized that
the library gave the most reasonable estimation of cross sections through comparison amongst libraries and measured data.
If data were not available in JENDL/HE-2007, they were mostly adopted from TENDL-2011 to complete the library.
A summary table of the proton sublibrary including an indication of the source library can be found in~\cref{tbl:proton-sublib} in the appendix.

The quasi-monoenergetic neutrons produced from $p+$Li can be useful for e.g., the shielding analysis of neutrons.
However, evaluated data for the high energy region were not available at the time.
For instance, ENDF/B-VII.0 only contained results of an R-matrix analysis up to about~5\,MeV.
Therefore, one of the authors, Satoshi Kunieda, performed a new evaluation for $^{6,7}$Li up to 200~MeV,
which combined an R-matrix analysis and the results of  the nuclear model code CCONE~\cite{Iwamoto2016}.
This new evaluation provided a reasonable description of the $(p,xn)$ double-differential cross section as included in
JENDL-4.0/HE~\cite{Kunieda2017}, which has been released in 2015.
This evaluation was also adopted later-on by FENDL-3.2b as these data are pertinent information for fusion research. 

\subsection{Incident deuteron data}
Evaluated deuteron data are indispensable for the design of neutron sources based on deuteron accelerators, such as IFMIF.
Data need not only to cover light elements employed as deuteron irradiation targets but also medium-heavy nuclides
occurring in structural materials.
Therefore, deuteron nuclear data for various nuclides up to at least 150 MeV were also included in FENDL-3.

The status of deuteron nuclear data as of 2012, when FENDL-3 was first released, was as follows.
ENDF/B-VII.1 contained $R$-matrix parameters for $^{2,3}$H, $^{3}$He, and $^{6,7}$Li up to 20\,MeV incident energy. 
TENDL-2011 provided deuteron data up to 200 MeV for 1160 nuclides from Li to Lr.
Evaluated data for neutron and $\gamma$-ray production for the $^{6,7}\mathrm{Li}+d$ reaction have been published by Pereslavtsev~\cite{Pereslavtsev2008}.
These data were bundled with the McDeLicious code~\cite{Simakov2012}, which has been successfully applied to the neutronics design of IFMIF.
However, the upper energy limit of the incident deuterons was 50 MeV and, moreover, the evaluated data were not compiled in the ENDF-6 format.
Some evaluated data related to medical radionuclide production were available~\cite{Tarkanyi2001}, but they were evaluated only for specific reaction channels and the information on the energy and angular distribution of the emitted particles was not provided.
Due to this situation regarding data availability, the first version of FENDL-3 adopted the TENDL-2011 data for the 175 major nuclides from Li to U.

After the first release of FENDL-3, the evaluated data for $^{7}$Li were updated in ENDF/B-VIII.0 but the maximum incident energy remained at 20 MeV.
Some updated data became available for deuteron-induced reaction data related to medical applications~\cite{Hermanne2018,Tarkanyi2019_1,Tarkanyi2019_2,Engle2019} but the evaluation was only performed for excitation functions.
In 2021, JENDL/DEU-2020~\cite{Nakayama2021}, a deuteron nuclear data library for $^{6,7}$Li, $^{9}$Be, and $^{12,13}$C up to 200 MeV, was released.
The data of the library were evaluated by employing the code system dedicated to deuteron-induced reactions called DEURACS~\cite{Nakayama2016}.
In~\cite{Nakayama2021} it is shown that JENDL/DEU-2020 reproduces measured neutron production data much better than TENDL in the incident energy range up to 200 MeV.
Also a benchmark test for $d$-Li thick target neutron yields~\cite{Nishitani2021} concludes that JENDL/DEU-2020 is useful for estimating neutron yields and the characterization of irradiation fields at IFMIF and similar facilities.
Given these circumstances, the evaluations for Li, Be and C isotopes in JENDL/DEU-2020 have been adopted in the latest version of FENDL. 
For the other nuclides, the data of the first version of FENDL, which are from TENDL-2011, were carried over.

Recent advances in deuteron reaction modeling in TALYS are expected to improve the deuteron evaluations in TENDL and will also be considered for adoption in FENDL. For example, an advanced break-up model by M. Avrigeanu~\cite{avrigeanuAdvancedBreakupnucleonEnhancement2022} has been implemented in the TALYS code but needs to be further tested before it can be applied on a wide-scale. At the same time, TALYS model parameters have been automatically adjusted to the available deuteron cross section data, improving at least the general description and prediction of the excitation functions. In addition, there is the possibility to adopt an explicit contribution from the break-up mechanism in the double-differential spectra of deuteron-induced reactions using an extension of the Kalbach parameterization. This has been implemented and validated by P. Sauvan with special adapted versions of NJOY and MCNP, resulting in a significant better description of integral shielding data~\cite{sauvanImplementationNewEnergyangular2017}.

\subsection{Comment on available covariance matrices}
Since the inception of the FENDL library project, the availability of covariance matrices has never been a design objective.
Because evaluations of library projects, such as JENDL, ENDF/B, JEFF and TENDL were fully or partially adopted in FENDL, the covariance information in the original files was also carried over.
However, many ENDF files in FENDL are combinations of evaluations from different library projects and additional modifications have sometimes been effected without updating the associated covariance data.
Therefore, the covariance information may not be fully consistent with the cross sections and distribution data.
Furthermore covariance data are missing for some important isotopes and hence full uncertainty quantification is sometimes not possible.
Some of those covariance data may even not be processable by any version of NJOY or other processing codes.
For these reasons, the validity and suitability of available covariance information cannot be guaranteed.
Users who still want to use it, should proceed with caution.

\section{PROCESSING}
\label{sec:processing}

\subsection{Processing nuclear data for incident neutrons}

\begin{figure}[ht]
    \centering
    \includegraphics[width=8.69cm,height=10cm]{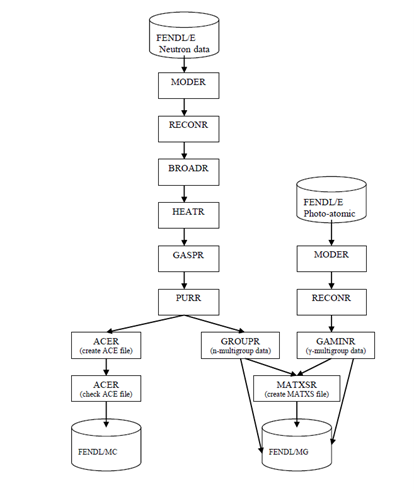}
    \caption{NJOY processing sequence for incident neutrons}
    \label{fig:neutronprocessing}
\end{figure}

The FENDL-3.2b evaluated nuclear data files for incident neutrons were processed using the modular code system NJOY-2016.60~\cite{NJOYReport} with local updates developed during the project. Four main kinds of updates are relevant for FENDL nuclear data processing:
\begin{enumerate}
\item Heating calculation: At the JAEA, the kinematic method was implemented as a new feature in the module HEATR/NJOY. In this approach, the heating values calculated using the energy-balance method are replaced by the upper kinematic limits. This choice is useful to avoid negative heating numbers when the evaluation does not show a good energy balance. The feature includes consistent treatment of partial kermas for resonant reactions, which are subject to self-shielding.
\item Alpha particle production: A patch was applied to GASPR/NJOY to take into account the production of two alpha particles when the residual nucleus is $^{8}$Li. $^{8}$Li decays in about 840 ms to $^{8}$Be, which decays almost instantaneously into two alpha particles. Therefore, it is assumed that $^{8}$Li disintegrates into two alphas.
\item Unresolved resonance table sampling: Due to the use of the single-level Breit-Wigner formalism in the unresolved resonance range, nonphysical too small values of the total cross section can be obtained during the sampling procedure. It produces an overestimation of the self-shielding effect in the unresolved resonance region. Therefore, the probability tables and the Bondarenko cross sections are incorrectly estimated, which are later used in Monte Carlo and deterministic codes, respectively. To overcome this problem, the module PURR/NJOY was patched limiting the minimum value of the sampled total cross section to a tenth of the potential cross section. The set of samples with a total cross section below this limit is ignored~\cite{DLAFENDL32PURR}.
\item Coding patches: These modifications include increased array sizes, the explicit de-allocation of some arrays, the correct reading of certain formatted data and rewinds of a few NJOY tapes, among others. These issues caused NJOY to crash or to produce incorrect results during the processing of evaluated nuclear data files in FENDL.
\end{enumerate}
It is worth noting that for a correct processing of the evaluated nuclear data files in FENDL, NJOY-2016.60 must be modified with patches developed in the frame of the FENDL project. The patched version is available on the NDS/IAEA GitHub site as well as the full processing pipeline to produce the application files, see the appendix for more information.
The processing sequence for generating the MATXS- and ACE-formatted files is shown in~\cref{fig:neutronprocessing}.
Processing details are described in~\cite{DLAFENDL21Processing,DLAFENDL30Processing}, but the key features are also summarized in \cref{tbl:tablesig0} in the appendix for the sake of completeness.

\subsection{Processing nuclear data for incident charged particles: protons and deuterons}

The evaluated nuclear data files for incident protons and deuterons were also processed using the NJOY modular system but with different versions than the one mentioned above for neutron and photo-atomic data. 

The evaluated nuclear data files coming from different versions of JENDL, such as $^{6}$Li and $^{7}$Li from the proton sub-library JENDL-4.0/HE~\cite{Shibata2011} and $^{6}$Li, $^{7}$Li, $^{9}$Be, $^{12}$C and $^{13}$C from the deuteron sub-library JENDL/DEU-2020~\cite{Nakayama2021}, adopted the laboratory angle-energy representation (denoted by LAW=7~\cite{ENDF_6_Format}) for the distribution of the outgoing particles. The same representation was used by the JENDL-2007/HE library~\cite{Watanabe2011}. The NJOY-2016.60 code internally converts the LAW=7 data to the continuous energy-angle distribution LAW=1 representation~\cite{ENDF_6_Format}, which is finally converted to the correlated energy-angle tabular distribution known as LAW=61~\cite{ACE_Format} in the ACE-formatted files. The current version of the MCNP code~\cite{mcnp62} can correctly work with the LAW=61 data for incident neutrons, but produces wrong results for incident charged particles prepared in this representation.

Therefore, NJOY99.259 was patched~\cite{SASA} for processing the JENDL-2007/HE evaluated nuclear data files to solve this issue. The patch converts the LAW=7 data into laboratory angle-energy representation LAW=67 in the ACE-formatted files, which can be correctly interpreted for incident charged particles by the MCNP code. An equivalent patch was developed for the NJOY-2016.49 code and the $^{6}$Li and $^{7}$Li evaluated files from JENDL-4.0/HE were processed with this patched version. The resulting ACE-formatted files are the same as those on the JAEA website~\cite{JAEAProton}. In the case of the isotopes $^{6}$Li, $^{7}$Li, $^{9}$Be, $^{12}$C and $^{13}$C coming from the deuteron sub-library JENDL/DEU-2020, a patched version of NJOY99.364 was applied to produce the ACE-formatted files, which are also the same as those on the JAEA website~\cite{JAEADeuteron}. 

For incident charged particle evaluations that represent the outgoing particle distributions with ENDF-laws different from the laboratory angle-energy distribution (LAW=7), such as Legendre polynomials (LAW=1, LANG=1) and tabulated angular data (LAW=1, LANG=11-15), the current version of NJOY-2016 with the patches from NDS/IAEA can be applied for producing the corresponding ACE-formatted files. Furthermore, the MCNP code is being updated to successfully treat LAW=61 for incident charged particles and photo-nuclear data.

It is worth mentioning that the evaluated nuclear data files for incident protons and deuterons should not include unresolved resonance parameters. Furthermore, the detailed heating numbers and gas production cross sections are not computed by HEATR and GASPR, respectively, for charged particles. Besides that, the MATXS-formatted files were not required. Therefore, the processing sequence for incident protons and deuterons is like the one for neutrons shown in \cref{fig:neutronprocessing}, but without calling modules HEATR, GASPR, PURR, GROUPR, GAMINR and MATXSR. The NJOY input parameters are similar for generating the ACE-formatted files.

\section{ACTIVATION}
\label{sec:activation}
For several decades, the European Activation File (EAF) has been the standard for activation analyses for fusion.
It was coupled to activation codes like FISPACT and relies on a simple, restricted ENDF format to cover only non-elastic 
partial cross sections including separation into ground state and isomers. The range of nuclides of EAF-2010 
(816 nuclides) was much wider than what is traditionally available in 
libraries for transport calculations such as ENDF/B, JEFF and JENDL.
Several years ago, the F4E and later EUROfusion nuclear data projects in collaboration with the JEFF project promoted the use
of TENDL as future reference neutron activation library, which was eventually adopted by the FENDL as activation library.
For that to happen, a lot of effort was invested with support from F4E and EUROfusion into the
TENDL project to make the quality of the activation cross sections, when compared to both differential and integral activation data, at least as 
good as EAF-2010 (the last released version of EAF).
For that effort, model parameters of TALYS were adjusted to get the best possible description of activation experimental data.
The data evaluation performed by Dzysiuk et al \cite{activation}, for many nuclides and reaction channels, finally brought TENDL at the 
level of EAF for the most important reaction channels. The added value of TENDL over EAF was then that it 
covers more nuclides, energies and projectiles.

Finally, we note that it is recommended~\cite{leichtleFENDLLibraryFusion2019} to use data from IRDFF-II for neutron-induced activation and data from the medical isotope database~\cite{gulChargedParticleCrosssection2001,betakNuclearDataProduction2011,Hermanne2018,Tarkanyi2019_1,Tarkanyi2019_2,Engle2019,hermanneUpgradeIAEARecommended2021} for charged particle-induced activation instead of TENDL-2017 if the required data are available in those libraries. \\

\section{VERIFICATION AND VALIDATION}
\label{sec:validation}
Verification and validation are an important part of the FENDL library development.  
This section will focus on validation of the ACE-formatted neutron
transport library in FENDL.  The first part will consist of
computational benchmarks and the second part will consist of
experimental benchmarks.

Validation using computational benchmarks
provides nuclear data evaluators with neutron response information
(e.g. neutron flux, dpa, gas production, heating, etc.) to indicate
impacts or changes due to the use of different cross section libraries.  The
computational benchmarks often represent realistic fusion reactor
geometries and materials to indicate potential impacts in actual designs or
systems.  The observed differences in responses can be used to rapidly provide
feedback to evaluators or identify obvious errors or omissions
introduced in the library creation 
process (e.g. processing errors).  The computational benchmarks
provide neutronics analysts with practical information to assess
potential differences in neutronics responses as newer and potentially
improved cross section libraries become available.  These differences
in results may necessitate design changes in the system being analyzed.
In contrast, the experimental benchmarks provide a direct assessment of the
accuracy of the nuclear data and transport codes in predicting various
neutron responses.  Typically experimental benchmarks are performed in
somewhat simple geometries as compared to actual fusion reactor designs.
\subsection{Computational Benchmarks}
\label{subsec:computationalbenchmarks}
\subsubsection{Leakage Sphere}
\label{subsubsec:leakagesphereFabbri}
Due to their complex production process, their extension and the involvement of multiple parameters, a nuclear data library must undergo systematic and extensive Verification and Validation (V\&V) procedures before it can be released. In order to help with this process by increasing the automation, standardization, and reproducible quality of these procedures, a Python based open-source tool named JADE~\cite{fabbri1jadedocs,fabbri2jadegithub} has been developed over the last few years. It includes a series of computational and experimental benchmarks whose inputs are automatically generated and run using the MCNP6 code \cite{mcnp62}.  The MCNP outputs are automatically post-processed in order to highlight differences between different libraries, agreement with experimental benchmark simulations, or consistency between different benchmarks with similar sensitivity profiles. JADE has been used to support the V\&V procedure of the new FENDL release with three of its benchmarks, the first of which is described in this section.

The Leakage Sphere benchmark is one of JADE’s computational benchmarks and a detailed description of it can be found in \cite{fabbri4jadefusengdes}. In brief, a 14 MeV neutron isotropic point source at the center of a sphere composed of a single isotope (or a typical fusion important material such as concrete, water, boron carbide, silica, etc.), where different quantities (integral values and spectra) are tallied and compared. Consistency checks are also performed to ensure that nuclear responses are physically valid and coherent. The main outcomes of this study are depicted hereinafter.

\begin{figure}[htp]
\includegraphics[width=8.69cm]{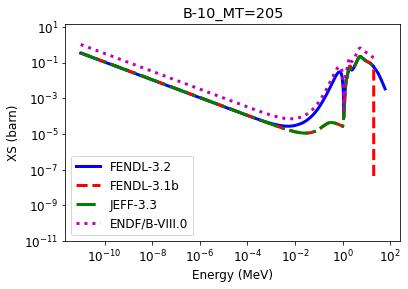}
    \caption{$^{10}$B MT=205 (triton production) cross sections from different libraries.}
\label{fig:fabbriFigure1b10mt205}
\end{figure}

The FENDL-3.2b materials/isotopes passed all JADE consistency checks: flux, helium production, dpa, neutron and gamma spectra and heating values are positive and coherent. The detailed comparison against the ENDF/B-VIII.0 data libraries highlights an overall good agreement for the main fusion-related elements. The higher (and also engineering relevant) differences in material/elements usually employed in the fusion field are found in the boron carbide tritium production (-164.40\%), seen in \cref{table:fabbrileakagesphereITERmatls}.  The FENDL-3.2b response exceeds the ENDF/B-VIII.0 response due to the double-counting in the $^{10}$B MT=205 reaction rate (\cref{fig:fabbriFigure1b10mt205}) and marginally by the limited alteration in the neutron flux spectra. It is worth highlighting that the difference spotted for $^{10}$B during the V\&V of the FENDL-3.2 beta \cite{fabbri5nucfus} has been addressed and now the heating values of the SS316L(N)-IG, concrete and water are aligned with the ones obtained using the ENDF/B-VIII.0 data.

\begin{table*}[htp]
\caption{Sphere leakage computational benchmark results comparison for typical ITER materials.}
\label{table:fabbrileakagesphereITERmatls}
\begin{tabular}{l | l c c c c c} \hline \hline
Material                 & FENDL-3.2b Vs & T prod.    & He prod.  & DPA      & Nheat (f6)  & Gammaheat (f6) \\ \hline
Water                    & FENDL-3.1d     &   -0.11\%  &  9.17\%   &  -0.39\% &  -0.12\%    &  -0.28\%    \\
Water                    & ENDF-VIII.0   &   62.51\%  &  0.10\%   &   0.01\% &  -0.22\%    &   0.12\%    \\
Ordinary Concrete        & FENDL-3.1d     &   -0.71\%  &  1.28\%   &   1.05\% &   0.14\%    &  -1.12\%    \\
Ordinary Concrete        & ENDF-VIII.0   &   -4.99\%  &  7.53\%   &   5.74\% &  -0.10\%    &   3.87\%    \\
Boron Carbide            & FENDL-3.1d     &   13.09\%  &  0.78\%   &   0.11\% &  -0.09\%    &   0.98\%    \\
Boron Carbide            & ENDF-VIII.0   &  164.40\%  & -7.37\%   &   3.83\% &  -0.42\%    &   3.59\%    \\
SS316L(N)-IG             & FENDL-3.1d     &    8.63\%  &  1.39\%   &  -0.17\% &  -8.78\%    &   3.05\%    \\
SS316L(N)-IG             & ENDF-VIII.0   &  -79.64\%  & -7.26\%   &   1.41\% &  -4.12\%    &  -0.89\%    \\
Nat. Silicon             & FENDL-3.1d     &  -         &  0.00\%   &   0.00\% &   0.00\%    &   0.00\%    \\
Nat. Silicon             & ENDF-VIII.0   &  -         &  0.00\%   &   0.01\% &  -0.15\%    &   0.09\%    \\
Polyethylene non-borated & FENDL-3.1d     &  -         &  0.00\%   &   0.01\% &   0.02\%    &   0.02\%    \\
Polyethylene non-borated & ENDF-VIII.0   &  -         &  4.61\%   &  -0.67\% &  -1.62\%    &  -0.48\%    \\ \hline \hline
\end{tabular}
\end{table*}

\begin{figure}[htp]
\includegraphics[width=8.69cm]{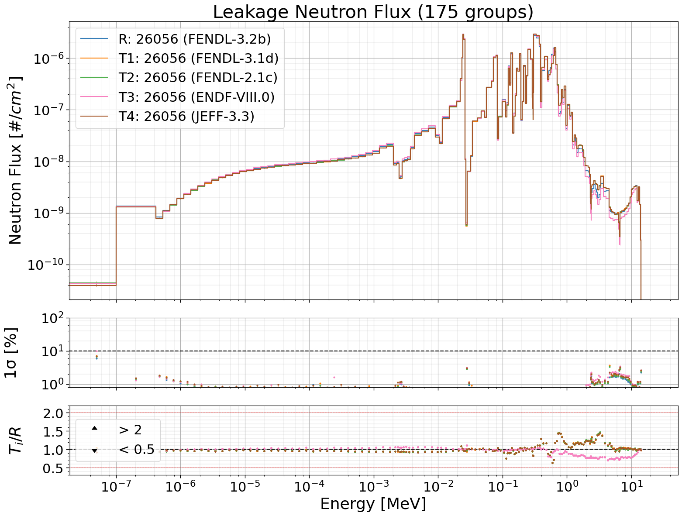}
    \caption{$^{56}$Fe leakage neutron flux.}
\label{fig:fabbriFigure2fe56flux}
\end{figure}

Moreover, the complete set of the neutron and gamma spectra leaking from the sphere have been compared against the results using JEFF-3.3, FENDL-2.1, FENDL-3.1d and ENDF/B-VIII.0 data. The plots summarizing these responses are collected in a dedicated atlas \cite{fabbri7figshare}. For each single isotope, a plot is provided comparing the energy binned flux (with the corresponding relative error) obtained using the different libraries and the ratio of the results from all libraries against the reference FENDL-3.2b library. As an example, the plot for $^{56}$Fe, which is important for fusion is shown in \cref{fig:fabbriFigure2fe56flux}.

Among the 192 FENDL isotopes, several discrepancies with the cross sections in the major evaluated data libraries are highlighted. For the sake of brevity, only a subset is discussed here, giving priority to the elements with experimental data available in the EXFOR library~\cite{fabbri6exfor}, high isotopic abundance, and a high engineering interest for fusion application (e.g., usage in main structural materials and breeding blanket). This limited subset of isotopes can be further subdivided in (a) evaluations which show a significant difference in the area in which the EXFOR data are available, (b) evaluations that differ in different libraries, and (c) a combination of both (a) and (b):
\begin{itemize}
\item Type (a): $^{46}$Ti, $^{52}$Cr, $^{62}$Ni, $^{92}$Mo, $^{94}$Mo, $^{107}$Ag, $^{109}$Ag
\item Type (b): $^{94}$Zr, $^{184}$W, $^{186}$W
\item Type (c): $^{54}$Cr, $^{57}$Fe, $^{182}$W
\end{itemize}

Following, we describe a few key isotopes of the three categories.

\begin{figure}[htp]
\includegraphics[width=8.69cm]{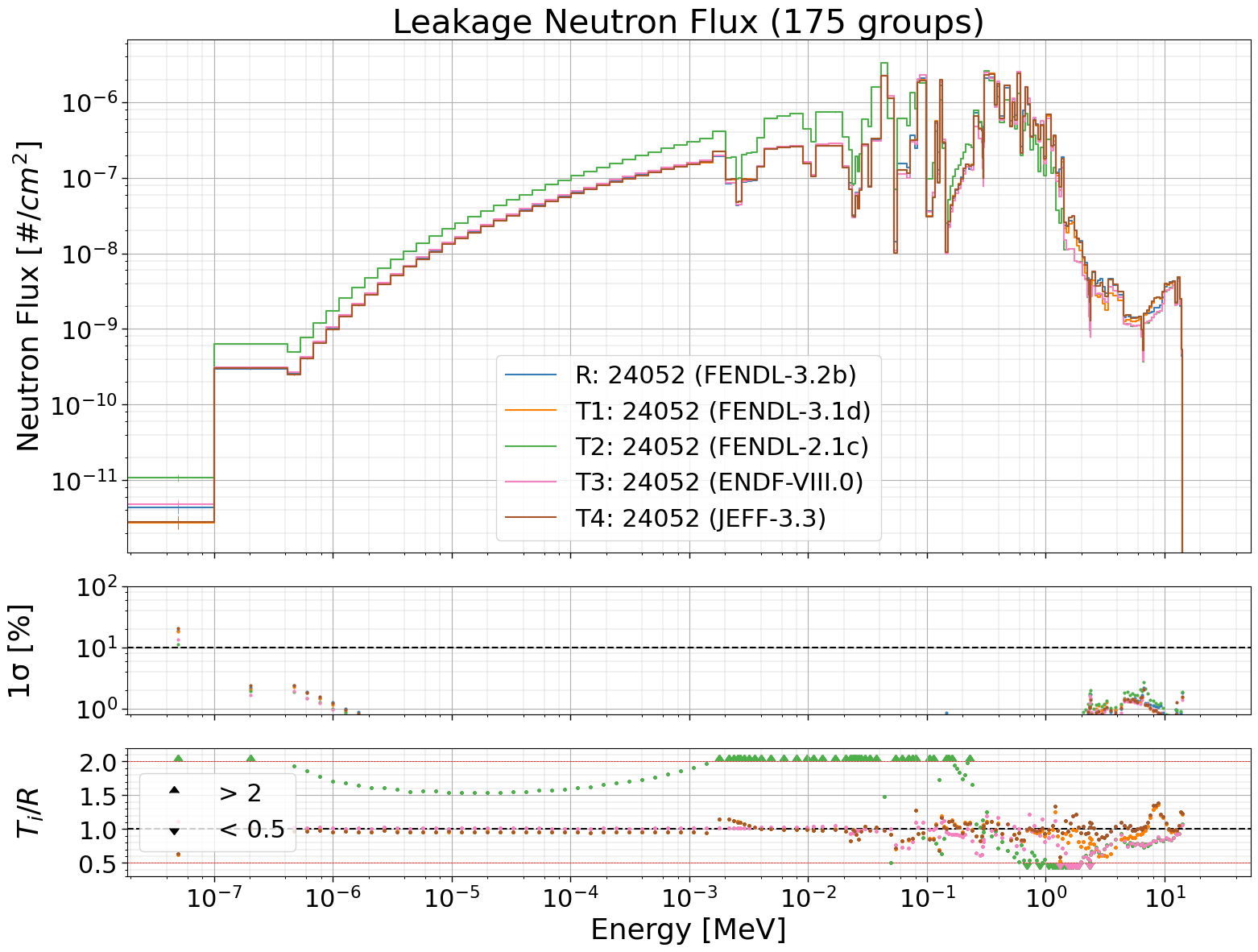}
    \caption{$^{52}$Cr leakage neutron flux.}
\label{fig:fabbriFigure3cr52flux}
\end{figure}

\begin{figure}[htp]
\includegraphics[width=8.69cm]{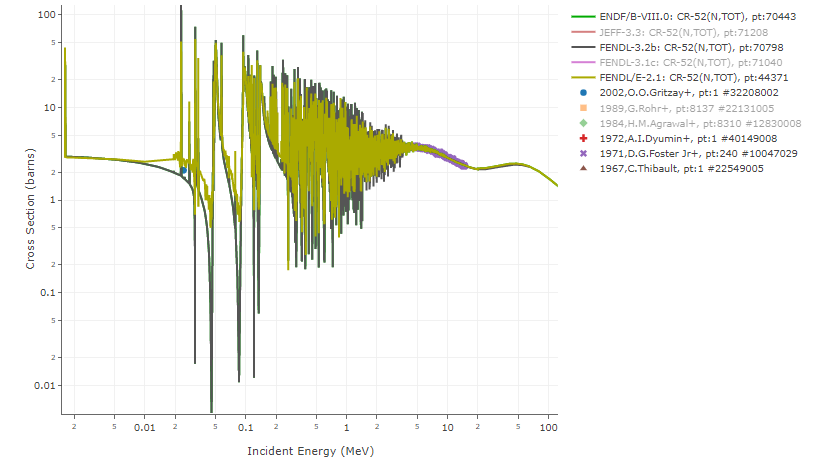}
    \caption{Comparison of the $^{52}$Cr(n,tot) cross section from different libraries and the EXFOR data.}
\label{fig:fabbriFigure4cr52sigtot}
\end{figure}

\textbf{Type (a)/$^{52}$Cr:} Apart from the superseded FENDL-2.1, which adopted the ENDF/B-VI.2 evaluation, all libraries produce a very similar leakage neutron flux, as shown in \cref{fig:fabbriFigure3cr52flux}. Comparing the (n,tot) reaction, we see that FENDL-2.1 has less structure in the resonance range, leading to an increase of the neutron flux at lower energies, see \cref{fig:fabbriFigure4cr52sigtot}. Unfortunately, the EXFOR database could not help in further judgments because data are only available at higher energies. The data in FENDL-3.2b are from a more recent evaluation and may therefore be considered more trustworthy. The users should be aware of the differences and their possible impact on simulations of integral experiments.

\begin{figure}[htp]
\includegraphics[width=8.69cm]{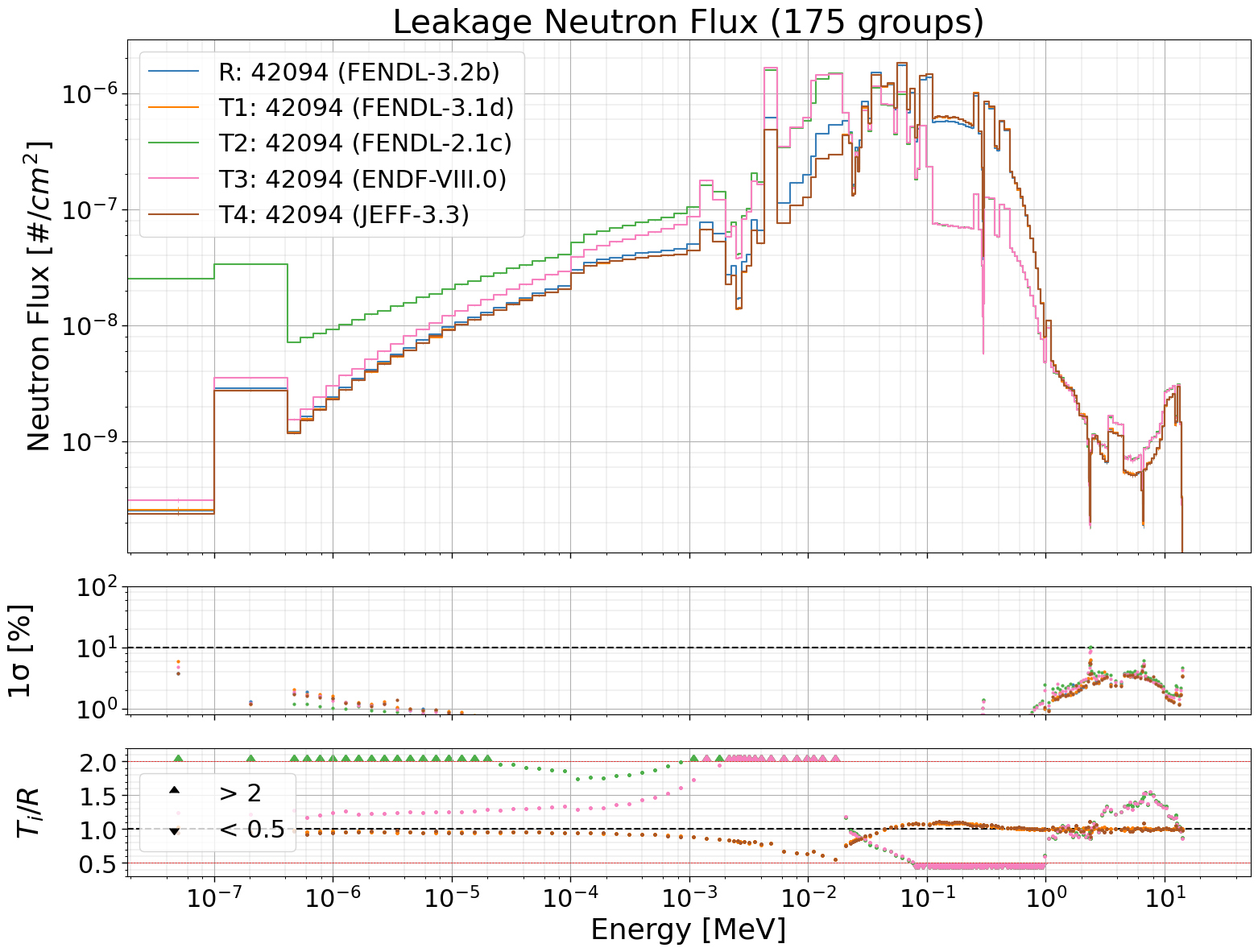}
    \caption{$^{94}$Mo leakage neutron flux.}
\label{fig:fabbriFigure5mo94flux}
\end{figure}

\begin{figure}[htp]
\includegraphics[width=8.69cm]{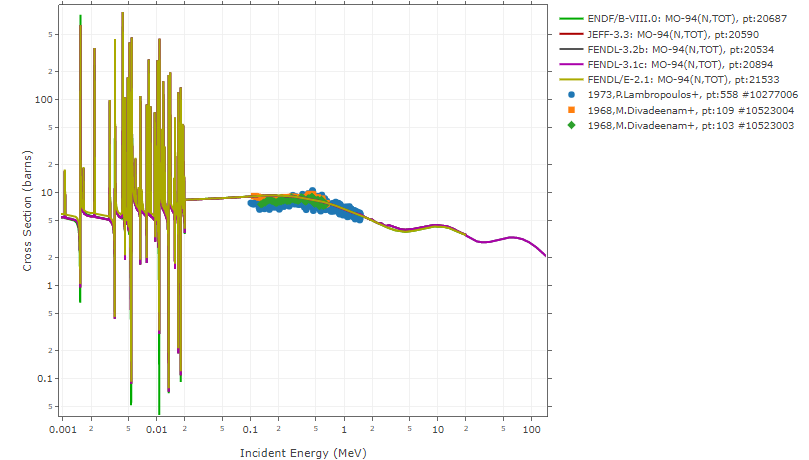}
    \caption{Comparison of the $^{94}$Mo(n,tot) cross section from different libraries and the EXFOR data.}
\label{fig:fabbriFigure6mo94sigtot}
\end{figure}

\textbf{Type (a)/$^{94}$Mo:} \Cref{fig:fabbriFigure5mo94flux} shows significant differences in the leakage neutron flux at about 0.1 MeV where the use of ENDF/B-VIII.0 data results in a lower flux, possibly caused by the larger amplitude of fluctuations in the resolved resonance region (RRR),~\cref{fig:fabbriFigure6mo94sigtot}. In the low energy region, the FENDL-2.1 result deviates the most compared to the others, perhaps due to more pronounced valleys below the low-lying resonance. Apart from these discrepancies, a good fit of experimental data is observed. As in the case of $^{52}$Cr, the more recent evaluation in FENDL-3.2b is considered better. We stress again that the users should be aware of the differences and the possible impact on the simulation of integral experiments.

\begin{figure}[htp]
\includegraphics[width=8.69cm]{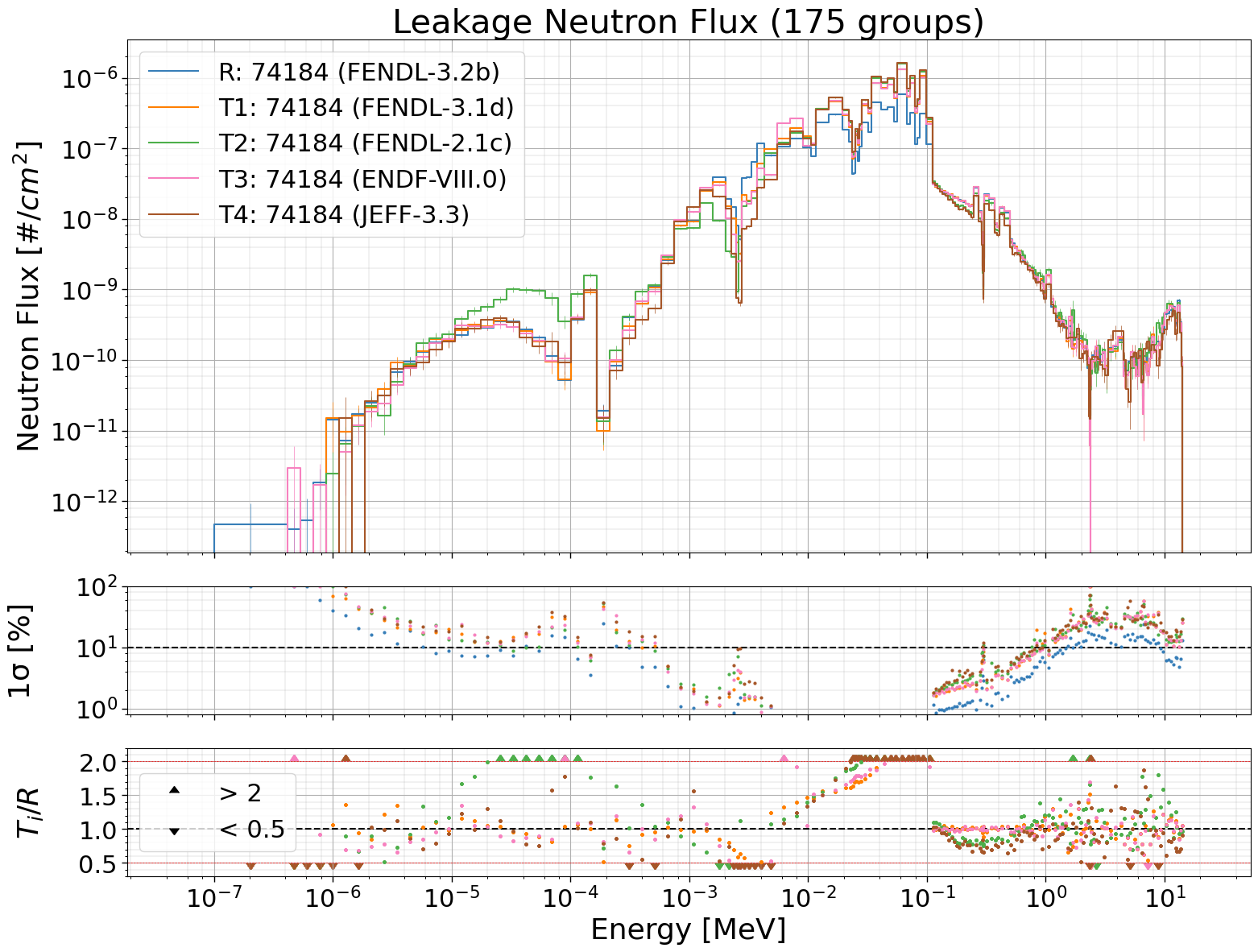}
    \caption{$^{184}$W leakage neutron flux.}
\label{fig:fabbriFigure7w184flux}
\end{figure}

\begin{figure}[htp]
\includegraphics[width=8.69cm]{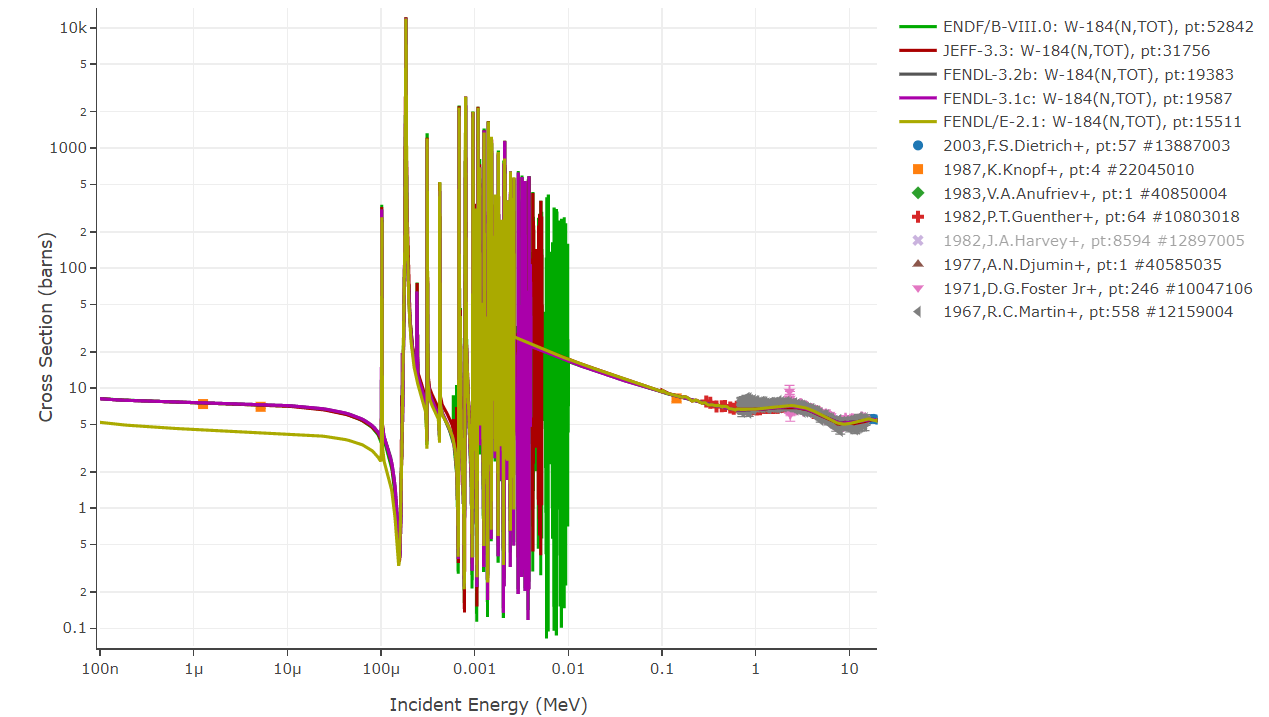}
    \caption{Comparison of the $^{184}$W(n,tot) cross section from different libraries and the EXFOR data.}
\label{fig:fabbriFigure8w184sigtot}
\end{figure}

\textbf{Type (b)/$^{184}$W:} The leakage neutron flux of JEFF-3.3, ENDF/B-VIII.0 and different versions of FENDL is quite similar as can be seen in~\cref{fig:fabbriFigure7w184flux}.
Only FENDL-2.1 deviates from the other libraries between $10^{-5}$ and $10^{-4}$ MeV,
which may be due to the fact that the two data points of K.~Knopf (EXFOR entry 22045010) below the RRR were not considered in the fitting of the total cross section, see~\cref{fig:fabbriFigure8w184sigtot}.
FENDL-3.2b seems to have a lower RRR range compared to ENDF/B-VIII.0 and JEFF-3.3, which may cause the variation around $10^{-1}$ MeV.

\begin{figure}[htp]
\includegraphics[width=8.69cm]{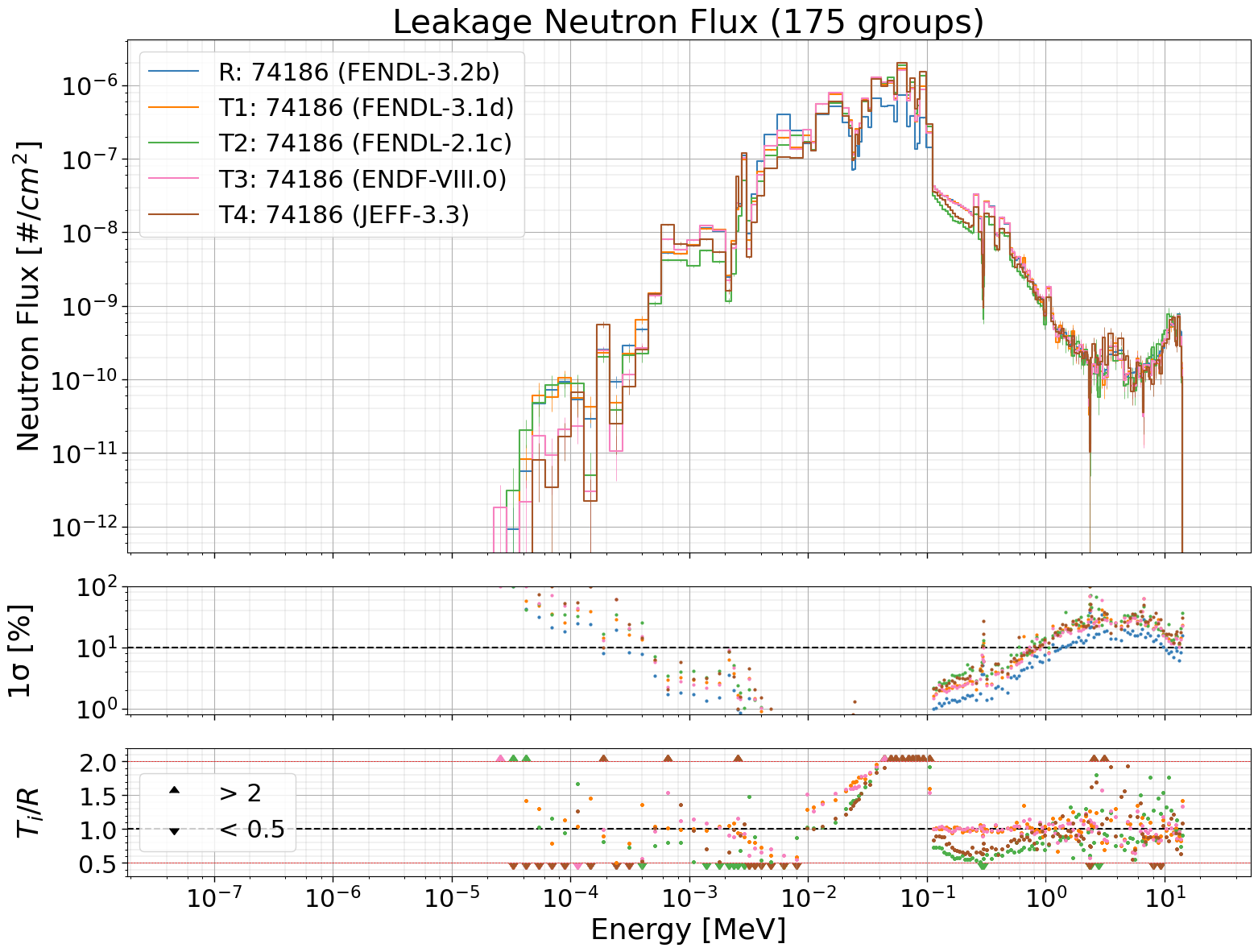}
    \caption{$^{186}$W leakage neutron flux.}
\label{fig:fabbriFigure9w186flux}
\end{figure}

\begin{figure}[htp]
\includegraphics[width=8.69cm]{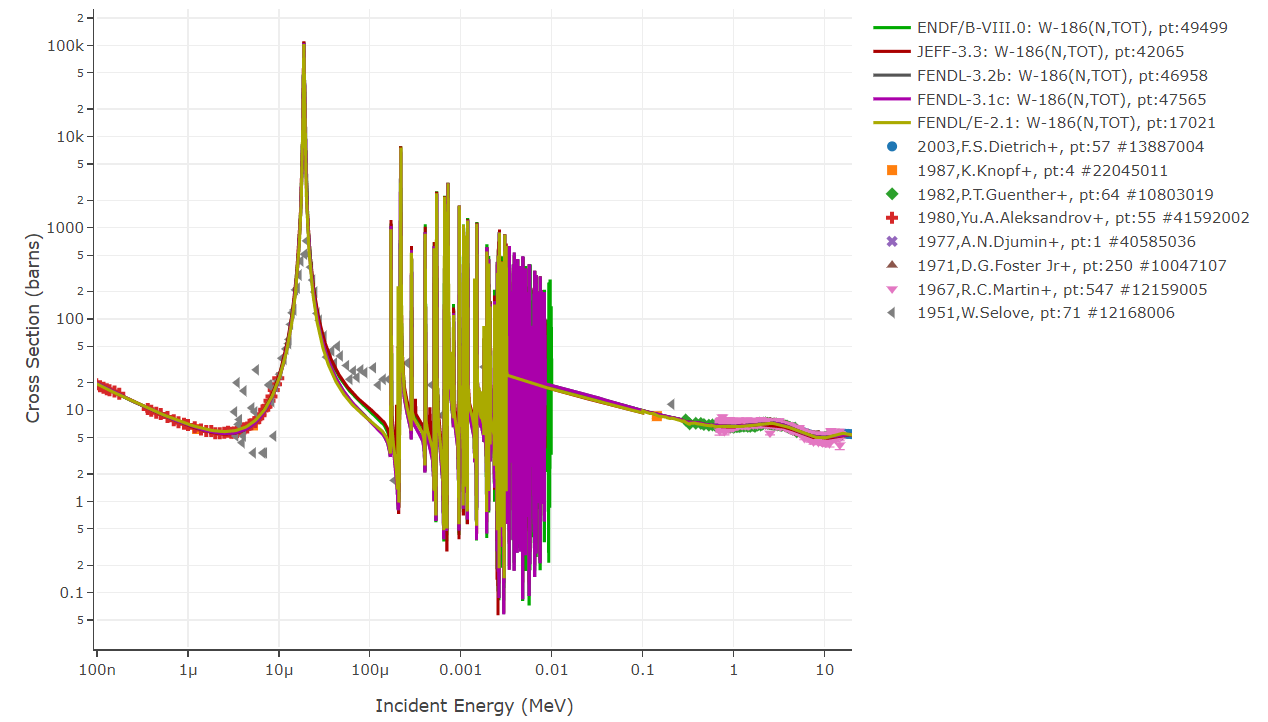}
    \caption{Comparison of the $^{186}$W(n,tot) cross section from different libraries and the EXFOR data.}
\label{fig:fabbriFigure10w186sigtot}
\end{figure}

\textbf{Type (b)/$^{186}$W:} Differences in the leakage neutron flux are small among the libraries, as shown in \cref{fig:fabbriFigure9w186flux}. The good agreement may be explained by similar good agreement of the (n,tot) cross section channel, see~\cref{fig:fabbriFigure10w186sigtot}. All considered libraries seem to employ the same comprehensive set of EXFOR data in a similar manner. Finally, we remark that ENDF-VIII.0 contains a more extended and pronounced RRR, which may explain the difference at about $10^{-2}$ MeV.

\begin{figure}[htp]
\includegraphics[width=8.69cm]{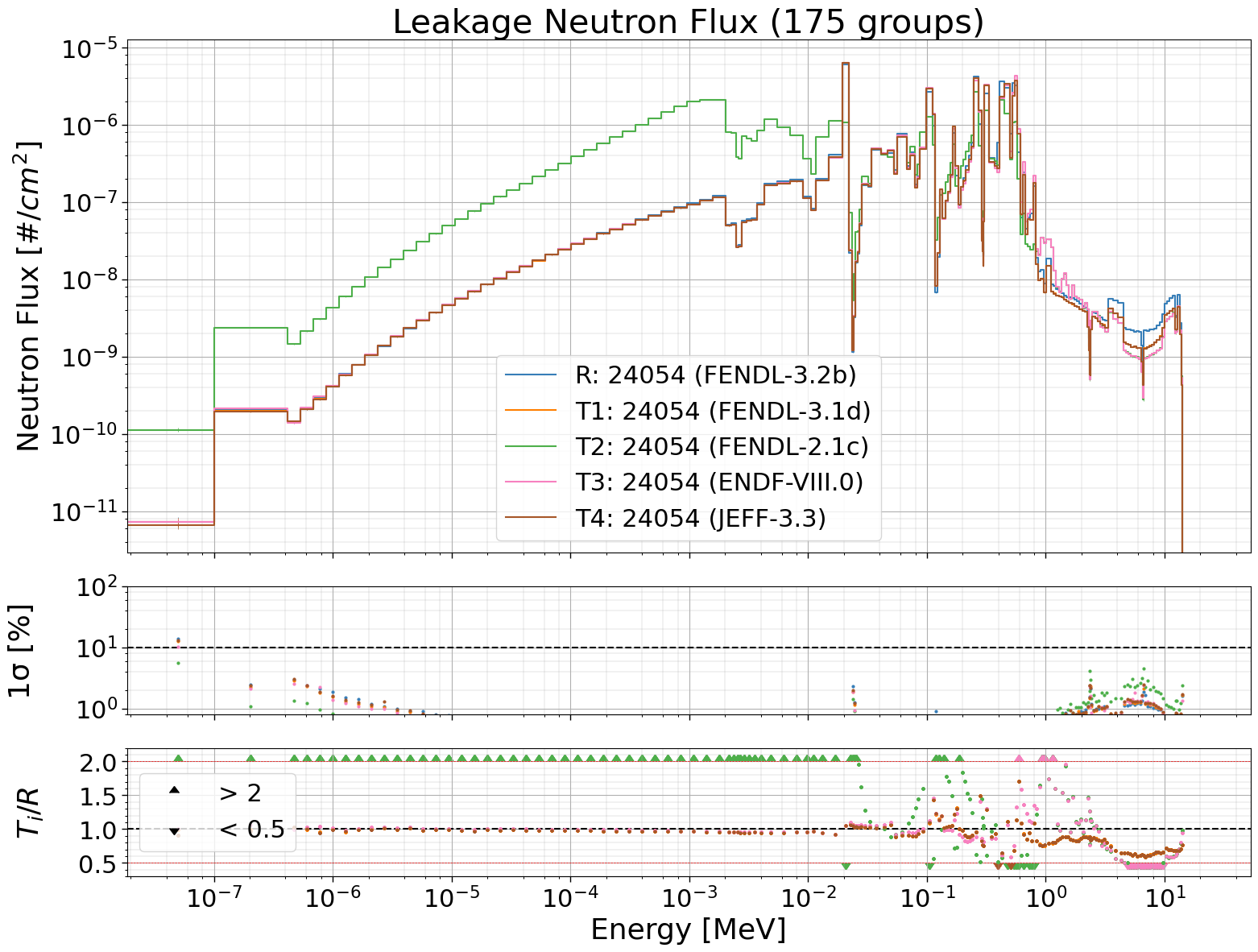}
    \caption{$^{54}$Cr leakage neutron flux.}
\label{fig:fabbriFigure11cr54flux}
\end{figure}

\begin{figure}[htp]
\includegraphics[width=8.69cm]{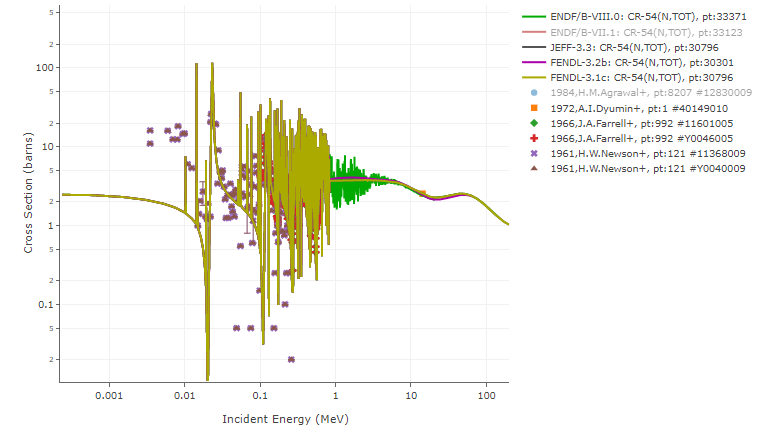}
    \caption{Comparison of the $^{54}$Cr(n,tot) cross section from different libraries and the EXFOR data.}
\label{fig:fabbriFigure12cr54sigtot}
\end{figure}

\textbf{Type (c)/$^{54}$Cr:} A substantial difference in neutron flux is computed for FENDL-2.1 below $10^{-2}$~MeV, see~\cref{fig:fabbriFigure11cr54flux}. Several cross sections appear very similar in this energy range, as shown in \cref{fig:fabbriFigure12cr54sigtot}. Therefore, the reason for the observed discrepancy may be found at higher energies. The FENDL-3.2b cross sections are similar in the amplitude of fluctuations representing the resonances, but the fluctuations are replaced by a smooth average above 1 MeV, while other evaluations have fluctuations up to 10 MeV, which is the most probable source of the observed discrepancies.

\begin{figure}[htp]
\includegraphics[width=8.69cm]{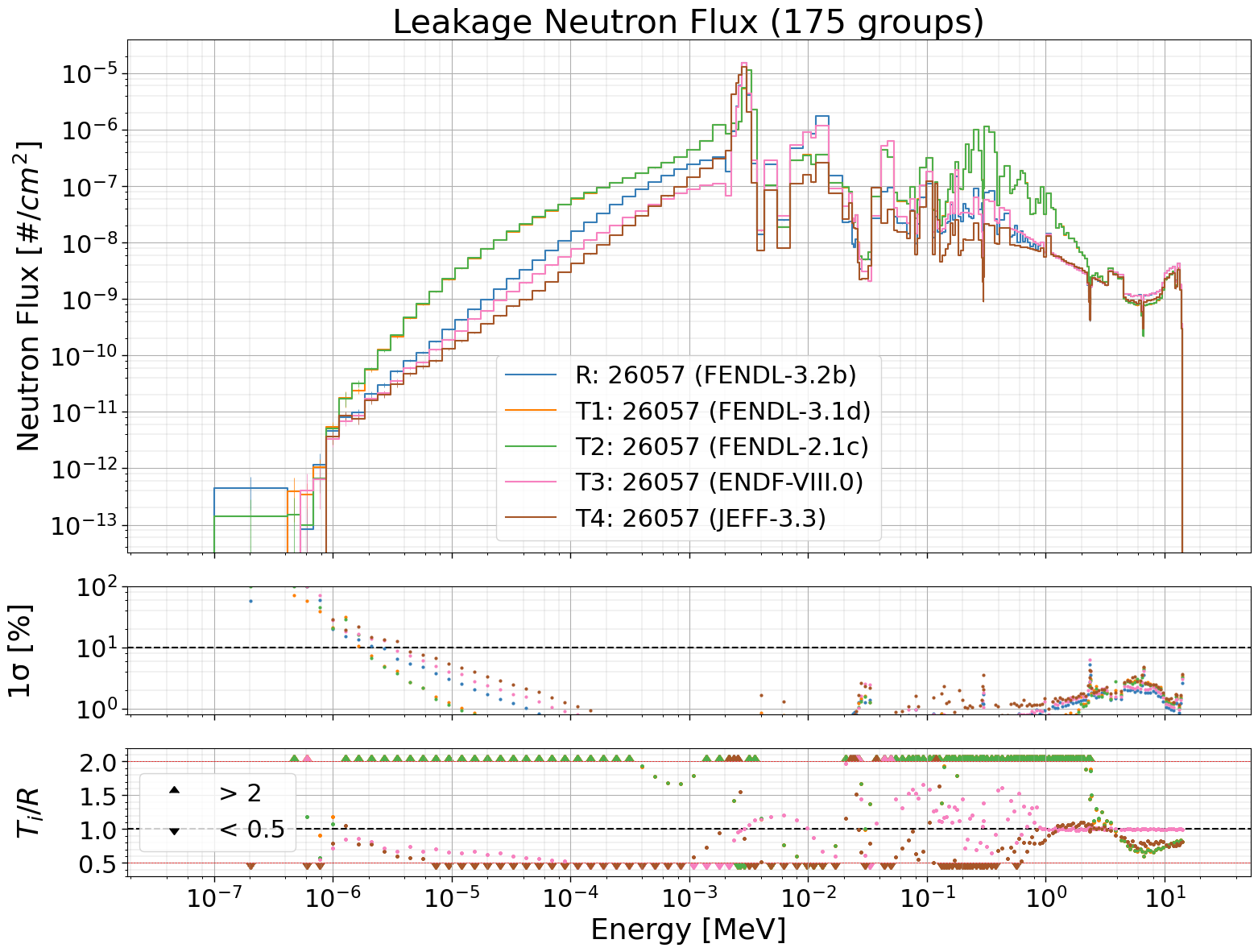}
    \caption{$^{57}$Fe leakage neutron flux.}
\label{fig:fabbriFigure13fe57flux}
\end{figure}

\begin{figure}[htp]
\includegraphics[width=8.69cm]{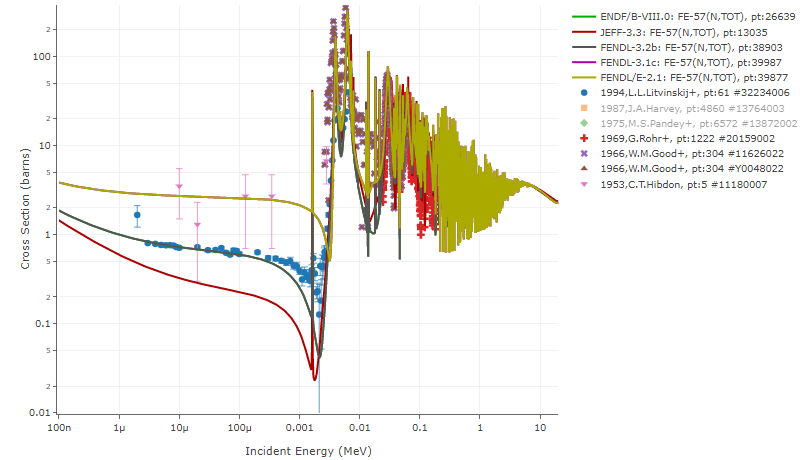}
    \caption{Comparison of the $^{57}$Fe(n,tot) cross section from different libraries and the EXFOR data.}
\label{fig:fabbriFigure14fe57sigtot}
\end{figure}

\textbf{Type (c)/$^{57}$Fe:}  Good agreement is generally obtained in the neutron flux as well as in the cross sections, as seen in \cref{fig:fabbriFigure13fe57flux} and \cref{fig:fabbriFigure14fe57sigtot}. The main discrepancies occur at energies below $10^{-2}$ MeV between previous FENDL releases (v2.1 and 3.1d) and the other major libraries. The reason may be some fine-tuning of the evaluations based on different EXFOR data causing a different lower-energy tail, namely the data of 1953 C.T. Hibdon (EXFOR number 11180007) versus 1994 L.L. Litvinskij (EXFOR number 32234006). Moreover, the FENDL-3.2b cross section seems to have resonance structure extending to higher energies compared to ENDF/B-VIII.0, as described in Section~\ref{subsubsec:jaea-evaluations}. Fluctuations in the cross sections in FENDL-2.1/3.1d extend to higher energies still, based on the total cross section measurements by Pandey (EXFOR number 13872002). The elastic cross section was adjusted to preserve unitarity. These differences may affect the flux difference around 1~MeV.

\subsubsection{ITER 1-D}
\label{subsubsec:iter1dBohm}
The ITER 1-D cylindrical benchmark is a calculational benchmark based
on an early ITER design and was developed for the FENDL evaluation
process in 1994~\cite{sawanITER1Dbenchmark1994}. The model includes the
inboard (IB), plasma, and outboard (OB) regions of ITER, and contains 26
IB and 26 OB layers.  Detailed schematic diagrams
for this model are shown in
reference~\cite{sawanITER1Dbenchmark1994}. Looking at these schematic
drawings, one can see that the blanket/shield region, between the first wall (FW)
and the vacuum vessel (VV), consists of alternating SS-316 and
water regions.  The VV consists of an Inconel 625 shell with a water
cooled SS316 VV filler.  The neutron
source for this benchmark is uniformly distributed in space in the
plasma region. Neutron
and photon fluxes, nuclear heating, dpa, and gas production are
tallied in the benchmark calculation. This benchmark was run with
MCNP-6.2 for all calculations shown in this work \cite{mcnp62}. For photon
transport, we are using the mcplib84 photon cross section library
distributed with MCNP \cite{mcplib84}.  The neutron source uses MCNP's built-in Muir velocity Gaussian D-T fusion neutron spectrum at 10 keV. 

Previous work examined the impact of some improved Cr, Fe, and O cross
sections that were preliminary candidates to consider for the FENDL-3.2
release~\cite{bohmfendlFST2021}. This previous work compared the
FENDL-2.1, FENDL-3.1d, ENDF/B-VII.1, and ENDF/B-VIII.0 libraries, as
well as the new candidate cross sections inserted into the FENDL-3.1d
and ENDF/B-VIII.0 libraries.  In this work, we use the latest release
of FENDL-3.2 (version 3.2b released Feb 15, 2022). 

\begin{figure}[htp]
\includegraphics[width=8.69cm]{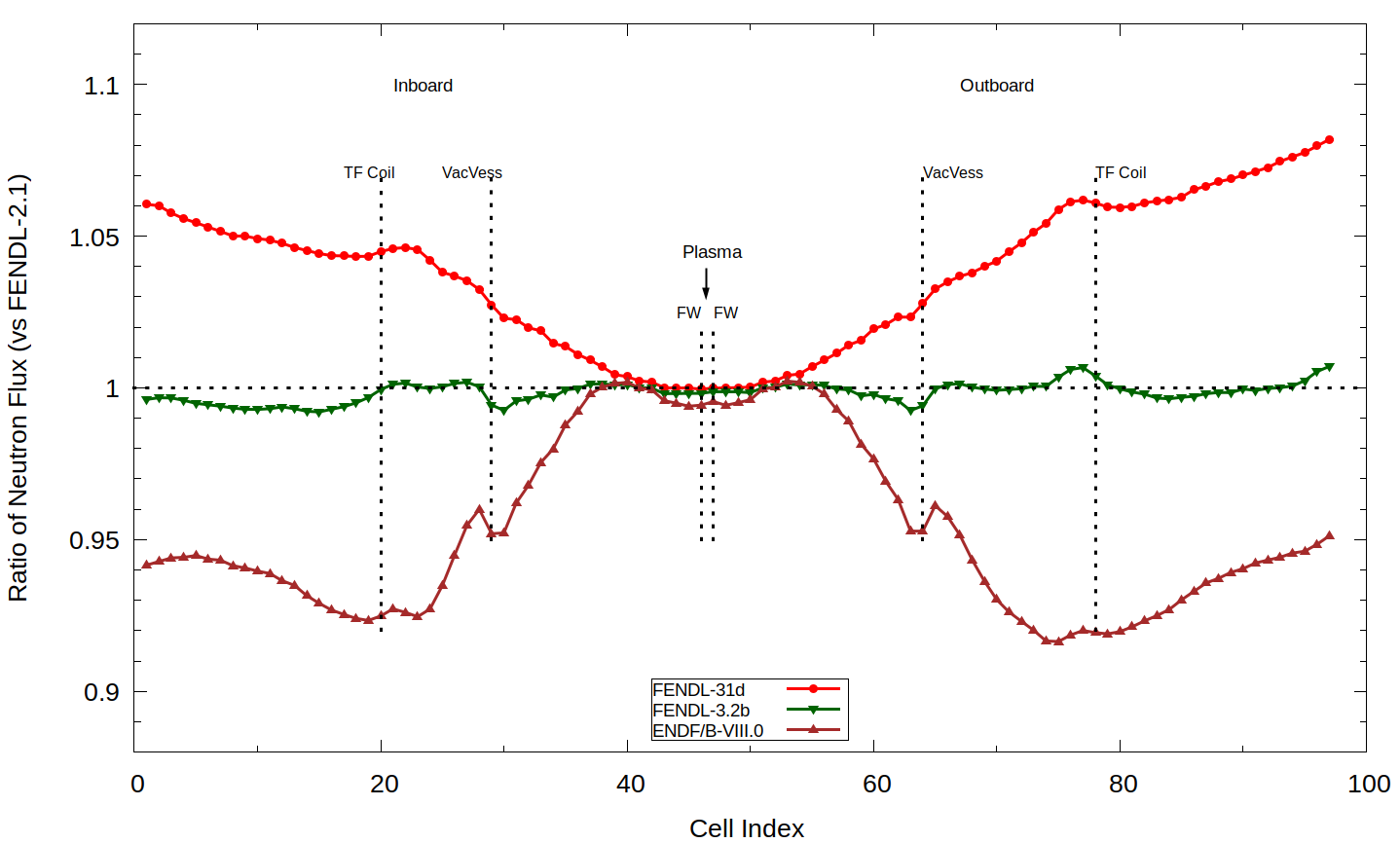}
    \caption{Ratio of neutron flux calculated with different neutron cross section data for the ITER 1-D cylindrical benchmark model.  The cell index is ordered radially.}
\label{fig:bohmiter1dnflux}
\end{figure}

\Cref{fig:bohmiter1dnflux} shows the ratio of neutron flux calculated with different
neutron cross section libraries for the ITER 1-D cylindrical benchmark
model.  The fluxes are compared against the flux calculated with the
FENDL-2.1 neutron library.  The cell index for the horizontal axis is
ordered radially and the inboard, plasma, and outboard sections are
labeled in the figure. Vertical dotted lines indicate the front
(closest to the plasma) location of some important components such as the FW
Be layer, VV, and toroidal field (TF) coil. The statistical uncertainty for the neutron
flux calculations has a maximum of 0.26\%.  The neutron
flux calculated with FENDL-3.1d is the highest of all libraries at
deeper locations (e.g. VV, TF Coil).  It is as much as 8\% higher than
the flux calculated with FENDL-2.1.  The neutron flux calculated with
ENDF/B-VIII.0 is the lowest of all the libraries shown at deeper locations.
It is as much as 8\% lower than the flux calculated with FENDL-2.1.
The flux calculated with FENDL-3.2b is within 1\% of that calculated
with FENDL-2.1.

The photon flux calculated using FENDL-3.2b and the mcplib84 photon library is within 2.5\% of values calculated using FENDL-2.1 and the mcplib84 photon library.  The photon flux calculated using FENDL-3.1d and the mcplib84 photon library is up to 10\% higher than values calculated using FENDL-2.1 and the mcplib84 photon library.

\begin{figure}[htp]
\includegraphics[width=8.69cm]{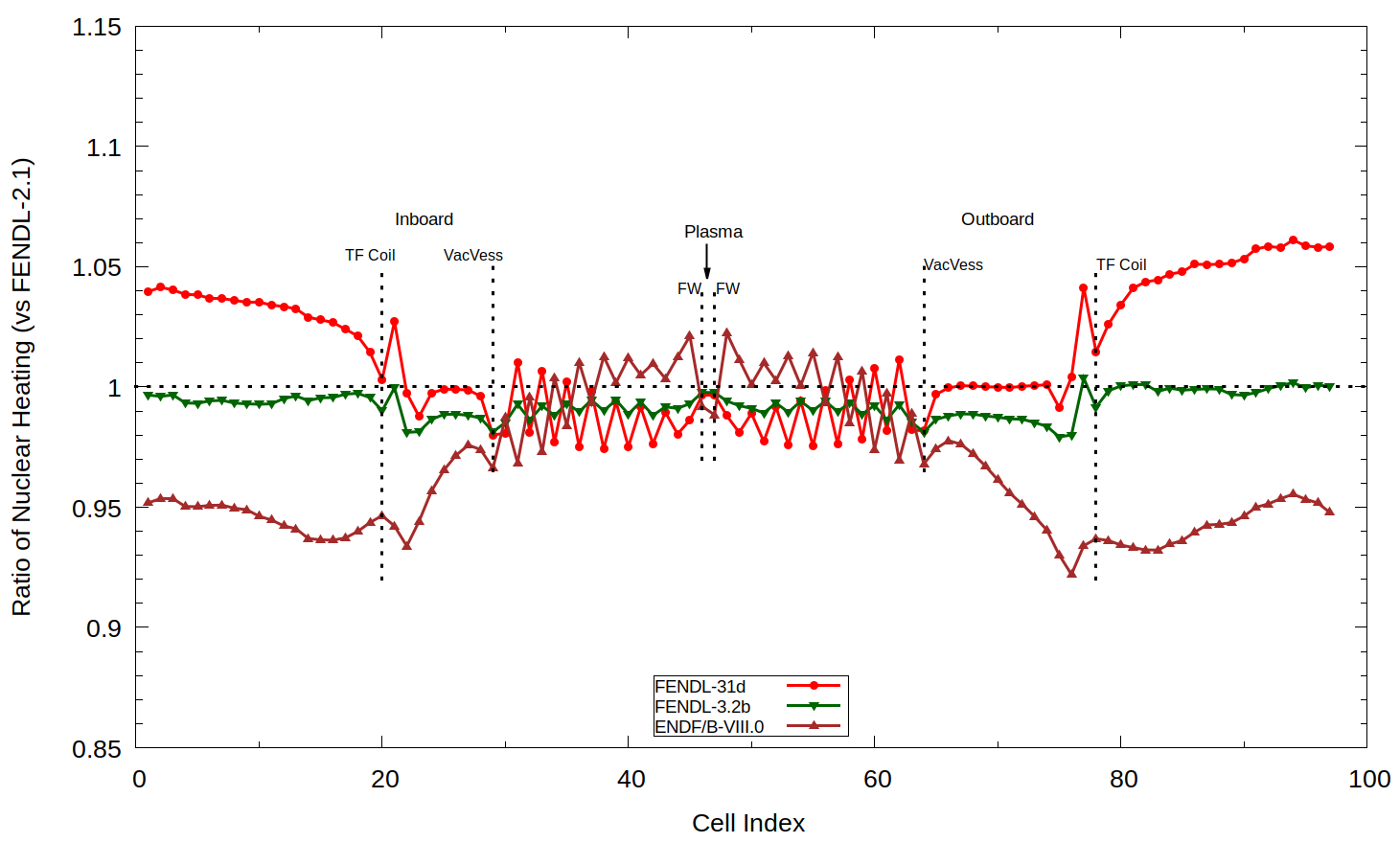}
    \caption{Ratio of total nuclear heating (n+gamma) calculated with different neutron cross section data for the ITER 1-D cylindrical benchmark model.  The cell index is ordered radially.}
\label{fig:bohmiter1dtotheat}
\end{figure}

\Cref{fig:bohmiter1dtotheat} shows the ratio of total nuclear heating (neutron+photon) calculated with different neutron cross section libraries for the ITER 1-D cylindrical benchmark.  The statistical uncertainty for the nuclear heating calculations has a maximum of 0.27\%.  The total nuclear heating calculated with FENDL-3.1d is the highest of all libraries at deep locations (TF Coil).  It is as much as 6\% higher than the heating calculated with FENDL-2.1.  The total nuclear heating calculated with ENDF/B-VIII.0 is the lowest of all the neutron libraries at deeper locations.  It is as much as 8\% lower than the heating calculated with FENDL-2.1.  Total nuclear heating calculated using FENDL-3.2b neutron cross sections is generally slightly lower than the heating calculated with FENDL-2.1 (up to 2\%). For FENDL-3.2b results, the oscillating behaviour in the FW to VV region is greatly reduced compared to the up to 12\% peaks observed with a preliminary version of FENDL-3.2 \cite{bohmCSEWG2021}.  This was traced to a problem with neutron heating in water, specifically in $^{16}$O, and corrected in newer versions of FENDL.

\Cref{table:bohmiter1ddpa} shows the ratio in dpa calculated with various libraries versus the dpa calculated with the FENDL-2.1 library.  The dpa is calculated at different locations in the model for the material indicated in the table.  The statistical uncertainty for these dpa calculations has a maximum of 0.15\%.  The dpa calculated with FENDL-3.1d is up to 6\% higher than values calculated with FENDL-2.1 and higher values occur at deeper locations.  The dpa calculated with ENDF/B-VIII.0 is up to 9\% lower than values calculated with FENDL-2.1.  Using the FENDL-3.2b library produces dpa values within 1\% of values calculated with FENDL-2.1.

\begin{table}[htp]
\vspace{-2mm}
\caption{Ratio of dpa calculated with various libraries versus the dpa calculated with the FENDL-2.1 library in the ITER-1D model.}
\label{table:bohmiter1ddpa}
\begin{tabular}{l | c | c c c c} \hline \hline
Component  & matl &  F-21 &   F-31d &  E-8.0 &  F-32b    \\ \hline
IB FW Cu   & Cu &    1.0000 &  0.9968 &  0.9884 &  0.9970 \\
IB FW SS   & Fe &    1.0000 &  0.9993 &  1.0015 &  0.9965 \\
IB VV Inc. & Ni &    1.0000 &  1.0303 &  0.9204 &  0.9988 \\
IB VV SS   & Fe &    1.0000 &  1.0274 &  0.9631 &  1.0110 \\
IB Mag.    & Cu &    1.0000 &  1.0452 &  0.9104 &  1.0041 \\ \hline
OB FW Cu   & Cu &    1.0000 &  0.9972 &  0.9896 &  0.9974 \\
OB FW SS   & Fe &    1.0000 &  0.9997 &  1.0010 &  0.9964 \\
OB VV Inc. & Ni &    1.0000 &  1.0307 &  0.9205 &  0.9980 \\
OB VV SS   & Fe &    1.0000 &  1.0282 &  0.9656 &  1.0115 \\
OB Mag.    & Cu &    1.0000 &  1.0630 &  0.9073 &  1.0095 \\ \hline \hline
\end{tabular}
\vspace{-3mm}
\end{table}

\Cref{table:bohmiter1dheprod} shows the ratio in helium production calculated with various libraries versus the helium production calculated with the FENDL-2.1 library. The statistical uncertainty for these helium production calculations has a maximum of 0.19\%.  The helium production values calculated with FENDL-3.1d are higher than those calculated with FENDL-2.1 (as much as 18\%).  The helium production values calculated with ENDF/B-VIII.0 are also higher than those calculated with FENDL-2.1 (up to 7\%) except at the magnet where values are lower (1-1.5\%) than those calculated with FENDL-2.1. The helium production values calculated using the FENDL-3.2b library are higher than values calculated with FENDL-2.1 (up to 20\% for the VV Inconel shell and 5-9\% higher for the SS-316 components).

Past work has shown that most of the differences observed here are due to missing gas production reactions in the various isotopes contained in the benchmark model’s materials in the cross section libraries (particularly FENDL-2.1) \cite{fischerINDC-631}.  Note that helium production means both He-3 and He-4 gas production (reaction numbers MT 206 and MT 207 respectively in ENDF teminology).  None of the libraries examined had any isotopes that were missing the He-4 production cross sections, however all of the libraries had some isotopes that were missing He-3 production cross sections.  FENDL-3.2b had the fewest number of missing cross sections.  Because of the large number of isotopes contained in the SS-316 material definition for this model, it is difficult to determine the exact cause of the observed differences in He production.  Portions of the differences are due to missing gas production as well as different values of the gas production cross sections themselves (when present in the libraries).  Note that determination of helium production is important in ITER to determine reweldability of components during maintenance.

\begin{table}[htp]
\vspace{-2mm}
\caption{Ratio of He production calculated with various libraries versus the He production calculated with the FENDL-2.1 library in the ITER-1D model.}
\label{table:bohmiter1dheprod}
\begin{tabular}{l | c c c c} \hline \hline
Component     & F-21 &   F-31d &  E-8.0 &  F-32b  \\ \hline
IB FW Be      & 1.0000 & 1.0004 & 1.0080 & 1.0008 \\
IB FW CuBeNi  & 1.0000 & 1.0039 & 1.0370 & 1.0048 \\
IB FWSS316    & 1.0000 & 1.0424 & 1.0616 & 1.0875 \\
IB VV Inconel & 1.0000 & 1.1796 & 1.0684 & 1.1970 \\
IB VV SS316   & 1.0000 & 1.0806 & 1.0126 & 1.0593 \\
IB Mag. (Cu)  & 1.0000 & 1.0502 & 0.9883 & 1.0148 \\ \hline
OB FW Be      & 1.0000 & 1.0007 & 1.0090 & 1.0010 \\
OB FW CuBeNi  & 1.0000 & 1.0047 & 1.0375 & 1.0053 \\
OB FWSS316    & 1.0000 & 1.0441 & 1.0698 & 1.0961 \\
OB VV Inconel & 1.0000 & 1.1798 & 1.0696 & 1.1952 \\
OB VV SS316   & 1.0000 & 1.0807 & 1.0149 & 1.0592 \\
OB Mag. (Cu)  & 1.0000 & 1.0686 & 0.9848 & 1.0201 \\ \hline \hline
\end{tabular}
\vspace{-3mm}
\end{table}

\Cref{table:bohmiter1dtritprod} shows the ratio in tritium production calculated with various libraries versus the tritium production calculated with the FENDL-2.1 library. The statistical uncertainty for these tritium production calculations has a maximum of 0.27\%.  The tritium production values calculated with FENDL-3.1d are higher than those calculated with FENDL-2.1 (up to a factor of 2.4) at the VV and magnet locations.  The tritium production values calculated with ENDF/B-VIII.0 are higher at most locations than those calculated with FENDL-2.1 (up to a factor of 3). Using FENDL-3.2b, tritium production is similar to that seen with FENDL-3.1d.  As discussed with the helium production, past work has shown that much of the differences observed is due to missing gas production reactions in the cross section libraries for the various isotopes contained in these benchmark materials \cite{fischerINDC-631}.  For example, the missing tritium production cross sections for Al-27 in FENDL-2.1 caused a large portion of the difference in the Inconel tritium production. In the current design for ITER, the vacuum vessel shell is composed of a special grade of SS-316 rather than the Inconel used in the early design upon which this benchmark is based.  Because of this, the wide range of values observed in Inconel are less concerning, but significant variations in tritium production are also seen for SS-316.  Future work will look to identify the exact cause of the different tritium production rate in the large number of isotope contained in the SS-316 material definition for this benchmark model.  Note that tritium production is important even for non-breeding materials for determining total tritium levels and potential environmental impacts during accidents and waste disposal activities. 

\begin{table}[htp]
\vspace{-2mm}
\caption{Ratio of tritium production calculated with various libraries versus the tritium production calculated with the FENDL-2.1 library in the ITER-1D model.}
\label{table:bohmiter1dtritprod}
\begin{tabular}{l | c c c c} \hline \hline
Component     & F-21 &   F-31d &  E-8.0 &  F-32b  \\ \hline
IB FW Be      & 1.0000 & 1.0000 & 0.9953 & 1.0007 \\
IB FW CuBeNi  & 1.0000 & 0.9998 & 1.1870 & 1.0007 \\
IB FWSS316    & 1.0000 & 0.9967 & 1.8999 & 0.9489 \\
IB VV Inconel & 1.0000 & 2.3708 & 2.9941 & 2.0296 \\
IB VV SS316   & 1.0000 & 1.0229 & 1.7551 & 0.9956 \\
IB Mag. (Cu)  & 1.0000 & 1.0584 & 1.1535 & 1.0171 \\ \hline
OB FW Be      & 1.0000 & 1.0003 & 0.9960 & 1.0008 \\
OB FW CuBeNi  & 1.0000 & 1.0003 & 1.1879 & 1.0010 \\
OB FWSS316    & 1.0000 & 0.9974 & 1.9079 & 0.9482 \\
OB VV Inconel & 1.0000 & 2.3827 & 3.0162 & 2.0435 \\
OB VV SS316   & 1.0000 & 1.0233 & 1.7576 & 0.9973 \\
OB Mag. (Cu)  & 1.0000 & 1.0793 & 1.1573 & 1.0244 \\ \hline \hline
\end{tabular}
\vspace{-6mm}
\end{table}

\subsubsection{ITER 3-D}
\label{subsubsec:iter3dBohm}
\begin{figure}[htp]
\includegraphics[width=8.69cm]{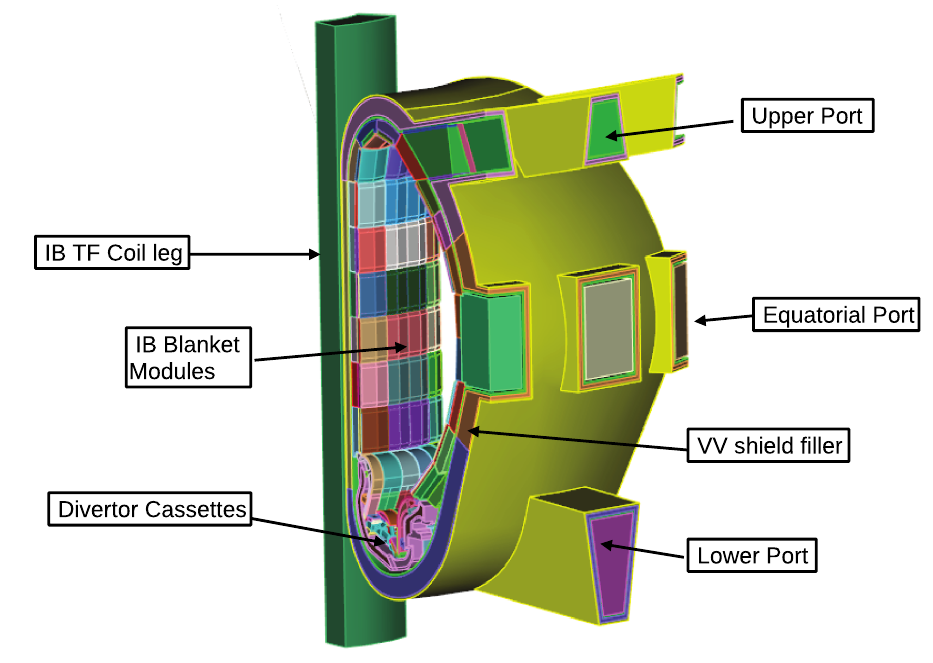}
    \caption{Overall view of ITER 3-D 40 degree benchmark model.}
\label{fig:bohmiter3dmodel}
\end{figure}

This benchmark model is a CAD model consisting of a partially homogenized 40 degree sector model based on the UW ITER Blanket Lite (BL-Lite) model \cite{uwITERbllite}.  A homogenized IB TF coil volume is added to the BL-Lite model to complete the benchmark model.  The model consists of a total of 723 volumes and is shown in \cref{fig:bohmiter3dmodel}.  Detailed CAD models of ITER that were current at the time of the BL-Lite model creation were used to determine the homogenized compositions.  Each blanket module (BM) consists of 5 volumes.  There are two versions of BM: a normal heat flux (NHF) module or an enhanced heat flux module (EHF).  The BM is composed of a water cooled FW and water cooled SS-316 Shield Block (SB).  The vacuum vessel is water cooled and consists of a SS-316 inner shell, a SS-316 outer shell, and four different filler regions (water cooled borated SS-304).  A more detailed description of this model is provided in previous work\cite{bohmfendlFST2021}.  This benchmark model was run with DAG-MCNP6.2 \cite{dagmcnp} for all calculations shown in this work. For photon transport, we are using the mcplib84 photon cross section library distributed with MCNP \cite{mcplib84}.

\Cref{table:bohmiter3dibtfcoilheat} shows the ratio of total nuclear heating of the IB TF Coil volume calculated with different neutron cross section libraries for the ITER 3-D benchmark model.  The ratio is taken with respect to the nuclear heating calculated with FENDL-2.1.  In these heating calculations, the maximum statistical uncertainty was 0.66\%.  For the FENDL v3.1d and v3.2b libraries, we see that the nuclear heating is nearly identical to that calculated with FENDL-2.1. The nuclear heating is ~8\% lower for ENDF/B-VIII.0 respectively as compared to FENDL-2.1.

\begin{table}[htp]
\caption{Ratio of total nuclear heating in the IB TF Coil volume calculated with different neutron cross section data for the ITER 3-D benchmark model.}
\label{table:bohmiter3dibtfcoilheat}
\begin{tabular}{l | c } \hline \hline
Library       & Ratio  \\ \hline
FENDL-2.1     &  1     \\  
FENDL-3.1d    &  1.004 \\
ENDF/B-VIII.0 &  0.925 \\
FENDL-3.2b    &  0.994 \\ \hline \hline
\end{tabular}
\end{table}

\begin{figure}[htp]
\includegraphics[width=8.69cm]{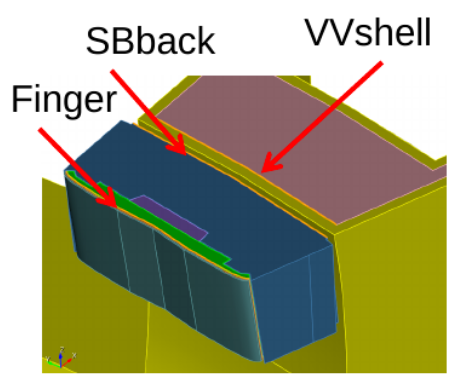}
    \caption{Location at BM14 for the surface tallies used in the ITER-3D benchmark model.}
\label{fig:bohmiter3dbm14tallies}
\end{figure}

Radiation damage in terms of dpa and helium production were calculated in three different locations near BM14 which is located on the OB side near the machine mid-plane.  \Cref{fig:bohmiter3dbm14tallies} shows the location of surface tallies used for these calculations.  Values of iron dpa at these locations were within 1\% of each other when calculated with the FENDL-2.1, FENDL-3.1d, and FENDL-3.2b cross section libraries.  The value of iron dpa calculated using ENDF/B-VIII.0 was 0\%, 3.2\% and 3.6\% lower than the value calculated with FENDL-2.1 at the FWfinger, SBback and VVshell locations respectively.  The maximum statistical uncertainty for these iron dpa calculations was 0.31\%.  

\Cref{table:bohmiter3dheprodbm14} shows the ratio of helium production calculated near BM14.  The maximum statistical uncertainty for these calculations is 0.30\%. Overall, the He production is 4-11\% higher when calculated with the various libraries tested as compared to that calculated with FENDL-2.1.  Helium production calculated with FENDL-3.2b has the highest values of the libraries examined.  Past work has shown that most of the difference in helium production was due to the different He production cross sections (mt=207) in the various libraries \cite{bohmfendlFST2021}.  More work will be done in the future to determine the details of these different helium production cross section values for the various isotopes contained in the SS-316L(N)-IG alloy used in ITER.

\begin{table}[htp]
\caption{Ratio of helium production calculated in different locations near BM14 for the ITER-3D benchmark model.}
\label{table:bohmiter3dheprodbm14}
\begin{tabular}{l | c c c c} \hline \hline
Component    &  F-21 & F-31d  &  E-8.0 &  F-32b  \\ \hline
BM14 finger  &  1.0  & 1.0419 & 1.0783 & 1.1081 \\
BM14 SBback  &  1.0  & 1.0562 & 1.0371 & 1.0578 \\
BM14 VVshell &  1.0  & 1.0572 & 1.0657 & 1.0877 \\ \hline \hline
\end{tabular}
\end{table}

\subsubsection{FNSF 3-D}
\label{subsubsec:fnsf3dBohm}
\begin{figure}[htp]
\includegraphics[width=8.69cm]{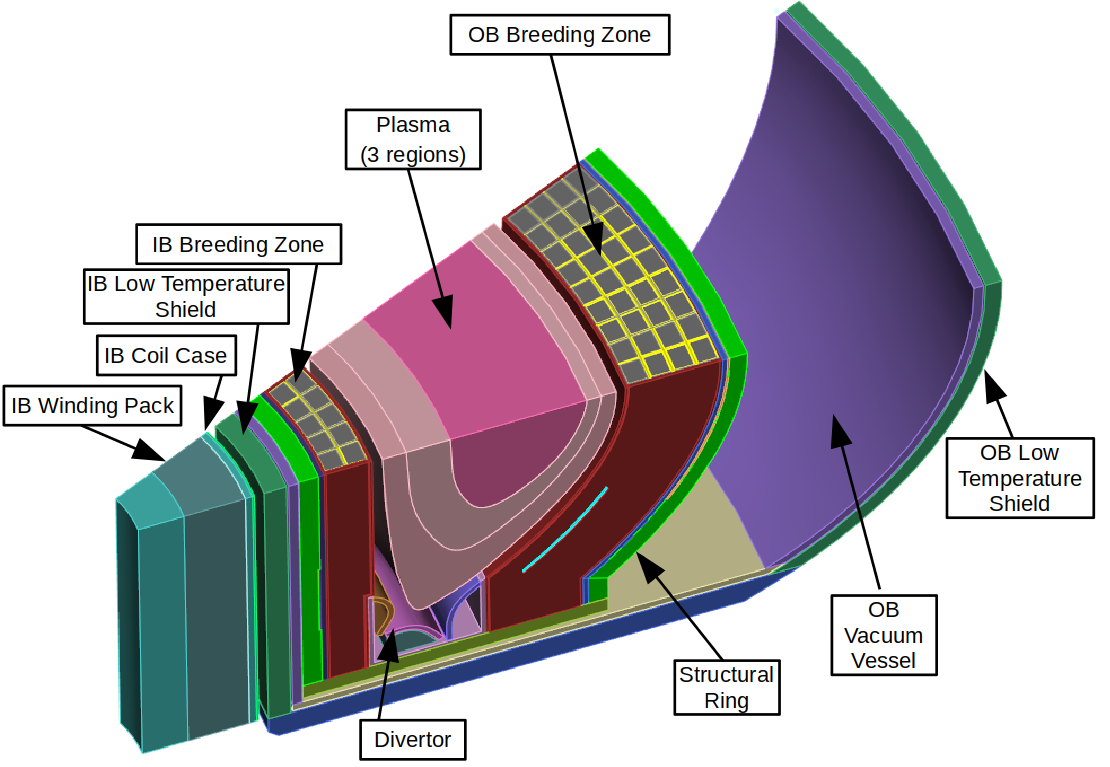}
    \caption{Overall view of the FNSF 3-D 22.5 degree benchmark model sliced at the mid-plane.}
\label{fig:bohmfnsf3dmodel}
\end{figure}

This benchmark model is a CAD model consisting of a partially homogenized 22.5 degree model of the Fusion Energy Systems Study (FESS) Fusion Nuclear Science Facility (FNSF) \cite{davisFNSFFusEngDes2018,bohmFNSFFST2019}.   A cut away view of this model taken at the mid-plane is shown in \cref{fig:bohmfnsf3dmodel}. This model has a 3 region plasma source, does not contain ports, and contains a detailed PbLi breeding zone (BZ) on both the IB and OB sides.  These BZ regions contain the thin SiC flow channel inserts lining the PbLi flow channel walls.  These walls are helium cooled low activation ferritic steel.  This benchmark model contains materials that are significantly different than ITER, mainly it has a large amount of PbLi (a eutectic of 84.3 atomic percent Pb and 15.7 atomic percent Li enriched to 90 wt. \% Li-6) in the BZ and its structural steel is a low activation modified ferritic steel (MF82H) which contains 90 wt. \% Fe, 7.5 wt. \% Cr, 2 wt. \% W, 0.2 wt. \% V, 0.02 wt. \% Ta and 0.1 wt. \% C.  The MF82H steel is also used for the face plates in the structural ring (SR).  A 3Cr-3WV Ferritic Steel (FS) (93 wt. \% Fe, 3 wt. \% Cr, 3 wt. \% W) is used for the face plates of the VV and the Low Temperature Shield (LTsh).  The main shielding for the super conducting magnets is provided by these three structures (SR, VV, LTsh) which contain shielding fillers.  The SR, VV, and LTsh fillers consist of WC for the IB SR, IB VV, and IB LTsh. For the OB fillers, borated MF82H, and borated ferritic steel are used. More details on component composition are provided in the FNSF 1-D schematic diagrams \cref{fig:bohmfnsf1dibschematic} and \cref{fig:bohmfnsf1dobschematic} as well as previous work~\cite{bohmfendlFST2021}.  Helium provides cooling for the FW, SR, and VV while water is used for cooling the LTsh.  Therefore in the FNSF design, most of the cooling is provided by He instead of water as is used in ITER.  This benchmark model was run with DAG-MCNP-5v160 for all calculations shown in this particular work.  For photon transport, we are using the mcplib84 photon cross section library distributed with MCNP \cite{mcplib84}.

\begin{figure}[htp]
\includegraphics[width=8.69cm]{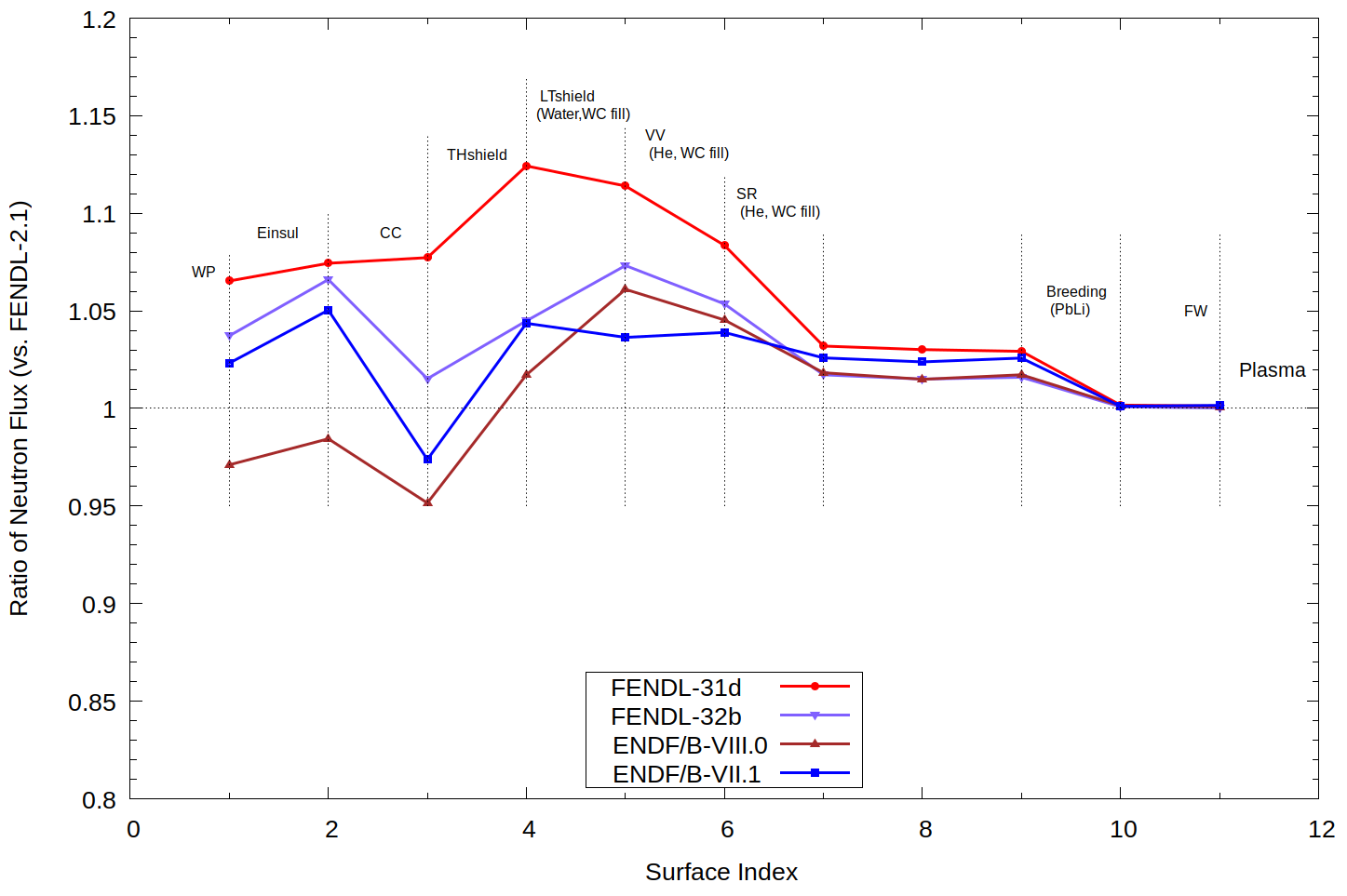}
    \caption{Ratio of neutron flux (vs. FENDL-2.1) calculated with different neutron cross section data for the FNSF 3-D benchmark model on the IB side.}
\label{fig:bohmfnsf3dnflux}
\end{figure}

\Cref{fig:bohmfnsf3dnflux} shows the ratio of neutron flux calculated at the front surface of each IB component with different neutron cross section libraries in the FNSF 3-D benchmark model. The major components are indicated on the plot with a dotted line and label. The maximum statistical uncertainty for these calculations is $<$0.1\% except in the deeper regions from the thermal shield (THshield) to the WP where the uncertainty is 2.0\% and 2.8\% respectively. The neutron flux calculated with the FENDL-3.1d cross section library is up to 12\% higher than the flux calculated with the FENDL-2.1 library. For calculations with the ENDF/B-VIII.0 library, the flux is as much as 5\% lower than that calculated with the FENDL-2.1 cross section library. Calculations of neutron flux with FENDL-3.2b agree more closely with those calculated for FENDL-2.1 than those calculated with FENDL-3.1d. In general, these neutron flux trends are similar to what was seen in the ITER-1D model.   

\Cref{table:bohmfnsf3dtbr} shows the ratio of TBR for the IB and OB breeding zones (BZ) calculated with different libraries as compared to that calculated with the FENDL-2.1 library.  Maximum statistical uncertainty was 0.01\% . For the libraries discussed in this report, the TBR for the IBBZ is higher and averages 1.3\% higher than that calculated with FENDL-2.1. The TBR for the OBBZ is also higher and averages 1.5\% higher for the libraries investigated in this work as compared to FENDL-2.1.

\begin{table}[htp]
\caption{Ratio of tritium breeding ratio (TBR) at the IBBZ and OBBZ compared to that calculated with FENDL-2.1 for the FNSF 3-D model.}
\label{table:bohmfnsf3dtbr}
\begin{tabular}{l | c c } \hline \hline
Library       &  Ratio IBBZ &  Ratio OBBZ \\ \hline    
FENDL-2.1     &   1         &   1         \\      
ENDF/B-VIII.0 &   1.0142    &   1.0160    \\    
FENDL-3.1d    &   1.0131    &   1.0171    \\    
ENDF/B-VII.1  &   1.0122    &   1.0130    \\    
FENDL-3.2b    &   1.0115    &   1.0151    \\ \hline \hline
\end{tabular}
\end{table}

\Cref{table:bohmfnsf3dvvccheat} shows the ratio of integrated total nuclear heating for the IB VV and the IB magnet coil case compared to that calculated using the FENDL-2.1 library. Maximum statistical uncertainty is 0.03\% and 1.3\% for the IB VV and IB coil case (CC) respectively. At the IB VV, heating values are ~5\% higher than those calculated with FENDL-2.1 when using ENDF/B-VIII.0 and FENDL-3.1d.
Using ENDF/B-VII.1 and FENDL-3.2b produces heating values that are ~3\% and 4\% higher respectively.  At the IBCC, the heating values are 8-14\% higher as compared to those calculated with FENDL-2.1. This is different than what was seen in the ITER-1D benchmark at deep locations.  There, some heating values were higher (e.g. those calculated with FENDL-3.1d) and some were lower (e.g. those calculated with ENDF/B-VIII.0) as compared to FENDL-2.1.   At this point, it is not exactly clear why the nuclear heating values at the coil case calculated with the newer libraries are higher than those calculated with FENDL-2.1.  The neutron fluxes at the coil case are higher when calculated with FENDL-3.2b and FENDL-3.1d but lower when calculated with ENDF/B-VIII.0 as compared to FENDL-2.1.  Comparing photon fluxes at the coil case location (plot not shown), shows higher photon fluxes (10-18\%) when calculated with all of these libraries as compared to FENDL-2.1.

The contribution from neutron heating to the total nuclear heating at the IBCC is small (4\%), so even though we have observed differences in the neutron flux and differences in neutron heating numbers amongst the libraries for the isotopes in SS-316, the overall effect on total nuclear heating is small. The nuclear heating at the IBCC is dominated by the contribution from photons (gammas) which is 96\% of the total. Recall that all calculations presented in this section use the same photon cross section library provided with MCNP (mcplib04) and therefore will use the same photon heating numbers.  Thus, the differences observed are due to differing photon flux levels (and possibly differing photon spectrum) which are determined by the photon production in the neutron libraries for the materials near the IBCC.  If one examines the neutron induced prompt photon production cross section for SS-316, one generally sees higher photon production cross sections in the newer libraries as compared to FENDL-2.1, but note that visual inspection is difficult due to the many isotopes and resonances.  More work will be done in the future to examine the exact cause of differences observed in total nuclear heating at this location as magnet heating is an important design driver for reactors.

\begin{table}[htp]
\caption{Ratio of integrated total nuclear heating for the IB VV and the IB magnet coil case (CC) compared to that calculated with FENDL-2.1 for the FNSF 3-D model.}
\label{table:bohmfnsf3dvvccheat}
\begin{tabular}{l | c c } \hline \hline
Library        &  Ratio IB VV &  Ratio IB CC \\ \hline   
FENDL-2.1      &  1           &  1                  \\ 
ENDF/B-VIII.0  &  1.0480      &  1.0783             \\ 
FENDL-3.1d     &  1.0539      &  1.1350             \\ 
ENDF/B-VII.1   &  1.0264      &  1.0762             \\ 
FENDL-3.2b     &  1.0431      &  1.1031             \\ \hline \hline 
\end{tabular}
\end{table}

For iron dpa at the IBFW and IBVV, FENDL-3.2b is 0\% and 8\% higher than FENDL-2.1 respectively.  Results for calculations using the other neutron cross section libraries for iron dpa were generally similar except for ENDF/B-VII.1 at the IBFW and were shown in the previous work in Table IX \cite{bohmfendlFST2021}.

For helium production at the IBFW and IBVV, FENDL-3.2b is 6\% and 2\% higher than FENDL-2.1 respectively. Results for calculations using the other neutron cross section libraries for helium production were shown in the previous work (in Table X) \cite{bohmfendlFST2021}.  Note that for FENDL-3.1d, helium production calculations showed little difference compared to calculations with FENDL-2.1 at these two locations. Calculations of helium production with ENDF/B-VIII.0 and ENDF/B-VII.1 libraries were 8-12\% higher than calculations with FENDL-2.1 at these two locations.

\Cref{table:bohmfnsf3dtprodfwvv} shows the ratio of tritium production calculated with various libraries compared to that calculated with FENDL-2.1 at the IBFW and IBVV. For tritium production at the IBFW and IBVV, FENDL-3.2b is 27\% and 65\% lower than FENDL-2.1 respectively.  For tritium production calculated with FENDL-3.1d, it is 43\% higher than FENDL-2.1 at the IBFW but 37\% lower than FENDL-2.1 at the IBVV.  As discussed in the results of tritium production in the ITER 1-D model, more work will have to be done to identify the cause of these large differences seen in the tritium production at these locations for the particular steel alloys used at the IBFW and IBVV (MF82H and 3Cr-3WV respectively).

\begin{table}[htp]
\caption{Ratio of tritium production for the IBFW and the IBVV compared to that calculated with FENDL-2.1 for the FNSF 3-D model.}
\label{table:bohmfnsf3dtprodfwvv}
\begin{tabular}{l | c c } \hline \hline
Library        &  Ratio IBFW &  Ratio IBVV \\ \hline   
FENDL-2.1      &  1          &  1          \\
ENDF/B-VIII.0  &  0.5000     &  0.2798     \\
FENDL-3.1d     &  1.4277     &  0.6308     \\
ENDF/B-VII.1   &  1.0501     &  0.5176     \\
FENDL-3.2b     &  0.7348     &  0.3494      \\ \hline \hline 
\end{tabular}
\end{table}

\subsubsection{FNSF 1-D}
\label{subsubsec:fnsf1dBohm}
\begin{figure*}[htp]
\includegraphics[width=17.75cm]{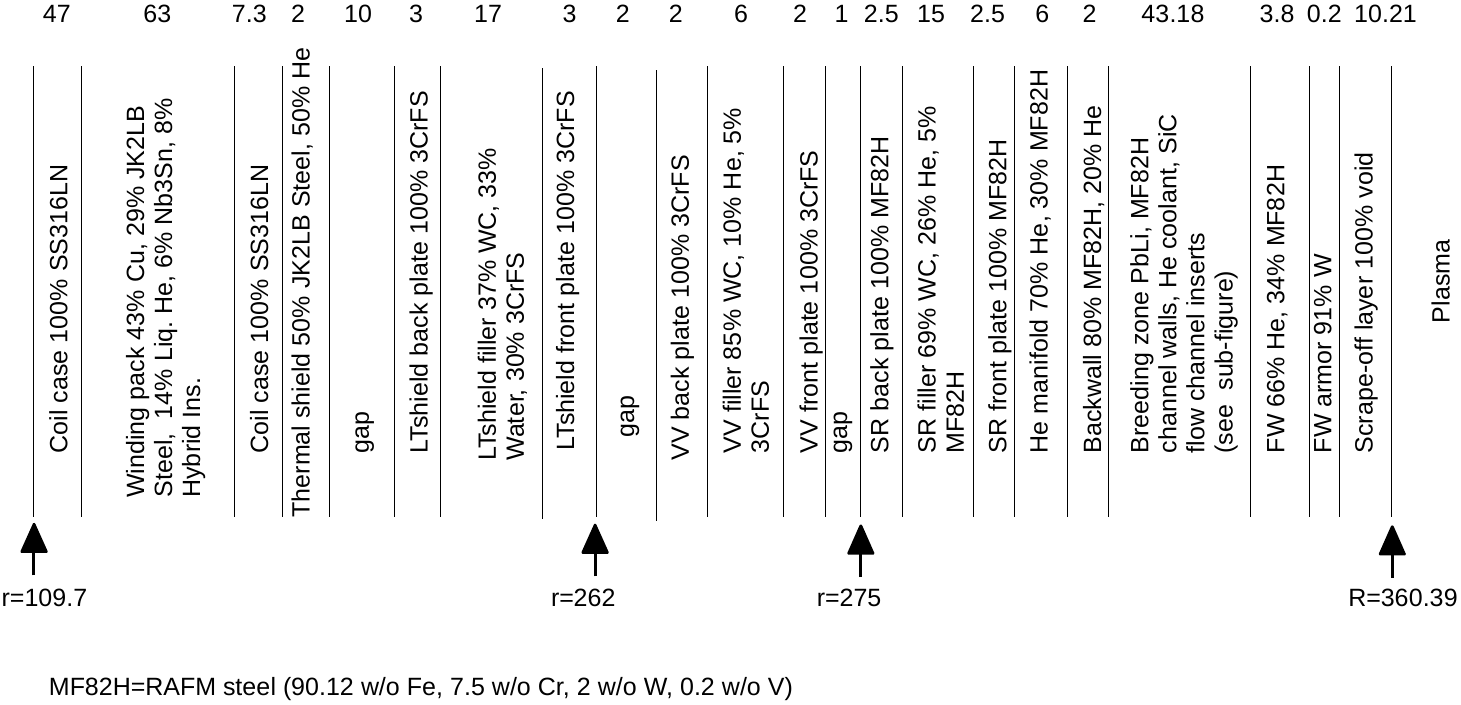}
    \caption{Schematic of the FNSF 1-D benchmark model IB portion. Compositions are volume percentages and numerical values along the top and bottom are zone thicknesses and radius in cm.}
\label{fig:bohmfnsf1dibschematic}
\end{figure*}

\begin{figure*}[htp]
\includegraphics[width=12.75cm]{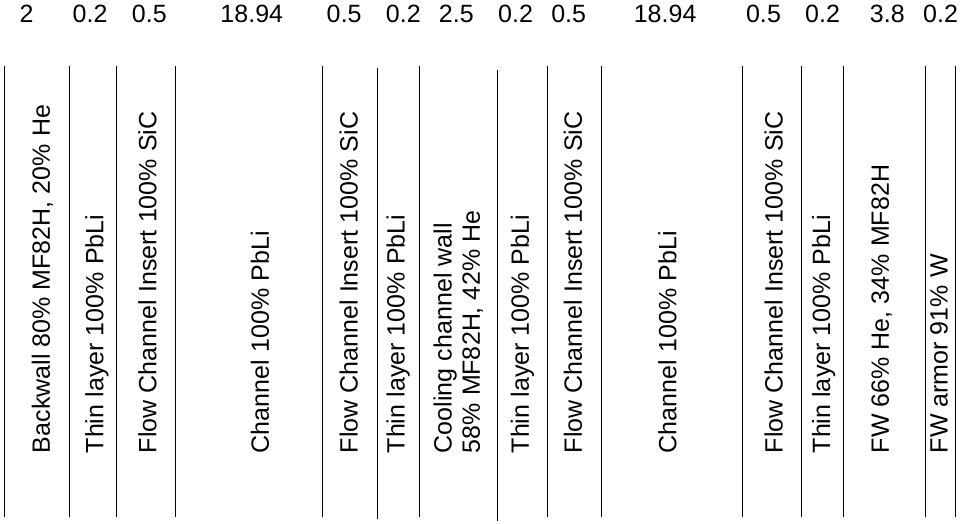}
    \caption{Schematic of the FNSF 1-D benchmark model IBBZ portion. Compositions are volume percentages and numerical values along the top and bottom are zone thicknesses and radius in cm.}
\label{fig:bohmfnsf1dibbz}
\end{figure*}

\begin{figure*}[htp]
\includegraphics[width=17.75cm]{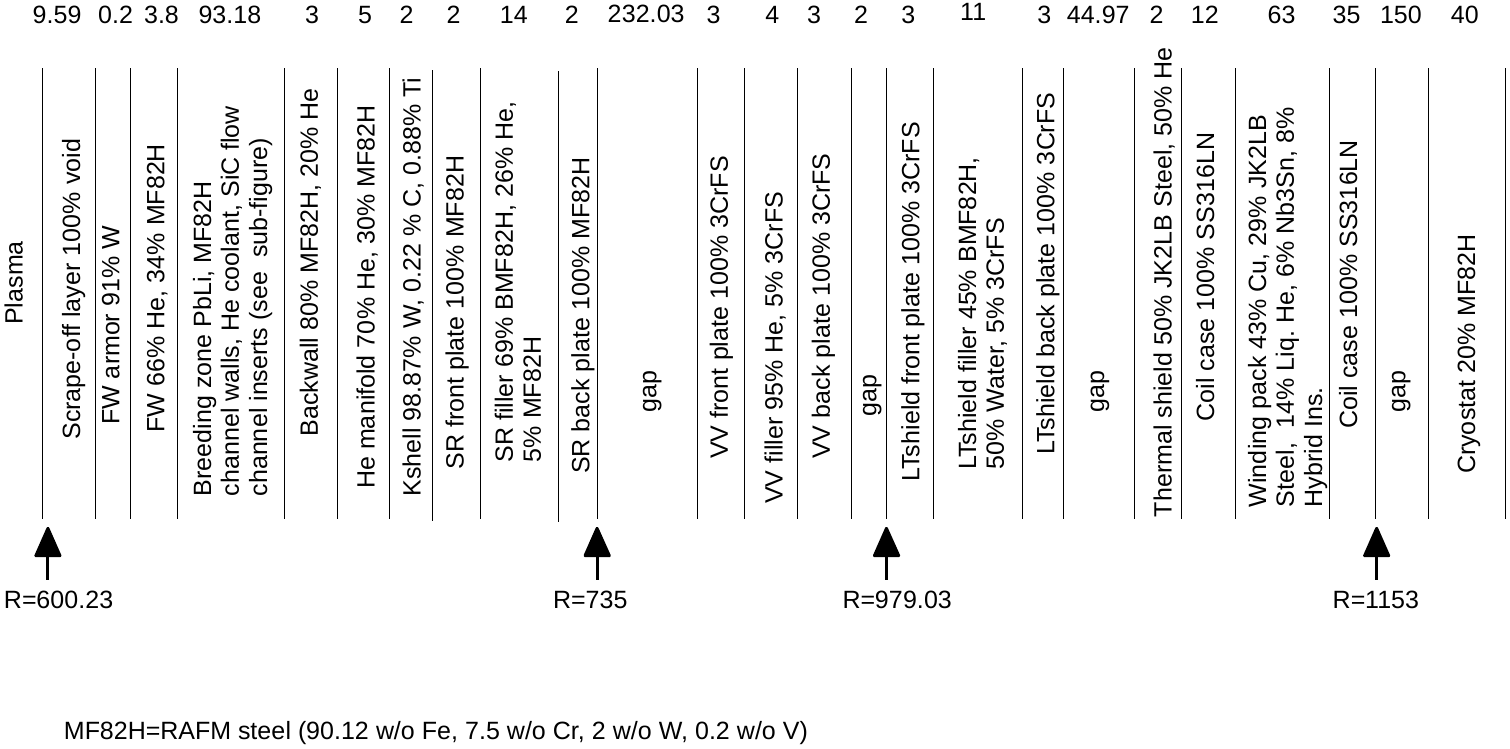}
    \caption{Schematic of the FNSF 1-D benchmark model OB portion. Compositions are volume percentages and numerical values along the top and bottom are zone thicknesses and radius in cm.}
\label{fig:bohmfnsf1dobschematic}
\end{figure*}

The FNSF 1-D cylindrical computational benchmark is based on the 3-D benchmark described above in section \ref{subsubsec:fnsf3dBohm}.  The model contains 85 radial zones and contains the details of the breeding zone which includes the PbLi flow channels, the SiC flow channel inserts, and the helium cooled MF82H flow channel dividers.  Some additional components were added to the 1-D model that were not present in the 3-D model and include the OB thermal shield, magnet coil case, magnet winding pack, and cryostat.  Also, the SR, VV, and LTsh were dehomogenized to specifically model the faceplates in these shielding structures.  \Cref{fig:bohmfnsf1dibschematic} shows the IB section and \cref{fig:bohmfnsf1dibbz} shows the details of the IB breeding zone in the 1-D model. Compositions are volume percentages and numerical values along the top and bottom are zone thicknesses and radius in cm. \Cref{fig:bohmfnsf1dobschematic} shows the OB section of the model.  The OBBZ is not shown but is similar to the IBBZ except it contains 4 PbLi flow channels rather than 2.

\begin{figure}[htp]
\includegraphics[width=8.69cm]{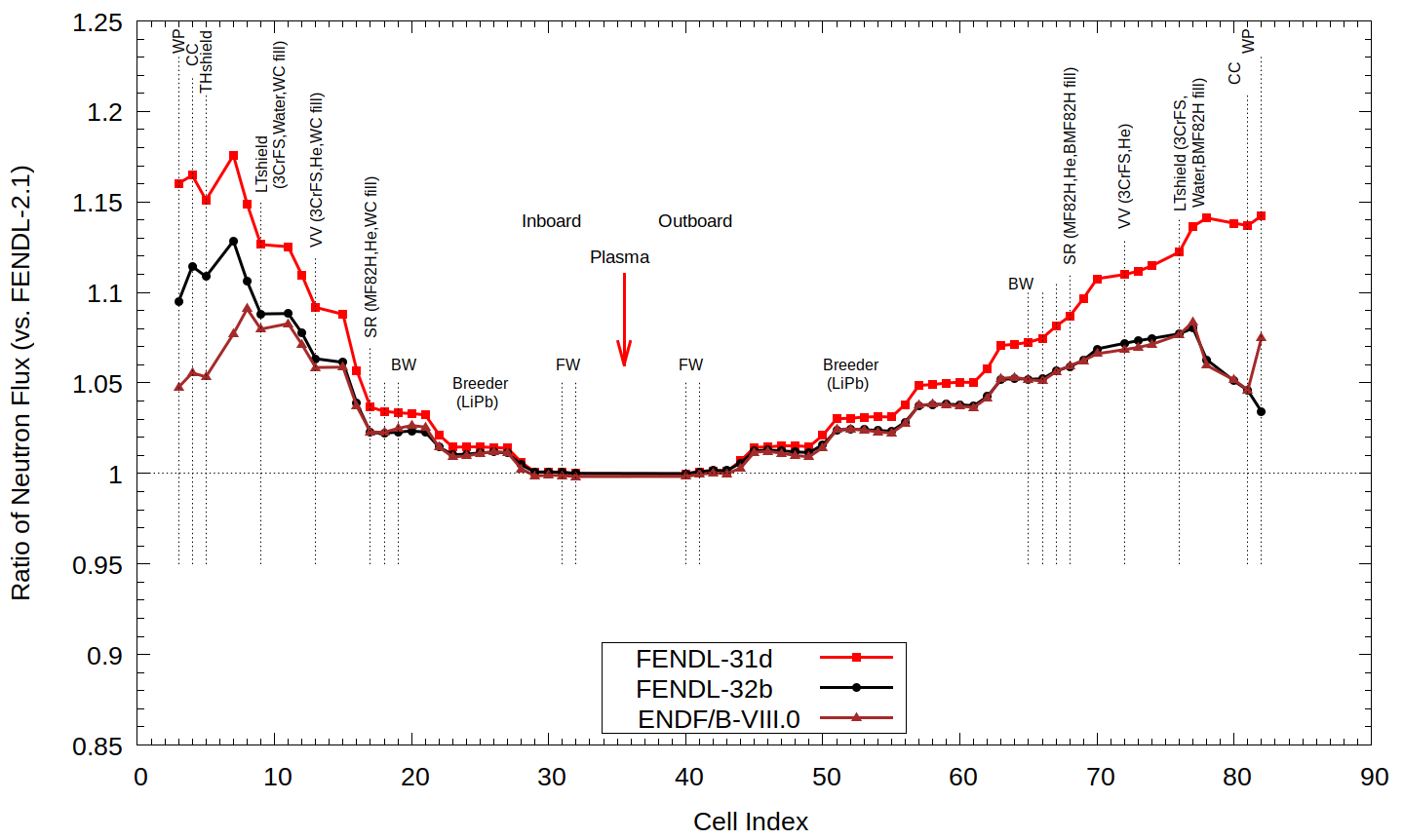}
    \caption{Ratio of neutron flux (vs. FENDL-2.1) calculated with different neutron cross section data for the FNSF 1-D benchmark model.  The cell index is ordered radially.}
\label{fig:bohmfnsf1dnflux}
\end{figure}

\Cref{fig:bohmfnsf1dnflux} shows the ratio of neutron flux calculated with different neutron cross section data for the FNSF 1-D cylindrical benchmark model. The fluxes are compared against the flux calculated with the FENDL-2.1 neutron library. The cell index for the horizontal axis is ordered radially. The inboard, plasma, and outboard sections are labeled in the figure. Also, vertical dotted lines indicate the front (closest to the plasma) location of some key components such as the FW, BW, SR, VV, and WP. The statistical uncertainty for the neutron flux calculations has a maximum $<$0.1\% except for the LT shield back plate to WP region which ranges from 0.8-1.7\% respectively for the IB, and 0.2\%-0.6\% on the OB.  Looking at this figure, fluxes calculated with all the libraries are higher than those calculated with FENDL-2.1 (this was different from what was seen with the ITER-1D benchmark where some calculations showed higher fluxes and some lower).  Fluxes calculated with FENDL-3.2b are as much as 13\% higher than those calculated with FENDL-2.1.    Fluxes calculated with FENDL-3.2b are close to those calculated with ENDF/B-VIII.0 up to the depth of the LT shield on the IB and up to the depth of the CC on the OB.  Fluxes calculated with FENDL-3.1d are the highest of all the libraries tested (this was also observed in the FNSF-3D model).  Further, looking at the figure, one can see the neutron flux ratio rises in comparison to FENDL-2.1 at the thick PbLi flow channels (19 cm) in the breeder zones, the thick WC shield filler regions in the IB SR, VV, and LTshield, and the thick borated MF82H shield filler regions in the OB SR, VV, and LTshield.  Examining the cross sections for these materials is difficult due to the large number of isotopes and resonances.  Small changes in the cross section will have a significant impact on the flux in these thicker regions.  More work will be done in the future to identify the exact cause of the increased flux compared to FENDL-2.1.

\begin{figure}[htp]
\includegraphics[width=8.69cm]{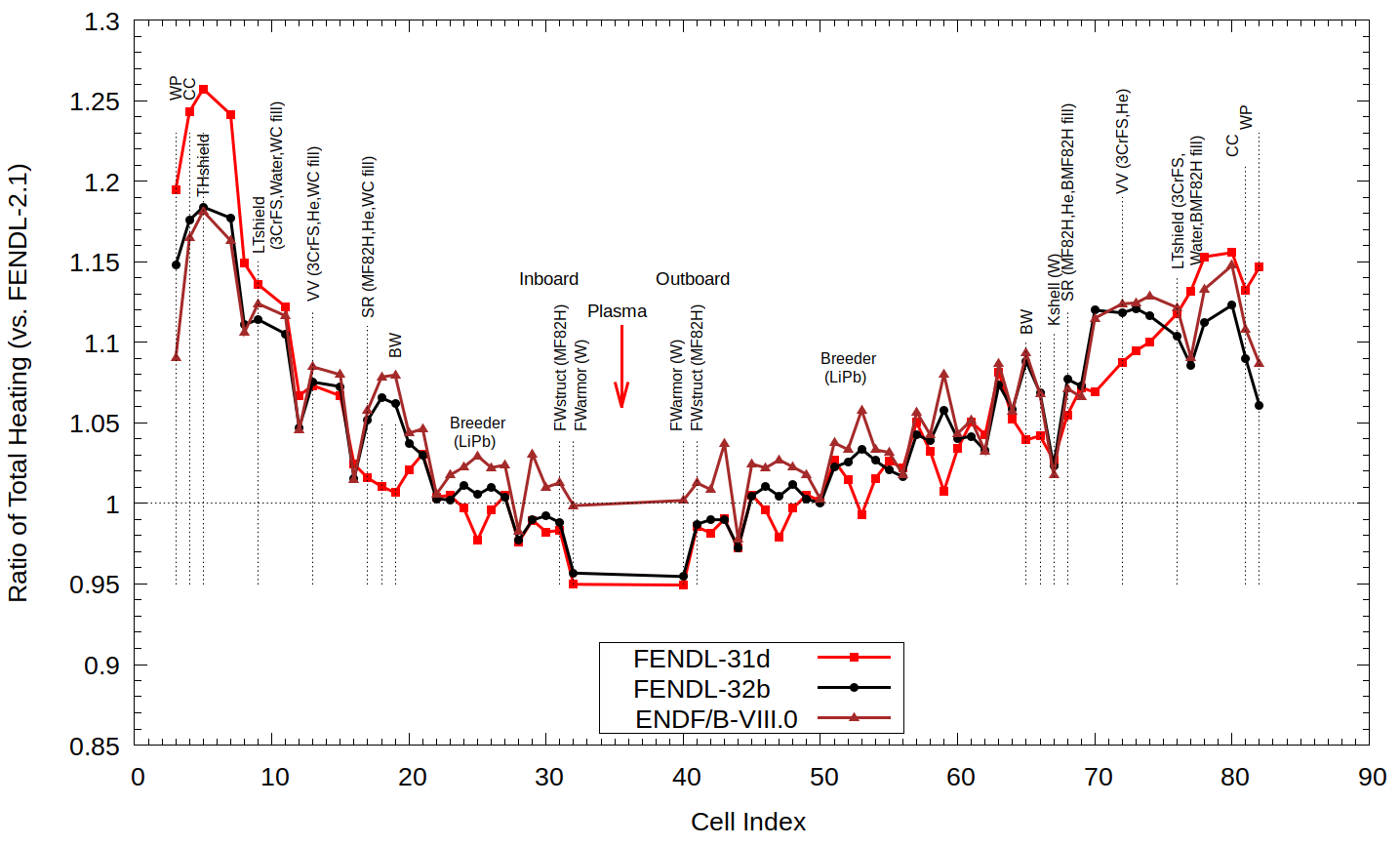}
    \caption{Ratio of total nuclear heating (vs. FENDL-2.1) calculated with different neutron cross section data for the FNSF 1-D benchmark model.  The cell index is ordered radially.}
\label{fig:bohmfnsf1dtotheat}
\end{figure}

\Cref{fig:bohmfnsf1dtotheat} shows the ratio of total nuclear heating (neutron+photon) calculated with different neutron cross section data for the FNSF 1-D cylindrical benchmark model.  The statistical uncertainty for the nuclear heating calculations is less than 0.1\% except at deep locations from the LTsh backplate to the WP where it ranges from 0.5\% to 1\% for the IB and 0.3\% to 0.8\% for the OB.  Looking at the figure, nuclear heating in tungsten at the FW is lower when calculated with FENDL-3.1d and FENDL-3.2b as compared to that calculated with FENDL-2.1.  Also, total heating is higher for all libraries at all depths beyond the breeder zone than that calculated with FENDL-2.1. Heating values calculated with FENDL-3.2b and ENDF/B-VIII.0 are in general agreement except at the FW tungsten and IB WP.  Heating calculated with FENDL-3.2b is up to 15\% higher at the IB WP and up to 6\% higher at the OB WP as compared to that calculated with FENDL-2.1.  

\Cref{table:bohmfnsf1ddpa} shows the ratio of iron dpa calculated with various libraries versus the iron dpa calculated with the FENDL-2.1 library.  The iron dpa is calculated at the front surface of various components in the model.  The statistical uncertainty for these dpa calculations has a maximum of 0.15\%. Looking at the table, the iron dpa calculated with various libraries is quite close at the FW.  However, at deeper locations, the iron dpa is 5-15\% higher when calculated with newer libraries as compared to when calculated with FENDL-2.1.  This is expected since the neutron fluxes calculated with these newer libraries were higher than FENDL-2.1 at these deeper locations (see \cref{fig:bohmfnsf1dnflux}).   

\begin{table}[htp]
\caption{Ratio of iron dpa calculated with various libraries versus the iron dpa calculated with the FENDL-2.1 library in the FNSF-1D model.}
\label{table:bohmfnsf1ddpa}
\begin{tabular}{l | c c c c} \hline \hline
Component   & F-21   & F-31d  & F-32b  & E-8.0    \\ \hline
IBFW        & 1.0000 & 1.0013 & 1.0076 & 1.0136    \\
IBBW        & 1.0000 & 1.0484 & 1.0619 & 1.0836    \\
IBSRface    & 1.0000 & 1.0505 & 1.0537 & 1.0594    \\
IBVVface    & 1.0000 & 1.1196 & 1.0988 & 1.0777    \\
IBLTshface  & 1.0000 & 1.1587 & 1.1187 & 1.0864    \\ \hline
OBFW        & 1.0000 & 0.9994 & 1.0026 & 1.0077    \\
OBBW        & 1.0000 & 1.1004 & 1.1045 & 1.1225    \\
OBSRface    & 1.0000 & 1.1177 & 1.1080 & 1.1158    \\
OBVVface    & 1.0000 & 1.1311 & 1.1022 & 1.0951    \\
OBLtshface  & 1.0000 & 1.1373 & 1.1019 & 1.0901    \\ \hline \hline
\end{tabular}
\end{table}

\Cref{table:bohmfnsf1dheprod} shows the ratio of helium production calculated with various libraries versus the helium production calculated with the FENDL-2.1 library.  The helium production is calculated at the front surface of various components in the model using the material at this location (as indicated in the table).  The statistical uncertainty for these helium production calculations has a maximum of 1\% on the IB (IBLTsh) but up to 4\% on the OB (OBLTsh).  For the components composed of steels, the helium production calculated with newer libraries is up to 15\% higher at IB locations and up to 27\% higher at OB locations as compared to FENDL-2.1.  However, for components composed of tungsten, it can be seen that helium production calculated with FENDL-3.1d and ENDF/B-VIII.0 is over 2x higher than that calculated with FENDL-2.1.  In order to better understand the cause of this difference, a separate calculation was run using FENDL-3.2b cross sections for transport, then helium production was determined in tungsten using the helium production cross section from the various libraries on the tally multiplier card (fm) of MCNP.  The results of this showed that most of the difference is due to the different helium production cross sections for tungsten in the various libraries.  Plotting these helium production cross sections from the various libraries clearly shows this (note that W-180 is not present in the FENDL-2.1 library so W-180 from FENDL-3.1d was used).   

\begin{table}[htp]
\caption{Ratio of helium production calculated with various libraries versus the helium production calculated with the FENDL-2.1 library in the FNSF-1D model.}
\label{table:bohmfnsf1dheprod}
\begin{tabular}{l | c | c c c c} \hline \hline
Component  & Material  & F-21   & F-31d  & F-32b  & E-8.0    \\ \hline
IBFWarmor  & Tungsten  & 1.0000 & 2.1038 & 1.0655 & 2.1065    \\
IBFW       & MF82H     & 1.0000 & 1.0027 & 1.0602 & 1.0940    \\
IBBW       & MF82H     & 1.0000 & 1.0183 & 1.0673 & 1.1205    \\
IBSRface   & MF82H     & 1.0000 & 1.0305 & 1.0706 & 1.1187    \\
IBVVface   & 3CrFS     & 1.0000 & 1.0238 & 1.0524 & 1.1403    \\
IBLTshface & 3CrFS     & 1.0000 & 1.0472 & 1.0667 & 1.1521    \\ \hline
OBFWarmor  & Tungsten  & 1.0000 & 2.0995 & 1.0632 & 2.1014    \\
OBFW       & MF82H     & 1.0000 & 1.0016 & 1.0588 & 1.0884    \\
OBBW       & MF82H     & 1.0000 & 1.0629 & 1.1063 & 1.1332    \\
OBKshell   & Tungsten  & 1.0000 & 2.5834 & 1.2984 & 2.2679    \\
OBSRface   & MF82H     & 1.0000 & 1.0711 & 1.1020 & 1.1167    \\
OBVVface   & 3CrFS     & 1.0000 & 1.1439 & 1.1758 & 1.1935    \\
OBLTshface & 3CrFS     & 1.0000 & 1.2679 & 1.1950 & 1.1763    \\ \hline \hline
\end{tabular}
\end{table}

\Cref{table:bohmfnsf1dtritprod} shows the ratio of tritium production calculated with various libraries versus the tritium production calculated with the FENDL-2.1 library.  The tritium production is calculated at the front surface of various components in the model using the material at this location (as indicated in the table).  The statistical uncertainty for these tritium production calculations has a maximum of 1\% on the IB (IBLTsh) but up to 5\% on the OB (OBLTsh).  Note the tritium production in tungsten is zero for this model.  For the components composed of MF82H, the tritium production calculated with FENDL-3.1d averages 33-53\% higher as compared to FENDL-2.1 but 9-31\% lower when calculated with FENDL-3.2b, and 45-53\% lower when calculated with ENDF/B-VIII.0. For the components composed of the 3CrFS, the tritium production calculated with newer libraries averages 40\% lower as compared to FENDL-2.1.  Further note that the tritium production ratios from this 1-D model of FNSF are consistent with that observed for the FNSF-3D model as shown in \cref{table:bohmfnsf3dtprodfwvv}.  

In order to better understand the cause of this difference, a separate calculation was run using FENDL-3.2b cross sections for transport, then tritium production was determined in MF82H and 3CrFS using the tritium production cross section from the various libraries on the tally multiplier card (fm) of MCNP.  The results of this showed that most of the difference is due to the different tritium production cross sections for these materials in the various libraries.  Plotting these tritium production cross sections from the various libraries clearly shows this.  More work will need to be done to determine which elements/isotopes are causing this effect.

\begin{table}[htp]
\caption{Ratio of tritium production calculated with various libraries versus the tritium production calculated with the FENDL-2.1 library in the FNSF-1D model.}
\label{table:bohmfnsf1dtritprod}
\begin{tabular}{l | c | c c c c} \hline \hline
Component  & Material  & F-21   & F-31d  & F-32b  & E-8.0    \\ \hline
IBFW       &   MF82H   & 1.0000 & 1.3283 & 0.6875 & 0.4680    \\
IBBW       &   MF82H   & 1.0000 & 1.4137 & 0.7962 & 0.5119    \\
IBSRface   &   MF82H   & 1.0000 & 1.4266 & 0.8079 & 0.5185    \\
IBVVface   &   3CrFS   & 1.0000 & 0.5921 & 0.3555 & 0.2723    \\
IBLTshface &   3CrFS   & 1.0000 & 0.5882 & 0.3583 & 0.2752    \\ \hline
OBFW       &   MF82H   & 1.0000 & 1.3257 & 0.6854 & 0.4665    \\
OBBW       &   MF82H   & 1.0000 & 1.5364 & 0.9140 & 0.5554    \\
OBSRface   &   MF82H   & 1.0000 & 1.5288 & 0.9111 & 0.5489    \\
OBVVface   &   3CrFS   & 1.0000 & 0.6268 & 0.3914 & 0.2803    \\
OBLTshface &   3CrFS   & 1.0000 & 0.7311 & 0.4206 & 0.2842    \\ \hline \hline
\end{tabular}
\end{table}

\subsubsection{ITER-1D HCPB and WCLL TBM}
\label{subsubsec:hcpbTBMFabbri}
Two 1-D ITER Test Blanket Module (TBM) MCNP models, namely the Helium Cooled Pebble Bed Test Blanket Module (HCPB-TBM) and Water Cooled Lithium Lead Test Blanket Module (WCLL-TBM) have been implemented in the JADE automated testing software (previously described in \Cref{subsubsec:leakagesphereFabbri}) to test neutronics responses obtained from different cross section libraries.  The HCPB-TBM and WCLL-TBM are two of the blanket options to be tested on the ITER equatorial port \cite{sumini1iterhcpbtbm}, \cite{sumini2iterwclltbm}. The main difference is the Lithium compound used for breeding and the neutron multiplier used to enhance breeding.  The WCLL will use the eutectic Pb-16\%Li (enriched in Li-6 at 90\%) as breeder and neutron multiplier, whereas the HCPB relies on Beryllium pebbles as neutron multiplier and Lithium based ceramics with analogous enrichment as the tritium breeder.

The 1-D ITER cylindrical model from Sawan \cite{sawanITER1Dbenchmark1994} has been modified in the outboard region.  Calculations were performed with MCNP6.2 as a computational benchmark with the JADE V\&V tool. Neutron fluence, heating, and tritium production were the focus of this comparison.  As the TBMs have been split into uniform and homogenized regions, different tallies have been used for estimating fluxes, tritium production, neutron, and gamma heating. The HCBP and the WCLL have been modeled by considering the radial inhomogeneities in the TBMs \cite{fabbri1jadedocs}.  For homogenizing each region, the source TBMs CAD model have been analyzed in terms of volumes, materials, and densities, as measured with SpaceClaim \cite{sumini3spaceclaim} for creating dedicated MCNP cards (the supporting structures which separate the breeders have not been considered). More specifically, regarding the HCPB-TBM, four regions have been identified: the first across the space separating the first wall and the lithium structures; the second as the lithium aligned with the z axis; the third as the radial lithium structure parts; the fourth as the containing pipework section. Also, in the case of the WCLL-TBM, different regions have been isolated: one between the first wall and the water structures; one containing the lithium-lead; one modeled as the azimuthal components of the water structures; another with the radial parts of the water structures; the latter represented the pipework section.

\begin{figure}[htp]
\includegraphics[width=8.69cm]{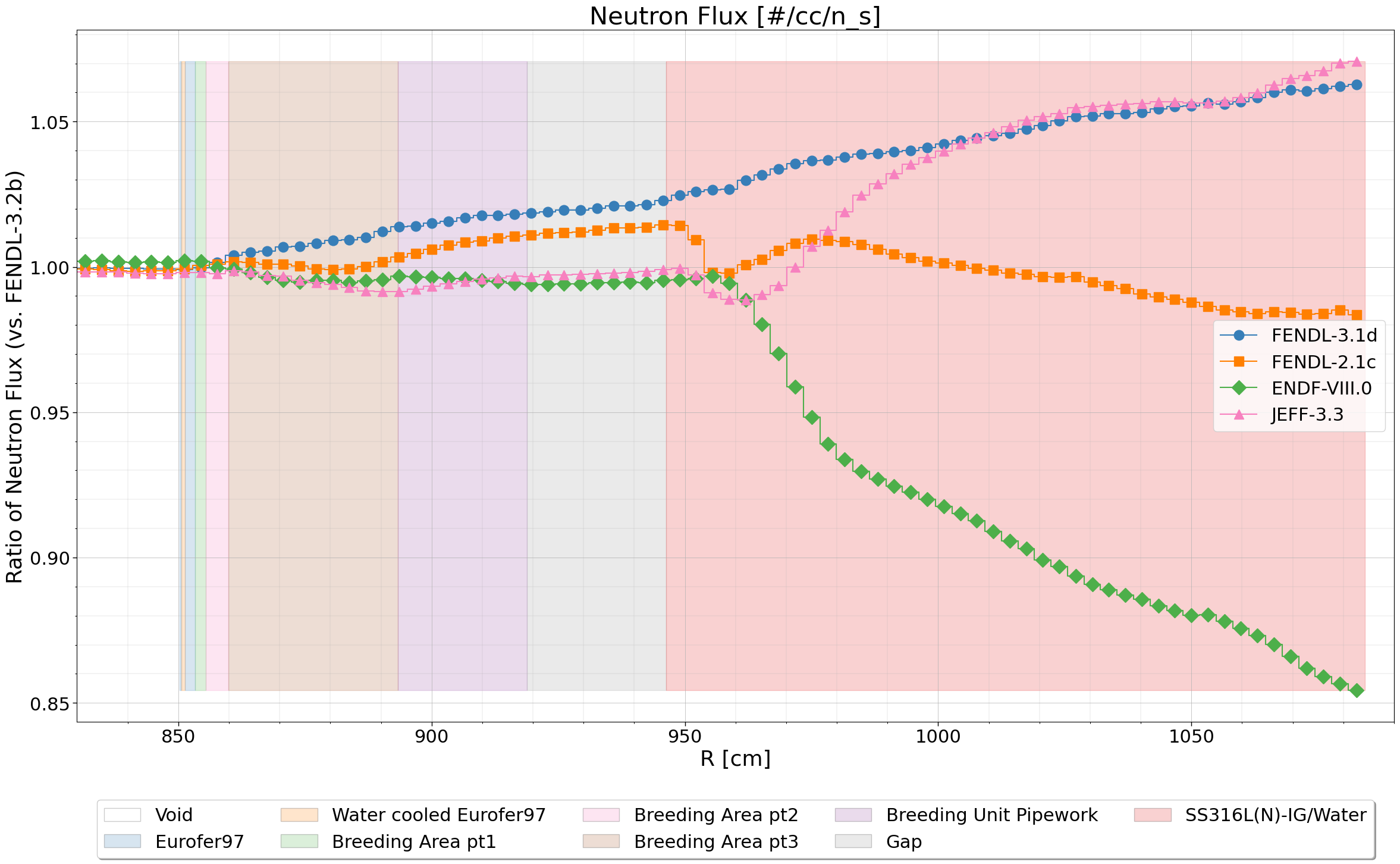}
    \caption{Ratio of neutron flux calculated with different neutron cross section data for the ITER 1-D HCPB TBM model.}
\label{fig:suminiFigure1hcpbnflux}
\end{figure}

\begin{figure}[htp]
\includegraphics[width=8.69cm]{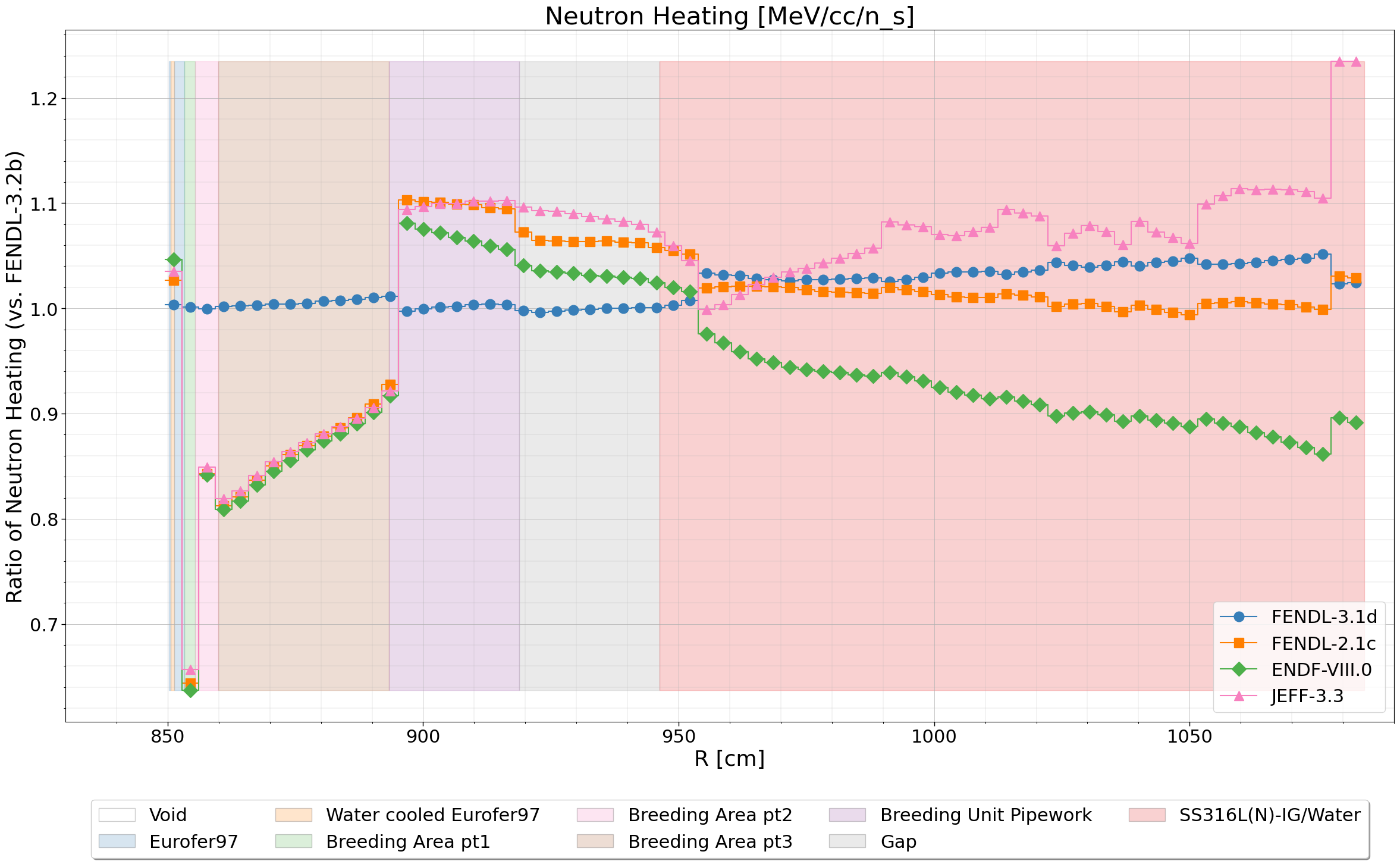}
    \caption{Ratio of neutron heating calculated with different neutron cross section data for the ITER 1-D HCPB TBM model.}
\label{fig:suminiFigure2hcpbnheat}
\end{figure}

\begin{figure}[htp]
\includegraphics[width=8.69cm]{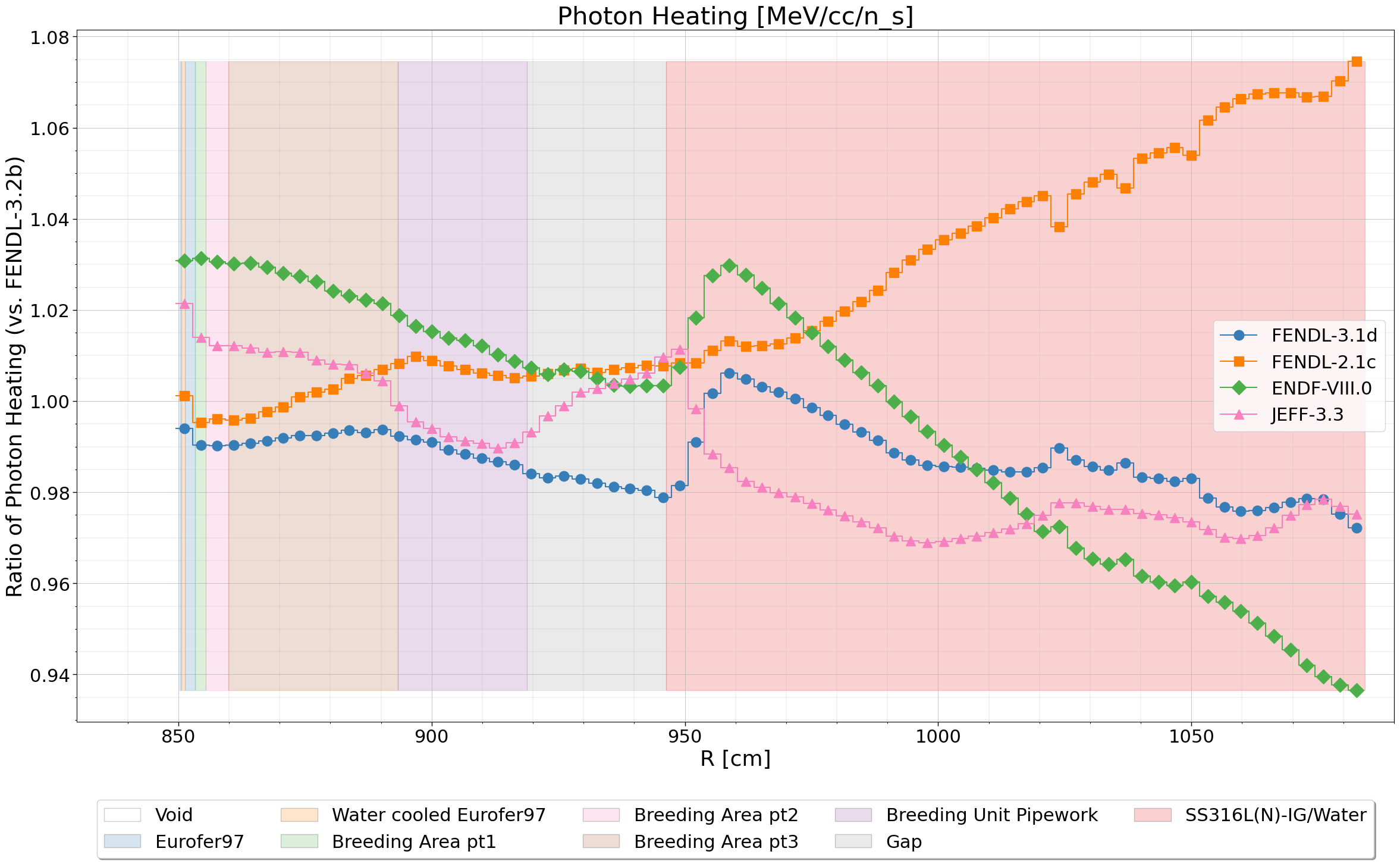}
    \caption{Ratio of photon heating calculated with different neutron cross section data for the ITER 1-D HCPB TBM model.}
\label{fig:suminiFigure3hcpbpheat}
\end{figure}

\begin{figure}[htp]
\includegraphics[width=8.69cm]{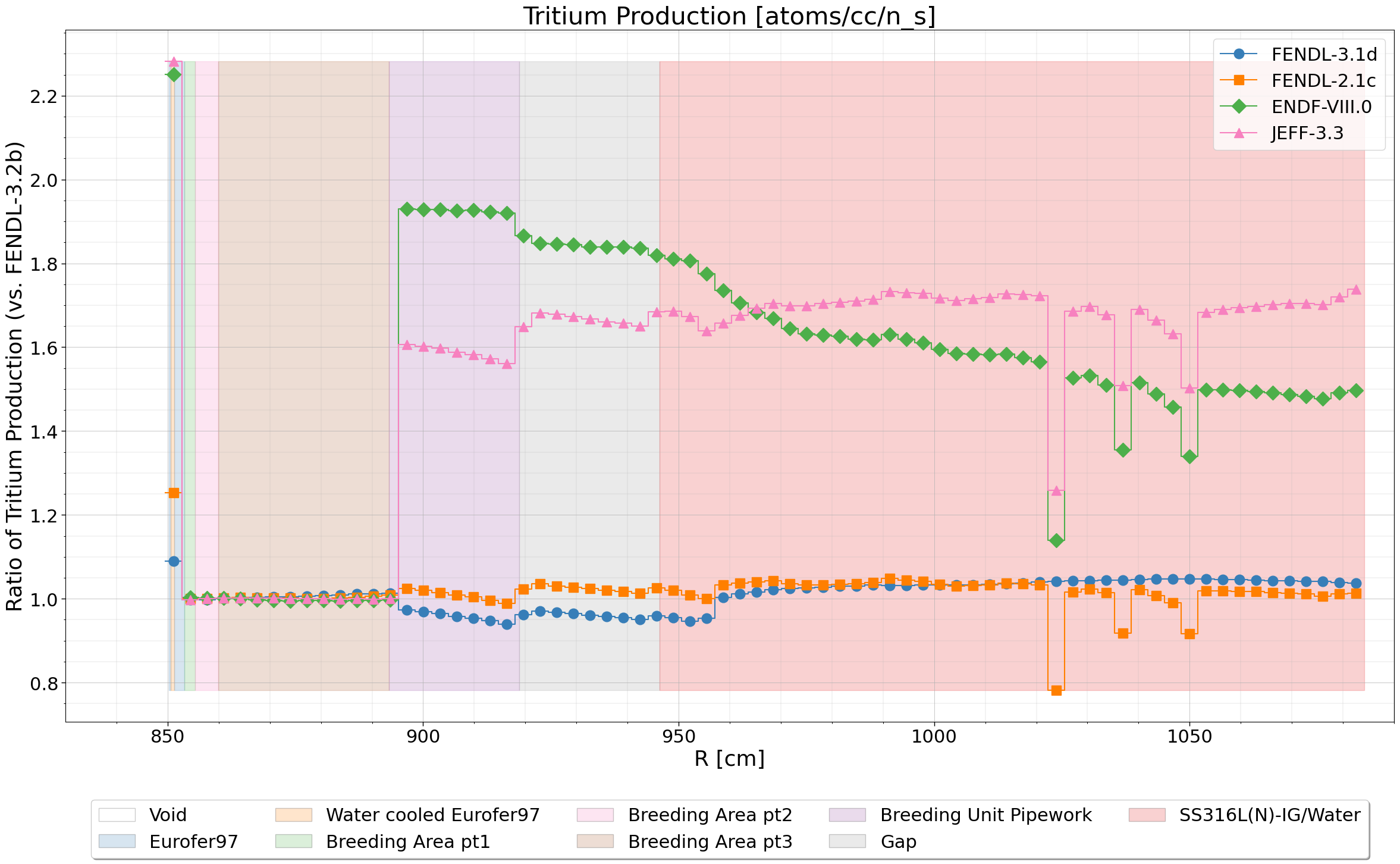}
    \caption{Ratio of tritium production calculated with different neutron cross section data for the ITER 1-D HCPB TBM model.}
\label{fig:suminiFigure4hcpbtprod}
\end{figure}

Neutron flux, neutron heating, photon heating, and tritium production ratios are shown in 
\cref{fig:suminiFigure1hcpbnflux,fig:suminiFigure2hcpbnheat,fig:suminiFigure3hcpbpheat,fig:suminiFigure4hcpbtprod} respectively for the HCPB TBM model.
The ratios are calculated versus that calculated with FENDL-3.2b.  
The results show good agreement in the comparison between FENDL 3.2b and the other cross section libraries apart from tritium production (MT=205).  As seen in \cref{fig:suminiFigure4hcpbtprod}, tritium production calculated with these different libraries is in good agreement in the breeding area, but over a factor of 2 different in cells outside the breeding sectors.

In order to identify the possible sources of these tritium productions differences, a map of the isotopes to be considered in the two reference structural materials, EUROFER97 \cite{sumini4eurofer97ukaea,sumini5eurofer97fusengdes} and INCONEL, has been analyzed.  Note that in these cells, very low tallied scores have been obtained, however all the MCNP statistical checks have been passed.  As has been shown analyzing the outputs of the sphere leakage benchmark, one possible source of these discrepancies could derive from the different cross section behavior with respect to the $^{10}$B isotope that must be considered in the EUROFER97 composition. 
A comparison of the two is shown in \cref{table:suminiEurofer97InconelcompPart1} and \cref{table:suminiEurofer97InconelcompPart2}, where also the specific isotope mass fractions (MF) implemented in the actual MCNP simulations in JADE have been considered.

\begin{table*}[htp]
\caption{Part 1: Compositions of the Reduced Activation Ferritic-Martensitic (RAFM) steel EUROFER97 and INCONEL in weight \%.
For EUROFER97, Fe is balanced and the reference composition bin is shown.
Alloying elements (AE) and Radiologically Undesired Elements (RUE), as defined for EUROFER97 (see ref.) are shown.
ALAP = As Low As Possible. N.A.= Not Available, N.AP.= Not Applicable, ‘*’= As+Sn+Sb+Zr = 0.5 as total max value.}
\label{table:suminiEurofer97InconelcompPart1}
\begin{tabular}{l|c|c|c|c|c|c|c|c|c|c|c} \hline \hline 
\multicolumn{3}{|c|}{ } & \multicolumn{7}{|c|}{EUROFER-97} & \multicolumn{2}{|c|}{Inconel} \\ \hline
Elem. & Z & A & Isot. MF (\%) & Elem. MF (\%) & AE & RUE & MIN (\%) & MAX (\%) & TARGET (\%) & ZAID MF(\%) & Isot. MF(\%) \\ \hline
B & 5 & 10 & 3.66E-04 & 2.00E-03 & X &  & ALAP & 2.00E-03 & N.A. &  & \\
  & 5 & 11 & 1.63E-03 &  &  &  &  &  &  &  & \\ \hline
C & 6 & 12 & 1.19E-01 & 1.19E-01 & X &  & 9.00E-02 & 1.20E-01 & 1.10E-01 & 5.02E-02 & 5.07E-02 \\ \hline 
 &  & 13 &  &  &  &  &  &  &  & 5.88E-04 & \\ \hline 
N & 7 & 14 & 4.48E-02 & 4.50E-02 & X &  & 1.50E-02 & 4.50E-02 & 3.00E-02 &  & \\ \hline 
 & 7 & 15 & 1.76E-04 &  &  &  &  &  &  &  & \\ \hline 
O & 8 & 16 & 9.99E-03 & 9.99E-03 & X &  & N.A. & 1.00E-02 & N.A. &  & \\ \hline 
Al & 13 & 27 & 1.00E-02 & 1.00E-02 &  & X & ALAP & 1.00E-02 & N.A. & 2.02E-01 & 2.02E-01 \\ \hline 
Si & 14 & 28 & 4.59E-02 & 5.00E-02 &  & X &  & 5.00E-02 & N.AP. & 2.32E-01 & 2.53E-01 \\ \hline 
 & 14 & 29 & 2.41E-03 &  &  &  &  &  &  & 1.22E-02 & \\ \hline 
 & 14 & 30 & 1.64E-03 &  &  &  &  &  &  & 8.34E-03 & \\ \hline 
P & 15 & 31 & 5.00E-03 & 5.00E-03 & X &  & N.A. & 5.00E-02 & N.A. &  & \\ \hline 
S & 16 & 32 & 4.73E-03 & 5.00E-03 & X &  & N.A. & 5.00E-02 & N.AP. & 7.73E-03 & 8.16E-03 \\ \hline 
 & 16 & 33 & 3.85E-05 &  &  &  &  &  &  & 6.29E-05 & \\ \hline 
 & 16 & 34 & 2.23E-04 &  &  &  &  &  &  & 3.64E-04 & \\ \hline 
 & 16 & 36 & 1.12E-06 &  &  &  &  &  &  & 1.83E-06 & \\ \hline 
Ti & 22 & 46 & 1.58E-03 & 2.00E-02 &  & X &  & 2.00E-02 & N.AP. & 1.61E-02 & 2.03E-01 \\ \hline 
 & 22 & 47 & 1.45E-03 &  &  &  &  &  &  & 1.48E-02 & \\ \hline 
 & 22 & 48 & 1.47E-02 &  &  &  &  &  &  & 1.50E-01 & \\ \hline 
 & 22 & 49 & 1.10E-03 &  &  &  &  &  &  & 1.12E-02 & \\ \hline 
 & 22 & 50 & 1.08E-03 &  &  &  &  &  &  & 1.09E-02 & \\ \hline 
V & 23 & 50 & 6.12E-04 & 2.50E-01 & X &  & 1.50E-01 & 2.50E-01 & N.A. &  & \\ \hline 
 & 23 & 51 & 2.49E-01 &  &  &  &  &  &  &  & \\ \hline 
Cr & 24 & 50 & 3.75E-01 & 9.00E+00 & X &  & 8.50E+00 & 9.50E+00 & 9.00E+00 & 8.99E-01 & 2.17E+01 \\ \hline 
 & 24 & 52 & 7.53E+00 &  &  &  &  &  &  & 1.82E+01 & \\ \hline 
 & 24 & 53 & 8.70E-01 &  &  &  &  &  &  & 2.12E+00 & \\ \hline 
 & 24 & 54 & 2.20E-01 &  &  &  &  &  &  & 5.42E-01 & \\ \hline 
Mn & 25 & 55 & 6.00E-01 & 6.00E-01 & X &  & 2.00E-01 & 6.00E-01 & 4.00E-01 & 2.53E-01 & 2.53E-01 \\ \hline
Fe & 26 & 54 & 5.00E+00 & 8.85E+01 & X &  & \multicolumn{3}{|c|}{Balance Value} & 1.42E-01 & 2.54E+00 \\ \hline
 & 26 & 56 & 8.14E+01 &  & &  & \multicolumn{3}{|c|}{Balance Value} & 2.33E+00 &  \\ \hline
 & 26 & 57 & 1.91E+00 &  & &  & \multicolumn{3}{|c|}{Balance Value} & 5.69E-02 &  \\ \hline
 & 26 & 58 & 2.59E-01 &  & &  & \multicolumn{3}{|c|}{Balance Value} & 7.90E-03 &  \\ \hline
Co & 27 & 59 & 1.00E-02 & 1.00E-02 &  & X & ALAP & 1.00E-02 & N.AP. &  & \\ \hline 
Ni & 28 & 58 & 6.71E-03 & 1.00E-02 &  & X & ALAP & 1.00E-02 & N.AP. & 4.14E+01 & 6.20E+01 \\ \hline 
 & 28 & 60 & 2.67E-03 &  &  &  &  &  &  & 1.65E+01 & \\ \hline 
 & 28 & 61 & 1.18E-04 &  &  &  &  &  &  & 8.36E-01 & \\ \hline 
 & 28 & 61 & 3.83E-04 &  &  &  &  &  &  & 2.35E+00 & \\ \hline 
 & 28 & 64 & 1.00E-04 &  &  &  &  &  &  & 8.10E-01 & \\ \hline 
Cu & 29 & 63 & 6.85E-03 & 1.00E-02 &  & X & ALAP & 1.00E-02 & N.AP. &  & \\ \hline 
 & 29 & 65 & 3.15E-03 &  &  &  &  &  &  &  & \\ \hline 
As & 33 & 75 & 1.25E-03 & 1.25E-03 &  & X &  & 5.00E-02* & N.AP. &  & \\ \hline 
Zr & 40 & 90 & 6.33E-04 & 1.25E-03 &  & X &  & * & N.AP. &  & \\ \hline 
 & 40 & 91 & 1.39E-04 &  &  &  &  &  &  &  & \\ \hline 
 & 40 & 92 & 2.15E-04 &  &  &  &  &  &  &  & \\ \hline 
 & 40 & 94 & 2.23E-04 &  &  &  &  &  &  &  & \\ \hline 
 & 40 & 96 & 3.67E-05 &  &  &  &  &  &  &  & \\ \hline 
Nb & 41 & 93 & 5.00E-03 & 5.00E-03 &  & X & ALAP & 5.00E-02 & N.AP. & 3.55E+00 & 3.55E+00 \\ \hline 
Mo & 42 & 92 & 7.10E-04 & 5.00E-03 &  & X & ALAP & 5.00E-02 & N.AP. & 1.29E+00 & 9.13E+00 \\ \hline 
 & 42 & 94 & 4.52E-04 &  &  &  &  &  &  & 8.27E-01 & \\ \hline 
 & 42 & 95 & 7.87E-04 &  &  &  &  &  &  & 1.43E+00 & \\ \hline 
 & 42 & 96 & 8.33E-04 &  &  &  &  &  &  & 1.52E+00 & \\ \hline 
 & 42 & 97 & 4.82E-04 &  &  &  &  &  &  & 8.81E-01 & \\ \hline 
 & 42 & 98 & 1.23E-03 &  &  &  &  &  &  & 2.24E+00 & \\ \hline 
 & 42 & 100 & 5.01E-04 &  &  &  &  &  &  & 9.16E-01 & \\ \hline 
\end{tabular}
\end{table*}

\begin{table*}[htp]
\caption{Part 2: Compositions of the Reduced Activation Ferritic-Martensitic (RAFM) steel EUROFER97 and INCONEL in weight \%.
For EUROFER97, Fe is balanced and the reference composition bin is shown.
Alloying elements (AE) and Radiologically Undesired Elements (RUE), as defined for EUROFER97 (see ref.) are shown.
ALAP = As Low As Possible. N.A.= Not Available, N.AP.= Not Applicable, ‘*’= As+Sn+Sb+Zr = 0.5 as total max value.}
\label{table:suminiEurofer97InconelcompPart2}
\begin{tabular}{l|c|c|c|c|c|c|c|c|c|c|c} \hline \hline 
\multicolumn{3}{|c|}{ } & \multicolumn{7}{|c|}{EUROFER-97} & \multicolumn{2}{|c|}{Inconel} \\ \hline
Elem. & Z & A & Isot. MF (\%) & Elem. MF (\%) & AE & RUE & MIN (\%) & MAX (\%) & TARGET (\%) & ZAID MF(\%) & Isot. MF(\%) \\ \hline
Sn & 50 & 112 & 1.14E-05 & 1.25E-03 &  & X &  & * & N.AP. &  & \\ \hline 
 & 50 & 114 & 7.91E-06 &  &  &  &  &  &  &  & \\ \hline 
 & 50 & 115 & 4.11E-06 &  &  &  &  &  &  &  & \\ \hline 
 & 50 & 116 & 1.77E-04 &  &  &  &  &  &  &  & \\ \hline 
 & 50 & 117 & 9.45E-05 &  &  &  &  &  &  &  & \\ \hline 
 & 50 & 118 & 3.00E-04 &  &  &  &  &  &  &  & \\ \hline 
 & 50 & 119 & 1.07E-04 &  &  &  &  &  &  &  & \\ \hline 
 & 50 & 120 & 4.11E-04 &  &  &  &  &  &  &  & \\ \hline 
 & 50 & 122 & 5.94E-05 &  &  &  &  &  &  &  & \\ \hline 
 & 50 & 124 & 7.55E-05 &  &  &  &  &  &  &  & \\ \hline 
Sb & 51 & 121 & 7.10E-04 & 1.25E-03 &  & X &  & * & N.AP. &  & \\ \hline 
 & 51 & 123 & 5.39E-04 &  &  &  &  &  &  &  & \\ \hline 
Ta & 73 & 181 & 1.40E-01 & 1.40E-01 & X &  & 1.00E-01 & 1.40E-01 & 1.20E-01 &  & \\ \hline 
W & 74 & 182 & 2.88E-01 & 1.10E+00 & X & X & 1.00E+00 & 1.20E+00 & 1.10E+00 &  &  \\ \hline 
 & 74 & 183 & 1.56E-01 &  &  &  &  &  &  &  &  \\ \hline 
 & 74 & 184 & 3.38E-01 &  &  &  &  &  &  &  &  \\ \hline 
 & 74 & 186 & 3.16E-01 &  &  &  &  &  &  &  &  \\ \hline 
\end{tabular}
\end{table*}

\begin{figure}[htp]
\includegraphics[width=8.69cm]{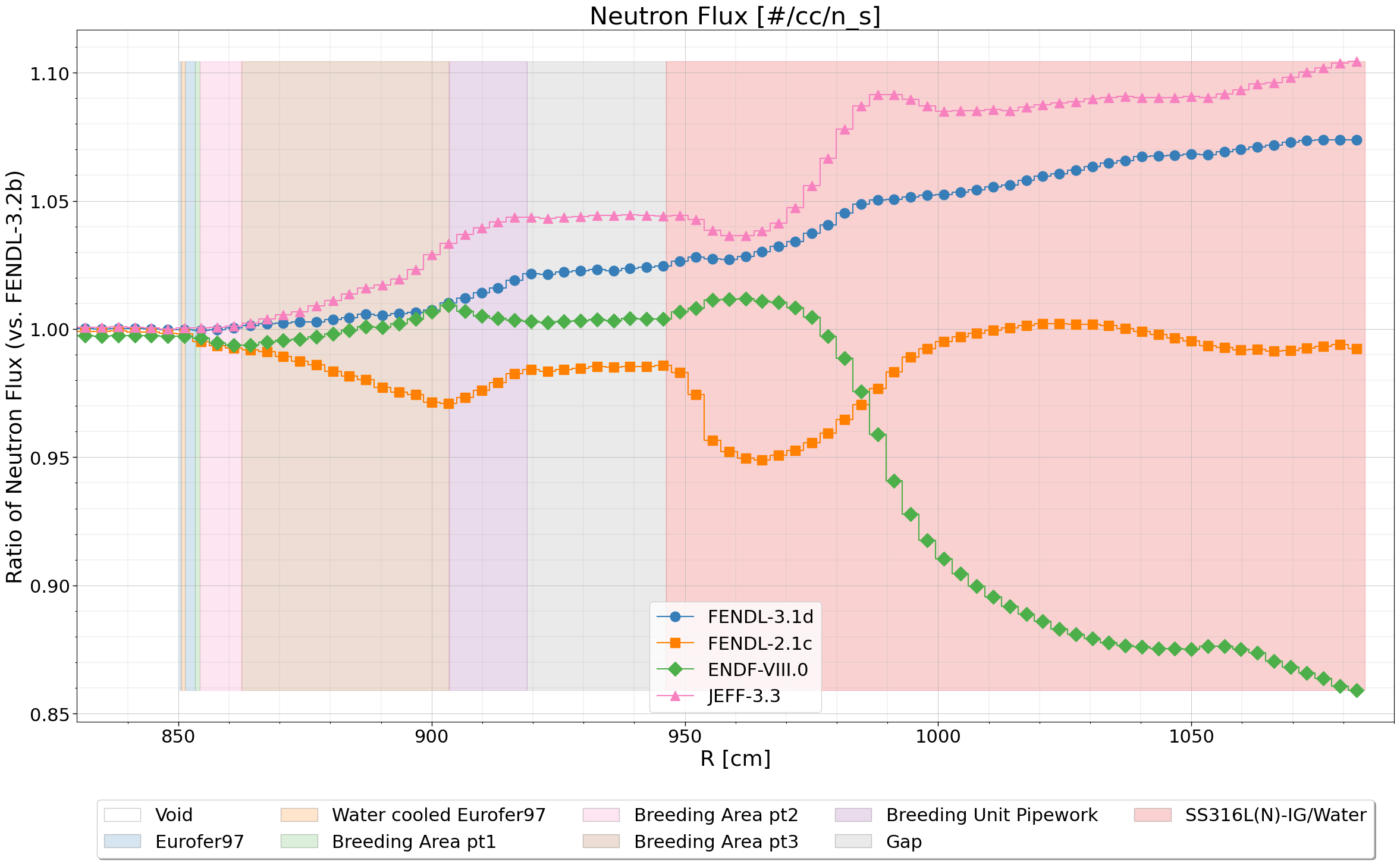}
    \caption{Ratio of neutron flux calculated with different neutron cross section data for the ITER 1-D WCLL TBM model.}
\label{fig:suminiFigure5wcllnflux}
\end{figure}

\begin{figure}[htp]
\includegraphics[width=8.69cm]{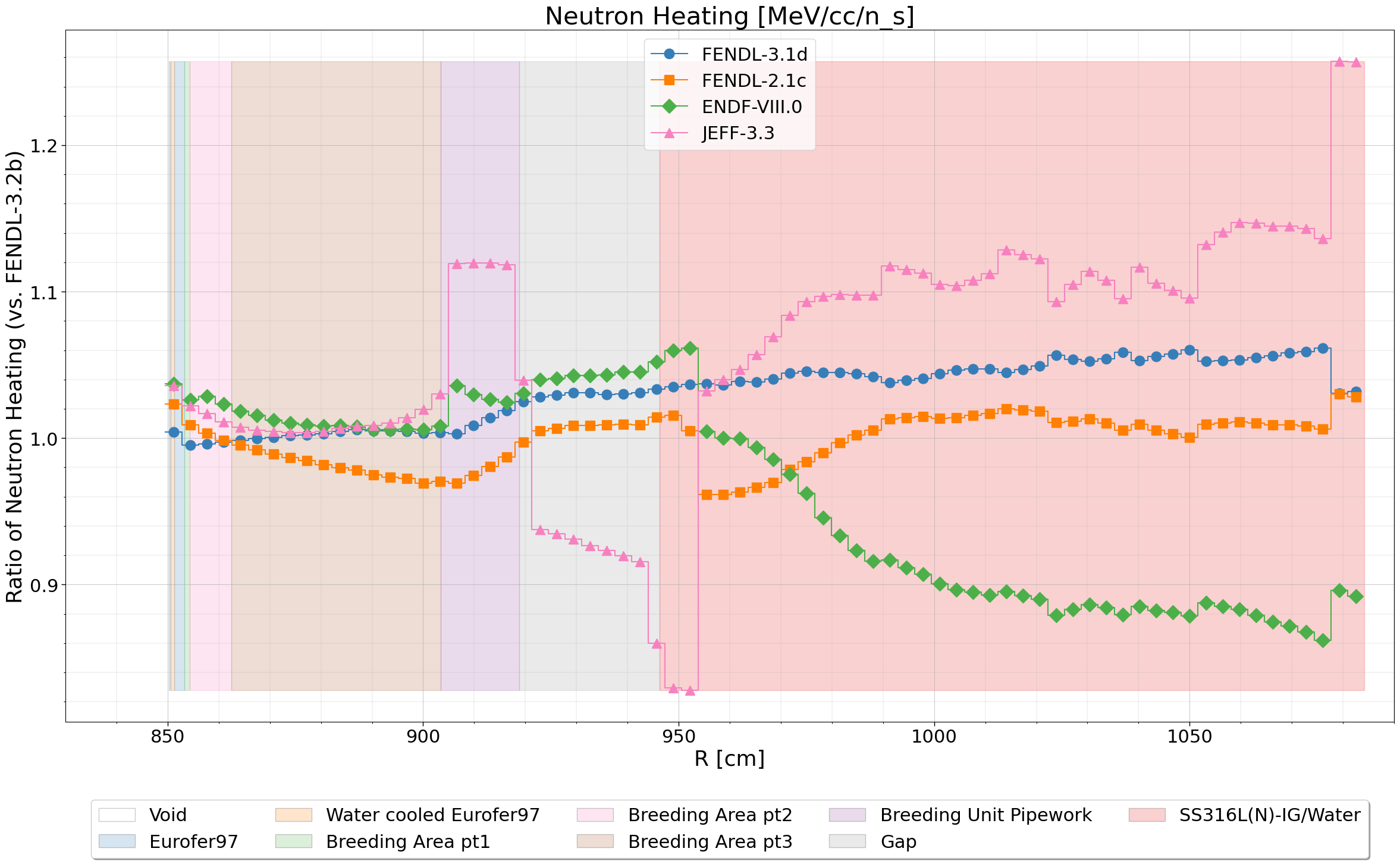}
    \caption{Ratio of neutron heating calculated with different neutron cross section data for the ITER 1-D WCLL TBM model.}
\label{fig:suminiFigure6wcllnheat}
\end{figure}

\begin{figure}[htp]
\includegraphics[width=8.69cm]{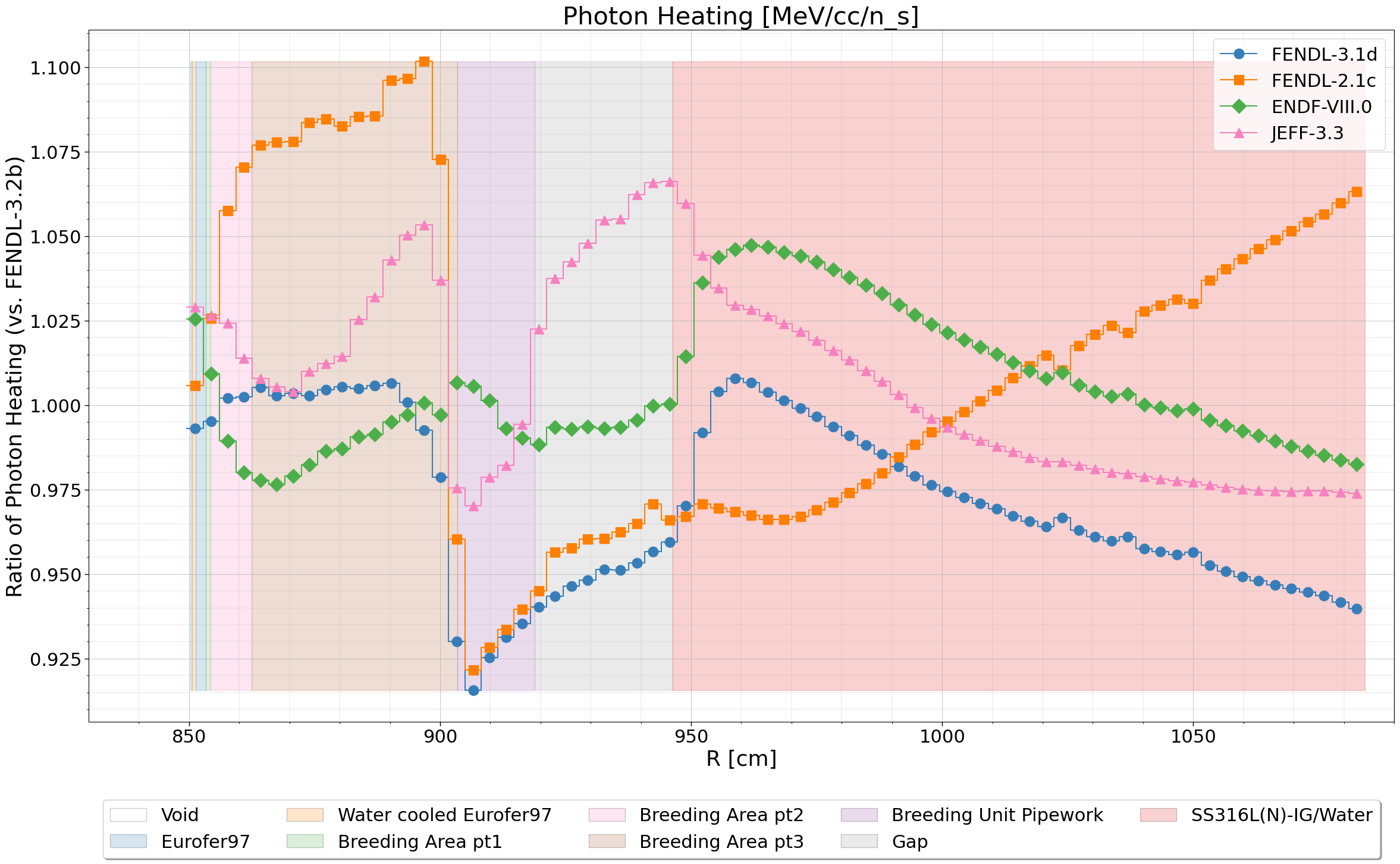}
    \caption{Ratio of photon heating calculated with different neutron cross section data for the ITER 1-D WCLL TBM model.}
\label{fig:suminiFigure7wcllpheat}
\end{figure}

\begin{figure}[htp]
\includegraphics[width=8.69cm]{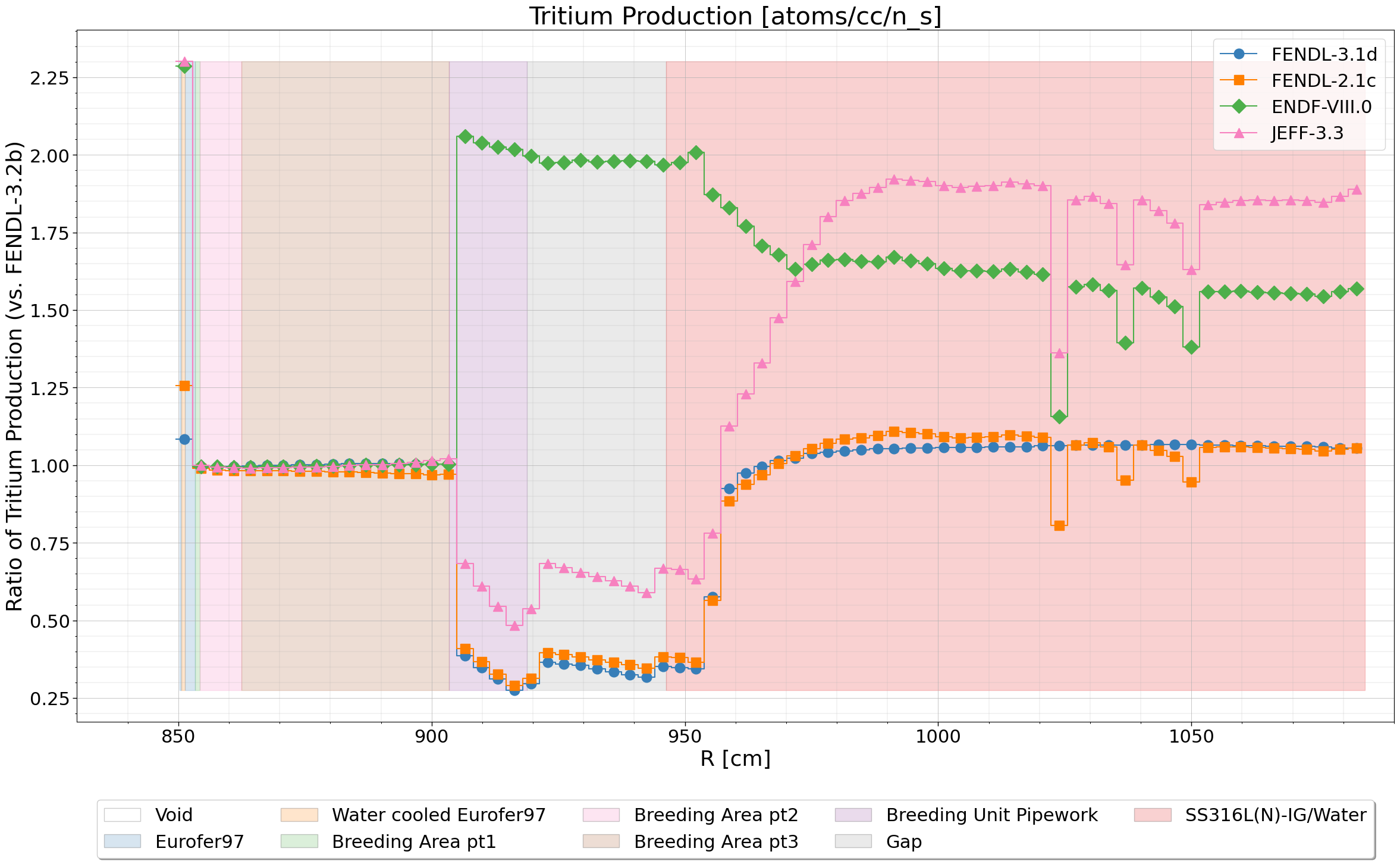}
    \caption{Ratio of tritium production calculated with different neutron cross section data for the ITER 1-D WCLL TBM model.}
\label{fig:suminiFigure8wclltprod}
\end{figure}

Neutron flux, neutron heating, photon heating, and tritium production ratios are shown in \cref{fig:suminiFigure5wcllnflux,fig:suminiFigure6wcllnheat,fig:suminiFigure7wcllpheat,fig:suminiFigure8wclltprod} respectively for the WCLL TBM model.  The ratios are calculated versus that calculated with FENDL-3.2b. The results are similar to that observed with the HCPB TBM model, i.e. good agreement in the comparison between FENDL 3.2b and the other cross section libraries apart from tritium production (MT=205).  With respect to tritium breeding, one can observe a spread that is slightly larger than the one in the HCPB in the breeding cells. This effect could be explained taking into account the more relevant fraction of materials due to the water cooling system and also considering that the WCLL, by design, should produce a smaller quantity of tritium (approximately 2/3) \cite{sumini1iterhcpbtbm}.

\subsubsection{EU DEMO-3D divertor}
\label{subsubsec:eudemo3dStankunas}
\begin{figure}[htp]
\includegraphics[width=8.69cm]{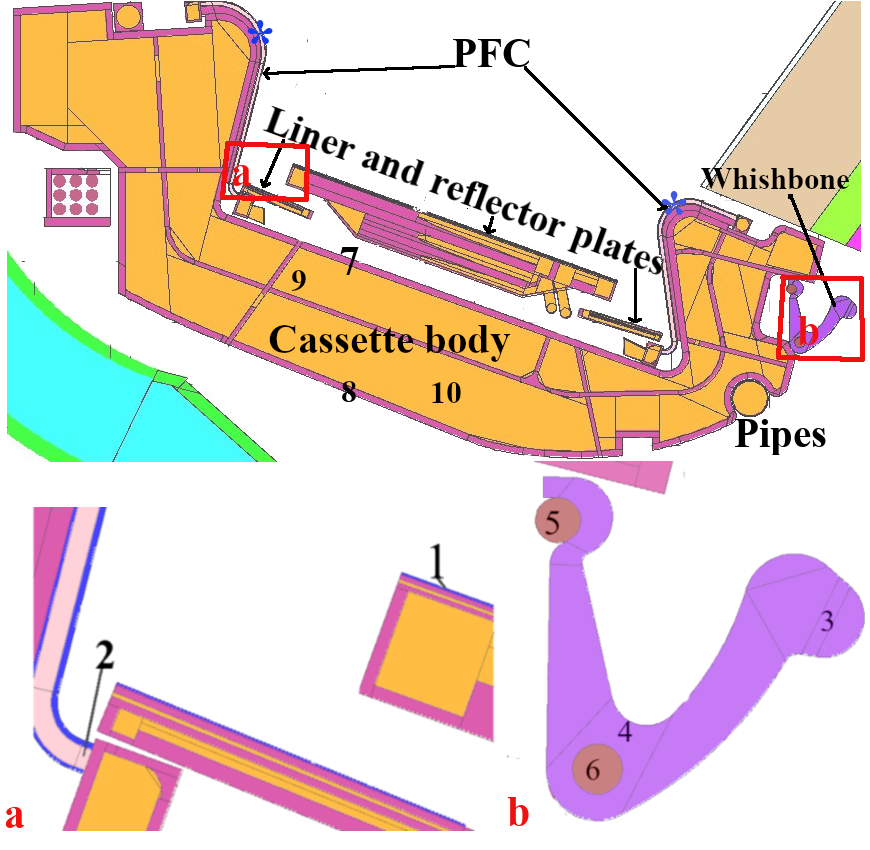}
    \caption{EU DEMO divertor model rendered with the MCNP geometry plotter.}
\label{fig:stankunasFig1divertor2D}
\end{figure}

\begin{figure}[htp]
\includegraphics[width=8.69cm]{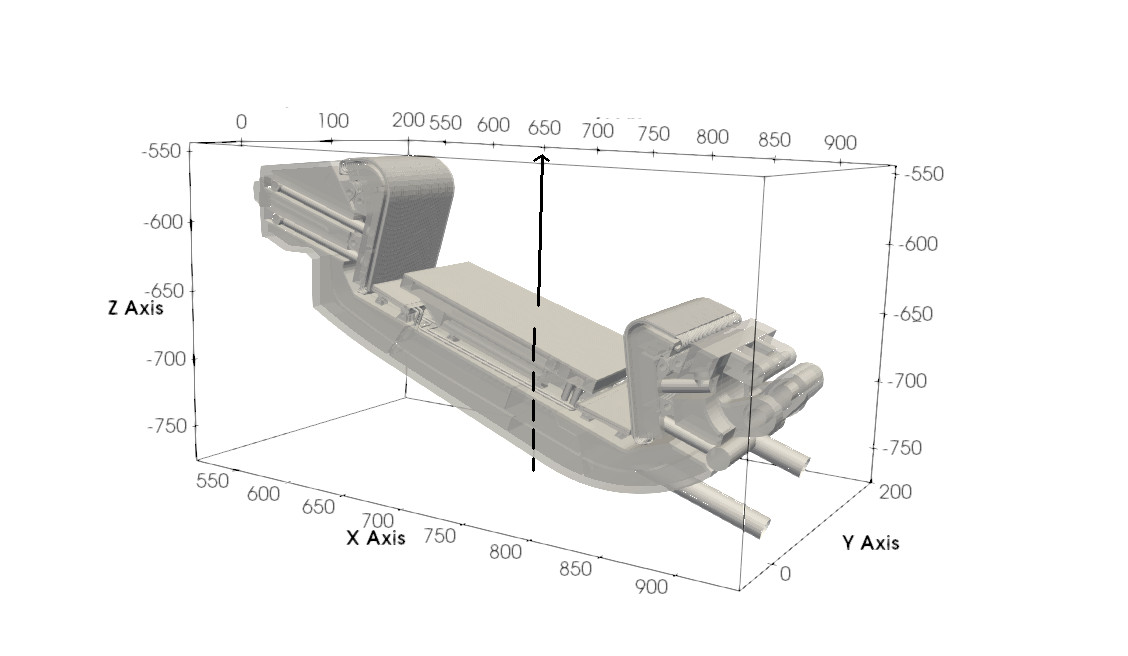}
    \caption{EU DEMO divertor model rendered in 3-D.  The line shown is the location where some results are shown.}
\label{fig:stankunasFig2divertor3D}
\end{figure}

For this 3-D calculational benchmark, an 11.25 degree toroidal sector of the EU DEMO 1 2017 model with a homogeneous Helium Cooled Pebble Bed (HCPB) \cite{stankunas1eudemohcpb2017} breeding blanket (BB) structure was used. For the divertor, the 2019 full heterogeneous configuration model~\cite{stankunas2eudemodivertor} has been used.  \Cref{fig:stankunasFig1divertor2D} shows a view of the divertor as rendered with the MCNP plotter and \cref{fig:stankunasFig2divertor3D} shows a 3-D view of the divertor model.

The MCNP6.2 code \cite{mcnp62} was used for neutron transport calculations and ADVANTG with FW-CADIS (Forward-Weighted Consistent Adjoint Driven Importance Sampling \cite{stankunas4advantg}) was used to generate weight windows in order to obtain accurate particle flux results and reduce statistical uncertainty.  The FENDL-2.1 and FENDL-3.2b cross section libraries were used for comparisons.  Nominally, neutron energy spectra were determined with 709 energy groups from $1.05\times10^{-11}$~ MeV to $10^{3}$~MeV \cite{stankunas5fispact}.  Results were normalized to a source strength of $7.09\times10^{20} n/s$.  Both mesh tallies and cell based tallies were used to determine flux distributions in the divertor region.

\begin{figure}[htp]
\includegraphics[width=8.69cm]{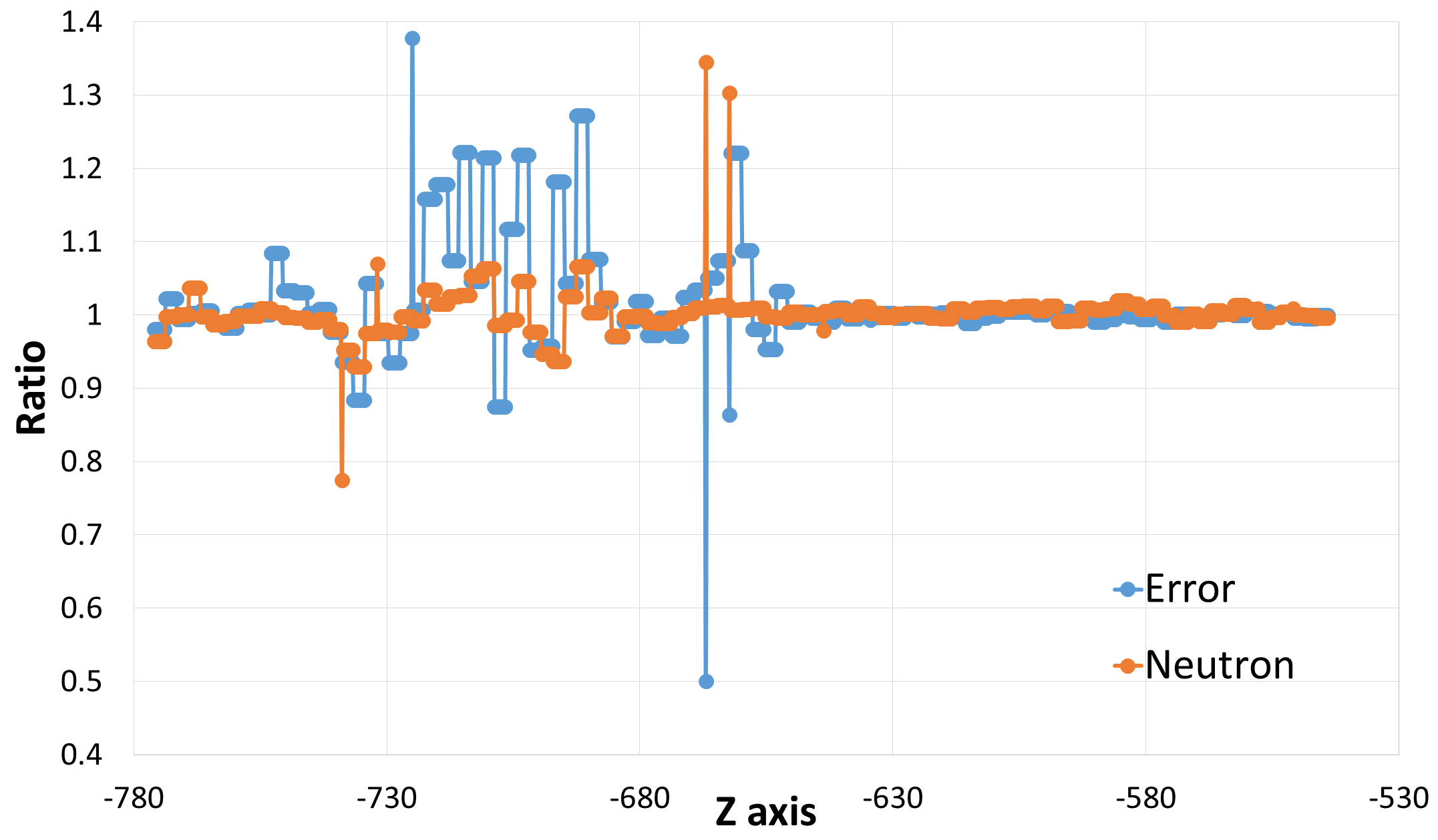}
    \caption{Ratio of neutron flux and ratio of statistical error (calculated with FENDL-3.2b to FENDL-2.1) along the line EU DEMO divertor mesh tally.}
\label{fig:stankunasFig6fluxratio}
\end{figure}

For the mesh tally in the divertor region, the flux calculated with both FENDL-2.1 and FENDL-3.2b revealed a marginal difference.  \Cref{fig:stankunasFig6fluxratio} shows a plot of the ratio of the neutron flux calculated with FENDL-3.2b to that calculated with FENDL-2.1 along the line shown in \cref{fig:stankunasFig2divertor3D}.   The flux ratio ranged from 0.774 to 1.3453.  Also shown in the plot is the ratio of the statistical error which ranged between 0.5003 and 1.3775.  The average ratio was 1.00193 for neutron flux and 1.0181 for the ratio of statistical error.

For the cell based tallies, neutron flux results from the pipes, plates, body, wishbone, and plasma facing component (PFC) parts of the divertor (as seen in \cref{fig:stankunasFig1divertor2D}) are discussed in the following paragraphs.  The neutron energy range from $10^{-8}$~MeV to $14.8$~MeV was analyzed since average statistical errors outside these energies exceeded 10\% in the neutron energy group structure for these tallies.

\begin{figure}[htp]
\includegraphics[width=8.69cm]{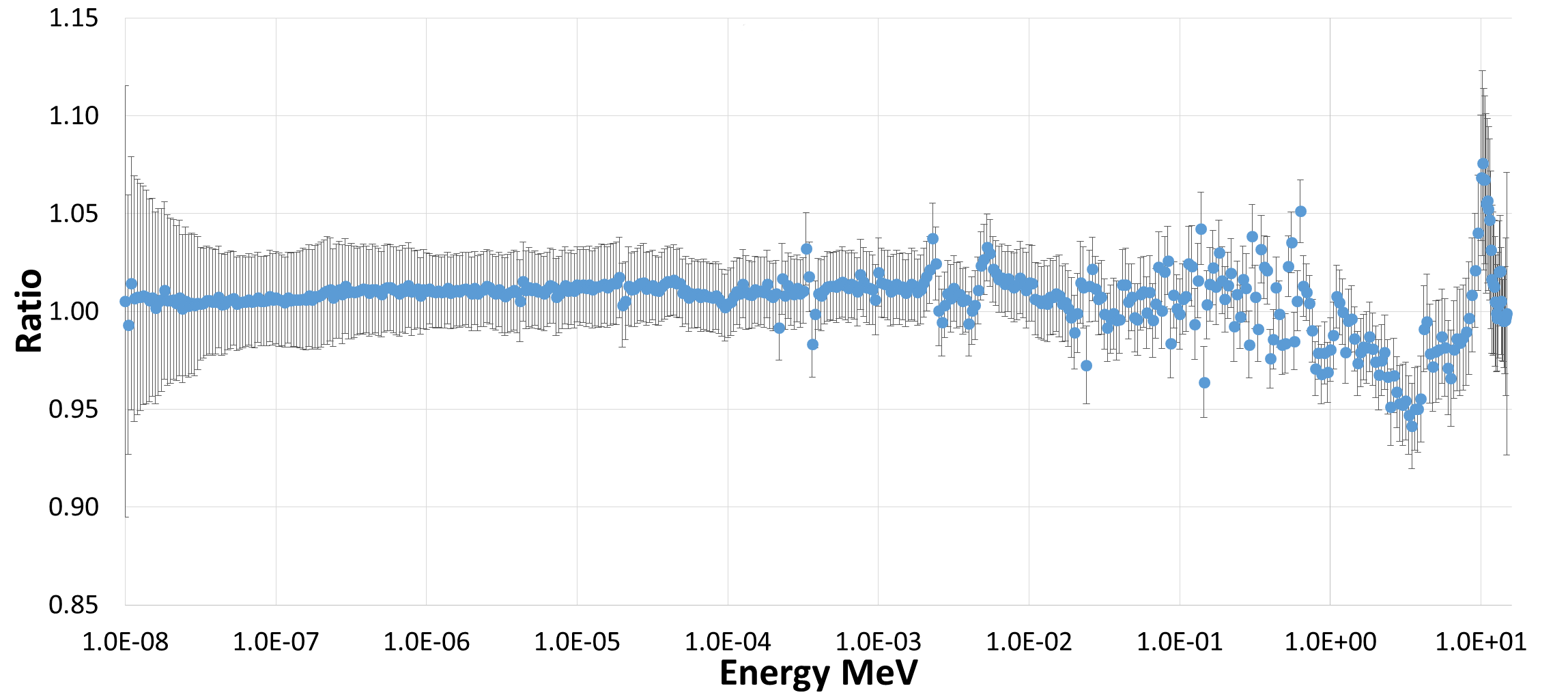}
    \caption{Ratio of neutron flux calculated with FENDL-3.2b to that calculated with FENDL-2.1 in the divertor pipe.}
\label{fig:stankunasFig8fluxratioPipe}
\end{figure}

The pipes are made of EUROFER97 and used to transport coolant. The total volume of one divertor cassette pipes is $1.71\times10^{5} cm^{3}$.  \Cref{fig:stankunasFig8fluxratioPipe} shows the ratio of neutron flux calculated with FENDL-3.2b versus that calculated with FENDL-2.1 in the energy range analyzed.  Calculations results showed that in the energy region of $10^{-8}$ – $10^{-2}$~MeV, the average flux ratio is 1.01.  In the higher energy regions the average ratio decreased to 1.0003, but with more variability in ratios above and below 1. The standard deviation of results showed an average of 0.024, while in the lower energy region it was 0.0054.

\begin{figure}[htp]
\includegraphics[width=8.69cm]{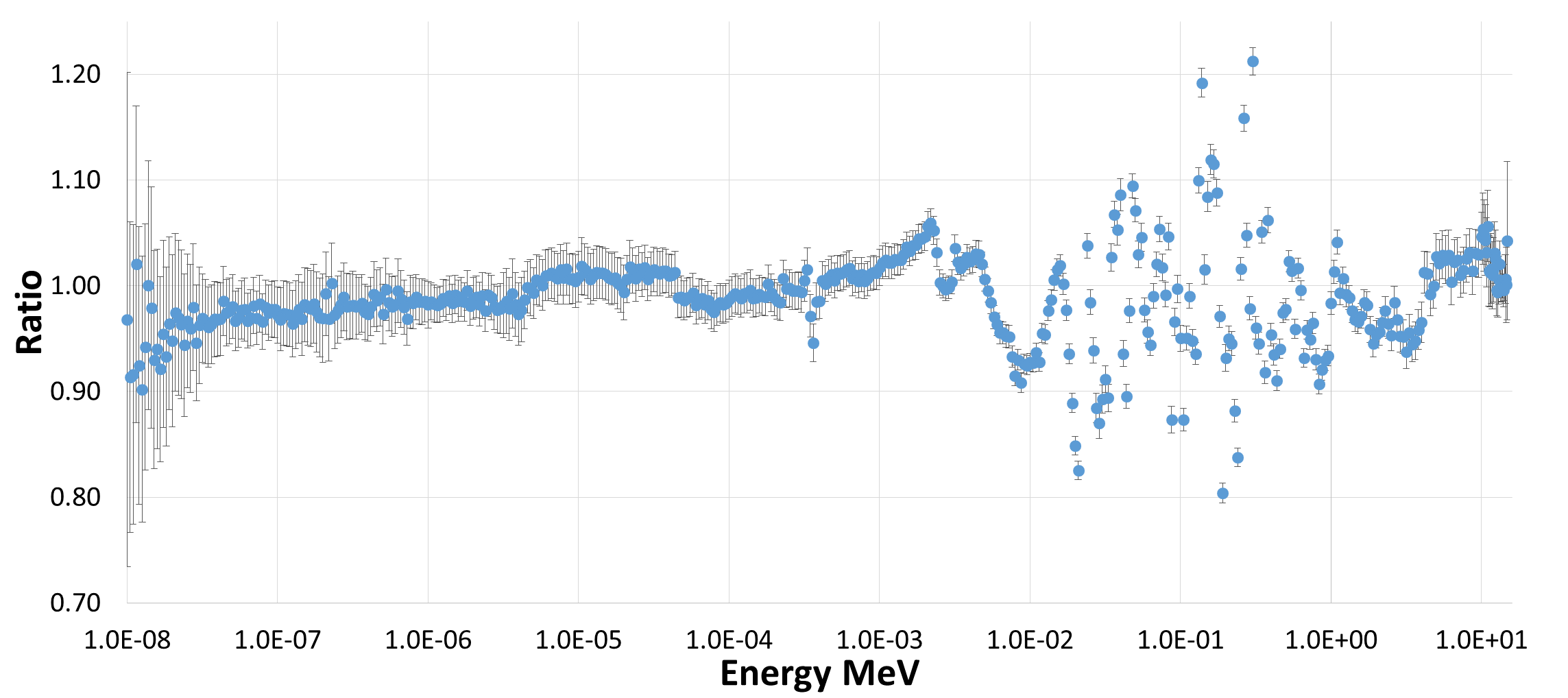}
    \caption{Ratio of neutron flux calculated with FENDL-3.2b to that calculated with FENDL-2.1 in the divertor wishbone.}
\label{fig:stankunasFig13fluxratioWishbone}
\end{figure}

\begin{figure}[htp]
\includegraphics[width=8.69cm]{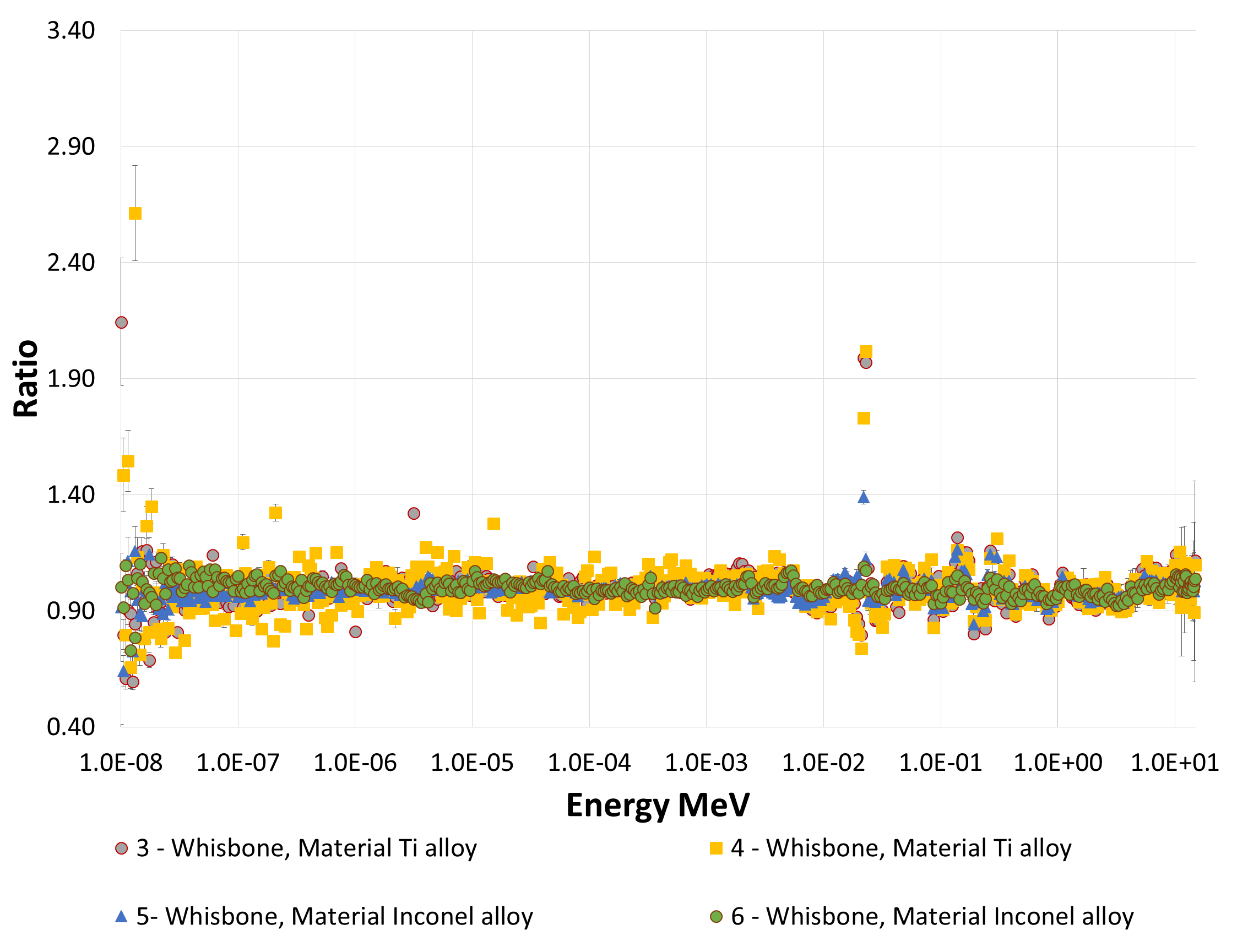}
    \caption{Ratio of neutron flux calculated with FENDL-3.2b to that calculated with FENDL-2.1 in the divertor wishbone cells (For cell location see \cref{fig:stankunasFig1divertor2D}).}
\label{fig:stankunasFig14fluxratioWishboneCells}
\end{figure}

The wishbone is made of Ti and Inconel alloy material and its main purpose is to fix the cassette body via pin connections to the Vacuum Vessel (VV) \cite{stankunas7eudemodivertorcad}. The total volume of one divertor cassette wishbone is $1.59 \times 10^{4} cm^{3}$. \Cref{fig:stankunasFig13fluxratioWishbone} shows the ratio of neutron flux calculated with FENDL-3.2b versus that calculated with FENDL-2.1 in the energy range analyzed.  The neutron flux ratio ranges from 0.804 to 1.21 and the average ratio value is 0.99.  The standard deviation in the lower energy region is 2 times lower than in the higher energy region (for $10^{-8}$ – $10^{-2}$~MeV it is 0.0254, while for $10^{-2}$ – $14.8$~MeV it is 0.0603).  Looking in more detail at the cells making up the wishbone (see \cref{fig:stankunasFig1divertor2D}), the maximum ratio values observed are rather high: cell 3 - 2.1444; cell 4 – 2.613; cell 5 – 1.39 and cell 6 - 1.128 as seen in \cref{fig:stankunasFig14fluxratioWishboneCells}.  Note however many of the neutron flux ratio values are much closer to 1 and recall the average flux ratio value was 0.99.  

\Cref{table:stankunasEUDEMOaveDivertorFlux} shows the average ratio of neutron flux calculated with FENDL-3.2b versus that calculated with FENDL-2.1 in the energy range analyzed for each of the components examined in the divertor.  As seen from the table, the average ratios are very close to 1 with a maximum difference of 1\% seen in the PFC. 

\begin{table}[htp]
\caption{Average ratio of neutron flux calculated with FENDL-3.2b to that calculated with FENDL-2.1 in different parts of the divertor. For part location see \cref{fig:stankunasFig1divertor2D}}
\label{table:stankunasEUDEMOaveDivertorFlux}
\begin{tabular}{l | c | c } \hline \hline
Component  & Averaged Ratio & Error (\%)  \\ \hline
Pipes      & 1.0064 & 2.442 \\
Wishbone   & 0.9923 & 2.943 \\
Plates     & 1.0062 & 2.434 \\
Body       & 1.0065 & 3.434 \\
PFC        & 1.0103 & 1.25  \\ \hline \hline
\end{tabular}
\end{table}

\subsection{Experimental Benchmarks}
\label{subsec:experimentalbenchmarks}
\subsubsection{Oktavian}
\label{subsubsec:oktavianLaghi}
OKTAVIAN is an experimental facility located at the Osaka University
which has been operative since 1981. It consists of an intense
deuterium-tritium (D-T) fusion neutron source (up to $3 \times 10^{12}$~n/s) that
has been used for many experiments on high energy
neutron transport. Among them, many Time Of Flight (TOF) experiments
were conducted and their results have been introduced in SINBAD
\cite{laghi1sinbad}. These experiments consist of placing the neutron source
inside a sphere composed only by a specific material of interest and measuring the leakage neutron and photon spectra exiting from such sphere with
the use of detectors. The particle energy measurement is performed
indirectly measuring the time of flight, which is then converted into
a velocity. Additional details on the experiments and the benchmark
definition can be found in \cite{laghi2oktavian}. Both MCNP inputs and experimental data for OKTAVIAN were extracted from CoNDERC \cite{laghi3conderc} and slightly modified in
order to be included in JADE’s benchmark suite
\cite{fabbri1jadedocs,fabbri2jadegithub}. The 11 studied materials 
are aluminium, cobalt, chromium, copper, lithium fluoride, manganese,
molybdenum, silicon, titanium, tungsten and zirconium. For the
zirconium benchmark, no experimental data were available for the photon
leakage current, hence, only the neutron benchmark was computed. Neutron
leakage current is tallied in 134 energy bins, while photons are
tallied in 57 energy bins. The nuclear data libraries considered for the benchmark
were FENDL-3.2b, FENDL-3.1d, FENDL-2.1, ENDF-VIII.0 and JEFF-3.3.

\begin{figure}[htp]
\includegraphics[width=8.69cm]{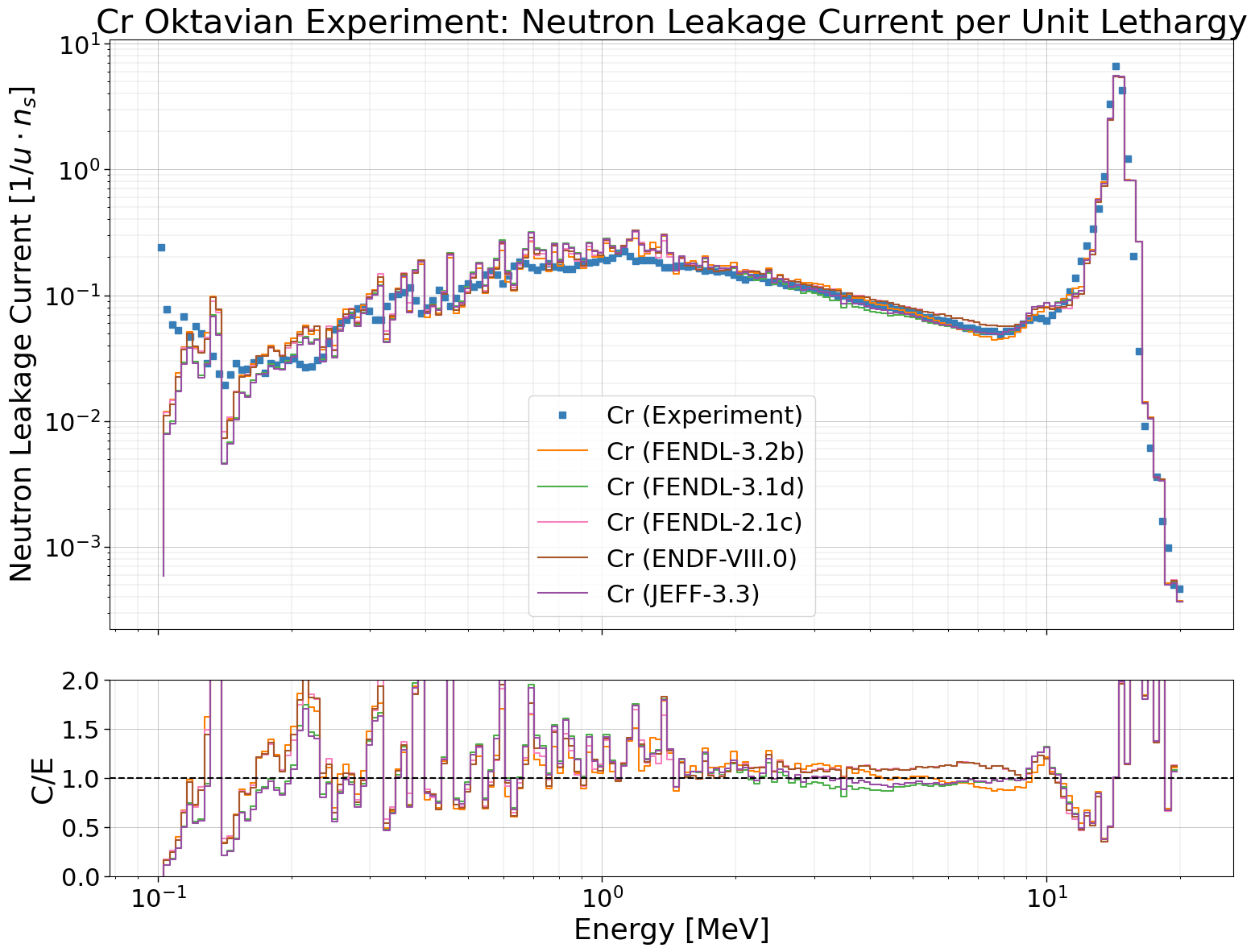}
    \caption{Neutron leakage current in the Cr Oktavian experimental benchmark.}
\label{fig:figure3CrNeutron}
\end{figure}

\begin{figure}[htp]
\includegraphics[width=8.69cm]{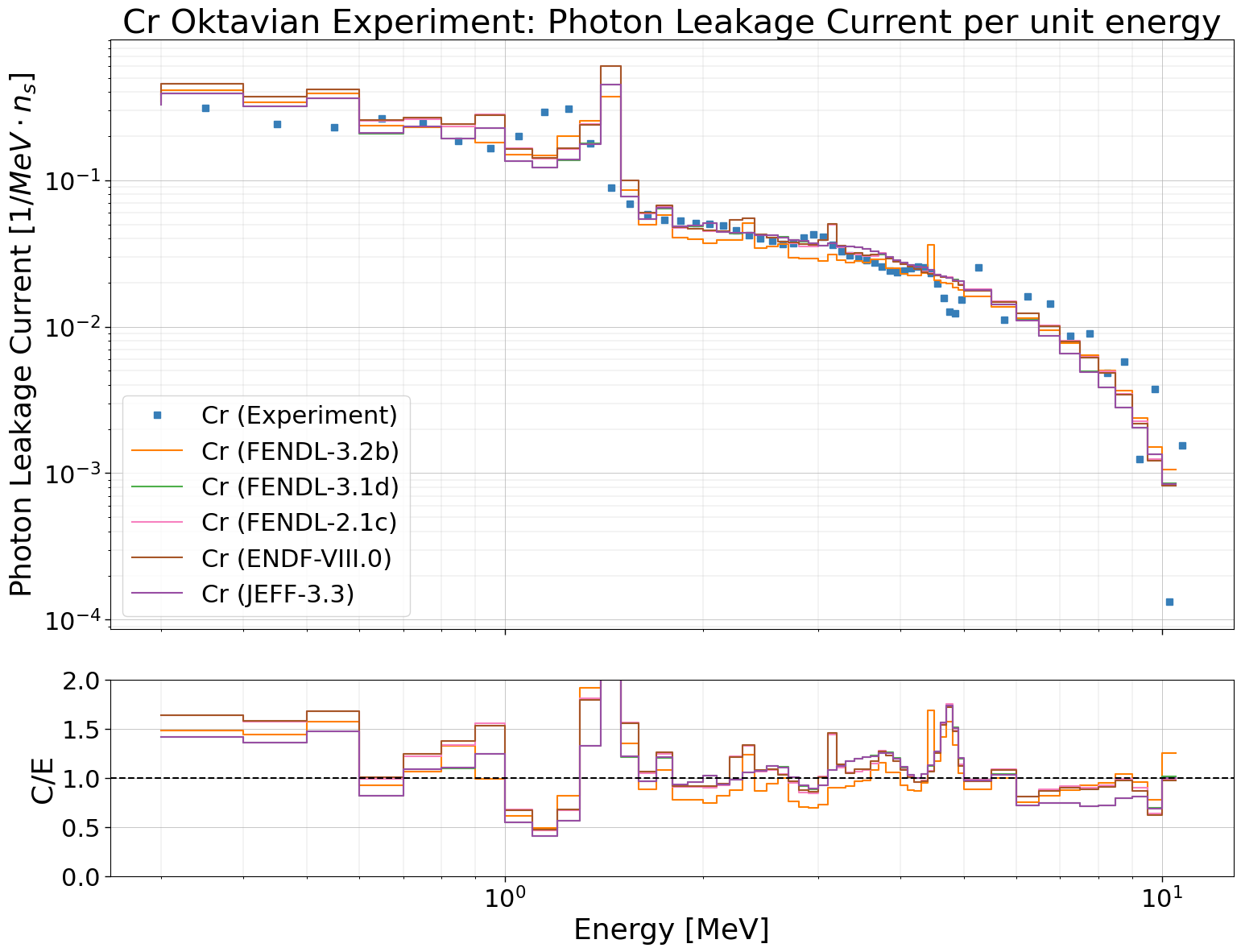}
    \caption{Photon leakage current in the Cr Oktavian experimental benchmark.}
\label{fig:figure3CrPhoton}
\end{figure}

JADE produces extensive post-processed outputs for all the benchmarks
that it includes. The complete set of outputs, which include raw data,
different C/E tables and plots can be found in a file share~\cite{laghi4figshareOktavian}.  \Cref{fig:figure3CrNeutron,fig:figure3CrPhoton} show sample plots of the neutron leakage and photon leakage for the Cr experiment.  For the sake of brevity, only the major results are discussed in this paper.   

A summary of computational to experimental (C/E) results averaged over all 11 materials examined is shown in \cref{table:laghiTable1summaryAll}.  This table is built by collapsing the leakage current results into 5 coarse energy bins and computing averages of C/E values. The C/E values averaged over all libraries are in column 3 and averaged for FENDL-3.2b only in column 4. 
In general, \cref{table:laghiTable1summaryAll} shows that the simulations are good at predicting the experimental results up to 10 MeV for neutrons and up to 5 MeV for photons with a mean C/E in the range of 0.95 to 1.05. At higher energies, i.e. larger than 10\,MeV for neutrons and larger than 5\,MeV for photons, the agreement is not as good with a mean C/E of almost 5 for neutrons and 0.6 for photons.  Mean C/E values calculated with FENDL-3.2b are similar to those calculated with the other libraries investigated as seen in column 4 of the table. 

\begin{table}[htp]
\caption{Summary of the mean C/E values for the leakage current in the Oktavian benchmark.} 
\label{table:laghiTable1summaryAll}
\begin{tabular}{l | c | c c} \hline \hline
Particle  &  (MeV) & mean C/E & mean C/E  \\ \hline
          &              & All Materials      & All Materials        \\
          &              & All Libs.          & FENDL-3.2b \\ \hline
Neutron   &  0.1 - 1.0   & 1.050870 & 1.058053 \\ 
          &  1.0 - 5.0   & 0.987454 & 0.984611 \\
          &  5.0 - 10.0  & 0.957024 & 0.947358 \\
          & 10.0 - 20.0  & 4.859306 & 4.854748 \\ \hline
Photon    &  0.1 - 1.0   & 1.009546 & 1.008698 \\
          &  1.0 – 5.0   & 0.990999 & 0.956824 \\
          &  5.0 – 10.0  & 0.763274 & 0.762119 \\
          & 10.0 – 20.0  & 0.582553 & 0.601473 \\ \hline \hline
\end{tabular}
\end{table}

\Cref{table:laghiTable2summaryByMatl} shows the mean C/E for neutron leakage in each of the 11 materials examined for the same coarse 5 energy bins discussed earlier.  The mean C/E values are shown averaged over all libraries examined in column 3, and averaged for FENDL-3.2b only in column 4.

\begin{figure}[htp]
\includegraphics[width=8.69cm]{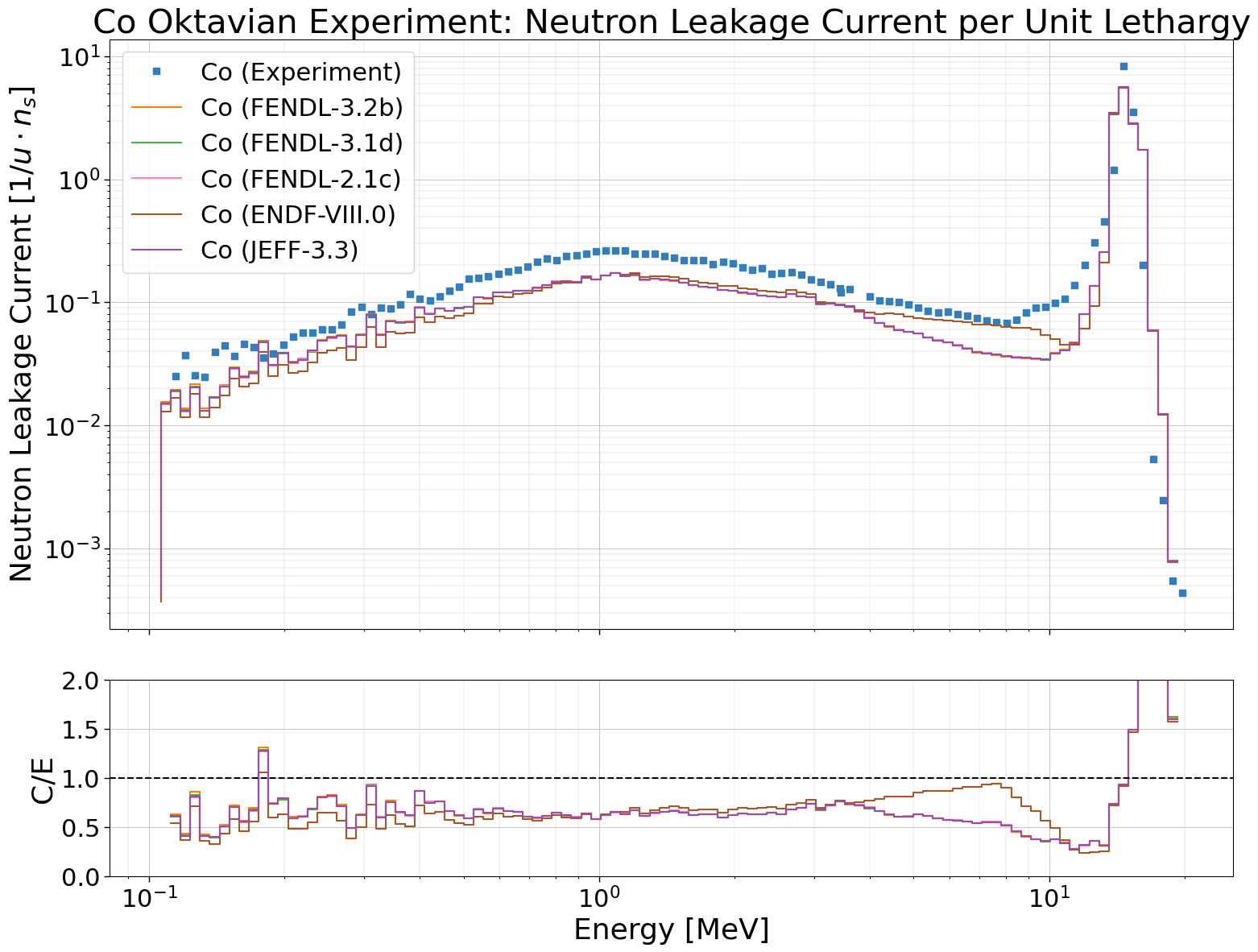}
    \caption{Neutron leakage current in the Co Oktavian experimental benchmark.}
\label{fig:figure2CoNeutron}
\end{figure}

Looking at \cref{table:laghiTable2summaryByMatl}, it can be observed that the obtained C/E values are quite different depending on the material. In particular, the simulations that present the largest discrepancies are for cobalt (Co) and molybdenum (Mo). For cobalt, the table shows that up to 10 MeV the mean C/E value obtained with FENDL-3.2b never surpasses 0.68, while from 10 MeV to 20 MeV it rises to 3.3.  These results are similar to the C/E values calculated with the other individual libraries with a minor difference in the range 5 MeV – 10 MeV where ENDF-VIII.0 results appear to be closer to the experimental measurements as seen in \cref{fig:figure2CoNeutron}. It should be noticed that the large discrepancies observed in the range above 10 MeV for all materials are most likely due to experimental limits and scarcity of data points. As an example, the clustered points just below 20 MeV observed in \cref{fig:figure5LiFNeutron} may introduce some bias in the computed C/E values.

\begin{figure}[htp]
\includegraphics[width=8.69cm]{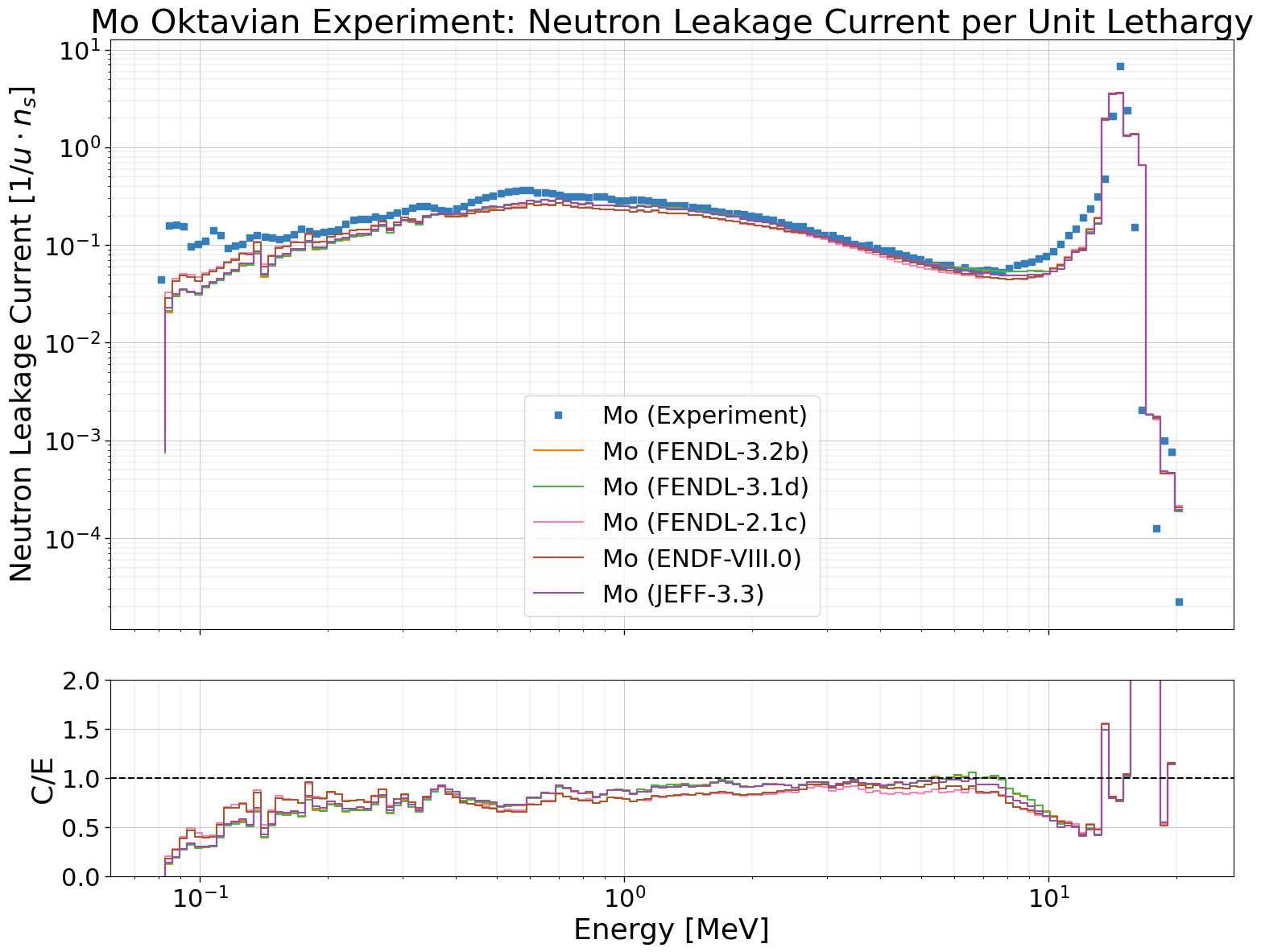}
    \caption{Neutron leakage current in the Mo Oktavian experimental benchmark.}
\label{fig:figure7MoNeutron}
\end{figure}

Regading the molybdenum results in \cref{table:laghiTable2summaryByMatl} in the 0.1 MeV – 1 MeV range and in the 10 MeV – 20 MeV range, the mean C/E values are 0.70 and 25 respectively. Once again, the FENDL-3.2b results are in line with the ones obtained with the other libraries (see also \cref{fig:figure7MoNeutron}).

\begin{figure}[htp]
\includegraphics[width=8.69cm]{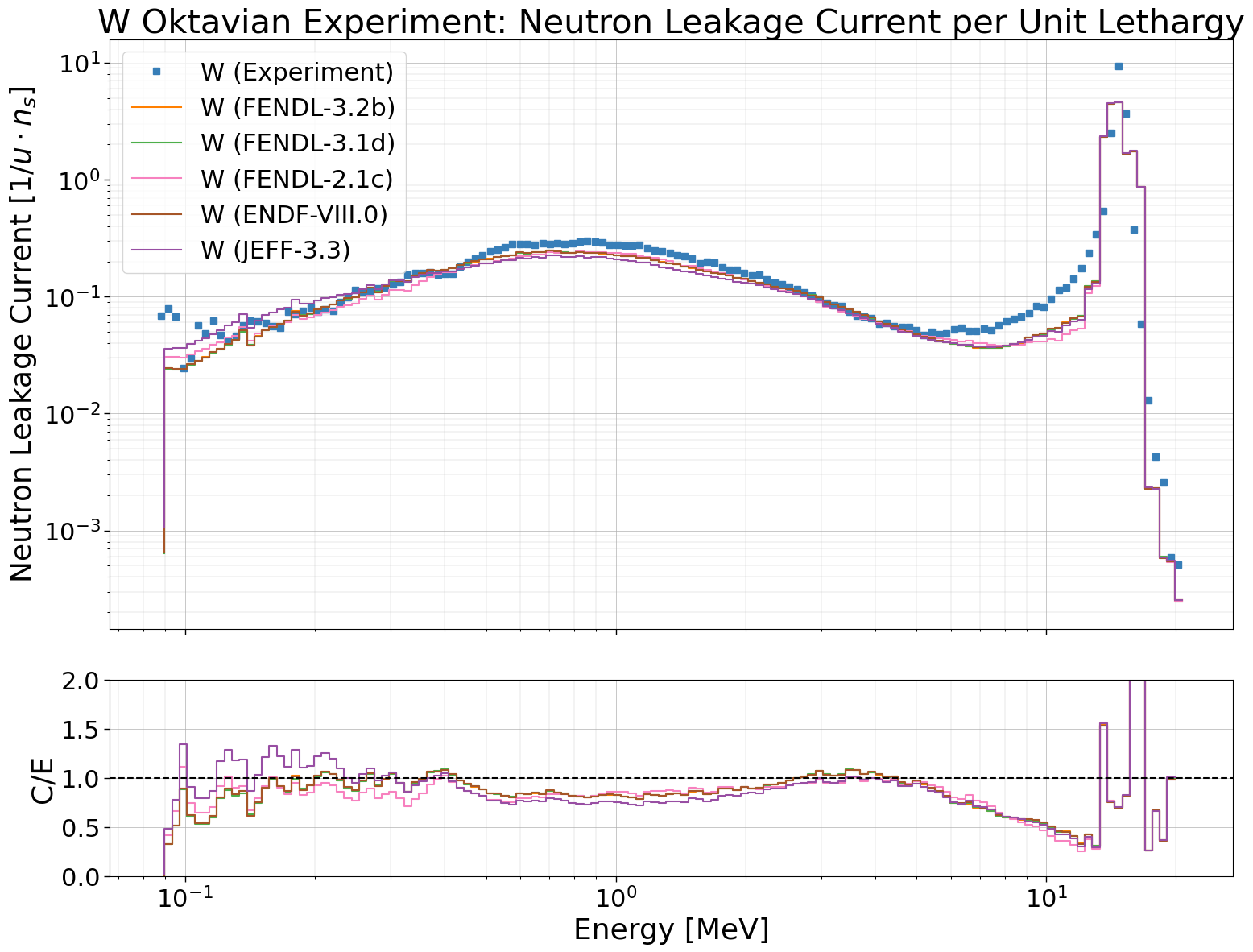}
    \caption{Neutron leakage current in the W Oktavian experimental benchmark.}
\label{fig:figure10WNeutron}
\end{figure}

\begin{figure}[htp]
\includegraphics[width=8.69cm]{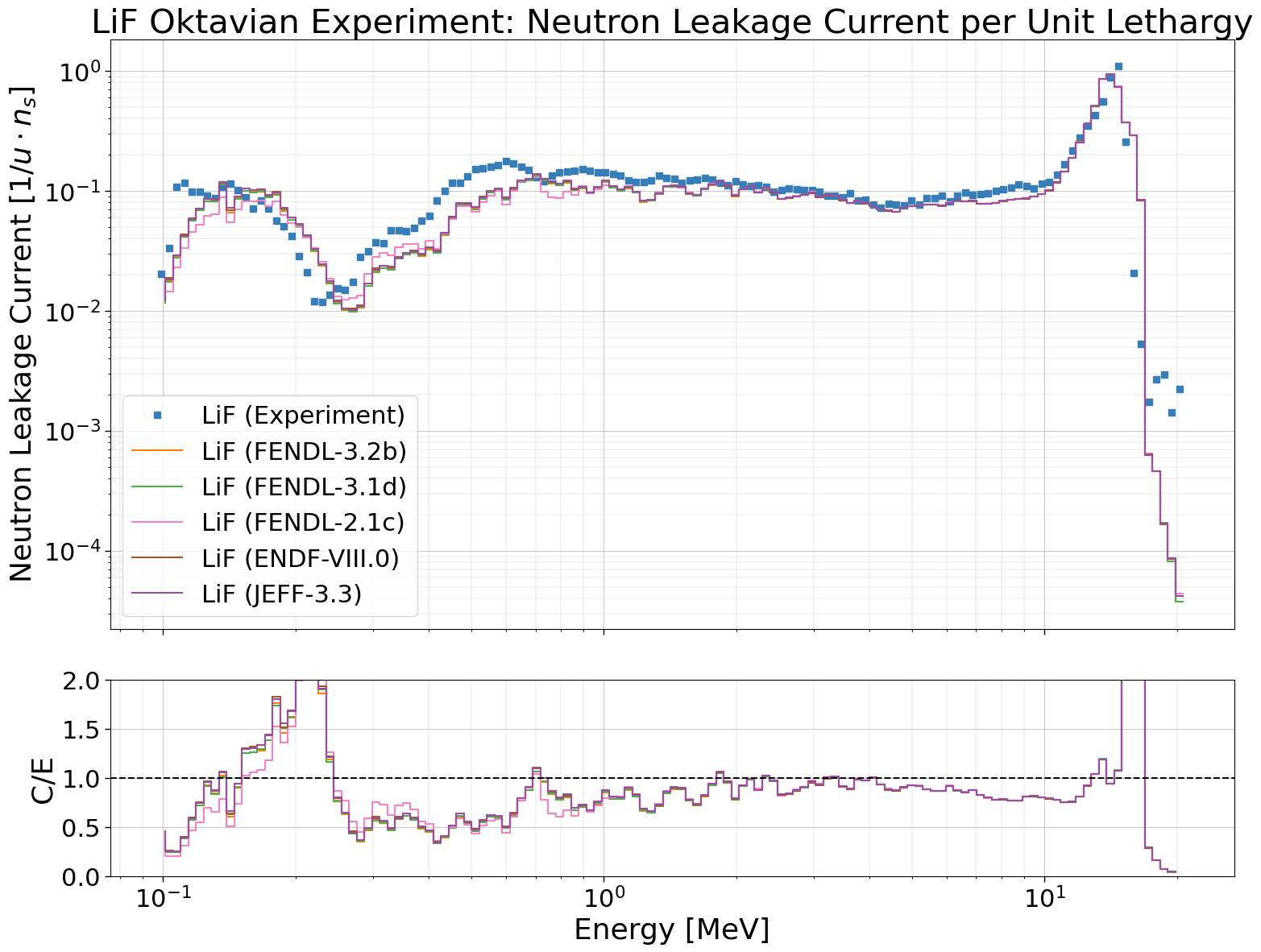}
    \caption{Neutron leakage current in the LiF Oktavian experimental benchmark.}
\label{fig:figure5LiFNeutron}
\end{figure}

\begin{figure}[htp]
\includegraphics[width=8.69cm]{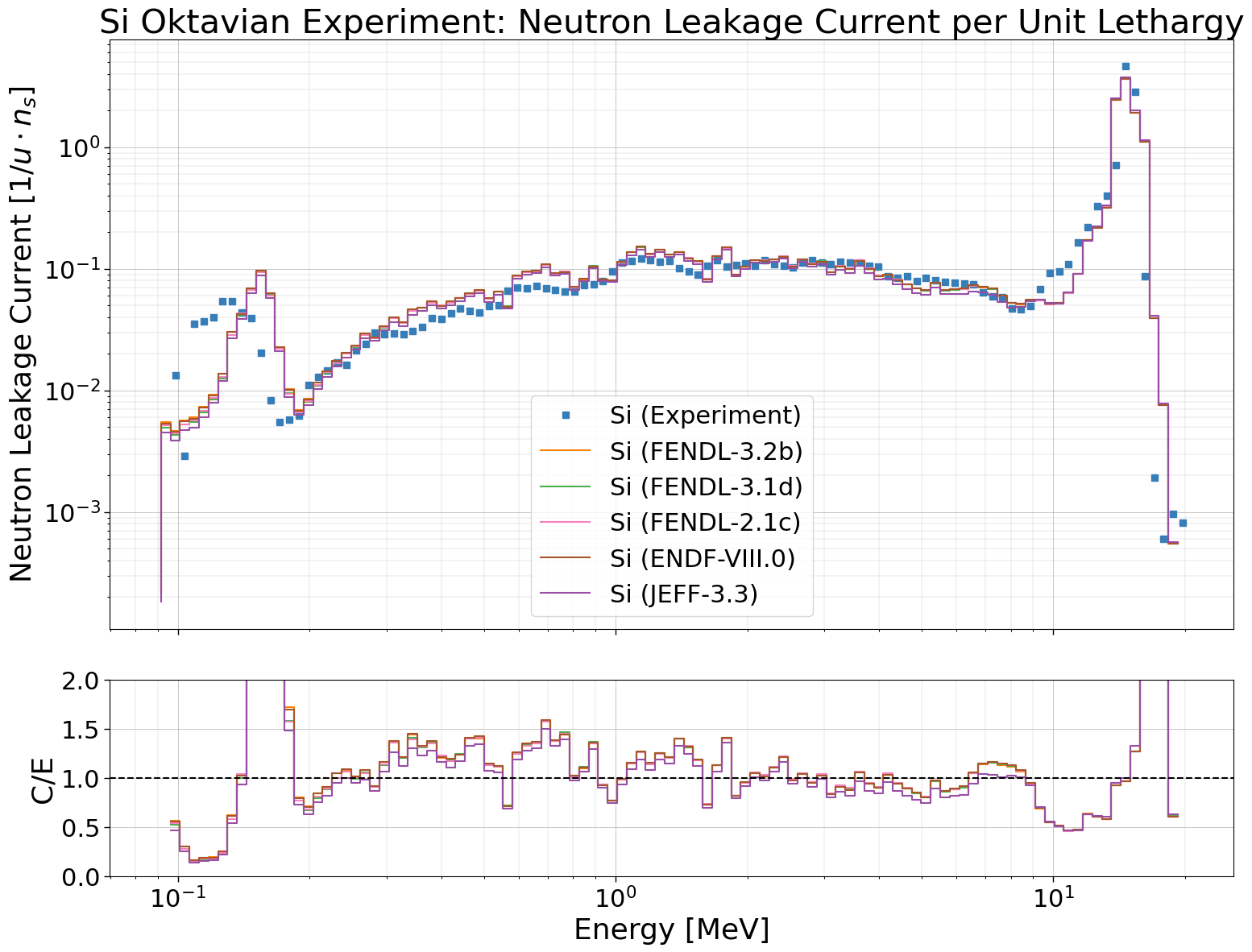}
    \caption{Neutron leakage current in the Si Oktavian experimental benchmark.}
\label{fig:figure8SiNeutron}
\end{figure}

Other results worth mentioning are the C/E discrepancy observed in tungsten (W) in the 5 MeV – 11 MeV range (see also \cref{fig:figure10WNeutron}) and the apparent shift between computational and experimental results of the low energy leakage current peak shown in lithium fluoride (\cref{fig:figure5LiFNeutron}) and silicon (\cref{fig:figure8SiNeutron}). 

\begin{table}[htp]
\caption{Summary of the mean C/E values for the neutron leakage current in the Oktavian benchmark.} 
\label{table:laghiTable2summaryByMatl}
\begin{tabular}{l | c |  c c} \hline \hline
Material  & Energy (MeV) & mean C/E & mean C/E  \\
          &              & All Libs.& FENDL-3.2b \\ \hline
Al  & 0.1 - 1 & 1.030898 & 1.029012 \\
    & 1 - 5 & 0.915619 & 0.898746 \\
    & 5 - 10 & 0.880712 & 0.901812 \\
    & 10 - 20 & 5.342176 & 5.335745 \\ \hline 
Co  & 0.1 - 1 & 0.65143 & 0.675886 \\
    & 1 - 5 & 0.667017 & 0.655133 \\
    & 5 - 10 & 0.598878 & 0.532827 \\
    & 10 - 20 & 3.33032 & 3.335812 \\ \hline 
Cr  & 0.1 - 1 & 1.1182 & 1.147679 \\
    & 1 - 5 & 1.113829 & 1.139045 \\
    & 5 - 10 & 1.033334 & 0.955579 \\
    & 10 - 20 & 1.797556 & 1.818001 \\ \hline 
Cu  & 0.1 - 1 & 1.070476 & 1.073474 \\
    & 1 - 5 & 1.136589 & 1.13987 \\
    & 5 - 10 & 1.421237 & 1.367503 \\
    & 10 - 20 & 1.796202 & 1.785743 \\ \hline 
LiF & 0.1 - 1 & 0.848191 & 0.840806 \\
    & 1 - 5 & 0.88751 & 0.88173 \\
    & 5 - 10 & 0.845356 & 0.84557 \\
    & 10 - 20 & 3.274919 & 3.269567 \\ \hline 
Mn  & 0.1 - 1 & 1.143793 & 1.158953 \\
    & 1 - 5 & 1.001599 & 0.939605 \\
    & 5 - 10 & 1.035322 & 1.011717 \\
    & 10 - 20 & 1.431198 & 1.421553 \\ \hline 
Mo  & 0.1 - 1 & 0.723993 & 0.705897 \\
    & 1 - 5 & 0.900606 & 0.931112 \\
    & 5 - 10 & 0.887087 & 0.948869 \\
    & 10 - 20 & 25.046755 & 25.02657 \\ \hline 
Si  & 0.1 - 1 & 1.394874 & 1.427578 \\
    & 1 - 5 & 1.033058 & 1.042741 \\
    & 5 - 10 & 0.968537 & 0.980885 \\
    & 10 - 20 & 5.215136 & 5.171156 \\ \hline 
Ti  & 0.1 - 1 & 1.471778 & 1.43928 \\
    & 1 - 5 & 1.18957 & 1.198541 \\
    & 5 - 10 & 1.112841 & 1.136364 \\
    & 10 - 20 & 2.300806 & 2.307081 \\ \hline 
W   & 0.1 - 1 & 0.892386 & 0.884962 \\
    & 1 - 5 & 0.916524 & 0.932336 \\
    & 5 - 10 & 0.729615 & 0.721748 \\
    & 10 - 20 & 2.308566 & 2.316111 \\ \hline 
Zr  & 0.1 - 1 & 1.213553 & 1.255059 \\
    & 1 - 5 & 1.100074 & 1.071864 \\
    & 5 - 10 & 1.014344 & 1.018061 \\
    & 10 - 20 & 1.608731 & 1.614884 \\ \hline \hline 
\end{tabular}
\end{table}

\subsubsection{FNS experiments}
\label{subsubsec:fnsKonno}
\begin{figure}[htp]
\includegraphics[width=8.69cm]{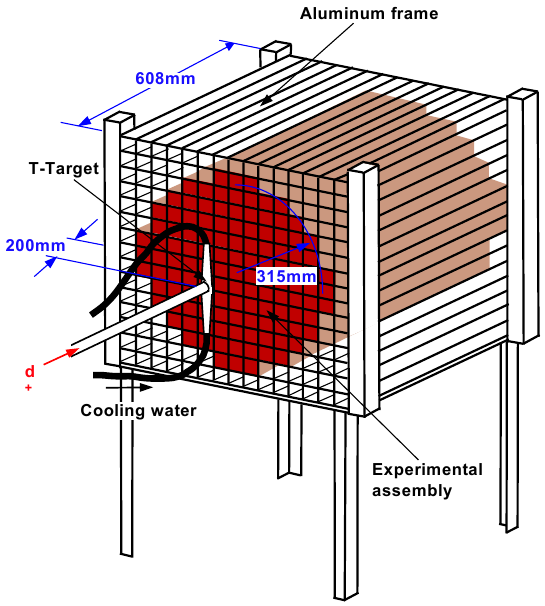}
\caption{Set-up of in-situ experiment.}
\label{fig:konnoFigure1}
\end{figure}

\begin{figure}[htp]
\includegraphics[width=8.69cm]{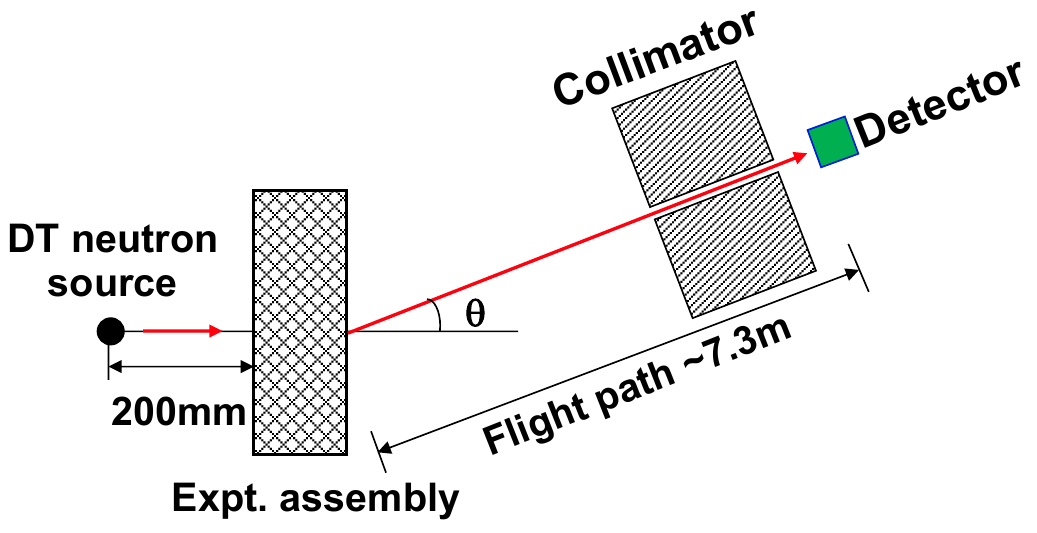}
\caption{Set-up of TOF experiment.}
\label{fig:konnoFigure2}
\end{figure}

\begin{figure}[htp]
\includegraphics[width=8.69cm]{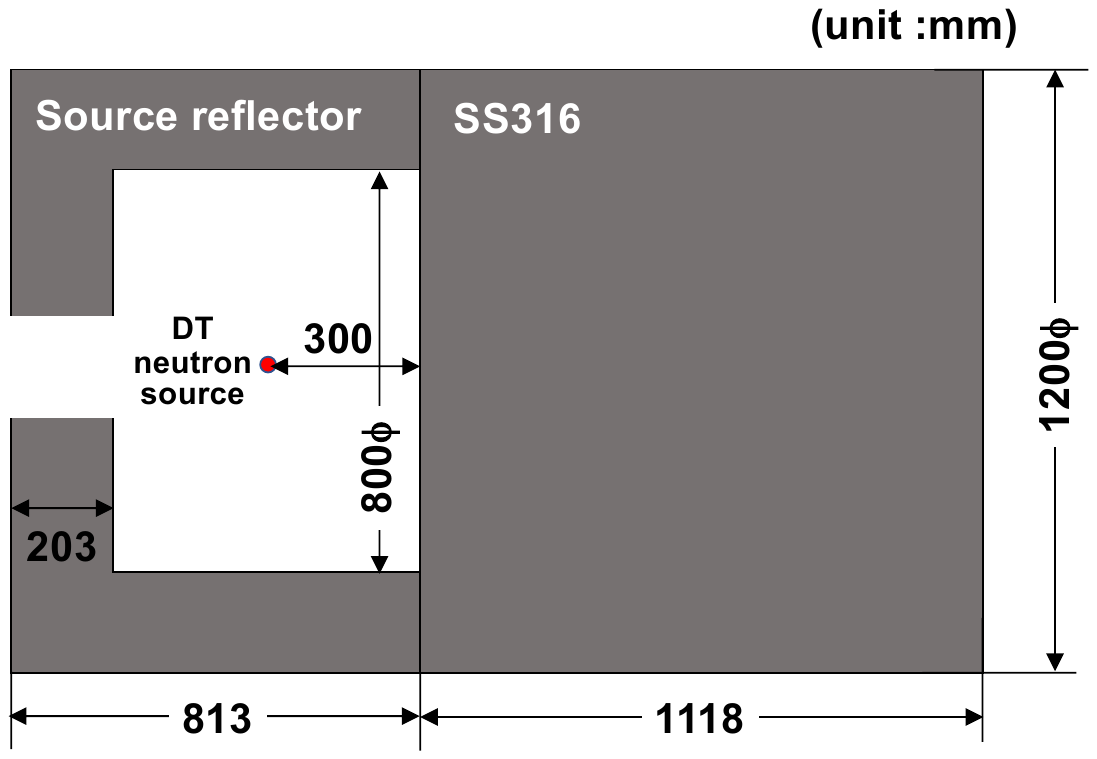}
\caption{Set-up of SS316 in-situ experiment.}
\label{fig:konnoFigure3}
\end{figure}

The \underline{F}usion \underline{N}eutronics \underline{S}ource (FNS) facility of \underline{J}apan \underline{A}tomic \underline{E}nergy \underline{A}gency (JAEA) had a DT neutron source, which was operated from 1981 to 2016. Two types of various integral benchmark experiments with the DT neutron source were performed for nuclear data validation there: (1) an in-situ benchmark experiment and (2) a Time-of-flight (TOF) experiment. \Cref{fig:konnoFigure1} shows a typical experimental configuration of the “in-situ” experiment. Reaction rates shown in \Cref{table:konnoTable1} and/or neutron spectra were measured inside the experimental assembly. \Cref{fig:konnoFigure2} shows a typical configuration of the TOF experiment. Angular neutron spectra leaked from the assembly were measured from 100 or 500 keV to 15 MeV at several angles. \Cref{table:konnoTable2} summarizes the experiments described in this paper. Note that the configuration of the type 316 stainless steel (SS316) in-situ experiment is special as shown in \cref{fig:konnoFigure3}.

\begin{table}[htp]
\caption{Reactions of measured reaction rates.}
\label{table:konnoTable1}
\begin{tabular}{c | c } \hline \hline
Reaction & Sensitive neutron energy \\ \hline
$^{93}$Nb(n,2n)$^{92m}$Nb & ${>}$ 10 MeV \\
$^{27}$Al(n,$\alpha$)$^{24}$Na & ${>}$ 4 MeV \\
$^{115}$In(n,n')$^{115m}$In & ${>}$ 0.3 MeV \\
$^{186}$W(n,$\gamma$)$^{187}$W & low energy \\
$^{197}$Au(n,$\gamma$)$^{198}$Au & low energy \\
$^{6}$Li(n,t)$^{4}$He & low energy \\
$^{235}$U(n,fission) & low energy \\
$^{238}$U(n,fission) & ${>}$ 1 MeV \\
\hline
\hline
\end{tabular}
\end{table}

\begin{table}[htp]
\caption{FNS experiment specification.}
\label{table:konnoTable2}
\begin{tabular}{ c | c | c | c  } \hline \hline
\multicolumn{2}{ c }{\multirow{2}{*}{Experiment}} & \multicolumn{2}{| c }{Assembly} \\ \cline{3-4}
\multicolumn{2}{ c |}{ } & Shape & Size \\ \hline
\multirow{4}{*}{Li$_{2}$O} & \multirow{2}{*}{in-situ} & Quasi & 630 mm in effective $\phi$ \\
  &  & cylinder & 610 mm in thickness \\ \cline{2-4}
  & \multirow{2}{*}{TOF} & Quasi & 630 mm in effective $\phi$ \\
  &  & cylinder & 51, 202, 405 in thickness \\ \hline
\multirow{4}{*}{Be} & \multirow{2}{*}{in-situ} & Quasi & 630 mm in effective $\phi$ \\
  &  & cylinder & 455 mm in thickness \\ \cline{2-4}
  & \multirow{2}{*}{TOF} & Quasi & 630 mm in effective $\phi$ \\
  &  & cylinder & 51, 152 in thickness \\ \hline
\multirow{4}{*}{Graphite} & \multirow{2}{*}{in-situ} & Quasi & 630 mm in effective $\phi$ \\
  &  & cylinder & 610 mm in thickness \\ \cline{2-4}
  & \multirow{2}{*}{TOF} & Quasi & 630 mm in effective $\phi$ \\
  &  & cylinder & 51, 202, 405 in thickness \\ \hline
\multirow{2}{*}{Liq. N$_{2}$} & \multirow{2}{*}{TOF} & Cylinder & 600 mm $\phi$ \\
  &  & tank & 200 mm in thickness \\ \hline
\multirow{2}{*}{Liq. O$_{2}$} & \multirow{2}{*}{TOF} & Cylinder & 600 mm $\phi$ \\
  &  & tank & 200 mm in thickness \\ \hline
SiC & in-situ & Rectangular & 457 x 457 x 711 mm \\ \hline
\multirow{4}{*}{Ti} & \multirow{4}{*}{in-situ} & \multirow{4}{*}{Rectangular} & 457 x 457 x 711 mm \\
  &  &  & covered with Li$_{2}$O \\
  &  &  & (front: 51 mm thick, side \\
  &  &  & and rear: 101 mm thick) \\ \hline
\multirow{3}{*}{V} & \multirow{3}{*}{in-situ} & \multirow{3}{*}{Rectangular} & 254 x 254 x 254 mm \\
  &  &  & covered with 50 mm \\
  &  &  & thick graphite \\ \hline
\multirow{5}{*}{Fe} & \multirow{2}{*}{in-situ} & \multirow{2}{*}{Cylinder} & 100 mm $\phi$ \\
  &  &  & 95 mm in thickness \\ \cline{2-4}
  & \multirow{3}{*}{TOF} & \multirow{3}{*}{Cylinder} & 100 mm $\phi$ \\
  &  &  & 50, 200, 400, 600 mm \\
  &  &  & in thickness \\ \hline
\multirow{2}{*}{SS316} & \multirow{2}{*}{in-situ} & \multirow{2}{*}{Cylinder} & 1200 mm $\phi$ \\
  &  &  & 1118 mm in thickness \\ \hline
\multirow{6}{*}{Cu} & \multirow{6}{*}{in-situ} &   & 630 mm in effective $\phi$ \\
  &  &  & 610 mm in thickness \\
  &  & Quasi & covered with Li$_{2}$O \\
  &  & cylinder & (front: 51 mm thick, \\
  &  &  & side: 51 mm thick, \\
  &  &  & rear: 153 mm thick) \\ \hline
\multirow{5}{*}{Mo} & \multirow{5}{*}{in-situ} & \multirow{5}{*}{Rectangular} & 253 x 253 x 354 mm \\
  &  &  & covered with Li$_{2}$O \\
  &  &  & (front: 51 mm thick, \\
  &  &  & side: 202 mm thick, \\
  &  &  & rear: 253 mm thick) \\ \hline
\multirow{2}{*}{W} & \multirow{2}{*}{in-situ} & Quasi & 575 mm in effective $\phi$ \\
  &  & cylinder & 507 mm in thickness \\ \hline
\multirow{6}{*}{Pb} & \multirow{4}{*}{in-situ} & \multirow{4}{*}{Rectangular} & 557 x 557 x 607 mm \\
  &  &  & covered with Li$_{2}$O \\
  &  &  & (front and side: 101 mm thick, \\
  &  &  & rear: 203 mm thick, \\ \cline{2-4}
  & \multirow{2}{*}{TOF} & Quasi & 630 mm in effective $\phi$ \\
  &  & cylinder & 202, 405 mm in thickness \\ \hline
\end{tabular}
\end{table}

These experiments were analyzed with the Monte Carlo code MCNP6.2 \cite{mcnp62} using the officially distributed ACE files of the nuclear data libraries FENDL-3.2b \cite{kwon7}, FENDL-2.1 \cite{konno1} and ENDF/B-VIII.0 \cite{konno2}. A target isotope file in FENDL-3.1d \cite{kwon6} was also used only when it was different from that in FENDL-3.2b. The JENDL dosimetry file 99 \cite{konno3} was adopted as the dosimetry reaction cross sections for reaction rate calculations. The thermal scattering law data for beryllium metal and graphite in ENDF/B-VIII.0 were used for the analyses of the beryllium and graphite in-situ experiments because FENDL-3.2b includes no thermal scattering law data.\\

\noindent {\it (1) Li$_{2}$O in-situ experiment\\}

\begin{figure}[htp]
\includegraphics[width=8.69cm]{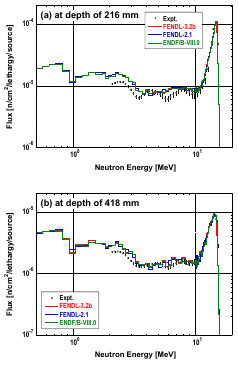}
\caption{Neutron spectra in Li$_{2}$O in-situ experiment.}
\label{fig:konnoFigure4}
\end{figure}

\begin{figure}[htp]
\includegraphics[width=8.69cm]{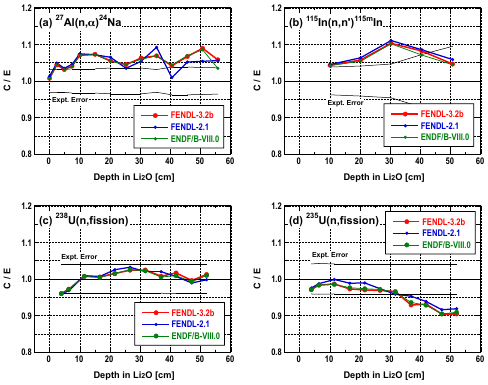}
\caption{C/E of reaction rates in Li$_{2}$O in-situ experiment.}
\label{fig:konnoFigure5}
\end{figure}

The neutron spectra above 2 MeV and several reaction rates were measured in this experiment \cite{konno4}. \Cref{fig:konnoFigure4} shows the neutron spectra at the depths of 216 and 418 mm. All the calculated neutron spectra are almost the same and those above 10 MeV agree with the measured ones well. On the contrary, there are large differences between the measured and calculated spectra below 10 MeV. The reason for this discrepancy can be very likely attributed to problems associated with the unfolding procedure to extract spectra from measurements as they are sometimes encountered in such experiments, e.g.~\cite{leeFastNeutronSpectrum1985}. \Cref{fig:konnoFigure5} shows the ratios of the calculated to the measured reaction rates (C/E) of typical reactions. Computed reaction rates are very close to each other and agree very well with the measured ones.\\

\noindent {\it (2) Li$_{2}$O TOF experiment\\}

\begin{figure}[htp]
\includegraphics[width=8.69cm]{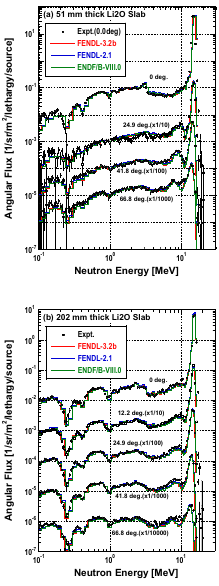}
\caption{Angular neutron spectra in Li$_{2}$O TOF experiment.}
\label{fig:konnoFigure6}
\end{figure}

The angular neutron spectra above 100 keV leaked from the Li$_{2}$O assemblies were measured with the TOF method in this experiment \cite{konno4}. \Cref{fig:konnoFigure6} shows the neutron spectra from the Li$_{2}$O assemblies of 51 and 202\,mm in thickness. All the calculated neutron spectra are almost the same and agree with the measured ones well.\\

\noindent {\it (3) Be in-situ experiment\\}

\begin{figure}[htp]
\includegraphics[width=8.69cm]{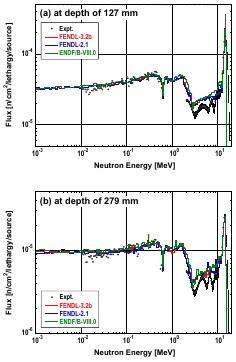}
\caption{Neutron spectra in Be in-situ experiment.}
\label{fig:konnoFigure7}
\end{figure}

\begin{figure}[htp]
\includegraphics[width=8.69cm]{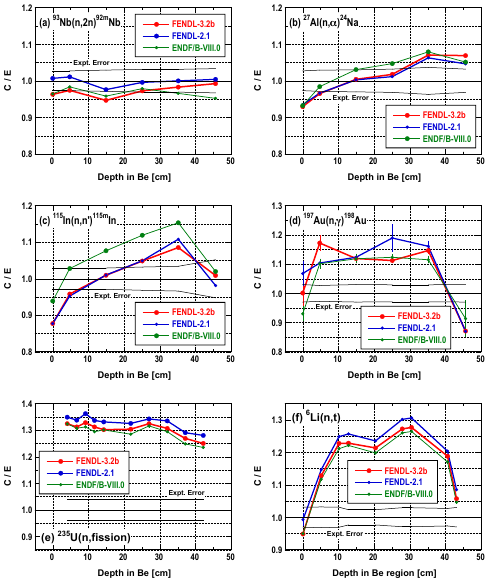}
\caption{C/E of reaction rates in Be in-situ experiment.}
\label{fig:konnoFigure8}
\end{figure}

The neutron spectra above 10 keV and several reaction rates were measured in this experiment \cite{konno4}. \Cref{fig:konnoFigure7} shows the neutron spectra at the depths of 127 and 279 mm. All the calculated neutron spectra are very close to each other and those above 10 MeV and from 10 keV to 1 MeV agree with the measured ones well.
Note that the strange oscillations in the measured spectra from 2 to 10 MeV are very likely artefacts of the unfolding procedure sometimes encountered in such experiments, e.g.~\cite{leeFastNeutronSpectrum1985}.
\Cref{fig:konnoFigure8} shows the C/E values of typical reactions. The calculated reaction rates are slightly scattered but they have a similar trend. The calculated reaction rates of the threshold reactions agree with the measured ones well, while those of the non-threshold reactions overestimate the measured ones. All the Be files may have some problems.\\

\noindent {\it (4) Be TOF experiment\\}

\begin{figure}[htp]
\includegraphics[width=8.69cm]{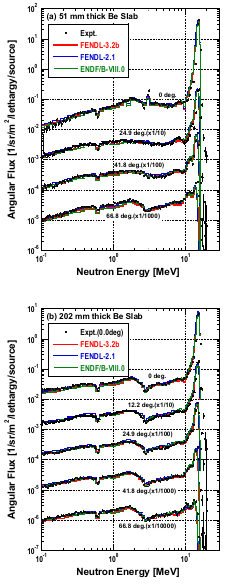}
\caption{Angular neutron spectra in Be TOF experiment.}
\label{fig:konnoFigure9}
\end{figure}

The angular neutron spectra above 100 keV leaked from the beryllium assemblies were measured with the TOF method in this experiment \cite{konno4}. \Cref{fig:konnoFigure9} shows the neutron spectra from the beryllium assemblies of 51 and 202 mm in thickness. All the calculated neutron spectra are almost the same and agree with the measured ones well.\\

\noindent {\it (5) Graphite in-situ experiment\\}

\begin{figure}[htp]
\includegraphics[width=8.69cm]{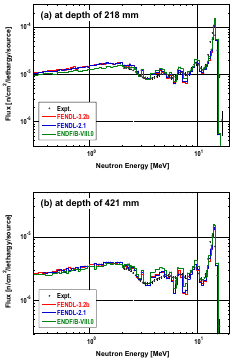}
\caption{Neutron spectra in graphite in-situ experiment.}
\label{fig:konnoFigure10}
\end{figure}

\begin{figure}[htp]
\includegraphics[width=8.69cm]{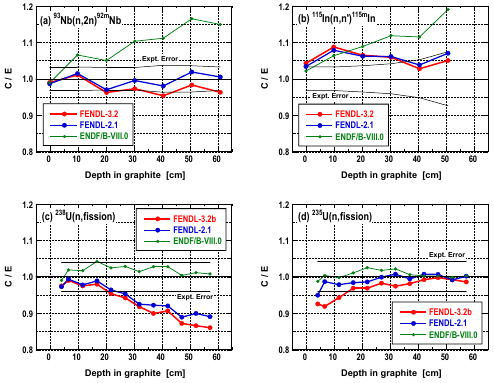}
\caption{C/E of reaction rates in graphite in-situ experiment.}
\label{fig:konnoFigure11}
\end{figure}

The neutron spectra above 2 MeV and several reaction rates were measured in this experiment \cite{konno4}. \Cref{fig:konnoFigure10} shows the neutron spectra at the depths of 218 and 421\,mm. The computed neutron spectra using FENDL-3.2b and FENDL-2.1 are very close to each other and very similar to the measured ones. In contrast to that, predictions based on ENDF/B-VIII.0 seem to overestimate the measurements. \Cref{fig:konnoFigure11} allows to compare the C/E values of common reactions. Calculated reaction rates based on FENDL-3.2b and FENDL-2.1 are similar to the measured ones with the exception of $^{238}$U(n,fission). Again, calculations relying on ENDF/B-VIII.0 tend to overestimate the measurements except for $^{238}$U(n,fission) and $^{235}$U(n,fission).\\

\noindent {\it (6) Graphite TOF experiment\\}

\begin{figure}[htp]
\includegraphics[width=8.69cm]{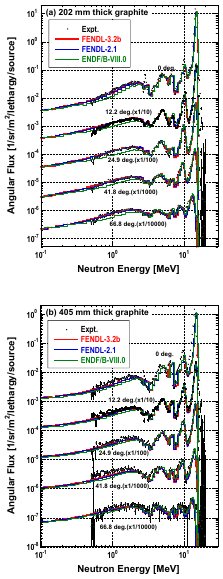}
\caption{Angular neutron spectra in graphite TOF experiment.}
\label{fig:konnoFigure12}
\end{figure}

The angular neutron spectra above 500 keV leaked from the graphite assemblies were measured with the TOF method in this experiment \cite{konno4}. \Cref{fig:konnoFigure12} shows the neutron spectra from the graphite assemblies of 202 and 405 mm in thickness. The computed neutron spectra with FENDL-3.2b and FENDL-2.1 are almost identical and compatible with the measured ones in the graphite assembly of 202 mm in thickness, while they differ slightly from the measured ones in the graphite assembly of 405 mm in thickness. Note that the calculated neutron spectra with ENDF/B-VIII.0 underestimate the measured ones around 1 MeV in the graphite assembly of 202 mm in thickness.\\

\noindent {\it (7) Liquid $N_{2}$ TOF experiment\\}

\begin{figure}[htp]
\includegraphics[width=8.69cm]{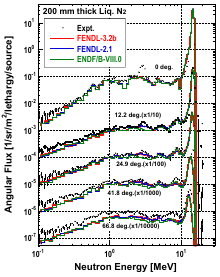}
\caption{Angular neutron spectra in liquid N$_{2}$ TOF experiment.}
\label{fig:konnoFigure13}
\end{figure}

The angular neutron spectra above 100 keV leaked from the liquid N$_{2}$ assembly were measured with the TOF method in this experiment \cite{konno5}. \Cref{fig:konnoFigure13} shows the neutron spectra from the liquid N$_{2}$ assembly of 200\,mm in thickness. All the calculated neutron spectra are almost the same and the differences from the measured ones are large at 66.8 degrees.\\

\noindent {\it (8) Liquid $O_{2}$ TOF experiment\\}

\begin{figure}[htp]
\includegraphics[width=8.69cm]{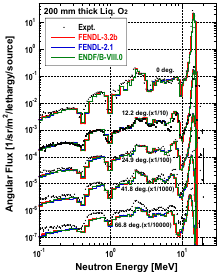}
\caption{Angular neutron spectra in liquid O$_{2}$ TOF experiment.}
\label{fig:konnoFigure14}
\end{figure}

The angular neutron spectra above 100 keV leaked from the liquid O$_{2}$ assembly were measured with the TOF method in this experiment \cite{konno5}. \Cref{fig:konnoFigure14} shows the neutron spectra from the liquid O$_{2}$ assembly of 200\,mm in thickness. All the calculated neutron spectra are nearly identical and deviations from measured ones increase with increasing leakage angle.\\

\noindent {\it (9) SiC in-situ experiment\\}

\begin{figure}[htp]
\includegraphics[width=8.69cm]{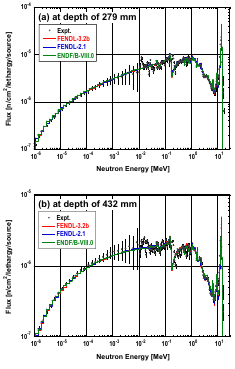}
\caption{Neutron spectra in SiC in-situ experiment.}
\label{fig:konnoFigure15}
\end{figure}

\begin{figure}[htp]
\includegraphics[width=8.69cm]{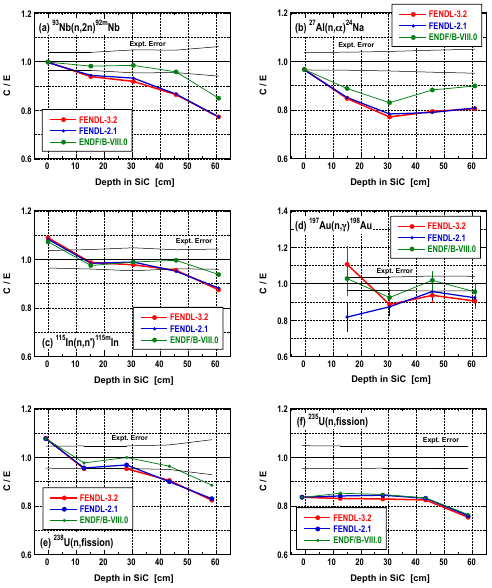}
\caption{C/E of reaction rates in SiC in-situ experiment.}
\label{fig:konnoFigure16}
\end{figure}

The neutron spectra over almost the whole energy range and several reaction rates were measured in this experiment \cite{konno6}. \Cref{fig:konnoFigure15} shows the neutron spectra at the depths of 279 and 432 mm. The calculated neutron spectra are very close to each other and follow measured ones well. \Cref{fig:konnoFigure16} depicts the C/E values of common reactions. The calculated and measured reaction rates are within 20$\%$ of each other. Note that those relying on ENDF/B-VIII.0 are slightly better than those relying on FENDL-3.2b and FENDL-2.1.\\

\noindent {\it (10) Ti in-situ experiment\\}

\begin{figure}[htp]
\includegraphics[width=8.69cm]{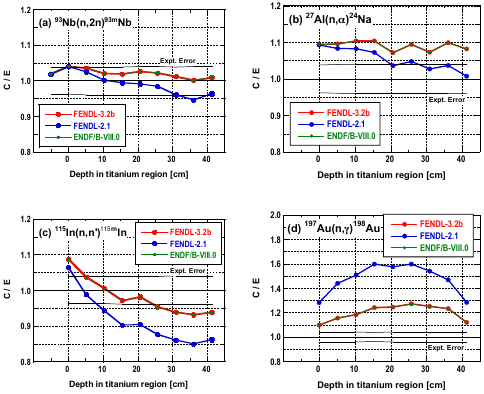}
\caption{C/E of reaction rates in Ti in-situ experiment.}
\label{fig:konnoFigure17}
\end{figure}

Only the reaction rates of the $^{93}$Nb(n,2n)$^{92m}$Nb, $^{27}$Al(n,$\alpha$)$^{24}$Na, $^{115}$In(n,n')$^{115m}$In, $^{186}$W(n,$\gamma$)$^{187}$W, $^{197}$Au(n,$\gamma$)$^{198}$Au, $^{235}$U(n,fission) and $^{238}$U(n,fission) reactions were measured in this experiment \cite{konno7}. \Cref{fig:konnoFigure17} shows several C/E plots. The calculated reaction rates with FENDL-3.2b are in better agreement with the measured ones than those with FENDL-2.1.
Calculations using ENDF/B-VIII.0 and FENDL-3.2b coincide because the Ti files in ENDF/B-VII.1 adopted by FENDL~\cite{konno8} were carried over from ENDF/B-VII.0.\\

\noindent {\it (11) V in-situ experiment\\}

\begin{figure}[htp]
\includegraphics[width=8.69cm]{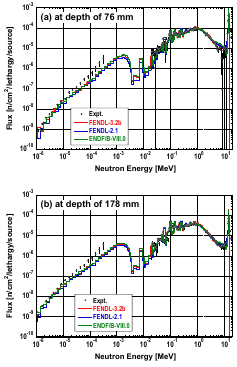}
\caption{Neutron spectra in V in-situ experiment.}
\label{fig:konnoFigure18}
\end{figure}

\begin{figure}[htp]
\includegraphics[width=8.69cm]{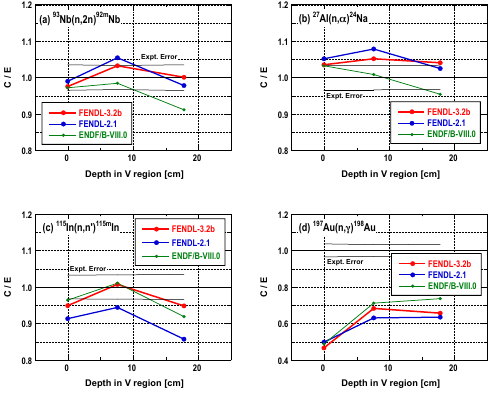}
\caption{C/E of reaction rates in V in-situ experiment.}
\label{fig:konnoFigure19}
\end{figure}

The neutron spectra covering almost the whole energy range and several reaction rates were measured in this experiment \cite{konno9}. \Cref{fig:konnoFigure18} shows the neutron spectra at the depths of 76 and 178 mm. Calculated neutron spectra are very similar and agree very well with the measured spectra above a few tens keV. Below a few hundred eV, they underestimate the measured ones. C/E values of common reactions are shown in \cref{fig:konnoFigure19}. There is good agreement between the calculated and measured spectra of threshold reactions but the measured reaction rate of $^{197}$Au(n,$\gamma$)$^{198}$Au is underestimated.\\

\noindent {\it (12) Fe in-situ experiment\\}

\begin{figure}[htp]
\includegraphics[width=8.69cm]{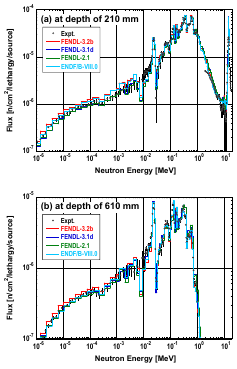}
\caption{Neutron spectra in Fe in-situ experiment.}
\label{fig:konnoFigure20}
\end{figure}

\begin{figure}[htp]
\includegraphics[width=8.69cm]{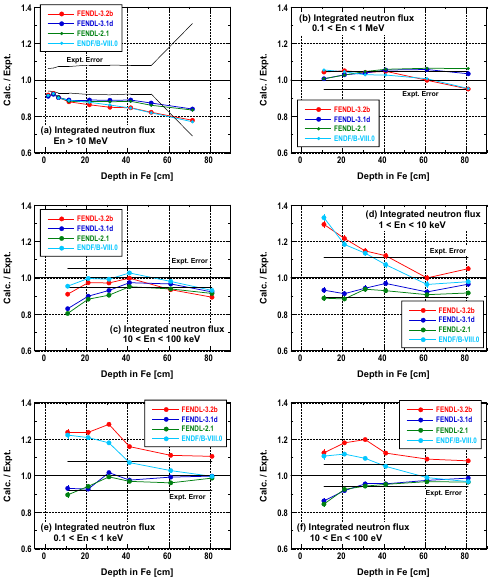}
\caption{C/E of integrated neutron fluxes for specified energy regions in Fe in-situ experiment.}
\label{fig:konnoFigure21}
\end{figure}

The neutron spectra over almost the whole energy range and several reaction rates were measured in this experiment \cite{konno9}. \Cref{fig:konnoFigure20} shows the neutron spectra at the depths of 210 and 610 mm. All the calculated neutron spectra are almost identical and in good agreement with the measured neutron fluxes above 100 keV. In contrast to that, the calculated neutron spectra below 100 keV based on FENDL-3.2b and ENDF/B-VIII.0 deviate from those based on FENDL-2.1 and FENDL-3.1d, which tend to overestimate the measured data. \Cref{fig:konnoFigure21} gives a clearer impression of the overestimation, illustrating the C/E values of integrated neutron fluxes for specified energy regions based on the measured and calculated neutron spectra.\\

\noindent {\it (13) Fe TOF experiment\\}

\begin{figure}[htp]
\includegraphics[width=8.69cm]{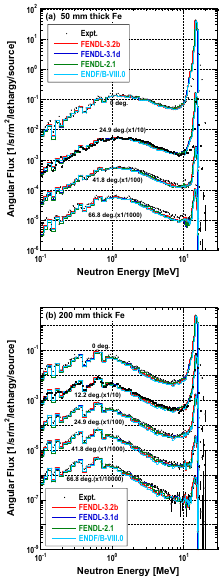}
\caption{Angular neutron spectra in Fe TOF experiment.}
\label{fig:konnoFigure22}
\end{figure}

The angular neutron spectra above 100 keV leaked from the iron assemblies were measured with the TOF method in this experiment \cite{konno9}. \Cref{fig:konnoFigure22} shows the neutron spectra from the iron assemblies of 50 and 200 mm in thickness. Good agreement between all the calculated and measured neutron spectra can be observed.\\

\noindent {\it (14) SS316 in-situ experiment\\}

\begin{figure}[htp]
\includegraphics[width=8.69cm]{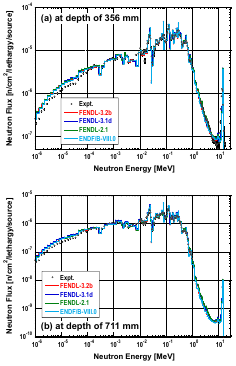}
\caption{Neutron spectra in SS316 in-situ experiment.}
\label{fig:konnoFigure23}
\end{figure}

\begin{figure}[htp]
\includegraphics[width=8.69cm]{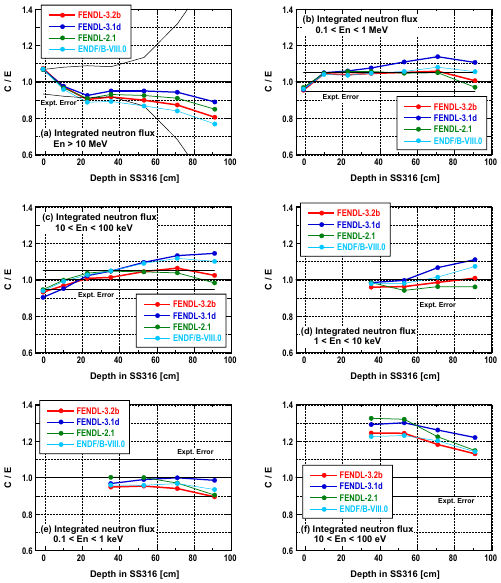}
\caption{C/E of integrated neutron fluxes for specified energy regions in SS316 in-situ experiment.}
\label{fig:konnoFigure24}
\end{figure}

The neutron spectra covering almost the entire energy range and several reaction rates were measured in this experiment \cite{konno10}. \Cref{fig:konnoFigure23} shows the neutron spectra at the depths of 356 and 711 mm. The calculated neutron spectra are very close to each other and very compatible with the measured neutron fluxes above 100 eV, while an overestimation below 100 eV can be observed. This tendency is clearly demonstrated in \cref{fig:konnoFigure24} which shows the C/E values of integrated neutron fluxes for specified energy regions from the measured and calculated neutron spectra.\\

\noindent {\it (15) Cu in-situ experiment\\}

\begin{figure}[htp]
\includegraphics[width=8.69cm]{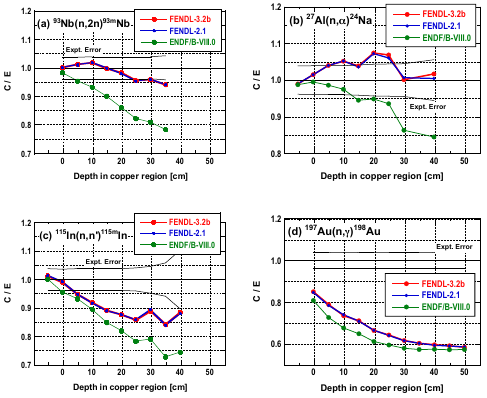}
\caption{C/E of reaction rates in Cu in-situ experiment.}
\label{fig:konnoFigure25}
\end{figure}

Only the reaction rates of the $^{93}$Nb(n,2n)$^{92m}$Nb, $^{27}$Al(n,$\alpha$)$^{24}$Na, $^{115}$In(n,n')$^{115m}$In, $^{186}$W(n,$\gamma$)$^{187}$W, $^{197}$Au(n,$\gamma$)$^{198}$Au, $^{235}$U(n,fission) and $^{238}$U(n,fission) reactions were measured in this experiment \cite{konno11}. The C/E plots are shown in \cref{fig:konnoFigure25}. The calculated reaction rates with FENDL-3.2b and FENDL-2.1 are almost the same because FENDL-3.2b and FENDL-2.1 only differ above 20 MeV for these isotopes. They are in better agreement with the measured ones than those with ENDF/B-VIII.0, though all the calculated reaction rates of $^{197}$Au(n,$\gamma$)$^{198}$Au drastically underestimate the measured ones.\\

\noindent {\it (16) Mo in-situ experiment\\}

\begin{figure}[htp]
\includegraphics[width=8.69cm]{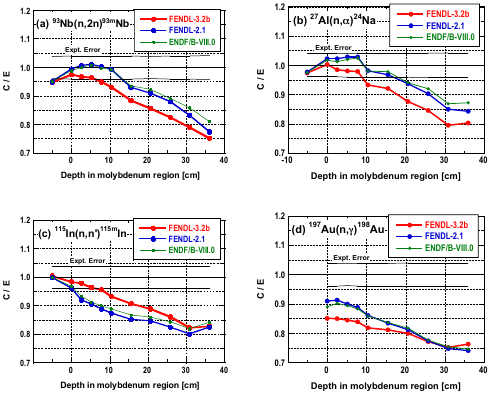}
\caption{C/E of reaction rates in Mo in-situ experiment.}
\label{fig:konnoFigure26}
\end{figure}

We measured the reaction rates of the $^{93}$Nb(n,2n)$^{92m}$Nb, $^{27}$Al(n,$\alpha$)$^{24}$Na, $^{115}$In(n,n')$^{115m}$In, $^{186}$W(n,$\gamma$)$^{187}$W, $^{197}$Au(n,$\gamma$)$^{198}$Au, $^{235}$U(n,fission) and $^{238}$U(n,fission) reactions in this experiment \cite{konno12}. The typical C/E plots are shown in \cref{fig:konnoFigure26}. The calculated $^{93}$Nb(n,2n)$^{92m}$Nb, $^{27}$Al(n,$\alpha$)$^{24}$Na, and $^{197}$Au(n,$\gamma$)$^{198}$Au reaction rates with FENDL-3.2b are slightly worse than those with FENDL-2.1 and ENDF/B-VIII.0, while the calculated $^{115}$In(n,n')$^{115m}$In reaction rate with FENDL-3.2b is slightly better than those with FENDL-2.1 and ENDF/B-VIII.0. Note that all the calculation results tend to underestimate the measured ones at larger depths.\\

\noindent {\it (17) W in-situ experiment\\}

\begin{figure}[htp]
\includegraphics[width=8.69cm]{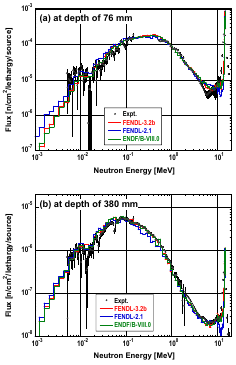}
\caption{Neutron spectra in W in-situ experiment.}
\label{fig:konnoFigure27}
\end{figure}

\begin{figure}[htp]
\includegraphics[width=8.69cm]{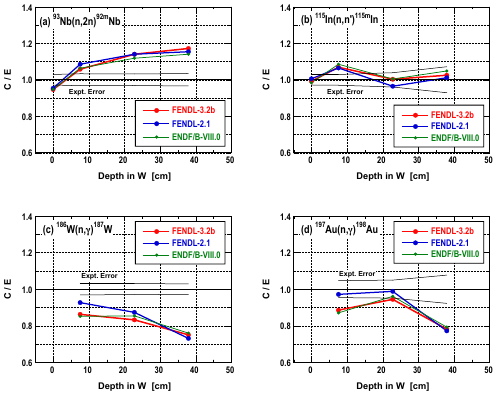}
\caption{C/E of reaction rates in W in-situ experiment.}
\label{fig:konnoFigure28}
\end{figure}

The neutron spectra above 5 keV and several reaction rates were measured in this experiment \cite{konno9}. \Cref{fig:konnoFigure27} visalizes the neutron spectra at the depths of 76 and 380\,mm. The calculated neutron spectra with FENDL-3.2b and ENDF/B-VIII.0 agree with the measured ones better than those with FENDL-2.1. The C/E values of common reactions presented in \cref{fig:konnoFigure28} indicate agreement within 20\% of calculated and measured reaction rates.\\

\noindent {\it (18) Pb in-situ experiment\\}

\begin{figure}[htp]
\includegraphics[width=8.69cm]{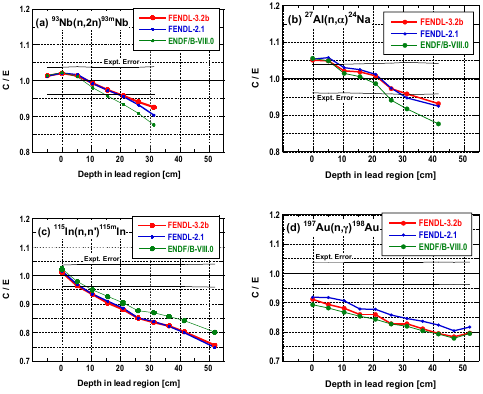}
\caption{C/E of reaction rates in Pb in-situ experiment.}
\label{fig:konnoFigure29}
\end{figure}

Only the reaction rates of the $^{93}$Nb(n,2n)$^{92m}$Nb, $^{27}$Al(n,$\alpha$)$^{24}$Na, $^{115}$In(n,n')$^{115m}$In, $^{186}$W(n,$\gamma$)$^{187}$W, $^{197}$Au(n,$\gamma$)$^{198}$Au, $^{235}$U(n,fission) and $^{238}$U(n,fission) reactions were measured in this experiment \cite{konno13}. We plot typical C/E values in \cref{fig:konnoFigure29}. The calculated $^{93}$Nb(n,2n)$^{92m}$Nb and $^{27}$Al(n,$\alpha$)$^{24}$Na reaction rates with FENDL-3.2b and FENDL-2.1 are slightly better than those with ENDF/B-VIII.0, while the calculated $^{115}$In(n,n')$^{115m}$In reaction rates with FENDL-3.2b and FENDL-2.1 are slightly worse than those with ENDF/B-VIII.0. The calculated $^{197}$Au(n,$\gamma$)$^{198}$Au reaction rates with FENDL-3.2b and ENDF/B-VIII.0 are smaller than that with FENDL-2.1. Note that all the calculation results also tend to underestimate the measured ones at larger depths.\\

\noindent {\it (19) Pb TOF experiment\\}

\begin{figure}[htp]
\includegraphics[width=8.69cm]{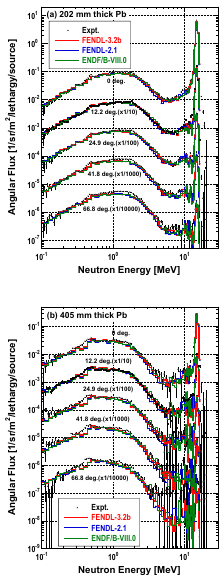}
\caption{Angular neutron spectra in Pb TOF experiment.}
\label{fig:konnoFigure30}
\end{figure}

The angular neutron spectra above 100 keV leaked from the lead assemblies were measured with the TOF method in this experiment \cite{konno4}. \Cref{fig:konnoFigure30} shows the neutron spectra from the lead assemblies of 202 and 405 mm in thickness. The calculated neutron spectra are nearly identical and in good agreement with the measured ones.\\

\subsubsection{TIARA shielding experiments}
\label{subsubsec:tiaraKwon}
\begin{figure}[htp]
\includegraphics[width=8.69cm]{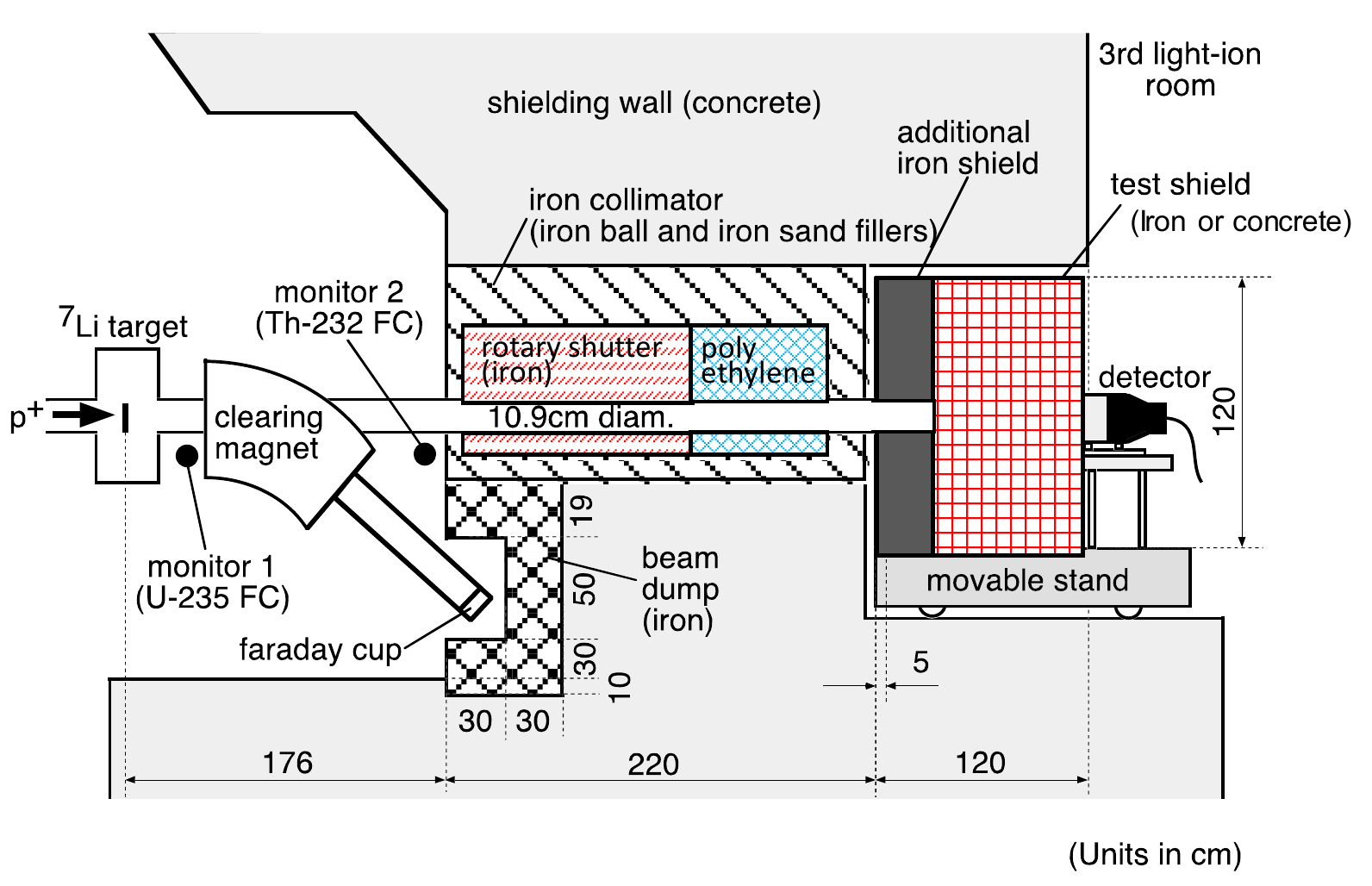}
    \caption{Experimental configuration of TIARA shielding experiments for iron and concrete.}
\label{fig:figure1}
\end{figure}

More than 20 years ago shielding experiments at the \underline{T}akasaki \underline{I}on \underline{A}ccelerators for Advanced \underline{R}adiation \underline{A}pplication, TIARA in National Institutes for \underline{Q}uantum \underline{S}cience and \underline{T}echnology, QST (former \underline{J}apan \underline{A}tomic \underline{E}nergy \underline{R}esearch \underline{I}nstitute, JAERI at that time) were carried out for iron and concrete with quasi-mono energetic neutrons produced by bombarding 43 or 68 MeV protons to a $^{7}$Li target. The generated neutrons of 40 or 65 MeV were collimated and entered into an iron or concrete test shield. The experimental configuration is shown in \cref{fig:figure1}. The neutron spectra above 10 MeV just behind the test shield on and off the beam axis were measured with a BC501A scintillation detector \cite{kwon1,kwon2}.

\begin{figure}[htp]
\includegraphics[width=8.69cm]{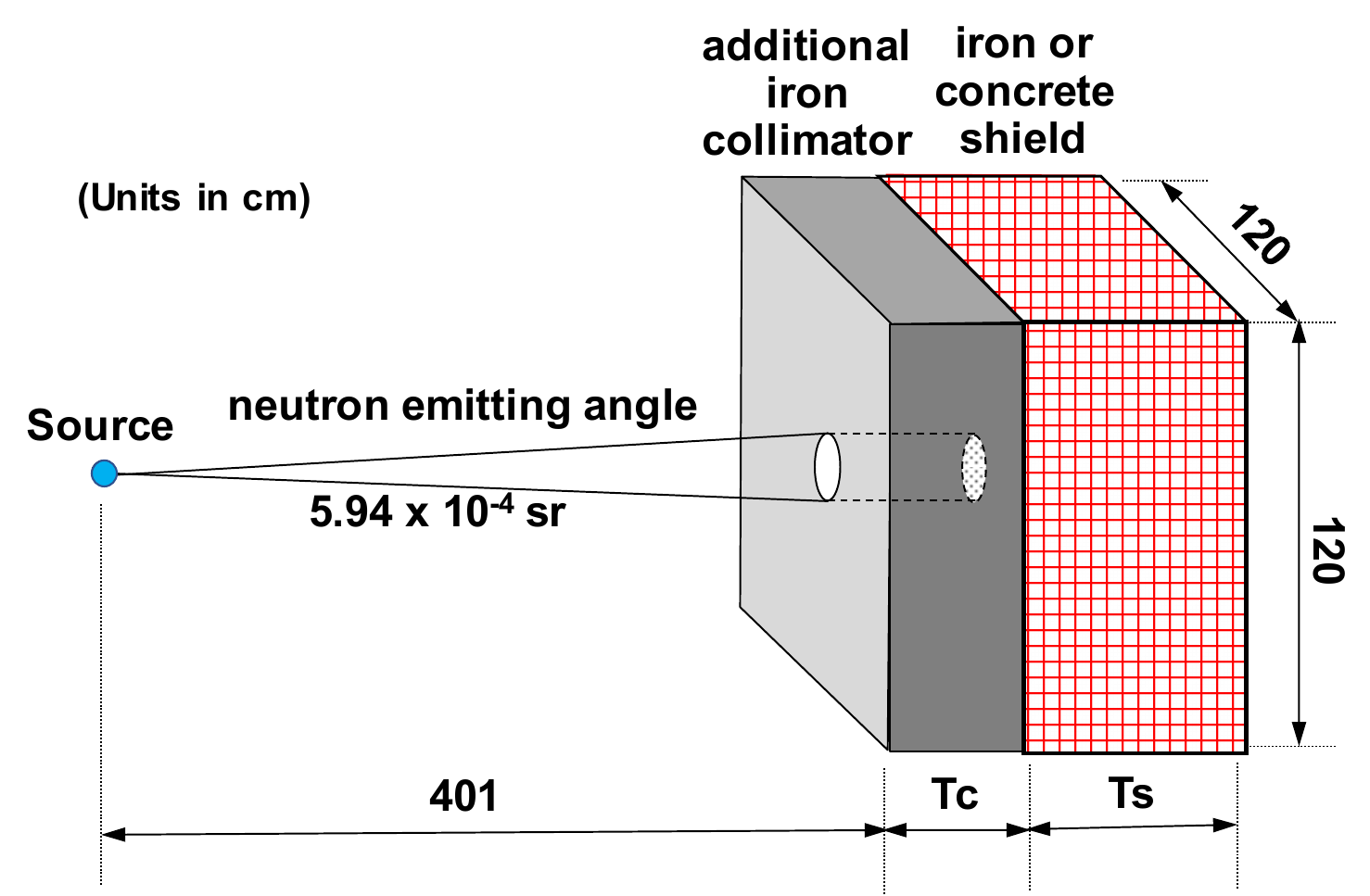}
    \caption{Calculation model of TIARA shielding experiments for iron and concrete.}
\label{fig:figure2}
\end{figure}

So far, several benchmark tests with the experiments have been performed for nuclear data libraries \cite{kwon3,kwon4,kwon5}. Here the experiments were analyzed by using the Monte Carlo code MCNP-6.2~\cite{mcnp62} with the officially distributed ACE files of ENDF/B-VIII.0~\cite{brownENDFBVIII8th2018}, FENDL-3.1d~\cite{kwon6} and FENDL-3.2b~\cite{kwon7}. Note that the $^{1}$H file of FENDL-3.2b was used in the analyses with ENDF/B-VIII.0 because the $^{1}$H file of ENDF/B-VIII.0 had no data above 20 MeV. FENDL-2.1 \cite{DLAFENDL21Processing} was not applied because it has no data above 20 MeV. \Cref{fig:figure2} shows the calculation model for the experiments.\\

\noindent {\it (1) Iron experiment with 40 MeV neutrons\\}

\begin{figure}[htp]
\includegraphics[width=8.69cm]{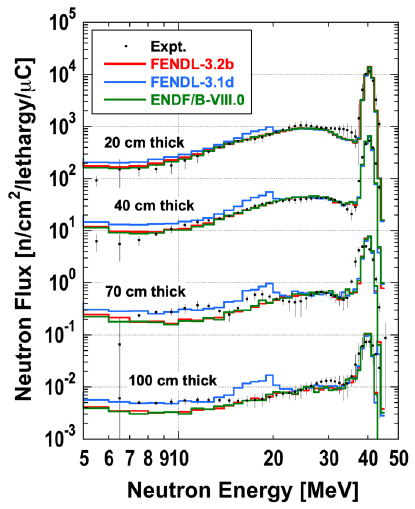}
    \caption{Neutron spectra in iron experiment with 40 MeV neutrons.}
\label{fig:figure3}
\end{figure}

\begin{figure}[htp]
\includegraphics[width=8.69cm]{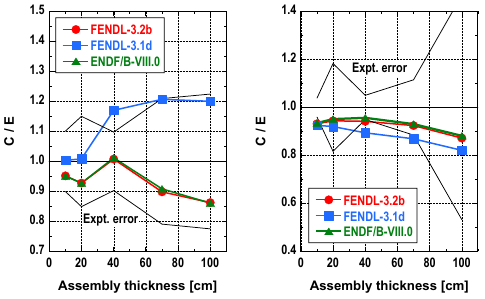}
    \caption{C/E of specified energy neutron fluxes in iron experiment with 40 MeV neutrons.  Left: continuum region (10-35 MeV), Right: peak region(35-45 MeV).}
\label{fig:figure4}
\end{figure}

The measured and calculated neutron spectra on the beam axis in the iron experiment with 40 MeV neutrons are shown in \cref{fig:figure3}. Note that the calculated neutron spectra with FENDL-3.1d unphysically increase below 20 MeV. This is due to wrong secondary neutron spectra just at 20 MeV as pointed out in Ref. \cite{kwon8}. The ratios of the calculated neutron fluxes to the experimental ones (C/E) in a continuum region (10-35 MeV) and a peak region (35-45 MeV) are shown in \cref{fig:figure4}. The overestimation tendency of the continuous neutron flux with FENDL-3.1d is improved in those with FENDL-3.2b. The calculated fluxes with FENDL-3.2b are almost the same as those with ENDF/B-VIII.0.\\

\noindent {\it (2) Iron experiment with 65 MeV neutrons\\}

\begin{figure}[htp]
\includegraphics[width=8.69cm]{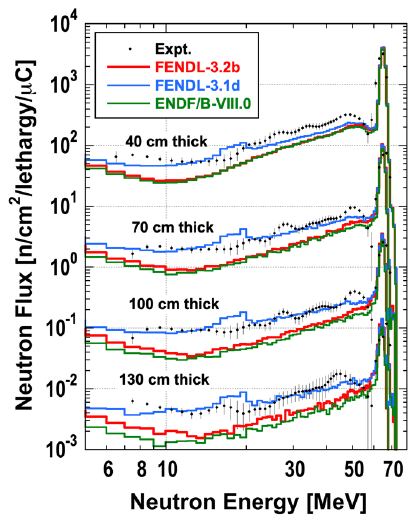}
    \caption{Neutron spectra in iron experiment with 65 MeV neutrons.}
\label{fig:figure5}
\end{figure}

\begin{figure}[htp]
\includegraphics[width=8.69cm]{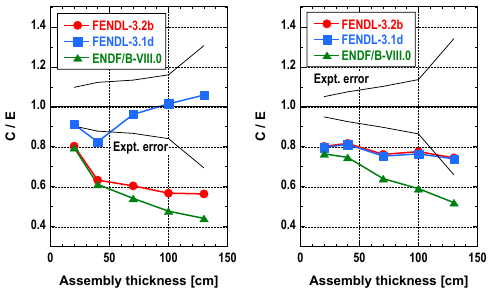}
    \caption{C/E of specified energy neutron fluxes in iron experiment with 65 MeV neutrons. Left: continuum region (10-60 MeV), Right: peak region (60-70 MeV).}
\label{fig:figure6}
\end{figure}

The measured and calculated neutron spectra on the beam axis in the iron experiment with 65 MeV neutrons are shown in \cref{fig:figure5}. The calculated neutron spectra with FENDL-3.1d agree with the measured ones well, though they also have the unphysical dip below 20 MeV due to the wrong secondary neutron data as was seen with the iron experiment with 40 MeV neutrons. On the contrary to calculations with FENDL-3.1d, those with FENDL-3.2b and ENDF/B-VIII.0 underestimate the measured ones below 60 MeV. These are clearly demonstrated in \cref{fig:figure6}, which shows the C/E results of neutron fluxes in a continuum region (10-60 MeV) and a peak region (60-70 MeV). The continuum and peak neutron fluxes with FENDL-3.1d agree with the measured ones. The peak neutron flux with FENDL-3.2b is almost the same as that with FENDL-3.1d, but the continuum neutron flux with FENDL-3.2b underestimates the measured one. The continuum and peak neutron fluxes with ENDF/B-VIII.0 have the poorest agreement with the measured ones.\\

\noindent {\it (3) Concrete experiment with 40 MeV neutrons\\}

\begin{figure}[htp]
\includegraphics[width=8.69cm]{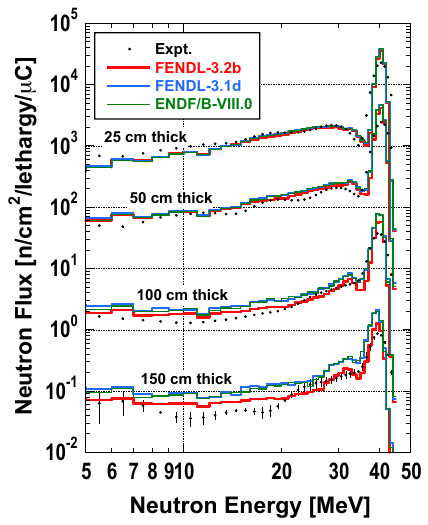}
    \caption{Neutron spectra in concrete experiment with 40 MeV neutrons.}
\label{fig:figure7}
\end{figure}

\begin{figure}[htp]
\includegraphics[width=8.69cm]{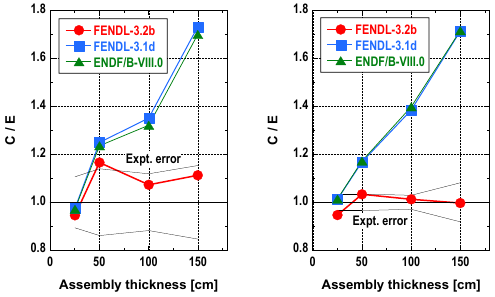}
    \caption{C/E of specified energy neutron fluxes in concrete experiment with 40 MeV neutrons. Left: continuum region (10-35 MeV), Right: peak region (35-45 MeV).}
\label{fig:figure8}
\end{figure}

\Cref{fig:figure7} shows the measured and calculated neutron spectra on the beam axis in the concrete experiment with 40 MeV neutrons. The calculated neutron spectra with FENDL-3.1d and ENDF/B-VIII.0 for the thicker concrete shields clearly overestimate the measured ones. It has been reported that the main reason of this overestimation is the $^{16}$O data above 20 MeV in FENDL-3.1d and ENDF/B-VIII.0 \cite{kwon3,kwon9}. The C/E results of neutron fluxes in the continuum region (10-35 MeV) and the peak region (35-45 MeV) are shown in \cref{fig:figure8}. In both regions, the calculated neutron fluxes with FENDL-3.2b improve the overestimation of those with FENDL-3.1d and ENDF/B-VIII.0.\\

\noindent {\it (4) Concrete experiment with 65 MeV neutrons\\}

\begin{figure}[htp]
\includegraphics[width=8.69cm]{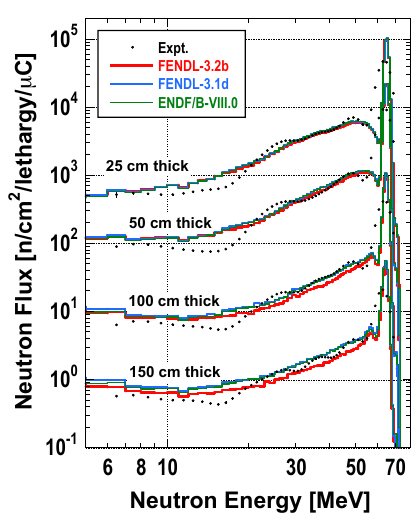}
    \caption{Neutron spectra in concrete experiment with 65 MeV neutrons.}
\label{fig:figure9}
\end{figure}

\begin{figure}[htp]
\includegraphics[width=8.69cm]{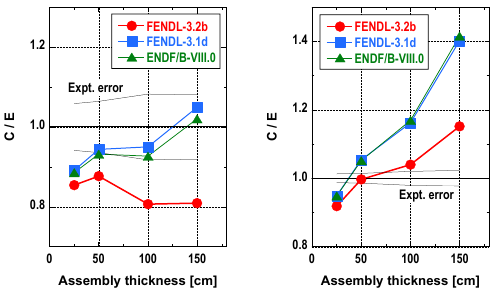}
    \caption{C/E of specified energy neutron fluxes in concrete experiment with 65 MeV neutrons. Left: continuum region (10-60 MeV), Right: peak region (60-70 MeV).}
\label{fig:figure10}
\end{figure}

\Cref{fig:figure9} shows the measured and calculated neutron spectra on the beam axis in the concrete experiment with 65 MeV neutrons. The overestimation of the calculated neutron spectra with FENDL-3.1d and ENDF/B-VIII.0 is smaller than that in the concrete experiment with 40 MeV neutrons. The C/E results of neutron fluxes in the continuum region (10-60 MeV) and the peak region (60-70 MeV) are shown in \cref{fig:figure10}. The calculated fluxes with FENDL-3.2b slightly underestimate the measured ones in the continuum region, while they show the best agreement with the measured ones in the peak region.\\

\subsubsection{FNG Cu, WCLL, W-SS-Water shield}
\label{subsubsec:fngAngelone}
The 14 MeV Frascati neutron generator (FNG) \cite{angelone1fng} has been
used since 1992 to 
run integral experiments (benchmark and/or mock-ups) mainly devoted to
validating nuclear data libraries and calculation tools to be used for
fusion neutronics studies (e.g. nuclear analysis of tokamaks). Owing
to the continuous effort to improve the nuclear data files, some of the FNG
experiments are re-analysed with the latest nuclear data files
in order to test them against experimental data. The newly released
files for FENDL-3.2 have been used to re-analyse three of the FNG
experiments: a benchmark of pure materials, a breeding blanket
mock-up, and the future tungsten shielding experiment (for which at
the moment we only have the comparison with other libraries).
The results of this new analysis are reported hereafter. A comparison
with previously used cross section libraries is also presented.

\begin{figure}[htp]
\includegraphics[width=0.90\columnwidth]{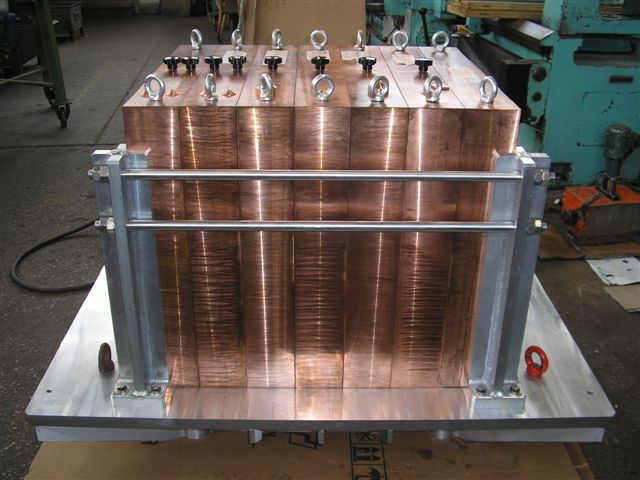}
    \caption{Picture of the Cu block in the FNG experimental benchmark.}
\label{fig:fig1aFNGCublockpicture}
\end{figure}

\begin{figure}[htp]
\includegraphics[width=0.90\columnwidth]{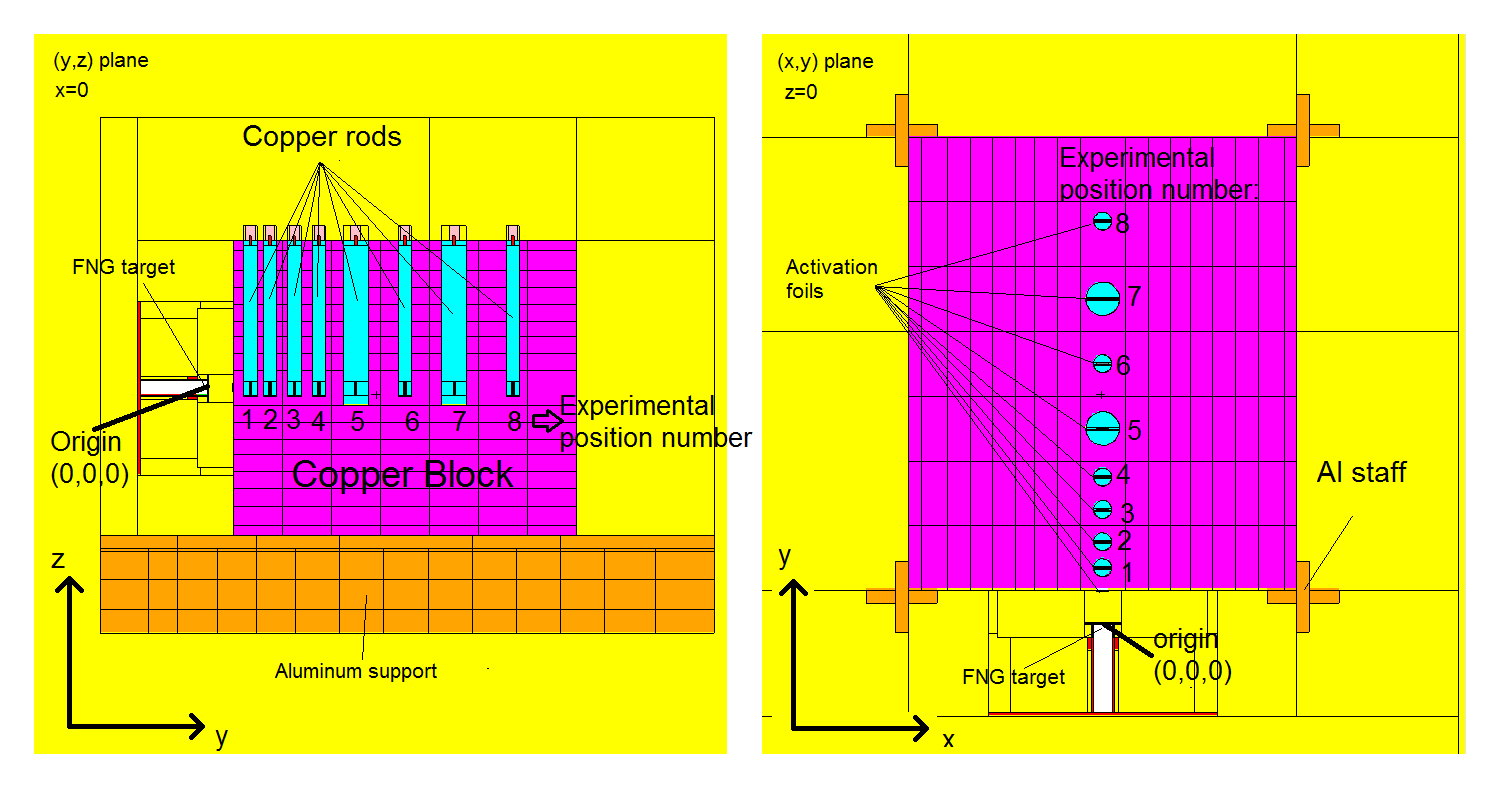}
    \caption{Schematic/MCNP model of Cu block in the FNG experimental benchmark.}
\label{fig:fig1bFNGCublockschem}
\end{figure}

\paragraph{Copper Benchmark Experiment}
A benchmark experiment using a block of pure copper was performed at FNG in 2015 \cite{angelone2fngCupre,angelone3fngCu}. The block dimensions were 600 mm $\times$ 600 mm $\times$ 700 mm (\cref{fig:fig1aFNGCublockpicture}). Eight experimental positions were available along the mid-plane of the block (\cref{fig:fig1bFNGCublockschem}). A set of activation foils was used to study the attenuation properties of copper across the entire neutron energy range relevant to fusion neutronics (from 14 MeV down to thermal energy).
The measured reaction rates were:
$ ^{197}Au(n,\gamma)^{198}Au$, $^{186}W(n,\gamma)^{187}W$, $^{115}In(n,n^{\prime})^{115}In$, $^{58}Ni(n,p)^{58}Co$, $^{27}Al(n,\alpha)^{24}Na$, $^{93}Nb(n,2n)^{92}Nb^{m}$.
The calculated reaction rates (C) were compared to the experimental reaction rates (E) in terms of C/E. Calculations were performed using the MCNP-5 code~\cite{mcnp5} and JEFF-3.1, JEFF-3.2, and subsequently with JEFF-3.3~\cite{plompenJointEvaluatedFission2020a} libraries for transport.
The reaction rates were calculated using the IRDFF version 1.05 dosimetry file~\cite{angeloneIRDFF105}.
The present analysis was performed using the available MCNP models and the FENDL-3.2a library.

The results in terms of the C/E ratio calculated with FENDL-3.2a are reported in \cref{table:angeloneFNGcuCoverE}.  \Cref{table:angeloneFNGcuCoverCpart1} and \cref{table:angeloneFNGcuCoverCpart2} reports a comparison of the calculated reaction rates obtained with FENDL-3.2a with those obtained using other libraries in terms of a ratio between the result using FENDL-3.2a and the result using the other library.
Looking at these tables, there is generally good agreement in C/E with the fast reaction activation foils, i.e. Nb, Ni and Al. The biggest discrepancies were in the thermal reaction rates, i.e. $ ^{197}Au(n,\gamma)^{198}Au$ and $^{186}W(n,\gamma)^{187}W$, with an average C/E of 0.7 and 0.5 respectively.
The comparison of reaction rates calculated with the various libraries in this experiment shows that in general, the calculations with the FENDL-3.2a library are in good agreement with FENDL-3.0 and JEFF-3.1.1.  Some relevant discrepancies are found with JEFF-3.2, in particular, the reaction rates calculated with FENDL-3.2a tend to be higher (up to 27\% for the $^{115}In(n,n^{\prime})$ reaction) than those calculated with JEFF-3.2.  The C/E values obtained using JEFF-3.3 closely reproduce those already discussed for JEFF-3.1.1 for all detectors.

\begin{table*}[htp]
\caption{C/E in activation foils inside the FNG copper benchmark experiment calculated with FENDL-3.2a. Relative errors (R.E.) take into account experimental and statistical uncertainty.}
\label{table:angeloneFNGcuCoverE}
\begin{tabular}{l | c c | c c | c c | c c | c c | c c | c c} \hline \hline
           & \multicolumn{2}{|c|}{$^{27}Al(n,\alpha)$} & \multicolumn{2}{|c|}{$^{58}Ni(n,p)$} & \multicolumn{2}{|c|}{$^{93}Nb(n,2n)$} & \multicolumn{2}{|c|}{$^{186}W(n,\gamma)$} & \multicolumn{2}{|c|}{$^{115}In(n,n^{\prime})$} & \multicolumn{2}{|c|}{$ ^{197}Au(n,2n)$} & \multicolumn{2}{|c}{$ ^{197}Au(n,\gamma)$} \\ \hline
Depth (cm) & C/E & R.E. & C/E & R.E. & C/E & R.E. & C/E & R.E. & C/E & R.E. & C/E & R.E. & C/E & R.E. \\ \hline 
3.5 & 0.95 & 4.00\% &  &  &  &  & 0.71 & 5.00\% & 0.87 & 5.00\% & 0.92 & 5.00\% & 0.91 & 7.00\% \\ \hline 
7.5 & 0.89 & 4.00\% & 1.00 & 5.00\% & 0.89 & 5.00\% & 0.64 & 5.00\% & 0.82 & 5.00\% & 0.86 & 5.00\% & 0.85 & 6.00\% \\ \hline 
12.5 & 0.9 & 4.00\% & 1.00 & 5.00\% & 0.87 & 5.00\% & 0.58 & 5.00\% & 0.8 & 5.00\% & 0.86 & 5.00\% & 0.79 & 5.00\% \\ \hline 
17.5 & 0.92 & 4.00\% & 1.04 & 5.00\% & 0.89 & 5.00\% & 0.54 & 5.00\% & 0.78 & 5.00\% & 0.84 & 5.00\% & 0.75 & 5.00\% \\ \hline 
24.9 & 0.84 & 5.00\% & 1.01 & 5.00\% & 0.88 & 5.00\% & 0.46 & 5.00\% & 0.77 & 5.00\% & 0.83 & 5.00\% & 0.69 & 5.00\% \\ \hline 
35 & 0.89 & 6.00\% & 1.06 & 5.00\% & 0.86 & 5.00\% & 0.43 & 5.00\% & 0.78 & 5.00\% &  &  & 0.65 & 5.00\% \\ \hline 
45 & 0.88 & 5.00\% & 1.02 & 9.00\% & 0.83 & 5.00\% & 0.41 & 5.00\% & 0.72 & 5.00\% &  &  & 0.61 & 5.00\% \\ \hline 
57 & 0.77 & 11.00\% & 0.38 & 12.00\% & 0.91 & 5.00\% & 0.4 & 5.00\% & 0.68 & 6.00\% &  &  & 0.58 & 5.00\% \\ \hline \hline
\end{tabular}
\end{table*}

\begin{table*}[htp]
\caption{Ratio between FENDL-3.2a and other libraries (FENDL-3.2a/[other lib.]) for the FNG copper benchmark experiment (Part 1).}
\label{table:angeloneFNGcuCoverCpart1}
\begin{tabular}{l | c c c | c c c | c c c | c c c } \hline \hline
     & \multicolumn{3}{|c|}{$^{27}Al(n,\alpha)$} & \multicolumn{3}{|c|}{$^{58}Ni(n,p)$} & \multicolumn{3}{|c|}{$^{93}Nb(n,2n)$} & \multicolumn{3}{|c}{$^{186}W(n,\gamma)$} \\ \hline
Depth (cm) & FENDL & JEFF & JEFF & FENDL & JEFF & JEFF & FENDL & JEFF & JEFF & FENDL & JEFF & JEFF \\
  & 3.0 & 3.1.1 &  3.2 &  3.0 &  3.1.1 &  3.2 &  3.0 &  3.1.1 &  3.2 &  3.0 &  3.1.1 &  3.2 \\ \hline
3.5 & 1.00 & 1.00 & 1.01 &  &  &  & 1.00 & 1.00 & 1.01 & 1.04 & 1.05 & 1.07 \\ \hline 
7.5 & 1.00 & 1.00 & 1.03 & 1.00 & 1.00 & 1.03 & 1.00 & 1.00 & 1.02 & 1.02 & 1.01 & 1.09 \\ \hline 
12.5 & 1.00 & 1.00 & 1.04 & 1.00 & 1.00 & 1.04 & 1.00 & 1.00 & 1.04 & 0.96 & 1.04 & 1.04 \\ \hline 
17.5 & 1.00 & 1.00 & 1.06 & 1.00 & 1.00 & 1.05 & 1.00 & 1.00 & 1.06 & 1.01 & 1.01 & 1.1 \\ \hline 
24.9 & 1.00 & 1.00 & 1.08 & 1.00 & 1.00 & 1.06 & 1.00 & 0.99 & 1.08 & 0.99 & 0.99 & 1.13 \\ \hline 
35 & 1.00 & 1.00 & 1.12 & 1.00 & 0.99 & 1.08 & 1.00 & 1.00 & 1.11 & 0.97 & 0.98 & 1.11 \\ \hline 
45 & 0.98 & 0.99 & 1.14 & 0.99 & 0.99 & 1.11 & 0.99 & 0.99 & 1.15 & 1.00 & 1.00 & 1.12 \\ \hline 
57 & 1.02 & 1.01 & 1.18 & 1.00 & 1.00 & 1.14 & 0.99 & 0.99 & 1.15 & 0.99 & 0.99 & 1.1 \\ \hline \hline
\end{tabular}
\end{table*}

\begin{table*}[htp]
\caption{Ratio between FENDL-3.2a and other libraries (FENDL-3.2a/[other lib.]) for the FNG copper benchmark experiment (Part 2).}
\label{table:angeloneFNGcuCoverCpart2}
\begin{tabular}{l | c c c | c c c | c c c } \hline \hline
     & \multicolumn{3}{|c|}{$^{115}In(n,n^{\prime})$} & \multicolumn{3}{|c|}{$ ^{197}Au(n,2n)$} & \multicolumn{3}{|c|}{$ ^{197}Au(n,\gamma)$} \\ \hline
Depth (cm) & FENDL & JEFF & JEFF & FENDL & JEFF & JEFF & FENDL & JEFF & JEFF \\
  & 3.0 & 3.1.1 & 3.2 & 3.0 & 3.1.1 & 3.2 & 3.0 & 3.1.1 & 3.2 \\ \hline
3.5 & 1.00 & 1.00 & 1.03 & 1.00 & 1.00 & 1.01 & 1.07 & 1.05 & 1.05 \\ \hline 
7.5 & 1.00 & 1.00 & 1.04 & 1.00 & 1.00 & 1.03 & 1.04 & 1.04 & 1.05 \\ \hline 
12.5 & 1.00 & 1.00 & 1.05 & 1.00 & 1.00 & 1.04 & 1.01 & 1.00 & 1.02 \\ \hline 
17.5 & 1.00 & 1.00 & 1.07 & 1.00 & 1.00 & 1.06 & 1.01 & 1.00 & 1.02 \\ \hline 
24.9 & 1.00 & 1.00 & 1.1 & 0.99 & 0.99 & 1.08 & 1.00 & 0.98 & 1.01 \\ \hline 
35 & 1.00 & 1.00 & 1.15 &  &  &  & 0.99 & 0.98 & 1.00 \\ \hline 
45 & 1.00 & 1.00 & 1.2 &  &  &  & 1.01 & 1.00 & 0.98 \\ \hline 
57 & 0.99 & 1.00 & 1.27 &  &  &  & 1.00 & 0.99 & 0.94 \\ \hline \hline
\end{tabular}
\end{table*}

\paragraph{DEMO WCLL blanket mock-up}
A mock-up of the DEMO Water Cooled Lithium Lead (WCLL) breeding blanket has been realized and irradiated at FNG~\cite{angelone4fngwcll,angelone4fngwcllfusengdes}. The WCLL mock-up is made of LiPb and Eurofer plus some SS-316 supporting structure (cage). The main features of the mock-up are:
\begin{itemize}
\item EUROFER was used in the breeding zone (11 slabs). SS-316 was used for modeling the external cage enveloping the LiPb breeding zone as well as the manifold region (at the bottom of the mock-up)
\item 110 LiPb bricks were used for modeling the breeding zone. All the experimental positions are located inside the LiPb region
\item Perspex (Polymethyl-methacrylate or PMMA, $(C_{5}O_{2}H_{8})_{n}$, nominal density = 1.19 $g/cm^{3}$) was used to mimic water
\item A 2 mm thick tungsten layer was added to the first wall (FW) of the mock-up to better match the expected design (W of comparable thickness is foreseen as plasma facing material in the WCLL-Breeding Blanket (BB) of DEMO)
\end{itemize}

\begin{figure}[htp]
\includegraphics[width=0.90\columnwidth]{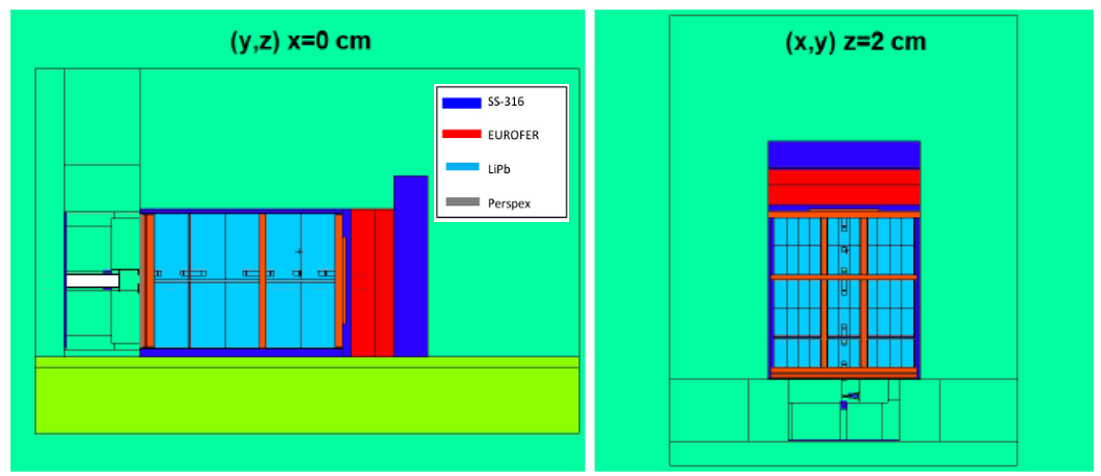}
    \caption{Schematic/MCNP model of DEMO WCLL mock-up in the FNG experimental benchmark.}
\label{fig:fig2FNGWCLLschem}
\end{figure}

The mock-up geometry exhibits strong heterogeneity and was modeled in detail using MCNP, see \cref{fig:fig2FNGWCLLschem}. Reaction rates were measured in seven positions located inside the LiPb zone.  The following reaction rates were measured:
$ ^{197}Au(n,\gamma)^{198}Au$, $^{186}W(n,\gamma)^{187}W$, $^{115}In(n,n^{\prime})^{115}In$, $^{58}Ni(n,p)^{58}Co$, $^{27}Al(n,\alpha)^{24}Na$, $^{93}Nb(n,2n)^{92}Nb^{m}$.
The calculated reaction rates (C) were compared to the experimental reaction rates (E) in terms of C/E. Calculations were performed using the MCNP-5 code and JEFF-3.3 and FENDL-3.2a library for transport.
The reaction rates were calculated using the IRDFF version 2 dosimetry file~\cite{IRDFFII}.

\begin{figure*}[htp]
\includegraphics[width=17.75cm]{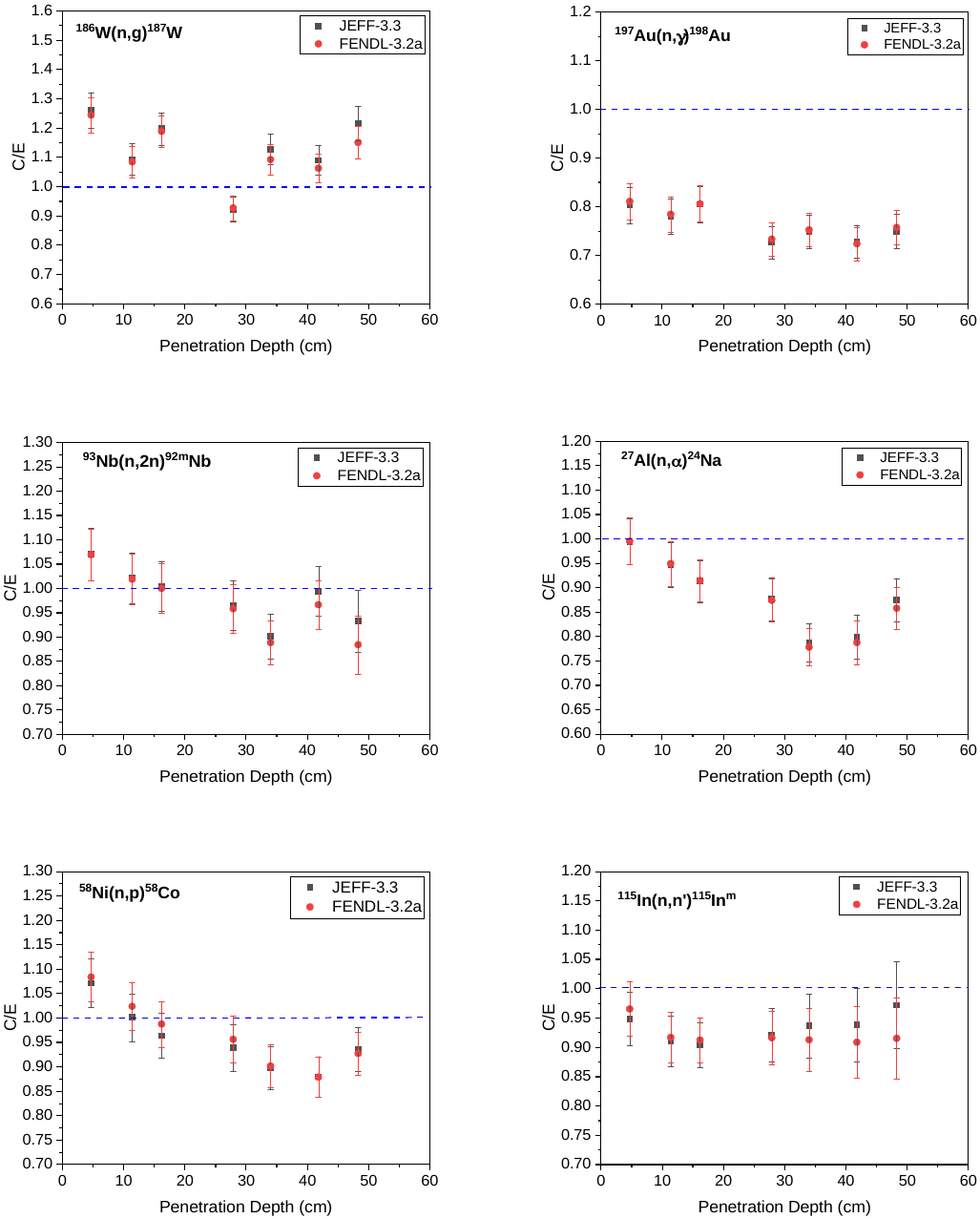}
    \caption{C/E of the FNG WCLL mock-up experiment for JEFF-3.3. and FENDL-3.2a as a function of the penetration depth.}
\label{fig:fig3FNGWCLLCoverE}
\end{figure*}

The results of the present analysis of the experiment in terms of C/E values for JEFF-3.3 and FENDL-3.2a are reported in \cref{table:angeloneFNGwcllCoverEpart1} and \cref{table:angeloneFNGwcllCoverEpart2}.  Additionally, \cref{fig:fig3FNGWCLLCoverE} shows plots of the C/E values for JEFF-3.3 and FENDL-3.2a as a function of the penetration depth for the reactions investigated. Values of C/E are generally good with the exception of thermal reactions, i.e. $ ^{197}Au(n,\gamma)^{198}Au$ and $^{186}W(n,\gamma)^{187}W$.
From these tables one can also see that C/E values calculated with FENDL-3.2a generally replicate those calculated with JEFF-3.3. Some minor discrepancies are found at large penetration depths, for example in tungsten where FENDL-3.2a seems to improve the agreement with the experiment by a few percent.  Furthermore, a discrepancy is observed at large penetration depths in the case of indium where the agreement between FENDL-3.2a and experimental data is slightly lower compared to JEFF-3.3.

The comparison of reaction rates calculated with FENDL-3.2a versus JEFF-3.3 is shown in \cref{table:angeloneFNGwcllCoverC}.  This table shows a general tendency to obtain lower reaction rates at large penetration depth for FENDL-3.2a versus JEFF-3.3. The discrepancies are in many cases statistically significant. 

\begin{table*}[htp]
\caption{C/E in activation foils inside the FNG WCLL mock-up calculated with JEFF-3.3 and FENDL-3.2a. Relative errors (R.E.) take into account experimental and statistical uncertainty (Part 1).}
\label{table:angeloneFNGwcllCoverEpart1}
\begin{tabular}{l | c c | c c | c c | c c | c c | c c } \hline \hline
 & \multicolumn{4}{|c|}{$^{93}Nb(n,2n)$} & \multicolumn{4}{|c|}{$ ^{197}Au(n,\gamma)$} & \multicolumn{4}{|c}{$^{115}In(n,n^{\prime})$} \\ \hline
Depth (cm) & JEFF-3.3 & R.E. & FENDL-3.2a & R.E. & JEFF-3.3 & R.E. & FENDL-3.2a & R.E. & JEFF-3.3 & R.E. & FENDL-3.2a & R.E. \\ \hline 
4.7 & 1.07 & 5.00\% & 1.07 & 5.00\% & 0.8 & 5.00\% & 0.81 & 5.00\% & 0.95 & 5.00\% & 0.97 & 5.00\% \\ \hline 
11.4 & 1.02 & 5.00\% & 1.02 & 5.00\% & 0.78 & 5.00\% & 0.78 & 5.00\% & 0.91 & 5.00\% & 0.92 & 5.00\% \\ \hline 
16.2 & 1.00 & 5.00\% & 1.00 & 5.00\% & 0.8 & 5.00\% & 0.81 & 5.00\% & 0.9 & 4.00\% & 0.91 & 4.00\% \\ \hline 
27.9 & 0.96 & 5.00\% & 0.96 & 5.00\% & 0.73 & 5.00\% & 0.73 & 5.00\% & 0.92 & 5.00\% & 0.92 & 5.00\% \\ \hline 
34 & 0.9 & 5.00\% & 0.89 & 5.00\% & 0.75 & 5.00\% & 0.75 & 5.00\% & 0.94 & 6.00\% & 0.91 & 6.00\% \\ \hline 
41.8 & 0.99 & 5.00\% & 0.97 & 5.00\% & 0.73 & 5.00\% & 0.72 & 5.00\% & 0.94 & 7.00\% & 0.91 & 7.00\% \\ \hline 
48.3 & 0.93 & 7.00\% & 0.88 & 7.00\% & 0.75 & 5.00\% & 0.76 & 5.00\% & 0.97 & 8.00\% & 0.92 & 8.00\% \\ \hline \hline 
\end{tabular}
\end{table*}

\begin{table*}[htp]
\caption{C/E in activation foils inside the FNG WCLL mock-up calculated with JEFF-3.3 and FENDL-3.2a. Relative errors (R.E.) take into account experimental and statistical uncertainty (Part 2).}
\label{table:angeloneFNGwcllCoverEpart2}
\begin{tabular}{l | c c | c c | c c | c c | c c | c c } \hline \hline
 & \multicolumn{4}{|c|}{$^{186}W(n,\gamma)$} & \multicolumn{4}{|c|}{$^{27}Al(n,\alpha)$} & \multicolumn{4}{|c}{$^{58}Ni(n,p)$} \\ \hline
Depth (cm) & JEFF-3.3 & R.E. & FENDL-3.2a & R.E. & JEFF-3.3 & R.E. & FENDL-3.2a & R.E. & JEFF-3.3 & R.E. & FENDL-3.2a & R.E. \\ \hline
4.7 & 1.26 & 5.00\% & 1.24 & 5.00\% & 0.99 & 5.00\% & 1.00 & 5.00\% & 1.07 & 5.00\% & 1.08 & 5.00\% \\ \hline 
11.4 & 1.09 & 5.00\% & 1.08 & 5.00\% & 0.95 & 5.00\% & 0.95 & 5.00\% & 1.00 & 5.00\% & 1.02 & 5.00\% \\ \hline 
16.2 & 1.2 & 5.00\% & 1.19 & 5.00\% & 0.91 & 5.00\% & 0.91 & 5.00\% & 0.96 & 5.00\% & 0.99 & 5.00\% \\ \hline 
27.9 & 0.92 & 5.00\% & 0.93 & 5.00\% & 0.88 & 5.00\% & 0.87 & 5.00\% & 0.94 & 5.00\% & 0.96 & 5.00\% \\ \hline 
34 & 1.13 & 5.00\% & 1.09 & 5.00\% & 0.79 & 5.00\% & 0.78 & 5.00\% & 0.9 & 5.00\% & 0.9 & 5.00\% \\ \hline 
41.8 & 1.09 & 5.00\% & 1.06 & 5.00\% & 0.8 & 6.00\% & 0.79 & 6.00\% & 0.88 & 5.00\% & 0.88 & 5.00\% \\ \hline 
48.3 & 1.22 & 5.00\% & 1.15 & 5.00\% & 0.88 & 5.00\% & 0.86 & 5.00\% & 0.94 & 5.00\% & 0.93 & 5.00\% \\ \hline \hline
\end{tabular}
\end{table*}

\begin{table*}[htp]
\caption{Ratio between reaction rates calculated with FENDL-3.2a and JEFF-3.3 for the FNG WCLL mock-up experiment.}
\label{table:angeloneFNGwcllCoverC}
\begin{tabular}{l | c c | c c | c c | c c | c c | c c} \hline \hline
   & \multicolumn{2}{|c|}{$^{93}Nb(n,2n)$} & \multicolumn{2}{|c|}{$ ^{197}Au(n,\gamma)$} & \multicolumn{2}{|c}{$^{115}In(n,n^{\prime})$} & \multicolumn{2}{|c|}{$^{186}W(n,\gamma)$} & \multicolumn{2}{|c|}{$^{27}Al(n,\alpha)$} & \multicolumn{2}{|c}{$^{58}Ni(n,p)$} \\ \hline
Depth (cm) & Ratio & R.E. & Ratio & R.E. & Ratio & R.E. & Ratio & R.E. & Ratio & R.E. & Ratio & R.E. \\ \hline 
4.7 & 1.00 & 0.06\% & 1.01 & 0.25\% & 1.02 & 0.09\% & 0.99 & 0.42\% & 1.00 & 0.09\% & 1.01 & 0.06\% \\ \hline 
11.4 & 1.00 & 0.10\% & 1.01 & 0.21\% & 1.01 & 0.13\% & 0.99 & 0.36\% & 1.00 & 0.09\% & 1.02 & 0.06\% \\ \hline 
16.2 & 1.00 & 0.14\% & 1.00 & 0.25\% & 1.01 & 0.19\% & 0.99 & 0.42\% & 1.00 & 0.09\% & 1.02 & 0.07\% \\ \hline 
27.9 & 0.99 & 0.51\% & 1.01 & 0.48\% & 1.00 & 0.37\% & 1.01 & 0.81\% & 1.00 & 0.17\% & 1.02 & 0.17\% \\ \hline 
34 & 0.99 & 0.76\% & 1.00 & 0.59\% & 0.97 & 0.51\% & 0.97 & 1.00\% & 0.99 & 0.24\% & 1.00 & 0.24\% \\ \hline 
41.8 & 0.97 & 1.18\% & 0.99 & 0.64\% & 0.97 & 0.68\% & 0.97 & 1.07\% & 0.99 & 0.37\% & 1.00 & 0.39\% \\ \hline 
48.3 & 0.95 & 1.77\% & 1.01 & 1.00\% & 0.94 & 1.01\% & 0.95 & 1.65\% & 0.98 & 0.56\% & 0.99 & 0.57\% \\ \hline \hline
\end{tabular}
\end{table*}

\begin{figure}[htp]
\includegraphics[width=0.90\columnwidth]{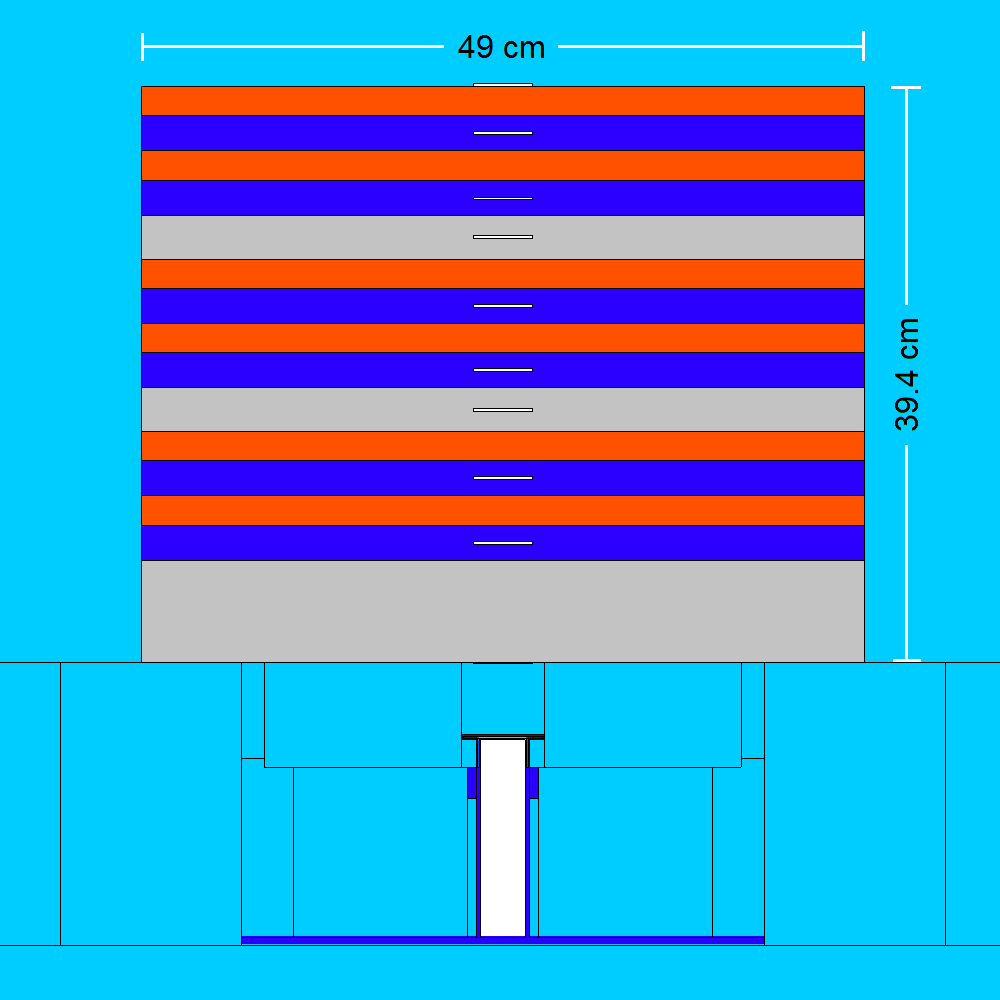}
    \caption{Schematic/MCNP model of the tungsten shielding mock-up in the FNG experiment. The tungsten alloy DENSIMET-180 is gray, SS-316 slabs are blue, and the Perspex slabs are red.}
\label{fig:fig4FNGWSSWaterschem}
\end{figure}

\paragraph{Tungsten-SS-Water shielding mock-up}
A tungsten shielding experiment is currently ongoing at the FNG and planned to be completed in summer 2023. The scope of the experiment is to study the neutron transport in a reactor-like shielding element made of tungsten, stainless steel, and water. A pre-analysis has been performed in order to optimize the experimental configuration~\cite{angelone5fngWSSWater}.
The selected configuration, shown in \cref{fig:fig4FNGWSSWaterschem} is a 49 cm $\times$ 39.4 cm $\times$ 45 cm block, composed of a 7 cm thick DENSIMET-180 (W alloy with 95\% W) slab (in gray), two 2.4 cm thick SS-316 slabs (in blue) alternated  with two 2 cm thick Perspex slabs (in red-orange). The sequence of these slabs is repeated two times to complete the block, except in these repeated structures, the DENISMET-180 slabs in the center of the block are 3 cm thick rather than 7 cm. The pre-analysis has been carried out using JEFF-3.3 nuclear data library for transport and IRDFF-II for dosimetry and it has been repeated with FENDL-3.2a for this work.

The ratios of reaction rates calculated with FENDL-3.2a to those calculated with JEFF-3.3 are shown in \cref{table:angeloneFNGWSSWaterCoverC} for the reactions investigated.  The location of the positions for the calculated reaction rates are indicated in \cref{fig:fig4FNGWSSWaterschem}.  In general, there is a close agreement between the reaction rates calculated with FENDL-3.2a and those calculated with JEFF-3.3.   A tendency to obtain slightly lower reaction rates with FENDL-3.2a is observed as was seen with the WCLL mock-up, which notably has significantly different materials.

\begin{table*}[htp]
\caption{Ratio between reaction rates calculated with FENDL-3.2a and JEFF-3.3 for the FNG Tungsten-SS-Water shielding mock-up experiment.}
\label{table:angeloneFNGWSSWaterCoverC}
\begin{tabular}{l | c c | c c | c c | c c | c c | c c | c c | c c } \hline \hline
Pos. & \multicolumn{2}{|c|}{$^{27}Al(n,\alpha)$} & \multicolumn{2}{|c|}{$ ^{197}Au(n,2n)$} & \multicolumn{2}{|c|}{$ ^{197}Au(n,\gamma)$} & \multicolumn{2}{|c|}{$^{115}In(n,n^{\prime})$} & \multicolumn{2}{|c|}{$^{93}Nb(n,2n)$} & \multicolumn{2}{|c|}{$^{58}Ni(n,2n)$} & \multicolumn{2}{|c|}{$^{58}Ni(n,p)$} & \multicolumn{2}{|c}{$^{186}W(n,\gamma)$} \\ \hline
1 & 0.99 & 0.02\% & 0.99 & 0.02\% & 0.98 & 4.08\% & 1.03 & 0.03\% & 0.99 & 0.02\% & 0.99 & 0.02\% & 1.00 & 0.02\% & 1.00 & 5.41\% \\ \hline 
2 & 0.99 & 0.07\% & 0.99 & 0.07\% & 0.98 & 0.74\% & 1.02 & 0.07\% & 0.99 & 0.07\% & 0.99 & 0.07\% & 1.00 & 0.07\% & 0.97 & 1.62\% \\ \hline 
3 & 0.99 & 0.11\% & 0.99 & 0.11\% & 0.99 & 0.19\% & 1.00 & 0.10\% & 0.99 & 0.11\% & 0.99 & 0.11\% & 0.99 & 0.11\% & 0.99 & 0.25\% \\ \hline 
4 & 0.97 & 0.16\% & 0.97 & 0.16\% & 0.98 & 1.16\% & 0.99 & 0.14\% & 0.97 & 0.16\% & 0.97 & 0.16\% & 0.97 & 0.16\% & 0.95 & 0.18\% \\ \hline 
5 & 0.98 & 0.20\% & 0.98 & 0.20\% & 1.01 & 0.77\% & 0.98 & 0.18\% & 0.98 & 0.20\% & 0.98 & 0.21\% & 0.97 & 0.19\% & 1.01 & 1.42\% \\ \hline 
6 & 0.98 & 0.28\% & 0.99 & 0.28\% & 1.03 & 0.72\% & 0.97 & 0.26\% & 0.99 & 0.29\% & 0.99 & 0.29\% & 0.97 & 0.27\% & 1.01 & 1.32\% \\ \hline 
7 & 0.96 & 0.41\% & 0.97 & 0.41\% & 0.98 & 2.70\% & 0.96 & 0.37\% & 0.97 & 0.42\% & 0.97 & 0.43\% & 0.95 & 0.40\% & 0.97 & 0.47\% \\ \hline 
8 & 0.99 & 0.76\% & 0.99 & 0.77\% & 1.02 & 1.58\% & 0.96 & 0.73\% & 0.99 & 0.78\% & 0.99 & 0.79\% & 0.97 & 0.74\% & 1.03 & 1.99\% \\ \hline 
9 & 0.98 & 0.97\% & 0.99 & 0.96\% & 1.03 & 4.64\% & 0.96 & 1.12\% & 0.99 & 0.97\% & 0.99 & 0.99\% & 0.98 & 1.05\% & 1.02 & 5.16\% \\ \hline \hline
\end{tabular}
\end{table*}

\subsubsection{Research Center Rez 10.7 and 12.7 MeV quasi-monoenergetic neutron source: Dosimetrical reactions}
\label{subsubsec:monoESchulc}
\begin{figure}[htp]
\includegraphics[width=0.90\columnwidth]{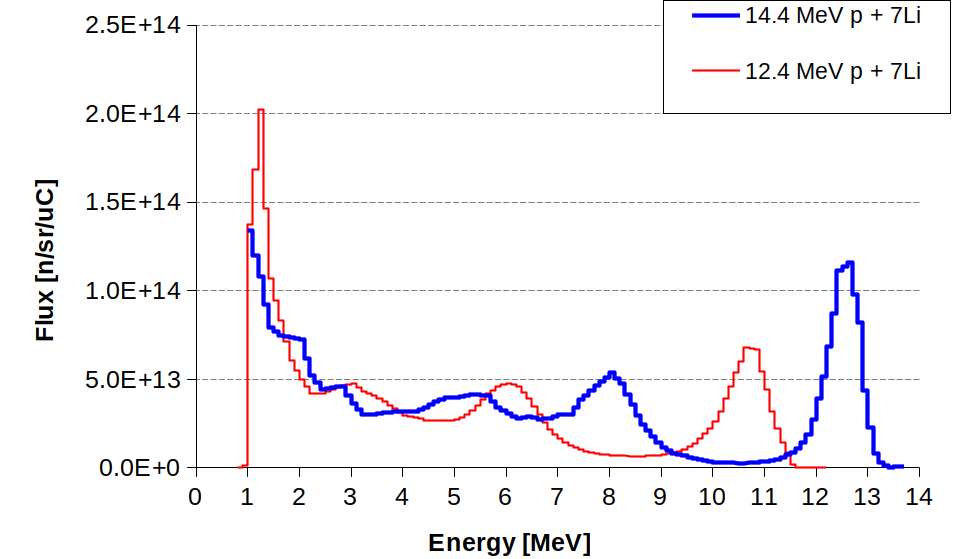}
    \caption{U120M cyclotron based neutron source spectra induced by 14.4 and 12.4 MeV protons interacting with a $^{nat.}Li$ target \cite{schulcMonoEmatej2022}.}
\label{fig:fig1srcNeutronSpectra}
\end{figure}

The dosimetric cross sections are fundamental quantities for the proper
determination of neutron fluxes using the activation method. In fusion applications, it is
important to validate this quantity, especially in regions close to a
fusion neutron energy spectrum. Activation reactions that are
dosimetric are taken from IRDFF-II as it is the recommended dosimetry
library for reaction dosimetry and fusion.
The validation of a set of
dosimeters was performed in a quasi-monoenergetic neutron
field established by proton irradiation of a thin Li target with protons of
energies 12.4 and 14.4 MeV. The neutron spectrum was measured in
position approximately 4 m from the neutron source given by a
$^\textrm{nat}$Li target irradiated by protons. The neutron spectra are shown in \cref{fig:fig1srcNeutronSpectra}.

The activation experiment was realized simultaneously with the
measurement of the neutron spectrum. To achieve a sufficient activation,
the experiment was performed in close distance to the target
assembly. Due to the relatively large volume of materials in vicinity
of the target assembly, the neutron spectrum in the irradiated foils
differs from the spectrum measured far from the target. The correction
factors given as ratios between the spectrum at the location of the measurement
and the spectrum in the target were determined by simulation.
These ratios were used to compute the neutron spectra at the position of
the activation foils (8.6 cm away from the Li target) based on the measured spectra
at the position of the scintillation detectors.
The flux was normalized
using Al and Ni monitors relying on the $^{27}Al(n,\alpha)$ and $^{58}Ni(n,p)$
reactions. These evaluated neutron fluxes in the target can be used
for the determination of evaluated reaction rates by folding
it with data from a nuclear data library. The details of
this methodology can be found in \cite{schulcMonoEmatej2022}.

The time evolution of the activation and the final activity measured
in the irradiation experiment allowed us to derive the reaction rates. These
experimentally obtained reaction rates were compared with evaluated
reaction rates obtained by folding the evaluated neutron spectra in the
target with a selected nuclear data library.
We employed this methodology to test a large set of commonly used dosimetry reactions
including data from IRDFF-II~\cite{IRDFFII}, ENDF/B-VIII~\cite{brownENDFBVIII8th2018},
JEFF-3.3~\cite{plompenJointEvaluatedFission2020a} and JENDL-4~\cite{JENDL4.0}.

\Cref{table:schulcMonoE12MeVpCoverE,table:schulcMonoE14MeVpCoverE} show the differences
between the calculated and the measured
reaction rates in terms of $C/E-1$ for the two neutron sources
investigated.  Uncertainties are also shown in the tables.
For common dosimetry reactions, 
the best agreement was obtained by using the IRDFF-II
evaluation. All of the considered IRDFF-II reactions are within two-sigma
uncertainty. The other libraries tested contain several reactions with
higher discrepancy. The most discrepant one is $^{19}F(n,2n)$ in JENDL-4
for which the discrepancy exceeds ten sigma. 

\begin{table*}[htp]
\caption{Calculated versus measured reaction rates in $^{7}Li(p,n)$ field from
  12.4 MeV protons in terms of C/E-1 \cite{schulcMonoEmatej2022}.} 
\label{table:schulcMonoE12MeVpCoverE}
\begin{tabular}{l | c | c c c c c} \hline \hline
Reaction & E50\% [MeV] & IRDFF-II & JEFF-3.3 & JENDL-4 & ENDF/B-VIII.0 & Uncertainty\\ \hline 
$^{54}Fe(n,p)$ & 6.78 & 1.90\% & -2.30\% & 5.20\% & 2.00\% & 4.20\% \\ \hline 
$^{47}Ti(n,p)$ & 7.43 & 7.80\% & 8.40\% &  \textbf{11.70\%} &  \textbf{23.60\%} & 3.70\% \\ \hline 
$^{46}Ti(n,p)$ & 9.61 & 11.70\% & 12.20\% & 9.80\% & 4.80\% & 8.30\% \\ \hline 
$^{59}Co(n,p)$ & 10.05 & 3.20\% & -7.10\% & -5.50\% & 3.20\% & 4.10\% \\ \hline 
$^{60}Ni(n,p)$ & 10.25 & 18.10\% & 18.80\% &  \textbf{20.90\%} & 18.10\% & 6.90\% \\ \hline 
$^{56}Fe(n,p)$ & 10.3 & -2.50\% & -4.00\% & 0.20\% & -2.50\% & 3.90\% \\ \hline 
$^{24}Mg(n,p)$ & 10.37 & 3.70\% & 8.40\% & 10.30\% & 7.70\% & 4.10\% \\ \hline 
$^{59}Co(n,\alpha)$ & 10.43 & -1.90\% & -4.60\% & 1.00\% & 1.70\% & 4.00\% \\ \hline 
$^{48}Ti(n,p)$ & 10.45 & -0.50\% & 0.10\% & -1.90\% & 10.60\% & 4.10\% \\ \hline 
$^{51}V(n,\alpha)$ & 10.57 & -1.70\% & -1.20\% & 8.40\% & 12.20\% & 5.30\% \\ \hline 
$^{197}Au(n,2n)$ & 10.6 & -3.40\% & -5.50\% &  \textbf{22.60\%} & -6.10\% & 4.30\% \\ \hline 
$^{58}Ni(n,x)^{57}Co$ & 10.89 & - &  \textbf{-20.90\%} & -10.60\% &  \textbf{-18.10\%} & 3.70\% \\ \hline 
$^{59}Co(n,2n)$ & 11.27 & -2.70\% & -2.70\% & 11.70\% & 14.00\% &
6.20\% \\ \hline \hline 
\end{tabular}
\end{table*}

\begin{table*}[htp]
\caption{Calculated versus measured reaction rates in $^{7}Li(p,n)$ field from
  14.4 MeV protons in terms of C/E-1 \cite{schulcMonoEmatej2022}.} 
\label{table:schulcMonoE14MeVpCoverE}
\begin{tabular}{l | c | c c c c c} \hline \hline
Reaction & E50\% [MeV] & IRDFF-II & JEFF-3.3 & JENDL-4 & ENDF/B-VIII.0 & Uncertainty \\ \hline 
$^{54}Fe(n,p)$ & 7.75 & -1.20\% & -3.00\% & 0.60\% & -1.00\% & 3.90\% \\ \hline 
$^{47}Ti(n,p)$ & 8.37 & 1.30\% & 1.30\% & 2.10\% & 15.90\% & 6.80\% \\ \hline 
$^{46}Ti(n,p)$ & 10.04 & 7.80\% & 7.80\% & 2.90\% & 4.20\% & 4.10\% \\ \hline 
$^{59}Co(n,p)$ & 12.13 & 7.20\% & -1.70\% & -1.30\% & 4.80\% & 4.60\% \\ \hline 
$^{60}Ni(n,p)$ & 12.24 & 3.80\% & 1.20\% & 7.90\% & 9.80\% & 5.30\% \\ \hline 
$^{54}Fe(n,\alpha)$ & 12.29 & -5.40\% & -17.40\% & -5.50\% &  \textbf{-61.70\%} & 6.40\% \\ \hline 
$^{24}Mg(n,p)$ & 12.31 & 3.00\% & 10.10\% & 10.10\% & 10.10\% & 5.00\% \\ \hline 
$^{56}Fe(n,p)$ & 12.31 & -1.70\% & -2.50\% & -3.90\% & -1.60\% & 5.00\% \\ \hline 
$^{59}Co(n,\alpha)$ & 12.4 & -2.90\% & -5.00\% & -3.20\% & -0.90\% & 5.30\% \\ \hline 
$^{48}Ti(n,p)$ & 12.41 & 2.20\% & 2.30\% & -3.90\% & 0.50\% & 5.40\% \\ \hline 
$^{51}V(n,\alpha)$ & 12.5 & -4.80\% & -5.00\% & -0.40\% & 1.50\% & 6.30\% \\ \hline 
$^{197}Au(n,2n)$ & 12.5 & 2.90\% & 4.20\% & 16.00\% & 4.20\% & 6.20\% \\ \hline 
$^{58}Ni(n,x)57Co$ & 12.57 & - & -7.70\% & 2.50\% & -10.20\% & 6.60\% \\ \hline 
$^{59}Co(n,2n)$ & 12.58 & -7.20\% & -6.40\% & -13.50\% & -4.00\% & 6.70\% \\ \hline 
$^{19}F(n,2n)$ & 12.64 & 10.70\% &  \textbf{40.10\%} &  \textbf{75.70\%} &  \textbf{40.10\%} & 7.30\% \\ \hline 
$^{55}Mn(n,2n)$ & 12.58 & -2.80\% & 4.90\% & -1.00\% & -3.90\% & 6.80\% \\ \hline \hline
\end{tabular}
\end{table*}

\subsubsection{Research Center Rez $^{252}$Cf(s.f.) source: Ni, Fe,
  Cu, stainless steel, and Pb leakage spectrum and
  dosimetrical reactions}
\label{subsubsec:Cf252Schulc}
The neutron transport in various materials was tested using a
$^{252}$Cf(s.f.) neutron source. The $^{252}$Cf(s.f.) reaction is a neutron
standard~\cite{IRDFFII}.
Uncertainties are therefore very low in the source spectrum and
and any discrepancies between calculated and measured neutron spectra after passing through a
material layer are the result of discrepancies in the material cross sections used in the neutron transport simulation.
The method for calculation and measurement using the $^{252}$Cf(s.f.) source in our
institute was summarized in detail in previous work \cite{schulcCf252schulc2019}.
The materials tested were copper, nickel, iron,
lead, and stainless steel AISI 321. Nickel and iron were in the form of a sphere
of diameter 50\,cm.
The copper and stainless steel were in the form of a cube. The
copper cube had dimensions of 48 cm $\times$ 49.5 cm $\times$ 49.5 cm,
while the dimensions of the stainless steel cube were 50.4 cm $\times$ 50.2 cm $\times$ 50.2 cm.
The lead block had
dimensions of 52.5 cm $\times$ 52.5 cm $\times$ 10 cm.  The validation process was
divided into two parts: the measurement of the neutron spectra in the
region of 1-12 MeV and the measurement of the reaction rates of the
activated materials. Reactions were carefully selected, i.e. only
IRDFF-II validated dosimetric reactions were used for the validation
of FENDL transport libraries. All calculations were performed using the
MCNP6.2 transport code \cite{mcnp62}.

\textbf{Neutron spectra measurement by proton recoil method}

\begin{figure}[htp]
\includegraphics[width=0.90\columnwidth]{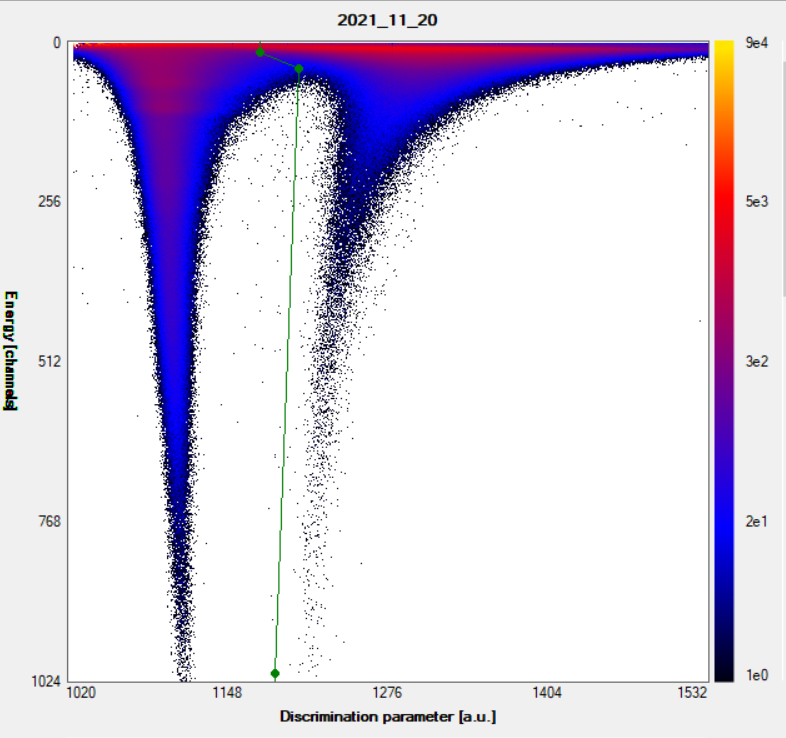}
    \caption{Neutron gamma discrimination in stainless steel cube
      measurement using the Research Center Rez $^{252}Cf(s.f.)$ source \cite{SchulcCf252Fe50kostal2018}.}
\label{fig:fig2neutgammaDiscrim}
\end{figure}

\begin{figure}[htp]
\includegraphics[width=0.90\columnwidth]{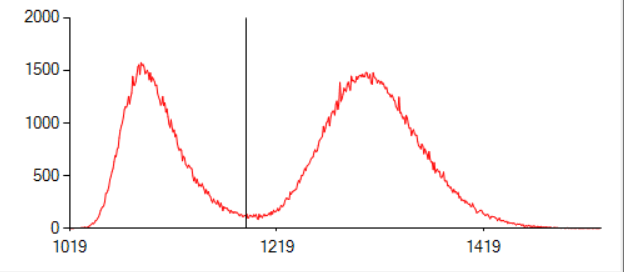}
    \caption{Discrimination between gammas (right) and neutrons (left)
      in 30 channel ($\sim$ 0.8 MeV) using the Research Center Rez $^{252}Cf(s.f.)$ source.}
\label{fig:fig3neutgammaDiscrimChann}
\end{figure}

The leakage neutron fluxes were monitored using a stilbene scintillation
detector of diameter 1\,cm and length 1\,cm. The spectra were
measured at a distance of 100\,cm from the center of the neutron
source. The spectrometric system that was used~\cite{schulcCf252matej2014,schulcCf252veskrna2014} features two-parameter data processing from scintillation
detectors in mixed gamma and neutron fields. In addition to the basic
data acquisition and parameter settings for individual experiments,
the control software contains mathematical evaluation algorithms, which
select the relevant data from the multi-parameter raw data \cite{schulcCf252cvachovec2008}. Using this method, it is possible to evaluate
individual components separately, i.e. neutrons and gammas. More
details concerning separation properties can be found in \cite{schulcCf252matej2021}. The plot of the pulse shape discrimination is shown in \cref{fig:fig2neutgammaDiscrim}. The  neutron and gamma signals are well separated because even in the region below 1 MeV, neutron and gamma pulses only overlap in the low 
energy region, see \cref{fig:fig3neutgammaDiscrimChann}.   

\textbf{Leakage validation by means of reaction rate measurements}

The leakage flux was also monitored by means of reaction rates in foils. The
reaction rates were derived from Net Peak Areas (NPA) associated with 
radioisotopes produced in activation foils attached to the block
surface or put inside the blocks. The NPAs were measured with well
characterized HPGe efficiency and energetic calibration. This accurate
description enables the computational determination of HPGe efficiency~\cite{kostalValidationZirconiumIsotopes2017}.
Relatively big foils (to compensate the low fluxes) were used and the measurement was performed in
HPGe end cap position. More details concerning the gamma spectrometry can
be found in \cite{schulcCf252kostal2017}. 

\textbf{Neutron leakage from the copper block:}

\begin{figure}[htp]
\includegraphics[width=0.90\columnwidth]{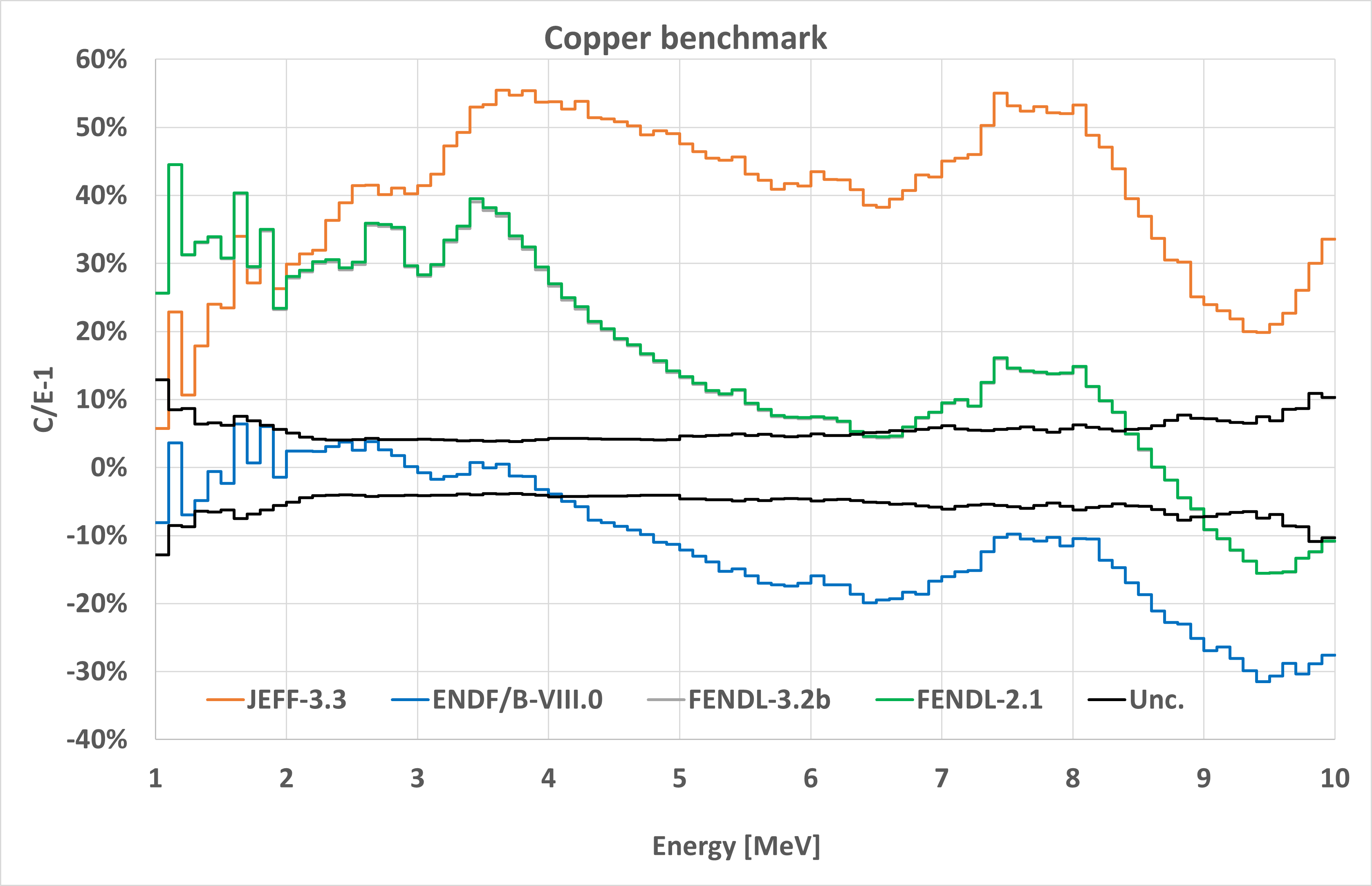}
    \caption{Cu benchmark neutron leakage spectra comparison using the Research Center Rez $^{252}Cf(s.f.)$ source.}
\label{fig:fig4culeakagespect}
\end{figure}

The copper block has dimensions of 48\,cm in length, 49.5\,cm in width, and 49.5\,cm in height. The neutron spectrum is measured at a distance of 100\,cm from the center of the source in the source plane.
\Cref{fig:fig4culeakagespect} shows the neutron leakage spectra calculated with various cross section libraries as compared to the measured spectrum in terms of $C/E-1$. Note that results calculated with FENDL-2.1, and FENDL3-.2b libraries are very similar.
For energies less than 5 MeV, agreement when using ENDF/B-VIII.0 cross sections is generally good, but a substantially higher flux is predicted when using the JEFF-3.3, FENDL-2.1, and FENDL-3.2b libraries.   
In higher energy regions (above 5 MeV), the agreement when using FENDL-2.1 and 3.2 are better than when using ENDF/B-VIII.0 and JEFF-3.3 cross sections.  More details can be found in~\cite{schulcCf252schulc2022a}.

The activation foils were attached to a thin aluminium foil and placed into the copper block at a distance of 16 cm from the $^{252}Cf(s.f.)$ neutron source center. The experiment specifically focuses on $^{58}Ni(n,p)^{58}Co$, $^{93}Nb(n,2n)^{92}Nb^{m}$, $^{197}Au(n,\gamma)^{198}Au$, and $ ^{55}Mn(n,\gamma)^{56}Mn$ dosimetry reactions.
\Cref{table:schulcCf252cufoilsCoverE} summarizes the results. Outliers over three-sigma uncertainty are highlighted by bold font in the table. Note that results calculated with FENDL-2.1 and FENDL-3.2b libraries are very similar.
Results calculated with the ENDF/B-VIII.0 library are similar to those calculated with the FENDL libraries except for $^{93}Nb(n,2n)^{92}Nb^{m}$ and $^{58}Ni(n,p)^{58}Co$ activation foils. More details concerning the experiment can be found in \cite{schulcCf252schulc2021}.

\begin{table*}[htp]
\caption{Calculated versus measured reaction rates in the Cu benchmark activation foils using the Research Center Rez $^{252}Cf(s.f.)$ source.} 
\label{table:schulcCf252cufoilsCoverE}
\begin{tabular}{l | c c c c} \hline \hline
Reaction & ENDF/B-VIII.0 & FENDL-3.2b & FENDL-2.1 & Exp. uncert. \\ \hline 
$^{197}Au(n,\gamma) I$ &  \textbf{-16.10\%} &  \textbf{-14.30\%} &  \textbf{-14.60\%} & 3.80\% \\ \hline 
$^{197}Au(n,\gamma) II$ & -7.80\% & -6.10\% & -6.10\% & 3.90\% \\ \hline 
$^{63}Cu(n,\gamma) I$ &  \textbf{-33.80\%} &  \textbf{-40.10\%} &  \textbf{-39.80\%} & 4.20\% \\ \hline 
$^{63}Cu(n,\gamma) II$ &  \textbf{-55.60\%} &  \textbf{-60.40\%} &  \textbf{-60.30\%} & 4.10\% \\ \hline 
$^{55}Mn(n,\gamma)$ & -4.80\% & -2.30\% & -0.80\% & 4.50\% \\ \hline 
$^{93}Nb(n,2n)$ & -13.69\% & -1.86\% & -1.86\% & 4.70\% \\ \hline 
$^{58}Ni(n,p)$ & 5.22\% &  \textbf{28.10\%} &  \textbf{28.10\%} & 4.10\% \\ \hline \hline
\end{tabular}
\end{table*}

\textbf{Neutron leakage from the nickel sphere:}

\begin{figure}[htp]
\includegraphics[width=0.90\columnwidth]{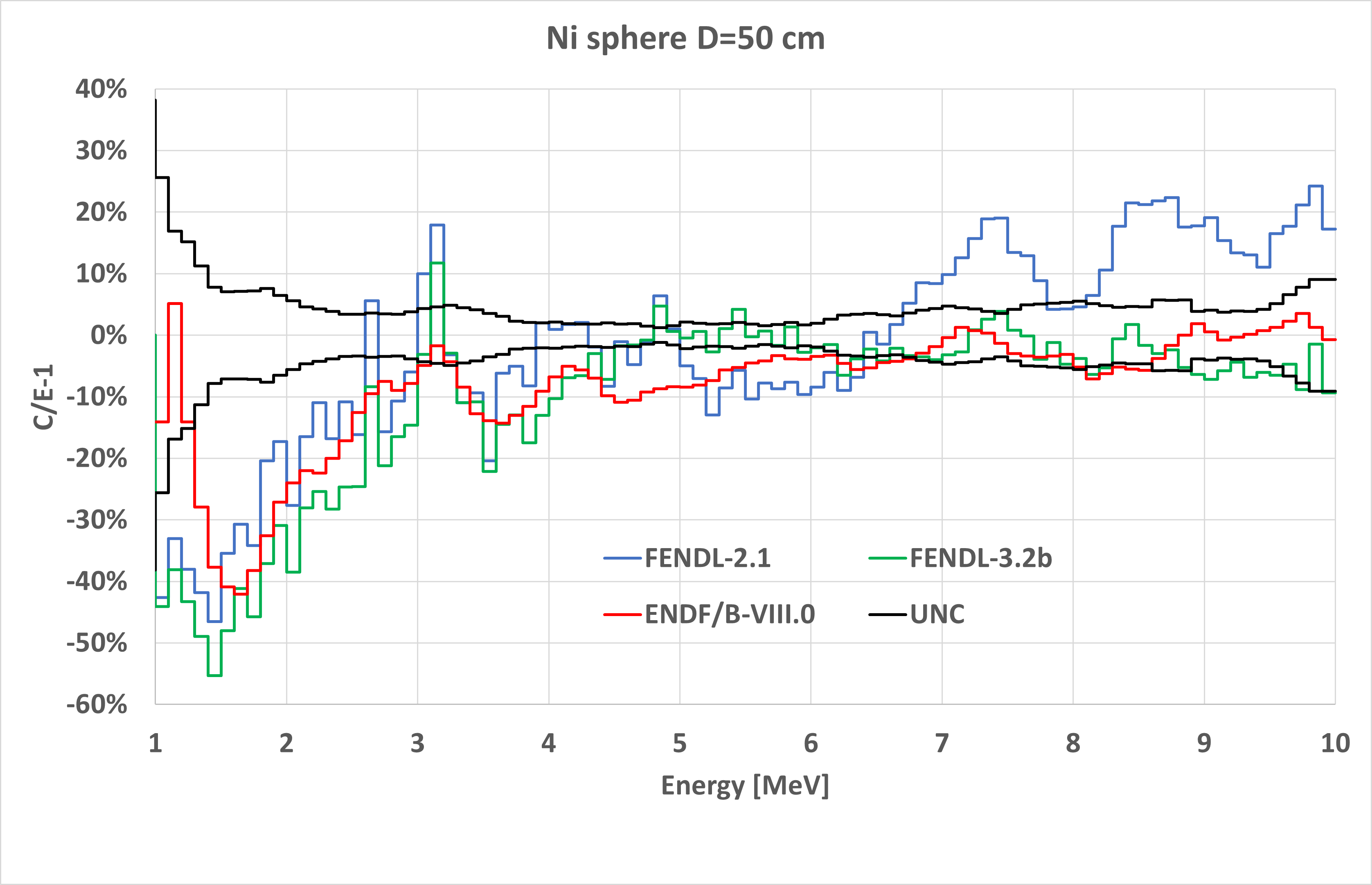}
    \caption{Ni benchmark neutron leakage spectra comparison using the Research Center Rez $^{252}Cf(s.f.)$ source.}
\label{fig:fig5nileakagespect}
\end{figure}

The fast leakage neutron spectra have been measured on spherical nickel benchmark assembly of diameter 50 cm. The $^{252}Cf(s.f.)$ neutron source was placed into the center of the sphere. Calculations using ENDF/B-VIII.0 and FENDL-3.2b agree with experimental results very well in the region above 4 MeV, as shown in \Cref{fig:fig5nileakagespect}. Performance with neutron energies below 4 MeV is much worse \cite{schulcCf252schulc2018}.  
\Cref{table:schulcCf252nifoilsCoverE} summarizes C/E-1 for the measured activation foils on the nickel sphere surface.   

\begin{table*}[htp]
\caption{Calculated versus measured reaction rates in the Ni benchmark activation foils using the Research Center Rez $^{252}Cf(s.f.)$ source.} 
\label{table:schulcCf252nifoilsCoverE}
\begin{tabular}{l | c c c c} \hline \hline
Reaction & ENDF/B-VIII.0 & FENDL-2.1 & FENDL-3.2b & Uncertainty \\ \hline 
$^{115}In(n,n')$ &  \textbf{-21.00\%} & 2.90\% & -5.80\% & 4.20\% \\ \hline 
$^{115}In(n,\gamma)$ & 10.00\% &  \textbf{19.80\%} &  \textbf{26.30\%} & 4.40\% \\ \hline 
$^{58}Ni(n,p)$ & -0.60\% & 13.70\% &  \textbf{19.30\%} & 5.00\% \\ \hline 
$^{55}Mn(n,\gamma)$ &  \textbf{-26.10\%} & 8.30\% & 13.40\% & 5.50\% \\ \hline \hline
\end{tabular}
\end{table*}

\textbf{Neutron leakage from the iron sphere:}

The fast leakage neutron spectra have been measured on the spherical iron benchmark assembly of diameter 50 cm. The $^{252}Cf(s.f.)$ neutron source was placed into the center of the sphere. \Cref{table:schulcCf252fefoilsCoverE} shows the results for activation foils on the iron sphere surface. Results calculated with FENDL-3.2b give better results than those calculated with FENDL-2.1 except for $^{115}In(n,\gamma)$. Overall, good agreement is not achieved with any of the libraries investigated \cite{SchulcCf252Fe50kostal2018}.

\begin{figure}[htp]
\includegraphics[width=0.90\columnwidth]{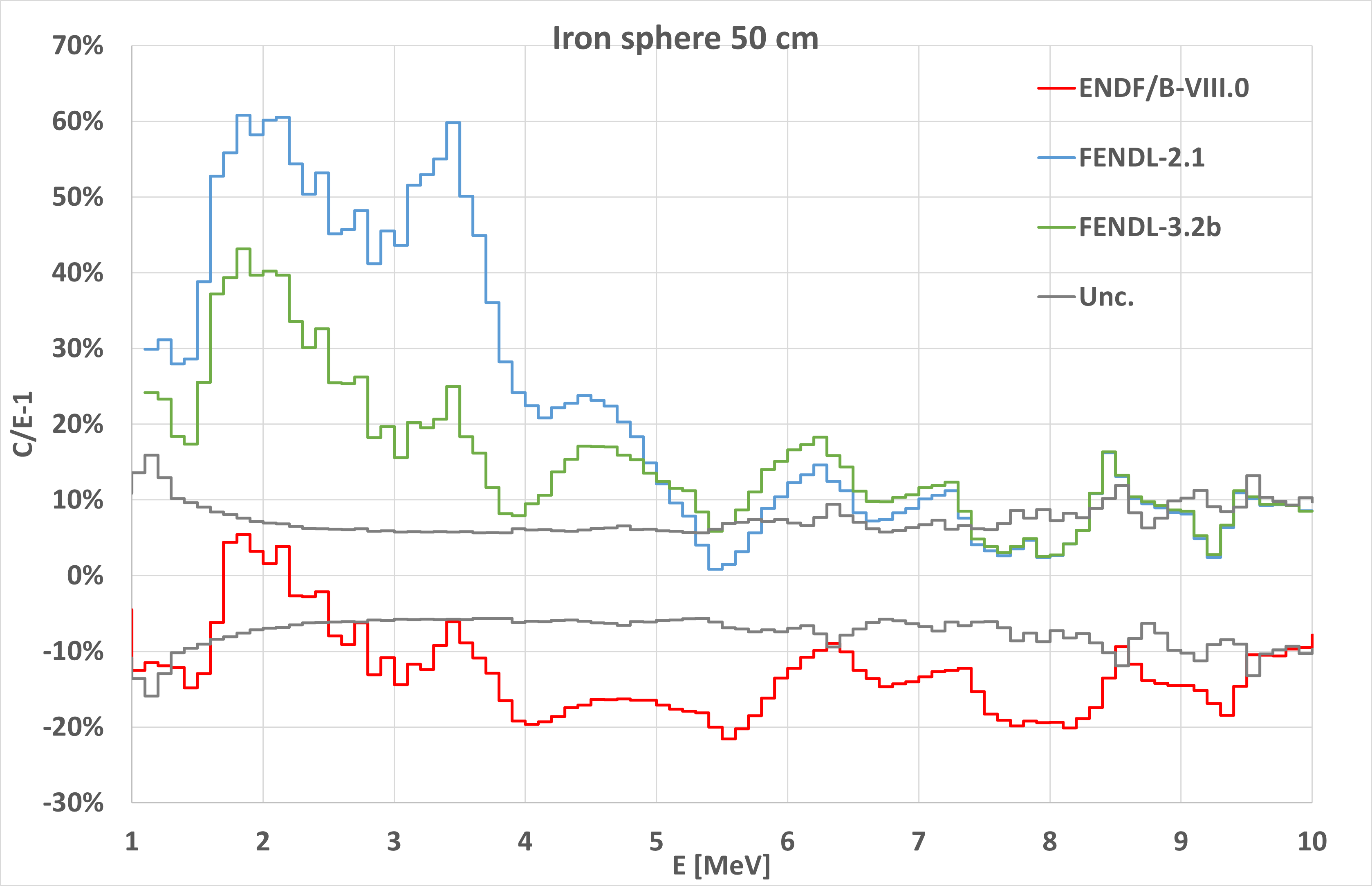}
    \caption{Fe benchmark neutron leakage spectra comparison using the Research Center Rez $^{252}Cf(s.f.)$ source.}
\label{fig:fig6feleakagespect}
\end{figure}

\Cref{fig:fig6feleakagespect} compares neutron leakage fluxes calculated with different libraries compared to the measured values. Calculations with the FENDL-3.2b library agree better with the experiment than fluxes calculated with FENDL-2.1. Reasonable agreement is achieved above 4 MeV.

\begin{table*}[htp]
\caption{Calculated versus measured reaction rates in the Fe benchmark activation foils using the Research Center Rez $^{252}Cf(s.f.)$ source.} 
\label{table:schulcCf252fefoilsCoverE}
\begin{tabular}{l | c c c c c} \hline \hline
Reaction & ENDF/B-VIII.0 & JEFF-3.3 & FENDL-3.2b & FENDL-2.1 & Unc.  \\ \hline 
$^{115}In(n,n')$ &  \textbf{-18.10\%} & -2.20\% & 1.70\% &  \textbf{13.90\%} & 4.00\% \\ \hline 
$^{115}In(n,\gamma)$ &  \textbf{62.80\%} &  \textbf{29.90\%} &  \textbf{50.20\%} &  \textbf{24.90\%} & 5.00\% \\ \hline 
$^{58}Ni(n,p)$ & -11.90\% & 15.00\% & 16.30\% &  \textbf{48.30\%} & 6.00\% \\ \hline 
$^{55}Mn(n,\gamma)$ &  \textbf{-28.30\%} &  \textbf{-43.70\%} &  \textbf{-26.30\%} &  \textbf{-41.90\%} & 6.00\% \\ \hline \hline
\end{tabular}
\end{table*}

\textbf{Neutron leakage from the stainless steel block:}

Stainless steel was used in the form of a block of dimensions of 50.4\,cm $\times$ 50.2\,cm $\times$ 50.2\,cm with a centrally located aperture for the neutron source. Measured were the neutron leakage spectra from the block and also the spatial distribution of reaction rates of dosimetrical reactions inside the block. The leakage spectra was measured at a distance of 1\,m from the block, two activation detectors were attached to the surface of the block and several inside in various depths.
The activation foils inside the block were placed between the various plates constituting the block so that 5\,cm, 10\,cm, 15\,cm and 20\,cm of steel was between the respective activation foil and the $^{252}$Cf(s.f.) source. 
The reactions measured were $^{58}Ni(n,p)^{58}Co$, $^{197}Au(n,2n)^{196}Au$, $^{63}Cu(n,\gamma)^{64}Cu$ and $^{181}Ta(n,\gamma)^{182}Ta$.  \Cref{table:schulcCf252ssfoilsCoverE} summarizes C/E-1 for the activation foil on the stainless steel block surface at the two locations.  The calculations agree with measurements except for the calculations using ENDF/B-VIII.0, which are 18\% lower than the measured values.

\begin{figure}[htp]
\includegraphics[width=0.90\columnwidth]{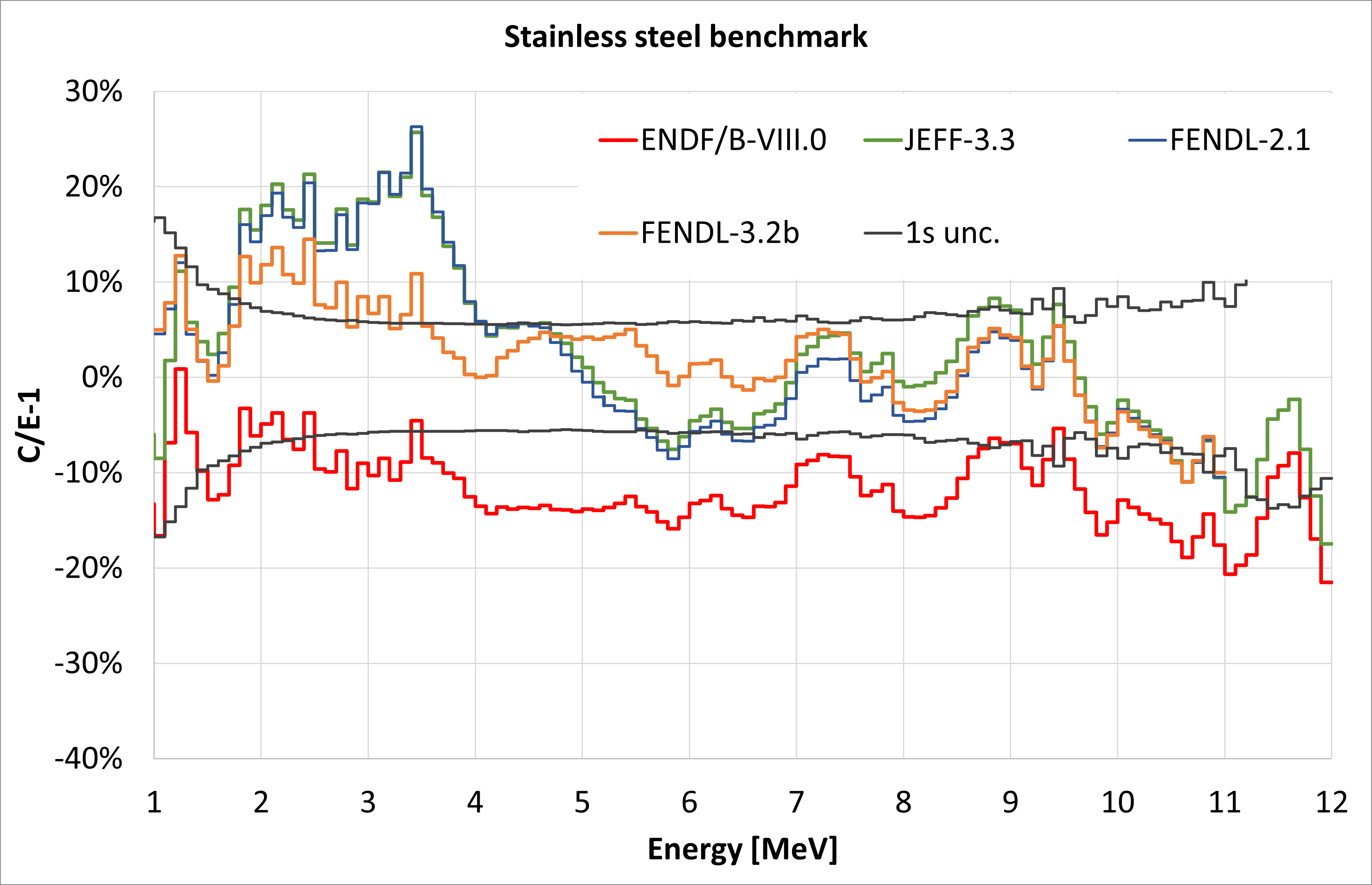}
    \caption{Stainless steel benchmark neutron leakage spectra comparison using the Research Center Rez $^{252}Cf(s.f.)$ source.}
\label{fig:fig7ssleakagespect}
\end{figure}

\Cref{table:schulcCf252ssfoilsCoverE2} summarizes the results for various thicknesses of steel. Calculations agree with measurements except for the calculations using ENDF/B-VIII.0, which significantly underpredict the trend with increasing steel thickness. It is worth noting that the results for the $^{58}Ni(n,p)^{58}Co$ reaction are consistent with validation results for neutron leakage spectra presented in \Cref{fig:fig7ssleakagespect}. About 10 - 20\% underprediction for ENDF/B-VIII.0 and a slight overprediction for JEFF-3.3 can be observed. Calculations using the FENDL-3.2b library agree well with the experiment above 4 MeV. More details concerning the experiment can be found in \cite{SchulcCf252SSschulc2022}.

\begin{table}[htp]
\caption{Calculated versus measured reaction rates in the stainless steel benchmark activation foil, $^{58}Ni(n,p)^{58}Co$, using the Research Center Rez $^{252}Cf(s.f.)$ source.} 
\label{table:schulcCf252ssfoilsCoverE}
\begin{tabular}{l | c c} \hline \hline
Library & Front side & Lateral side \\ \hline 
ENDF/B-VIII.0 &  \textbf{-18.40\%} &  \textbf{-18.50\%} \\ \hline 
JEFF-3.3 & 2.30\% & 5.50\% \\ \hline 
JENDL-4 & -5.40\% & -3.40\% \\ \hline 
FENDL-2.1 & -0.70\% & -0.70\% \\ \hline 
FENDL-3.2b & -3.80\% & -4.90\% \\ \hline 
Unc.  & 4.57\% & 3.24\% \\ \hline \hline
\end{tabular}
\end{table}

\begin{table*}[htp]
\caption{Calculated versus measured reaction rates in the stainless steel benchmark activation foils using the Research Center Rez $^{252}Cf(s.f.)$ source.} 
\label{table:schulcCf252ssfoilsCoverE2}
\begin{tabular}{l l | c c c c c c} \hline \hline
Steel thickness & Reaction & ENDF/B-VIII.0 & JEFF-3.3 & FENDL-2.1 & FENDL-3.2b & TENDL-2013 & Unc.  \\ \hline 
5.04 cm & $^{197}Au(n,\gamma)$ & 4.40 \% & -2.80\% & 6.39\% & 1.16\% &  \textbf{-8.40\%} & 2.60\% \\ \hline 
10.08 cm & $^{197}Au(n,\gamma)$ & 3.90\% & -3.70\% & 6.61\% & -0.19\% &  \textbf{-11.04\%} & 3.00\% \\ \hline 
15.12 cm & $^{197}Au(n,\gamma)$ & 3.70\% & -2.60\% & 7.21\% & -0.06\% &  \textbf{-11.31\%} & 3.10\% \\ \hline 
20.16 cm & $^{197}Au(n,\gamma)$ & 5.40\% & 0.00\% &  \textbf{9.90\%} & 2.42\% &  \textbf{-9.72\%} & 3.20\% \\ \hline \hline
5.04 cm & $^{58}Ni(n,p$ & -8.60\% & -3.40\% & -3.93\% & -5.47\% & -8.16\% & 7.10\% \\ \hline 
10.08 cm & $^{58}Ni(n,p)$ & -8.40\% & 1.30\% & 0.67\% & -2.25\% & -6.84\% & 3.30\% \\ \hline 
15.12 cm & $^{58}Ni(n,p)$ & -9.10\% & 5.50\% & 4.54\% & 0.71\% & -6.89\% & 4.00\% \\ \hline 
20.16 cm & $^{58}Ni(n,p)$ &  \textbf{-12.90\%} & 5.30\% & 3.24\% & 0.10\% & -8.39\% & 3.70\% \\ \hline \hline
5.04 cm & $^{181}Ta(n,\gamma)$ & 2.10\% & -3.90\% & 1.97\% & -1.58\% &  \textbf{-11.17\%} & 3.80\% \\ \hline 
10.08 cm & $^{181}Ta(n,\gamma)$ & 4.80\% & -3.30\% & 5.85\% & 0.93\% & -9.15\% & 4.00\% \\ \hline 
15.12 cm & $^{181}Ta(n,\gamma)$ & 4.80\% & 0.00\% & 13.22\% & 3.78\% & -6.93\% & 5.40\% \\ \hline 
20.16 cm & $^{181}Ta(n,\gamma)$ & 11.30\% & 7.10\% &  \textbf{16.88\%} & 9.62\% & -2.23\% & 5.30\% \\ \hline \hline
5.04 cm & $^{63}Cu(n,\gamma)$ &  \textbf{25.70\%} &  \textbf{19.60\%} &  \textbf{21.54\%} &  \textbf{20.96\%} &  \textbf{15.24\%} & 3.70\% \\ \hline 
10.08 cm & $^{63}Cu(n,\gamma)$ &  \textbf{27.90\%} &  \textbf{22.10\%} &  \textbf{28.16\%} &  \textbf{21.93\%} &  \textbf{14.95\%} & 4.80\% \\ \hline 
15.12 cm & $^{63}Cu(n,\gamma)$ &  \textbf{21.50\%} &  \textbf{19.70\%} &  \textbf{19.07\%} &  \textbf{16.77\%} & 9.79\% & 4.70\% \\ \hline 
20.16 cm & $^{63}Cu(n,\gamma)$ &  \textbf{25.00\%} &  \textbf{26.10\%} &  \textbf{28.56\%} &  \textbf{24.59\%} & 16.30\% & 6.20\% \\ \hline \hline
\end{tabular}
\end{table*}

\textbf{Neutron leakage from the lead block:}

For this benchmark setup, the samples were placed behind 5 and 10 cm of the lead material. The reactions measured were $^{197}Au(n,2n)^{196}Au$, $^{58}Ni(n,p)^{58}Co$, $^{93}Nb(n,2n)^{92}Nb^{m}$, $^{115}In(n,n')^{115}In^{m}$, $^{115}In(n,\gamma)^{116}In^{m}$, $^{197}Au(n,\gamma)^{198}Au$, and $^{63}Cu(n,\gamma)^{64}Cu$.
\Cref{table:schulcCf252pbfoilsCoverE} summarizes the comparison of calculated results versus the measurements in terms of C/E-1 for the libraries investigated. The measured values agree reasonably well with the calculations. However, the experimental data suggest a higher thermal neutron flux behind lead bricks.  More details are provided in~\cite{schulcCf252schulc2022b}.

\begin{table*}[htp]
\caption{Calculated versus measured reaction rates in the Pb benchmark activation foils using the Research Center Rez $^{252}Cf(s.f.)$ source.} 
\label{table:schulcCf252pbfoilsCoverE}
\begin{tabular}{l | c | c c c c} \hline \hline
Reaction & Pb thickness & FENDL-2.1 & FENDL-3.2 & ENDF/B-VIII.0 & Exp. Unc. \\ \hline 
$^{58}Ni(n,p)^{58}Co$ & 10 cm & -2.16\% & 1.71\% & 6.60\% & 3.20\% \\ \hline 
$^{93}Nb(n,2n)^{92}Nb^{m}$ & 10 cm & 6.09\% & 0.49\% & 2.90\% & 4.20\% \\ \hline 
$^{197}Au(n,\gamma)^{198}Au$ & 5 cm & -13.90\% & -15.20\% & -15.80\% & 2.90\% \\ \hline 
$^{197}Au(n,\gamma)^{198}Au$ & 10 cm & -51.61\% & -51.80\% & -51.40\% & 3.10\% \\ \hline 
$^{197}Au(n,2n)^{196}Au$ & 5 cm & -0.45\% & 4.94\% & -0.50\% & 4.10\% \\ \hline 
$^{197}Au(n,2n)^{196}Au$ & 10 cm & 5.54\% & -1.45\% & 2.40\% & 4.30\% \\ \hline 
$^{63}Cu(n,\gamma)^{64}Cu$ & 5 cm & 0.76\% & -1.91\% & -1.10\% & 3.70\% \\ \hline 
$^{115}In(n,n')^{115}In^{m}$ & 10 cm & -3.95\% & -1.81\% & 0.20\% & 3.40\% \\ \hline 
$^{115}In(n,\gamma)^{116}In^{m}$ & 10 cm & -38.23\% & -38.07\% & -37.90\% & 3.10\% \\ \hline \hline 
\end{tabular}
\end{table*}

\subsubsection{LLNL Pulsed Sphere}
\label{subsubsec:llnlpulsedsphereNeudecker}
The LLNL Pulsed Sphere measurement program was undertaken at Lawrence Livermore National Laboratory from the 1960s up to the late 1990s, see e.g.,~\cite{neudecker1Wong:1972,neudecker2Marchetti:1998}.
The purpose of these measurements was to validate neutron-transport codes and nuclear data for well-defined materials in a simple geometry.
To this end, neutrons of 13-15 MeV were produced by deuterons hitting a tritiated target in the center of a sphere of a specific material; the outgoing neutrons of the neutron-leakage spectra were then measured as a function of time-of-flight (TOF) with different neutron detectors and at angles and flight path lengths.
These experiments query scattering, and for actinides, also fission nuclear data.
Reference~\cite{neudecker3:2021pulsedsphere} showed that the peak of the spectrum is mostly sensitive to elastic and discrete inelastic scattering, while the valley and shoulder can be influenced by fission (only for actinides), discrete and continuum inelastic scattering (dependent on the level scheme of the isotopes).
Thin pulsed spheres mostly query nuclear data from 12-15 MeV, while thicker pulsed spheres can query nuclear data down to 6 MeV.

\begin{figure}[htp]
\includegraphics[width=0.90\columnwidth]{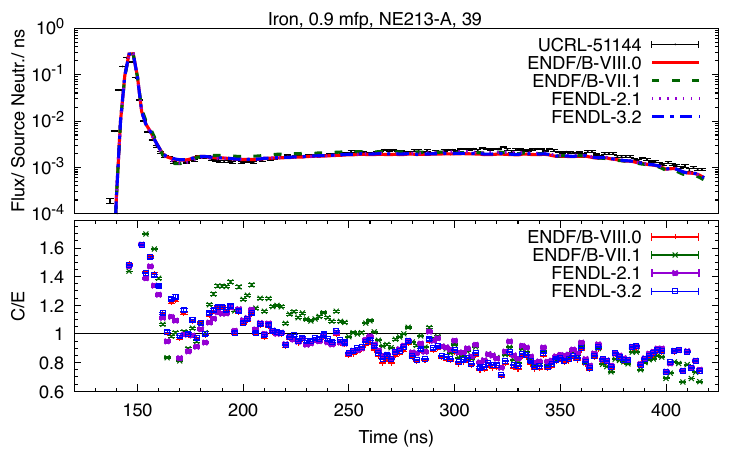}
    \caption{LLNL Pulsed Sphere experimental data compared to MCNP
      simulations using FENDL-2.1, FENDL-3.2, ENDF/B-VII.1 and
      ENDF/B-VIII.0 evaluated data for iron (0.9 mfp, detector NE213-A).}
\label{fig:fe0.9mfpAllnlpulsedsphere}
\end{figure}

\begin{figure}[htp]
\includegraphics[width=0.90\columnwidth]{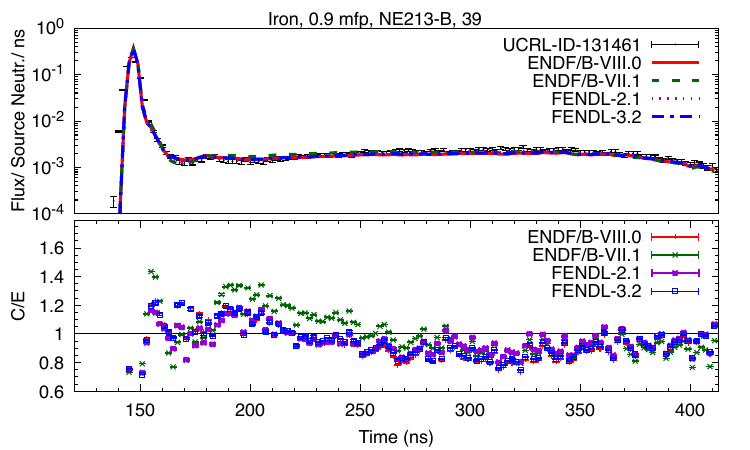}
    \caption{LLNL Pulsed Sphere experimental data compared to MCNP
      simulations using FENDL-2.1, FENDL-3.2, ENDF/B-VII.1 and
      ENDF/B-VIII.0 evaluated data for iron (0.9 mfp, detector NE213-B).}
\label{fig:fe0.9mfpBllnlpulsedsphere}
\end{figure}

\begin{figure}[htp]
\includegraphics[width=0.90\columnwidth]{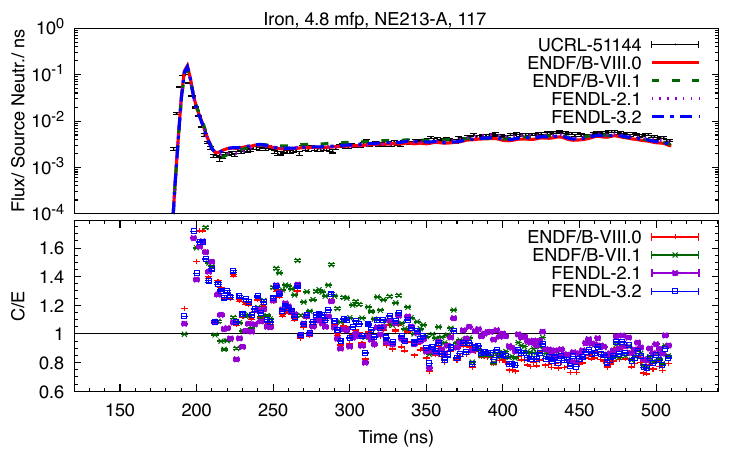}
    \caption{LLNL Pulsed Sphere experimental data compared to MCNP
      simulations using FENDL-2.1, FENDL-3.2, ENDF/B-VII.1 and
      ENDF/B-VIII.0 evaluated data for iron (4.8 mfp, detector NE213-A).}
\label{fig:fe4.8mfpAllnlpulsedsphere}
\end{figure}

\begin{figure}[htp]
\includegraphics[width=0.90\columnwidth]{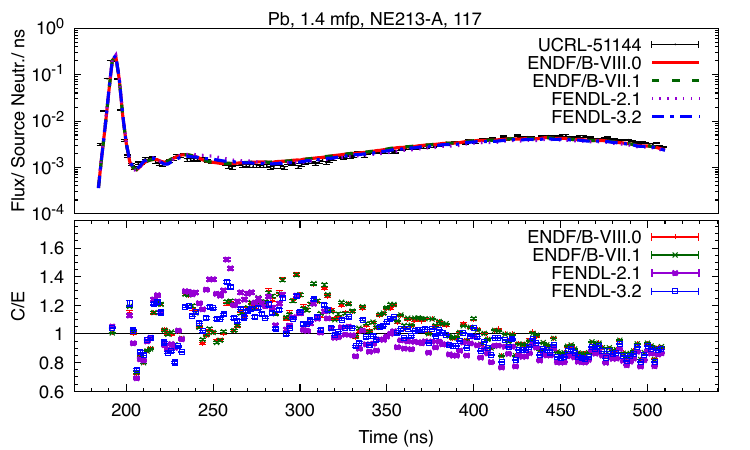}
    \caption{LLNL Pulsed Sphere experimental data compared to MCNP
      simulations using FENDL-2.1, FENDL-3.2, ENDF/B-VII.1 and
      ENDF/B-VIII.0 evaluated data for lead (1.4 mfp, detector NE213-A).}
\label{fig:pb1.4mfpAllnlpulsedsphere}
\end{figure}

While several (3 $^6$Li, 3 $^7$Li, 3 $^9$Be, 1 polyethylene, 2 carbon,
3 iron, and 1 lead sphere) pulsed-sphere neutron-leakage were
simulated in~\cite{neudecker4:FENDL}, only results for iron and
lead pulsed sphere change noticeably between FENDL-2.1 and FENDL-3.2.  Note that these calculations were run using MCNP-6.1.1.
\Cref{fig:fe0.9mfpAllnlpulsedsphere,fig:fe0.9mfpBllnlpulsedsphere} show the results for the 0.9 mfp
iron sphere (detector A and B respectively).
\Cref{fig:fe4.8mfpAllnlpulsedsphere} shows the results for the 4.8 mfp
  iron sphere (detector A).  
\Cref{fig:pb1.4mfpAllnlpulsedsphere}  shows the results for the 1.4 mfp
  lead sphere (detector A). 
The iron data in the valley, around 180 ns, are slightly better
described by FENDL-3.2 than FENDL-2.1 for the thin (0.9 Mfp) spheres,
while FENDL-2.1 performs better in describing the 4.8 mfp sphere in
the valley and at later times.
Continuum inelastic scattering plays an important role for describing the valley and later times of these pulsed spheres.
FENDL-3.2 describes the lead LLNL pulsed sphere measured spectra better than  FENDL-2.1 data above the peak.
Changes in discrete and continuum inelastic scattering cross sections and angular distributions are likely responsible for these changes.
ENDF/B-VII.1 and ENDF/B-VIII.0 data describe the spectra better from 200-260 ns but lead to worse predictions at later TOFs.
 
\subsubsection{JET Activation Foils}
\label{subsubsec:jetfoilsZohar}
This section summarizes and builds on work comparing experimental measurements of foil activation at the JET tokamak with calculations published in \cite{ANIMMAzohar} using the FENDL-3.2b nuclear data library.

\begin{figure*}[!htb]
    \centering
    \includegraphics[width=0.52\textwidth]{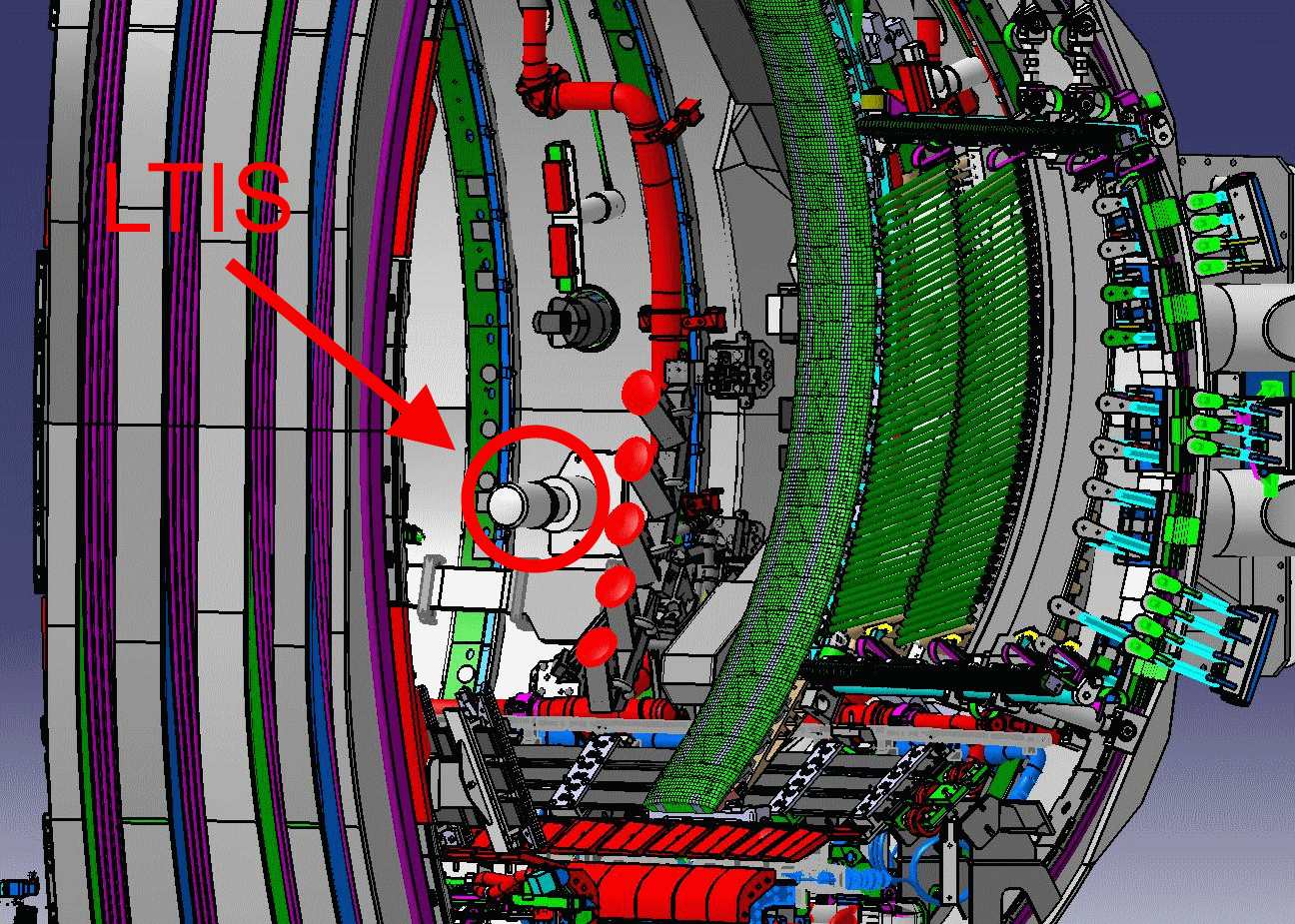}
    \includegraphics[width=0.358\textwidth]{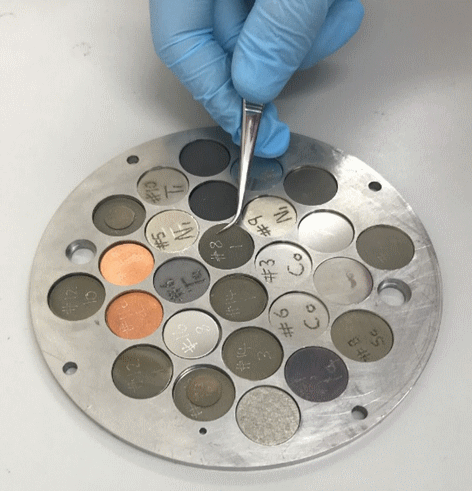}
    \caption{\textit{Left}: CAD model of the LTIS position in the JET tokamak, which located in the 7\textsuperscript{th} octant. \textit{Right}: Photography of the LTIS sample holder with ITER materials and dosimetry foils. Both figures were taken from \cite{ANIMMAzohar}.}
    \label{fig:JETltis}
\end{figure*}

In preparation for the JET D-T campaign, a C38 Deuterium-Deuterium campaign was performed at JET in 2019 \cite{DT1}. In the experimental campaign several different ITER-relevant materials and dosimetry foils were irradiated in a specially designed long-term irradiation station (LTIS) presented in \cref{fig:JETltis} \cite{ANIMMAzohar} with underpinning experimental characterisation in similar irradiation stations at JET using dosimetry foil reactions in~\cite{PACKER20171150, Packer_2018}. The ITER materials in LTIS included poloidal field (PF) coil jacket samples and toroidal field coil radial closure plate steels, EUROFER 97-2 steel, W and CuCrZr materials from the divertor, Inconel~718, CuCrZr and 316L stainless steel for blanket modules and vacuum vessel forging samples. A full list of materials is provided in \cite{Packer_2021} while the dosimetry foils included foils of cobalt, nickel, iron, scandium and yttrium to cover different neutron energy ranges. During the campaign, there were 101 days of irradiation and 46 days of no experimental irradiation. The neutron irradiation during plasma discharges was short (10 s to 20 s per plasma discharge) with typically several minutes to hours cooling time in between discharges during JET experimental campaign periods. The total neutron yield of the experimental campaign was 3.15$\cdot$10\textsuperscript{19} neutrons while the neutron fluence the samples in LTIS were exposed to was 1.9$\cdot$10\textsuperscript{14} n/cm\textsuperscript{2}.

The main goal of the experiment was testing activation of ITER materials by fusion neutrons. The simultaneous irradiation of dosimetry foils allowed the measurements of activated ITER materials to be compared to the measurements of well characterized activated dosimetry foils. Additionally, the experiment allowed calculations of dosimetry foils activation to be compared with the measurements of the biggest currently operating tokamak in the world. In this section, the comparison between experimental measurements of long-lived isotope activities in the activated dosimetry foils and calculated activities will be presented.

The measurement analysis of the activated dosimetry foils were performed by different European laboratories: Department of Fusion and Technologies of Nuclear Safety and Security, ENEA C. R. Frascati (ENEA), The Henryk Niewodniczanski Institute of Nuclear Physics, Polish Academy of Sciences (IFJ-PAN) and the National Centre for Scientific Research “Demokritos” (NCSRD). Further measurements and analysis were conducted at the UKAEA, which are covered in~\cite{Packer_2021}. All laboratories performed high resolution gamma ray measurements of samples with similar methodologies for calibration, analysis and reporting of the sample activity results. The reported measured activity of all isotopes in the samples were decay-corrected to the end of the JET irradiation period.

The computational support to the experiment was performed in two steps. In the first step, the Monte Carlo N-particle transport code (MCNP version 6.2 \cite{mcnp62}) was used to calculate the neutron spectra and reactions rates for all sample positrons in LTIS. For the neutron transport, the cross sections were taken from the FENDL-3.2b nuclear data library while the reactions rates were calculated using cross sections from the IRDFF-II nuclear data library \cite{IRDFFII}. For the analysis, the Doppler broadening of the cross sections was not considered. Despite the heating of the reactor, the reference nuclear data for JET calculations is the FENDL nuclear library as available on the IAEA web page without any modifications.

The MCNP model used for the validation is an existing MCNP model of the tokamak JET, which has been extensively used for JET neutronics studies, aiding in the design and installation of neutron diagnostic systems \cite{Batistoni_2017}, analysis of neutron detector response and their calibration \cite{Cufar_2018, BATISTONI_2016}, the heating of the tokamak’s walls and other vital components \cite{LENGAR20161011} and neutron streaming through JET structure and torus hall penetrations \cite{Batistoni_2015, KOS_2019}. The model has thus been used for neutron transport calculations inside and outside the tokamak's vacuum vessel as well as in the tokamak's torus hall penetrations. The model contains all larger components inside the vacuum vessel. The outside of the tokamak structure is partly defined with a low density mixture of Inconel with additions, representing the equipment surrounding the torus (homogenized surrounding region) and some of the ex-vessel diagnostic equipment. To properly support the neutron activation experiment, a detailed model of the LTIS with all samples was added to the validated MCNP model of the JET tokamak. 

\begin{figure}[!htb]
    \centering
    \includegraphics[width=8.69cm,height=6cm]{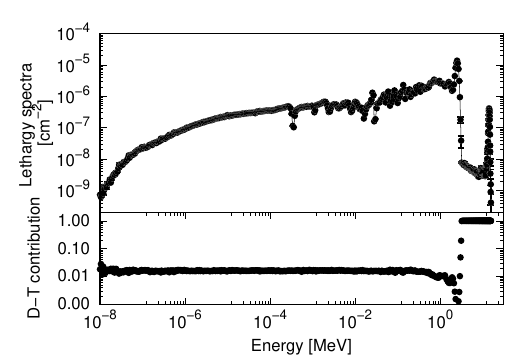}
    \caption{Calculated neutron lethargy spectra for D-D (2.45 MeV) and D-T (14.1 MeV) neutrons at the LTIS position in the JET tokamak.}
    \label{fig:JETnSpectra}
\end{figure}

The neutron source used for the analysis is based on the assumption that the plasma is in thermal equilibrium. The neutron emission profile is not described with an analytical model, but comes from a precomputed TRANSP plasma transport simulation \cite{TRANSP3}. The poloidal plasma profile of the source is divided into 386 rectangular zones over which the neutron emission probability is averaged. In a toroidally symmetric model, these poloidal rectangles form rings of equal neutron emissivity. The neutron spectrum for the D-D and D-T plasma is a Maxwellian one, computed for a plasma with average ion temperatures of T\textsubscript{th} = 20 keV. All of the neutrons are born isotropically in direction \cite{ANIMMAzohar}. The calculated neutron spectra at the LTIS positions, which are derived by combining D-D (2.45 MeV) neutron spectra with 1.4 \% D-T (14.1 MeV) neutron spectra, is presented in \cref{fig:JETnSpectra}.

In the second step of the analysis, the activities of long-lived isotopes in the activated dosimetry foils were calculated using the calculated reaction rates, neutron yield and irradiation time. As isotopes $^{58}$Co (half-life 70.86 days) and $^{46}$Sc (half-life 83.79 days) have a relatively short half-life in respect to the irradiation time, the irradiation and cooling days of the experimental campaign were merged on a weekly basis---22 weeks of irradiation. For more accurate calculations, the last day of experimental irradiation was simulated without any merger of pulses. The activity of the long lived isotopes was calculated using the equation:
\begin{equation}
    A_n = R(1-\mathrm{e}^{-\lambda t_{irr}})\mathrm{e}^{-\lambda t_{cool}}+A_{n-1}\mathrm{e}^{-\lambda t_w}
\end{equation}
where R is the total reaction rate obtained using the calculated MCNP reaction rate multiplied with the total neutron yield of the irradiation time, $\lambda$ is the decay constant of the activated isotope, $t_{irr}$ is the irradiation time in week, $t_{cool}$ is the cooling time in the week, $n$ is the number of the week and $t_w$ is the time in one week. The procedure is equal to the procedure used by the FISPACT code with the addition that continuous neutron spectra were used for the calculation of reaction rates whereas FISPACT uses group wise neutron spectra~\cite{FISPACT}.

\begin{figure*}[!htb]
    \centering
    \includegraphics[width=1\textwidth]{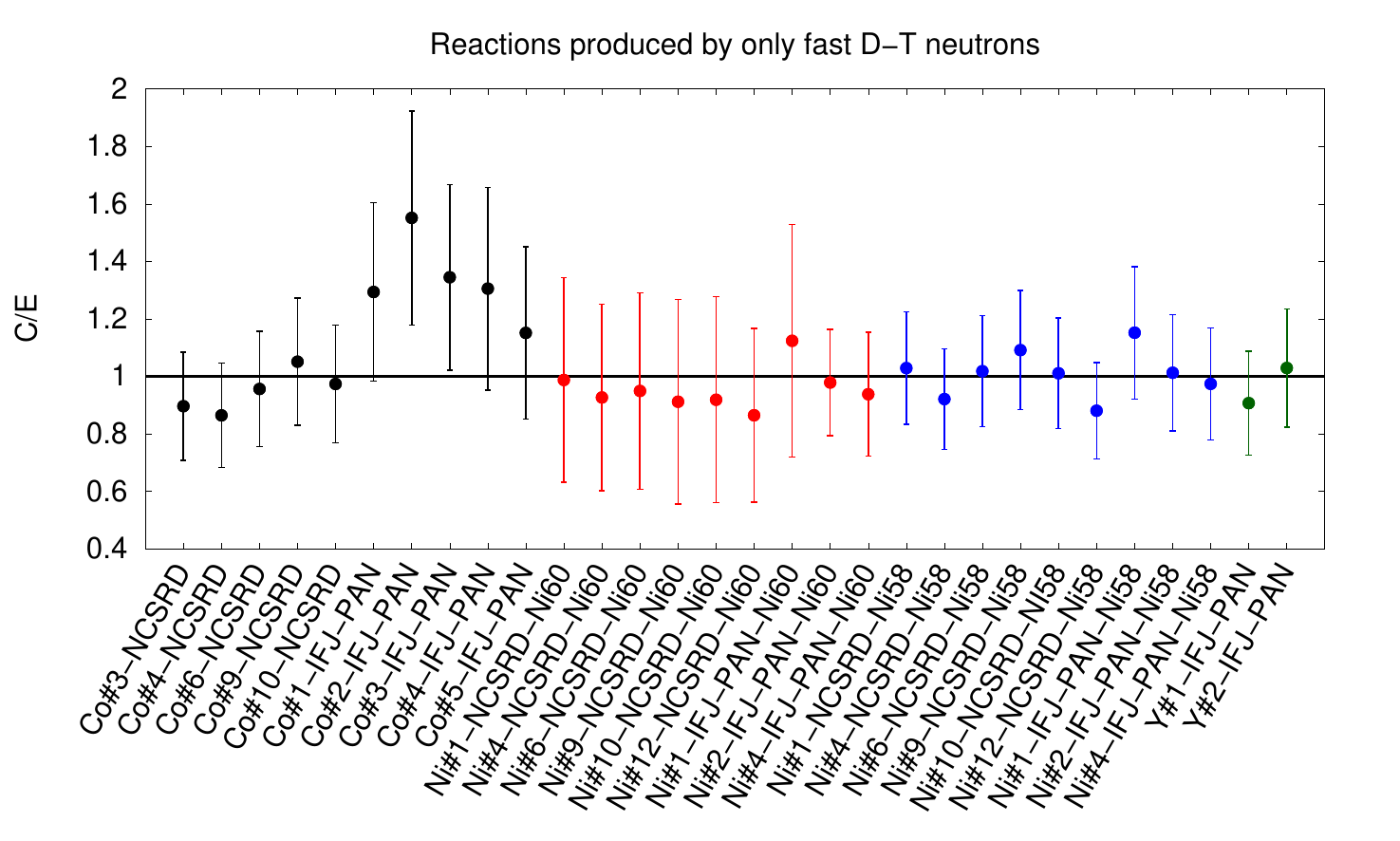}
    \caption{Comparison of the calculated and measured activity for isotopes in dosimetry foils induced by fast neutrons. The labels on the x-axis include the material of the dosimetry foil, the number of the foil in the irradiation station and the abbreviation of the institute that provided the dosimetry foil and performed the activation measurements.}
    \label{fig:fastJET}
\end{figure*}

The reactions $^{59}$Co(n,2n)$^{58}$Co, $^{60}$Ni(n,p)$^{60}$Co, $^{58}$Ni(n,np)$^{57}$Co and $^{89}$Y(n,2n)$^{88}$Y in the dosimetry foils have a threshold reaction above the D-D neutron energy (2.45 MeV) and can only be produced by D-T neutrons (14.1 MeV) due to tritium production in the D-D plasma. The analysis of these reactions allows the determination of the D-T neutron fraction throughout the experimental campaign. The analysis showed that a neutron fraction of 1.4 \% DT produces the calculated-to-measured activity ratios closest to 1 and are presented in~\cref{fig:fastJET}.

The uncertainties in the C/E results shown in~\cref{fig:fastJET} were calculated from the two sigma statistical uncertainties of the MCNP calculations, the uncertainties of the measured sample activity and the uncertainty of the neutron yield. The statistical uncertainty for all reaction rates calculated with MCNP was less than 2 \%. Thus, most of the uncertainty in the C/E values is due to the measurement uncertainties and the uncertainty in the neutron yield.

The C/E values vary between 0.9 and 1.6 for reactions sensitive to fast DT neutrons. The largest difference between the calculations and the measurements is observed for the cobalt samples measured by IFJ. The dispersion can be attributed to the different measurement methods used by different institutes and the different size of the samples. Despite the variations in the C/E values, the calculated activities agree quite well with the activities measured by different institutes.

\begin{figure*}[!htb]
    \centering
    \includegraphics[width=1\textwidth]{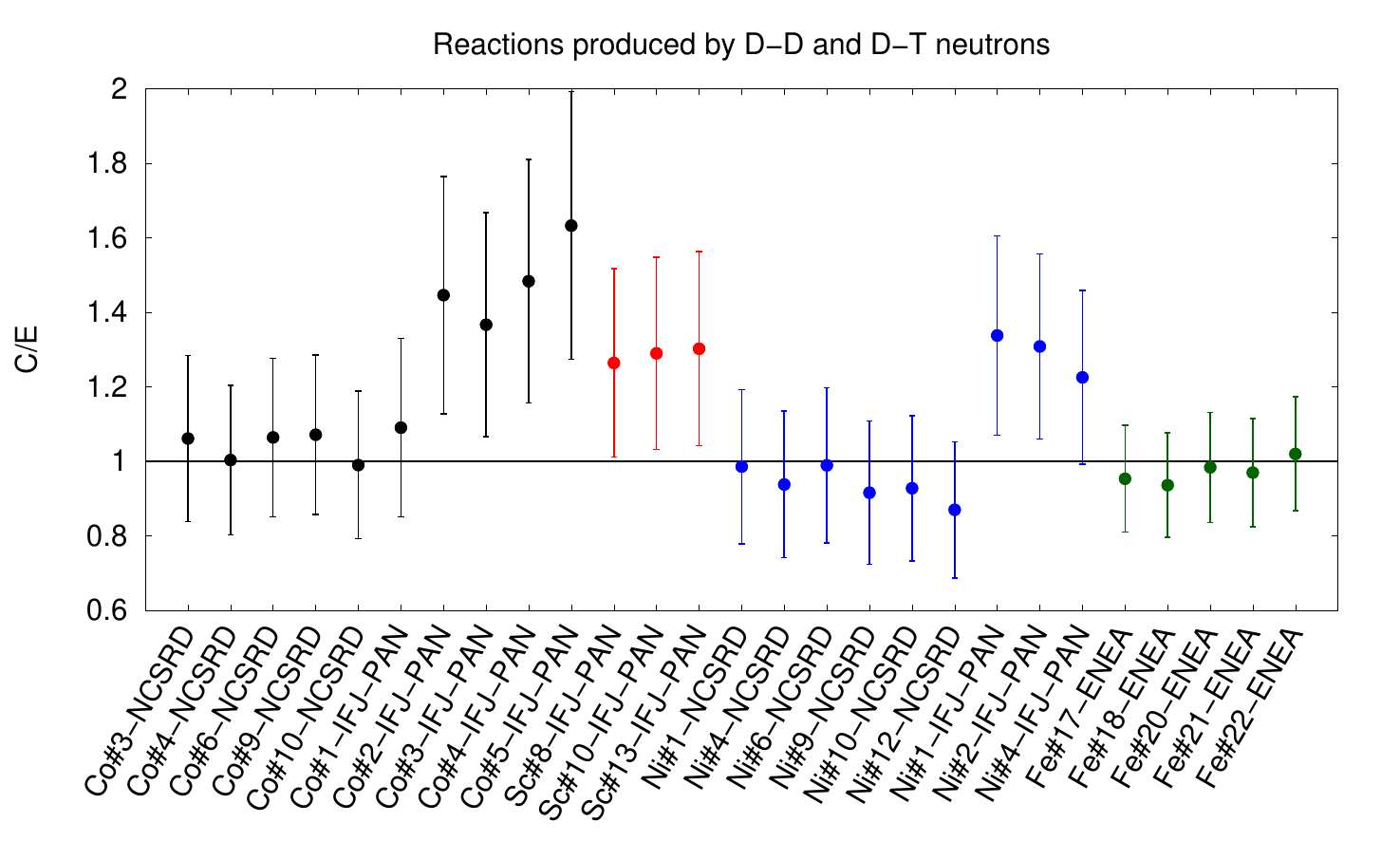}
    \caption{Comparison of the calculated activity with measured isotopes in dosimetry foils produced by D-D and D-T neutrons. The labels on the x-axis represent the material of the dosimetry foil, the number of the foil in the irradiation station and the abbreviation of the institute that provided the dosimetry foil and perform activation measurements.}
    \label{fig:thermalJET}
\end{figure*}

The reactions studied in dosimetry foils activated by both D-D and D-T neutrons were $^{59}$Co(n,$\gamma$)$^{60}$Co, $^{58}$Ni(n,p)$^{58}$Co, $^{54}$Fe(n,p)$^{54}$Mn and $^{45}$Sc(n,$\gamma$)$^{46}$Sc. The reactions $^{58}$Ni(n,p)$^{58}$Co and $^{54}$Fe(n,p)$^{54}$Mn are threshold reactions with threshold values at 0.8 MeV and 0.7 MeV, respectively, while the reactions $^{59}$Co(n,$\gamma$)$^{60}$Co and $^{45}$Sc(n,$\gamma$)$^{46}$Sc are predominantly produced by thermal neutrons. The C/E values for these reactions are shown in~\cref{fig:thermalJET}. The uncertainties of the C/E values shown in~\cref{fig:thermalJET} were determined in the same way as the C/E values for pure D-T neutron reactions.

For reactions predominantly produced by thermal neutrons, C/E values vary for cobalt samples measured by two different institutions. For the samples measured by NCSRD, the C/E values are around 1, while the cobalt samples measured by IFJ have higher C/E values of around 1.5. The Sc samples have C/E values of around 1.3. As with the D-T neutron activation reactions, the Co samples measured by NCSRD and IFJ were of different sizes. For the Ni-58(n,p)Co-58 reaction measured by NCSRD, the C/E values are around 1. The C/E values for the nickel samples measured by IFJ are 1.4. The C/E values for iron samples measured by ENEA are around 1. The dispersion in the C/E values in the results can be attributed to the differences in the measurement methodology of the different institutions and the different sizes of the samples (Co samples) measured by institutions.

Despite the variations in the C/E values, the calculated activities agree quite well with the activities measured by different institutions, thus indicating the correctness of the JET MCNP model and FENDL-3.2b used in the transport simulation.

Additionally, the initial analysis was performed with the FENDL-3.1d nuclear data library and the results are presented in \cite{ANIMMAzohar}. There are some small difference between both libraries, but they are within the uncertainty of the results (at most 3\% difference and the average difference being 1.4\%).

\def\figpath#1{figures/validation/experimentalbenchmarks/sinbadKodeli/#1}

\def\figpath#1{figures/validation/experimentalbenchmarks/sinbadKodeli/#1}

\subsubsection{SINBAD benchmarks}

The SINBAD database \cite{Kodeli2021,SINBAD} currently contains compilations and evaluations for 102 fission, fusion and accelerator relevant shielding problems and continues to represent an important experimental database for validating nuclear data, codes and nuclear design. Several of these benchmarks were used to validate the consistency and the predictive power of the FENDL-3.2a cross section library.  
The results are also representative for FENDL-3.2b as the only change between FENDL-3.2a and 3.2b concerned an update to the double-differential spectra of emitted tritons from $^{10}$B. 
Experiments used in the verification and validation (V \& V) process at UKAEA are listed in Table~\ref{tab:sinbad}. The benchmarks were selected according to their relevance for fusion nuclear data validation and the completeness and quality of benchmark experiment evaluations. Materials covered include: Fe/stainless steel, Pb, Li, SiC, W. 
The SINBAD benchmark evaluations used here went through a rigorous quality review and were found to be of good benchmark quality and ``valid for nuclear data and code benchmarking''.
Deterministic or probabilistic (Monte Carlo) radiation transport computer models used for the interpretation of the experiment are also provided in SINBAD in computer readable form. 
In addition, the CIAE iron slab experiment, not yet included in the official SINBAD distribution, was used for iron data benchmarking. The experiment was presented and discussed in the scope of the WPEC subgroup 47 activities \cite{Kodeli2020} and is under review for inclusion in SINBAD. Note that the FNG benchmarks analyzed in this section are different from those analyzed in section~\ref{subsubsec:fngAngelone}.

The results reported here were obtained using the MCNP5 code inputs provided in SINBAD. In some cases, minor modifications, such as in neutron source description and material compositions, were introduced in the original input files.
The ENDF ACE format cross sections of the FENDL-3.2a and FENDL-2.1 libraries were used and the detector response functions were taken from the IRDFF-II evaluation. Differences between the FENDL-3.2a and -2.1 libraries are evaluated and compared with the measured values to conclude on the progress achieved in the new release with respect to the FENDL-2.1 library.

\begin{table*}
\begin{center}
\caption{SINBAD shielding benchmarks used in this study at UKAEA. The benchmarks used are classified ``of benchmark quality'' (***) in SINBAD. Note that the FNG benchmarks analyzed here are different from those analyzed in section~\ref{subsubsec:fngAngelone}.}
\label{tab:sinbad}
\begin{tabular}{l|l|l|l}
\hline\noalign{\smallskip}
{\small Benchmark} & {\small Shielding material} & Detectors & {\small Computer code input}  \\
	\noalign{\smallskip}\hline\noalign{\smallskip}
{\small ASPIS Iron 88 (***)} & {\small steel 67 cm} & {\small Au, Rh, In, S, Al foils} & {\small MCBEND, DORT, TORT,} \\
 	& && {\small MCNPX/-5, (SERPENT)} \\
{\small PCA REPLICA (***)} & {\small H$_2$O /Fe shield} & {\small Mn, Rh, In, S, $^{235}$U foils,} & {\small TORT, TRIPOLI-3, -4,} \\
 &  & {\small SP-2, NE213 scintillator} & {\small MCNPX/-5/-6.1} \\
{\small FNG SiC (***)} & {\small SiC} & {\small Au, Al, Nb, Ni foils, TLD} & {\small MCNP5, DORT, TWODANT}  \\
	& {\small 45.72x45.72x71.12 cm$^3$} &&\\
{\small FNG Tungsten (***)} & {\small W block} & {\small Au, Mn, In, Ni, Fe, Al, Zr,} & {\small MCNP5, DORT,} \\
 & {\small 42-47 x 46.85 x 49 cm$^3$} & {\small Nb foils, TLD} & TWODANT \\
{\small FNG HCPB (***)} & {\small metallic Be with 2 layers} & {\small Au, Ni, Al, Nb foils, Li$_2$CO$_3$} & {\small MCNP5, DORT-TORT} \\
 & {\small of Li$_2$CO$_3$} & {\small pellets, TLD-300} & \\
{\small CIAE Fe} & {\small Fe slabs 5, 10, 15 cm} & {\small ongoing SINBAD evaluation} & MCNP \\
	\noalign{\smallskip}\hline
\end{tabular}
\end{center}
\end{table*}
	
\underline{ASPIS Iron88}

The ASPIS Iron88 benchmark was performed in 1988 at AEA Technology, Winfrith, UK using the fission plate neutron source. 

The ASPIS Iron-88 benchmark \cite{Wright1993,Milocco2015} consists of a 67-cm thick iron block irradiated with $^{235}$U fission neutrons. Several reaction rates were measured and calculated using the MCNP code: $^{27}$Al(n,$\alpha$), $^{103}$Rh(n,n’), $^{115}$In(n,n’), $^{32}$S(n,p), $^{197}$Au(n,$\gamma$). ASPIS-Iron88 was among the first benchmarks to be included in the SINBAD database around 1997. 
Although the neutron source and neutron spectra are not representative for fusion radiation environments, the benchmark, due to large iron shield thicknesses of 67 cm and high-quality measurements, proved useful for the validation of iron cross sections including for fusion applications.
ASPIS Fe88 represents one of the most severe and reliable validation test for iron transport cross section.

The ASPIS Iron-88 benchmark proved useful for the validation of iron cross sections, starting from JEF-2.2 in the 1980s to the recent data validations, pinpointing to the deficiencies of the recent JEFF-3.3 and ENDF/B-VIII.0 evaluations \cite{Kodeli2018}. 

\begin{figure*}[tbp]
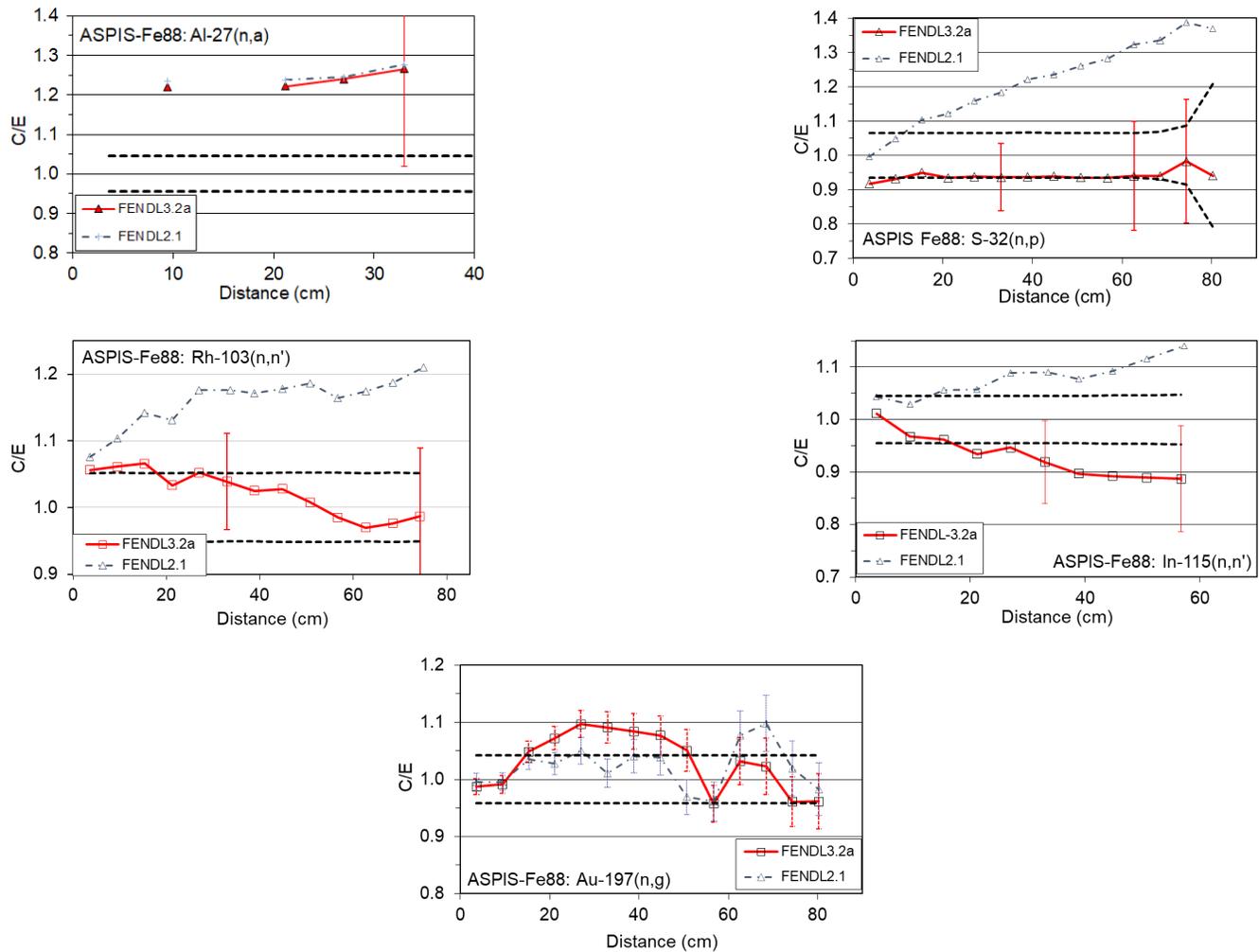

	\begin{minipage}[h]{7.cm}
		\begin{center}
			\resizebox{1.0\textwidth}{!}{%
			\includegraphics[width=6.5cm,clip]{\figpath{fe88_al.png}}
			}
		\end{center}      
	\end{minipage}
	\hfill 
	\begin{minipage}[h]{7.cm}
		\begin{center}
			\resizebox{1.0\textwidth}{!}{%
				\includegraphics[width=6.5cm,clip]{\figpath{fe88_s.png}} 	}
		\end{center}
	\end{minipage}
	\hfill 
	\begin{minipage}[h]{7.cm}
		\begin{center}
			\resizebox{1.0\textwidth}{!}{%
				\includegraphics[width=6.5cm,clip]{\figpath{fe88_rh.png}}	}
		\end{center}      
	\end{minipage}
	\hfill 
	\begin{minipage}[h]{7.cm}
		\begin{center}
			\resizebox{1.0\textwidth}{!}{%
				\includegraphics[width=6.5cm,clip]{\figpath{fe88_in.png}}	}
		\end{center}
	\end{minipage}
	\hfill
	\begin{minipage}[h]{7.cm}
	\begin{center}
		\resizebox{1.0\textwidth}{!}{%
			\includegraphics[width=6.5cm,clip]{\figpath{fe88_au.png}}	}
	\end{center}      
\end{minipage}
\hfill 
\caption{C/E comparison between FENDL-3.2 and FENDL-2.1 reults for ASPIS Iron88 benchmark. The uncertainty bars represent 1$\sigma$ nuclear data uncertainties obtained using ENDF/B-VIII.0 cross section covariances (Al, S, In and Rh detectors) and computational Monte Carlo uncertainty for Au detectors.}
	\label{fig-fe88}
\end{figure*}

The measured and calculated reaction rates are compared in~\cref{fig-fe88} in terms of C/E values. The results demonstrate progress achieved with the new FENDL-3.2 evaluation comparing to FENDL-2.1. Agreement with the measurement is improved in the high and intermediate energy ranges covered by S, Rh and In activation foils. FENDL-3.2 results agree with the measurements within about one-sigma experimental uncertainty also for high penetrations, with a slight underestimation observed for In detectors.
Trends in thermal energies are similar for the two libraries.
The new evaluation represents therefore also a net improvement comparing to ENDF/B-VIII.0. 

\underline{PCA Replica}

Due to smaller iron block thicknesses, PCA Replica experiments represent a less severe test of iron cross sections. Combining the two analyses allows to confirm or invalidate the Fe88 measurements and validation conclusions, and thus to increase the confidence in the conclusions based on Fe88 results.

\begin{figure*}
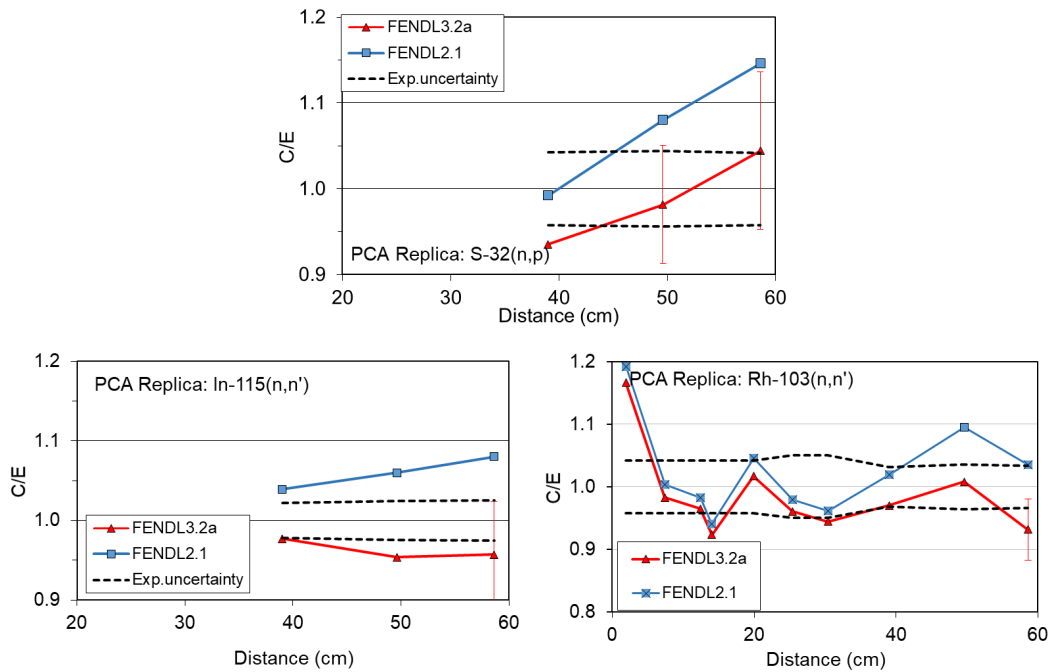

\centering
\includegraphics[width=7cm,clip]{\figpath{pca_s.png}}

\includegraphics[width=7cm,clip]{\figpath{pca_in.png}}
\includegraphics[width=7cm,clip]{\figpath{pca_rh.png}}
\caption{C/E comparison between FENDL-3.2 and FENDL-2.1 reults for ASPIS PCA Replica benchmark. The uncertainty bars represent 1-$\sigma$ nuclear data uncertainties obtained using ENDF/B-VIII.0 cross section covariances.}
\label{fig-pca}
\end{figure*}

As shown in~\cref{fig-pca}, the PCA Replica results indeed perfectly confirm the conclusions drawn from the Fe88 analysis. The measured and calculated reaction rates are again within about one-sigma experimental uncertainty but also exhibit a slight underestimation observed for In detectors.

\underline{CIAE iron slabs}

A series of shielding benchmarks measuring neutron leakage spectra from Fe, Be, SiC, graphite, U, polyethylene, W and multi-slab samples were performed at the China Institute of Atomic Energy (CIAE) Pulse Neutron Generator (CPNG) Cockcroft-Walton type accelerator, which can operate in DC or pulsed mode. A 25 GBq tritium-titanium (T-Ti) source target with an active diameter of 1.6 cm was used. Tritium gas is absorbed in a thin titanium layer (2.6 $\mu$m) evaporated on a 2.2 cm in diameter and 0.05 cm thick Molybdenum backing plate.

The CIAE leakage spectra measurements from iron slabs with D-T neutrons \cite{Ding2019} were presented at the meetings of the Working Party on International Nuclear Data Evaluation Co-operation Subgroup 47 (WPEC SG47) entitled ``Use of Shielding Integral Benchmark Archive and Database for Nuclear Data Validation'' \cite{Kodeli2020} and the SINBAD evaluation of the benchmark is under preparation. 
TOF leakage spectra from iron samples with a diameter of 13 cm and thicknesses of 5, 10 and 15 cm were measured with BC501A liquid scintillation and BaF$_2$ scintillation detectors. The background was estimated by measuring the same configuration without the presence of iron samples. Spectra were measured at two angles, 60$^\circ$ and 120$^\circ$ with respect to the D$^+$ beam direction. Measured spectra are provided in the time domain. MCNP input models are also provided in the proposed SINBAD evaluation.

\begin{figure*}
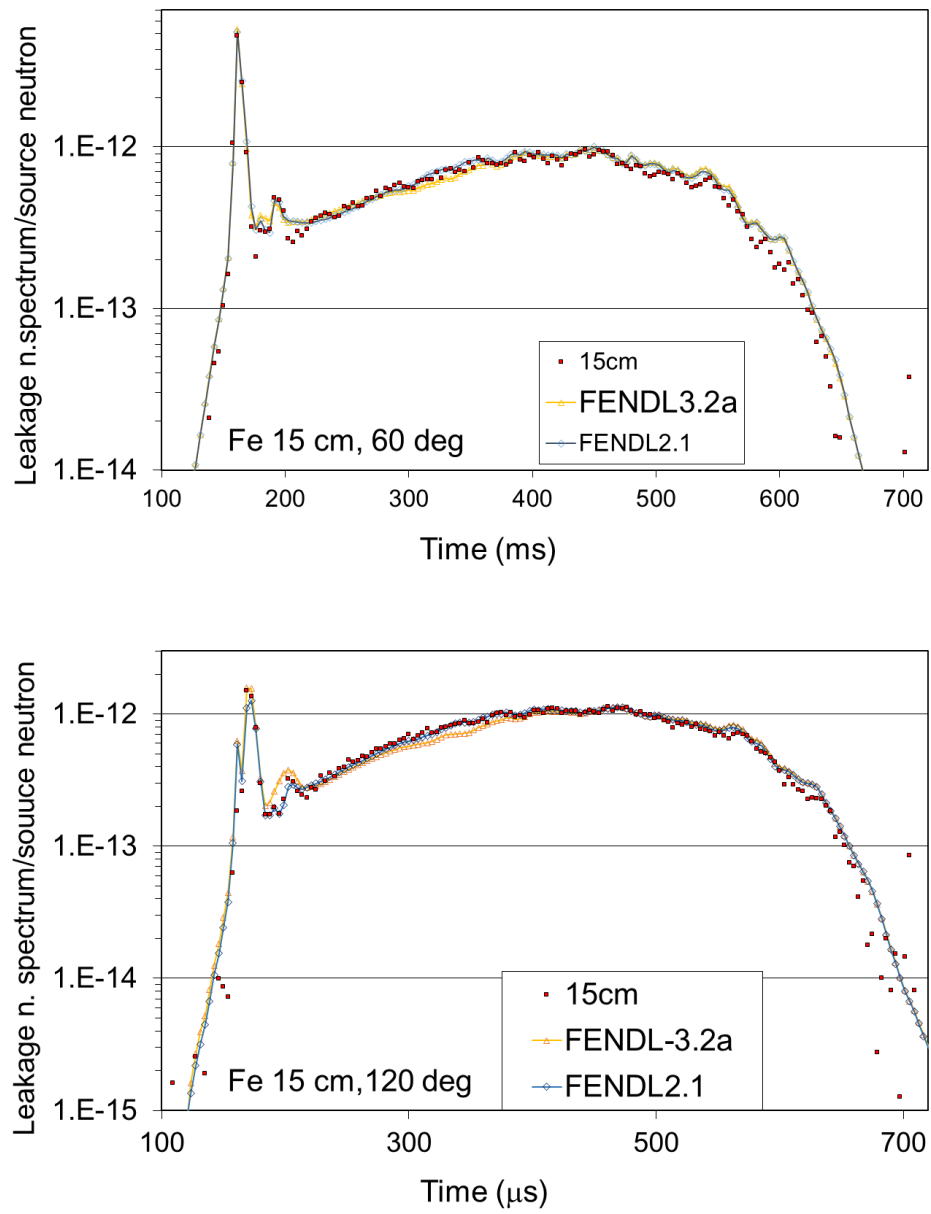

	\centering
	\includegraphics[width=13cm,clip]{\figpath{ciae-60.png}}
	\includegraphics[width=13cm,clip]{\figpath{ciae-120.png}}
	\caption{Comparison of the measured neutron leakage spectra from CIAE iron slabs with the calculated using FENDL-3.2 and FENDL-2.1 cross sections.}
	\label{fig-ciae}
\end{figure*}

Figure~\ref{fig-ciae} compares the measured and calculated neutron TOF spectra for the 15-cm thick iron slabs and the two angles, 60$^\circ$ and 120$^\circ$. Surprisingly, FENDL-2.1 shows better agreement with measurements at some energy ranges, which may point to possible deficiencies of the new iron evaluation in the angular distributions (or/and error compensation problems).

\underline{FNG-SiC benchmark}

The FNG-SiC experiment \cite{Angelone2002} was performed in 2001 to validate the cross sections of Si and C in the European Fusion File (EFF), as the SiC, in the form of a ceramic matrix (SiC-fiber/SiC), is a candidate structural material for advanced fusion reactor designs. 

The ENEA FNG 14-MeV-fusion-neutron source was used to perform measurements of neutron penetration within an experimental setup consisting of a block of sintered SiC (45.72 cm x 45.72 cm, 71.12 cm in thickness). Four experimental positions were available at different penetration depths inside the block for various detectors (activation foils, TLD holders, active spectrometers).

Four different activation foils were used to measure the reactions: $^{197}$Au(n,$\gamma$), $^{58}$Ni(n,p), $^{27}$Al(n,$\alpha$) and $^{93}$Nb(n,2n), covering neutron energies from thermal up to the fusion neutron peak. The reaction rates were measured at four experimental positions, 10.41 cm, 25.65 cm, 40.89 cm and 56.13 cm respectively from the block surface, using the absolutely calibrated HPGe detectors. The overall 1$\sigma$ measurement uncertainties were below 5\%, coming from the HPGe calibration (±2\%), measured activity ($\leq$±3\%) and total neutron yield (±3\%).



\begin{figure*}[tbp]
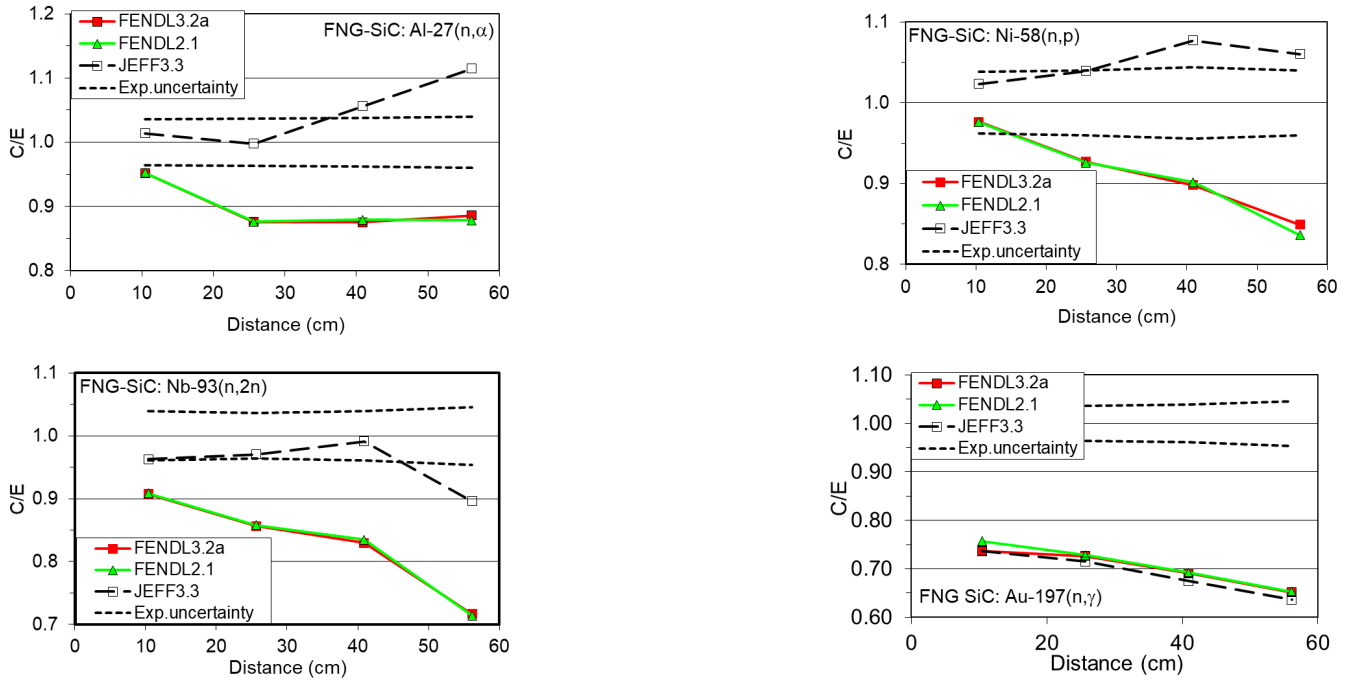

	\begin{minipage}[h]{7.cm}
		\begin{center}
			\resizebox{1.0\textwidth}{!}{
				\includegraphics[width=6.5cm,clip]{\figpath{sic_al.png}	}}
		\end{center}      
	\end{minipage}
	\hfill 
	\begin{minipage}[h]{7.cm}
		\begin{center}
			\resizebox{1.0\textwidth}{!}{
				\includegraphics[width=6.5cm,clip]{\figpath{sic_ni.png} 	}}
		\end{center}
	\end{minipage}
	\hfill 
	\begin{minipage}[h]{7.cm}
		\begin{center}
			\resizebox{1.0\textwidth}{!}{
				\includegraphics[width=6.5cm,clip]{\figpath{sic_nb.png}	}}
		\end{center}      
	\end{minipage}
	\hfill 
	\begin{minipage}[h]{7.cm}
		\begin{center}
			\resizebox{1.0\textwidth}{!}{
				\includegraphics[width=6.5cm,clip]{\figpath{sic_au.png}	}}
		\end{center}
	\end{minipage}
	\hfill 
	\caption{C/E comparison between FENDL-3.2, FENDL-2.1 and JEF-3.3 results for the FNG SiC benchmark.}
	\label{fig-sic}
\end{figure*}

The measured and the calculated neutron reaction rates using FENDL-3.2, -2.1 and JEFF-3.3 libraries at the four detector positions are compared in~\cref{fig-sic}, which indicates little progress of the new library. The comparison also suggests that the JEFF-3.3 SiC nuclear data may be a better choice.

\vspace{2ex}
\underline{FNG-W benchmark}

The FNG Benchmark Experiment on Tungsten \cite{Batistoni2003,Kodeli2004} is one in a series of the high quality fusion relevant benchmarks performed using the FNG 14 MeV neutron source. It was performed in 2002 in order to validate tungsten cross sections in the European Fusion File. Tungsten is a candidate material to be used as high heat flux component in fusion reactors. The mock-up consisted of a block of tungsten alloy with a size of about 42-47 cm wide, 46.85 cm high and 49 cm in thickness. The neutron flux was measured using $^{27}$Al(n,$\alpha$), $^{93}$Nb(n,2n), $^{90}$Zr(n,2n), $^{56}$Fe(n,p), $^{58}$Ni(n,2n), $^{58}$Ni(n,p), $^{115}$In(n,n'), $^{55}$Mn(n,$\gamma$) and $^{197}$Au(n,$\gamma$) activation foil reactions. 

The comparison of the measured and the calculated neutron reaction rates at the four detector positions is shown in~\cref{fig-w} and demonstrates good agreement, both for FENDL-3.2a and -2.1 to predict high threshold reaction rates (such as $^{93}$Nb(n,2n)) and epithermal and thermal reaction rates ($^{58}$Ni(n,p) and $^{197}$Au(n,$\gamma$)). 

\begin{figure*}
	\centering
	\includegraphics[width=7cm,clip]{\figpath{w_nb.png}}
	\includegraphics[width=7cm,clip]{\figpath{w_ni.png}}
	\includegraphics[width=7cm,clip]{\figpath{w_au.png}}
	\caption{C/E comparison between FENDL-3.2 and FENDL-2.1 results for the FNG W benchmark.}
	\label{fig-w}
\end{figure*}

\vspace{2ex}
\underline{FNG HCPB benchmark}

\begin{figure*}[t]
\centering
\includegraphics[width=8cm,clip]{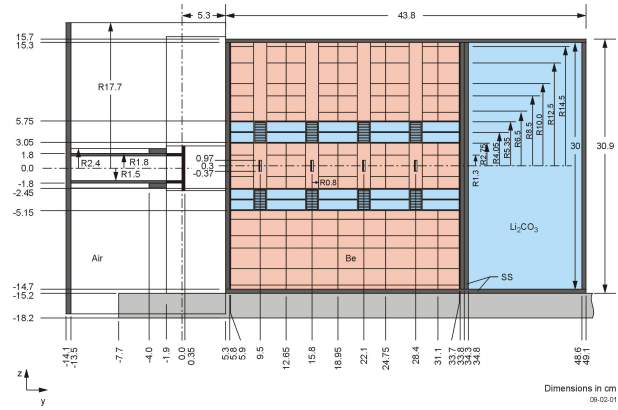}
\caption{View of the computational model for the FNG HCPB benchmark.}
\label{fig-hcpb-1}
\end{figure*}

The objective of the FNG-HCPB experiment \cite{Batistoni2012} performed in 2005 at the FNG facility in Frascati was to study the tritium breeding ratio and other nuclear quantities in a breeder blanket. The experiment consisted of a metallic beryllium setup with two double layers of breeder material (Li$_2$CO$_3$ pebbles) (see~\cref{fig-hcpb-1}). The reaction rate measurements included the Li$_2$CO$_3$ pellets (tritium breeding ratio), activation foils, and neutron and gamma spectrometers inserted at several axial and lateral locations in the block. 

\begin{figure*}[tbp]
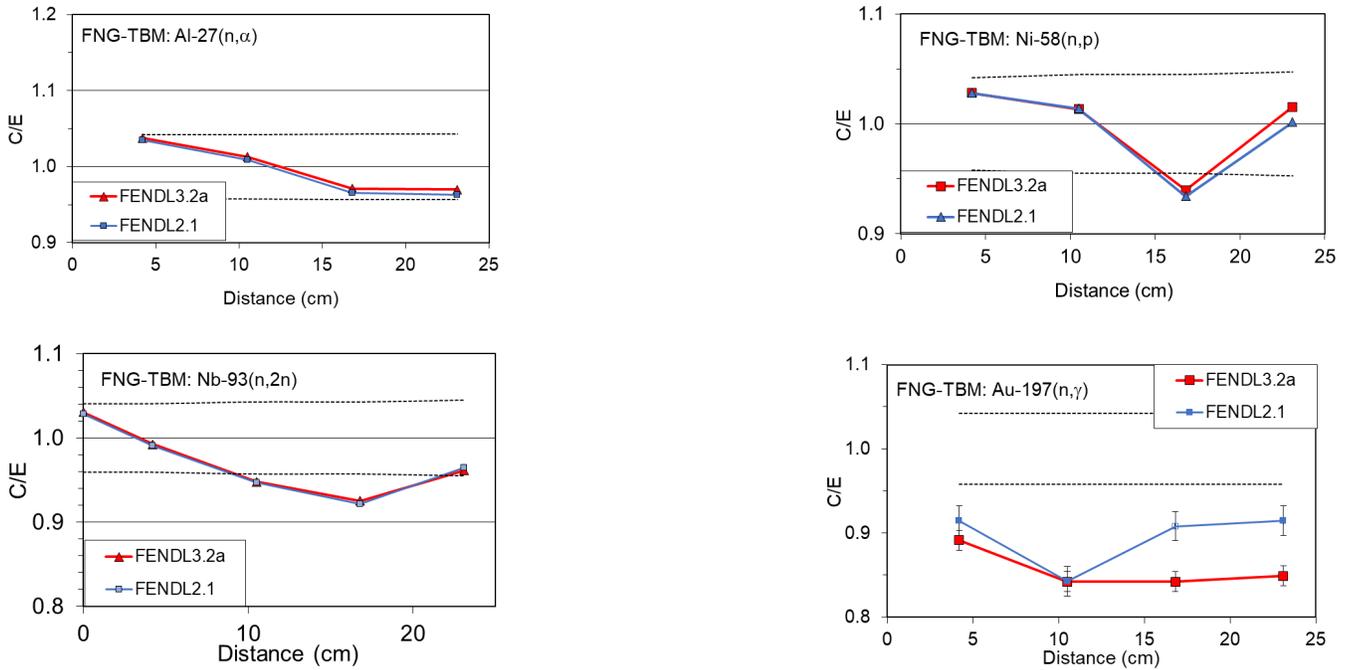

\begin{minipage}[h]{7.cm}
	\begin{center}
		\resizebox{1.0\textwidth}{!}{
			\includegraphics[width=6.5cm,clip]{\figpath{hcpb_al.png}}	}
	\end{center}      
\end{minipage}
\hfill 
\begin{minipage}[h]{7.cm}
	\begin{center}
		\resizebox{1.0\textwidth}{!}{
			\includegraphics[width=6.5cm,clip]{\figpath{hcpb_ni.png} 	}}
	\end{center}
\end{minipage}
\hfill 
\begin{minipage}[h]{7.cm}
	\begin{center}
		\resizebox{1.0\textwidth}{!}{
			\includegraphics[width=6.5cm,clip]{\figpath{hcpb_nb.png}	}}
	\end{center}      
\end{minipage}
\hfill 
\begin{minipage}[h]{7.cm}
	\begin{center}
		\resizebox{1.0\textwidth}{!}{
			\includegraphics[width=6.5cm,clip]{\figpath{hcpb_au.png}	}}
	\end{center}
\end{minipage}
\hfill 
\caption{C/E comparison between FENDL-3.2 and FENDL-2.1 results for the FNG HCPB benchmark.}
\label{fig-hcpb}
\end{figure*}

Good agreement between measured and calculated neutron reaction rates at the five detector positions can be observed in~\cref{fig-hcpb}.

%

\FloatBarrier

\section{CONCLUSION}
\label{sec:conclusion}
This paper summarizes the content and current status of the FENDL
library, which is currently at version 3.2b. The
evaluations in the FENDL library are the result of an international
collaboration in support of fusion research and technology
development.  FENDL contains sub-libraries for incident neutron, proton,
and deuteron cross sections and photo-atomic data for transport
calculations. Recommendations of libraries to be used for activation
and dosimetry are also provided as part of FENDL.  

A comprehensive validation exercise for the neutron-induced sublibrary
was presented in this work and included computational as well as
experimental benchmarks.  Recall that computational benchmarks can
be used to rapidly provide feedback to evaluators or identify obvious
errors introduced in the library whereas experimental benchmarks provide
a direct assessment of the accuracy of the nuclear data and transport codes.

The validation work presented here using computational benchmarks
showed generally good agreement among the tested neutron cross section
libraries for neutron flux, nuclear heating, and primary displacement
damage (dpa).  Also, the tritium production in breeding materials
showed good agreement.  However, gas production (hydrogen, helium, and
tritium) in non-breeding materials (e.g. structural materials) showed
substantial differences.  Note the amount of H/He gas production is
important in determining radiation damage and lifetime in structural
materials e.g. ~\cite{BhattacharyaZinkle2022}.  The amount of tritium
production in non-breeding materials is important in determining tritium
inventory for safety and licensing concerns.  More work will need to
be done to understand the differing gas production among different
libraries. 

Extensive results of the experimental validation effort of the neutron
sub-library for FENDL-3.2 were also presented.  The reader is
encouraged to examine details of the validation results for the
materials of greatest interest in their particular application. 

In general, it was demonstrated that the performance of FENDL-3.2b is
at least as good and in most cases better than FENDL-2.1, which was
the ITER reference library.  An important reason for the improved
performance are the new evaluations for the structural materials iron,
silicon and tungsten.  In particular, the quality of the iron
evaluation has been significantly improved and validation results
indicate that silicon and tungsten are of similar or better quality
compared to FENDL-2.1.  Copper is of similar quality as in FENDL-2.1
but not performing particularly well.  Looking in more detail,
comparison of calculated results and measurements at low energies tend
to have poorer agreement (e.g. FNG benchmarks
in \cref{subsubsec:fngAngelone}) although the Oktavian benchmark also
showed poor agreement above 10 MeV (see \cref{subsubsec:oktavianLaghi}).  

Very limited validation work was performed for high energy neutrons
(only iron and concrete were examined using 40 MeV and 65 MeV
neutrons) in section \cref{subsubsec:tiaraKwon}.  Thus future work should provide
additional validation benchmarks at high energies and with more
materials if possible. 

Validation of IRDFF-II dosimetry reaction data was presented in
\cref{subsubsec:monoESchulc} and shown to produce the best results of
all the major cross section libraries for the most commonly used
reactions. 

Apart from new evaluations of
structural materials, minor glitches in many evaluations and also
errors in the processing have been removed. Therefore, the current version of the
library, FENDL 3.2b, is recommended for fusion research and technology
developments. The FENDL library is available on the NDS website at: \url{https://nds.iaea.org/fendl/} and a GitHub
repository that is kept in sync with the website at: 
\url{https://github.com/IAEA-NDS}.

\section{FUTURE WORK}
\label{sec:future-data-needs}
The advances in computational possibilities for the design and
development of fusion facilities will put an even higher demand on the
quality of the evaluated data in the library being used.  Thus there are several
areas of future work for the FENDL library.

The availability of consistent, reliable and
comprehensive uncertainty information in the FENDL ENDF-formatted files will be
essential to avoid overly conservative safety margins and to reduce
construction costs of fusion facilities.  
Currently, the available covariances in the FENDL-3.2b ENDF files should not be
used because many ENDF files
were assembled from different libraries.  The cross sections and angular
distributions were modified but the covariance matrices were not updated to
reflect these changes. This was because proper uncertainty quantification was
not a priority in the early days of the FENDL project.
Inclusion of trustworthy uncertainty information will be very
important in the near future as fusion research and technology is
rapidly maturing, pending a comprehensive assessment of target accuracies for various important nuclear responses for design and safety.
Such values, together with associated sensitivity and uncertainty analyses, will be important to drive future nuclear data improvements.

Currently, new neutron evaluations are being developed and reviewed within the
INDEN network, which are expected to perform better than the currently
available ones in the FENDL transport library. These evaluations are
for neutron-induced reactions of oxygen, copper, fluorine and tungsten
isotopes. In addition, new improved evaluations are expected
from upcoming versions of international libraries (e.g., ENDF/B, JEFF, JENDL, CENDL, etc.).
In particular, the improvement of neutron data standards will have an impact on lithium and boron isotopes.
These improved evaluations will need to be considered for future versions of FENDL. 

For the deuteron general purpose files, many evaluations from TENDL-2011 were adopted within the IAEA coordinated research project 
between 2008 and 2012. More recent advancements in deuteron reaction modeling, such as described in~\cite{avrigeanuAdvancedBreakupnucleonEnhancement2022,sauvanImplementationNewEnergyangular2017,Nakayama2016}, are expected to provide more reliable nuclear data and should be reviewed for the FENDL project.  

Considering current best practices, the validation process for FENDL should be
automated as far as possible in order to reduce human effort and to
capture problems in evaluations or the processing at an early stage. 
The JADE verification and validation tool was successfully applied for
a part of the validation presented in this work and demonstrated the
advantage of such systems. 
Further development and the application of JADE and similar tools will
play a very important role in the automated validation for future
releases of the FENDL library.
For instance, integral responses should be computed and compared for the IFMIF-DONES test module~\cite{krolasIFMIFDONESFusionOriented2021}.
Further, once reliable uncertainty information
is available in the evaluated 
ENDF files, uncertainty propagation should also be considered in such
automated validation systems for an even more comprehensive assessment
of the library performance.

For dosimetry reactions, the IRDFF-II library is recommended as it has been
extremely well validated in integral measurements and the evaluations are mainly
based on assessed experimental data.
Note in particular, it is recommended to use data in IRDFF-II for gas
production if the reaction is available, which is the case for lithium and boron isotopes.

From the undertaken validation it was identified that the gas production
in structural materials (e.g. steel) needs further work.
Significant differences in gas production cross sections were observed
between different libraries.
For activation calculations, the TENDL-2017 library is usually recommended.
There are two important exceptions (including many evaluated reactions):
\begin{itemize}
    \item for neutron-induced activation, cross section data available in IRDFF-II
    should be given preference over TENDL-2017,
    \item for charged-particle induced activation, available data in the charged
    particle cross section database for medical isotope production and monitor
    reactions~\cite{gulChargedParticleCrosssection2001,betakNuclearDataProduction2011,Hermanne2018,Tarkanyi2019_1,Tarkanyi2019_2,Engle2019,hermanneUpgradeIAEARecommended2021}
    are recommended to be used instead of TENDL-2017.
\end{itemize}

It will be investigated how these libraries can be combined with data
from TENDL to form an improved activation library for FENDL.
We also note that TENDL-2017 ENDF files are general purpose files
and include both transport and activation data. It may be helpful to users
to create a dedicated activation library derived from the TENDL-2017
library for FENDL with the corresponding replacement of IRDFF and
charged-particle activation data mentioned above.

Additionally for future work, the photon production and material
activation data will need
to be validated as it is essential in the assessment of
radiation protection, shielding, decay heat, activation, and
shutdown dose rates. Similarly, the validation of the proton- and
deuteron-induced sublibraries needs to be addressed to support fusion
research and technology developments. 

The necessary developments described above will also require the
development of more computational and experimental benchmarks for the library
validation process.  This will be particularly important for some of
the newly proposed fusion facilities which might use newly developed
materials or materials that were not tested in the current validation process.

\vspace{1ex}

\noindent
\textbf{Acknowledgments}
We would like to thank Denise Neudecker and Maurizio Angelone for fruitful discussions and valuable input.
We also express our gratitude to the referee for their constructive feedback on the manuscript and pointing out additional opportunities for the FENDL project to contribute to the broader nuclear data community.

\bibliography{main}

\clearpage
\appendix

\section{Additional technical information}
\label{apx:add-tech-info}

\begin{minipage}{\linewidth}

The lists and tables in the appendix provide additional details
about the contents of the ENDF and application files in the
FENDL library and their meaning is described here.

\textbf{Application files}, such as ACE files and MATXS files
were produced from the ENDF files by making use of
NJOY2016 with a few additional patches. As reproducibility
is important, an Apptainer (formerly called Singularity) definition file has
been created to produce an Apptainer image file. Both the Apptainer definition
and image file are bundled together with the processed files at 
\url{https://github.com/IAEA-NDS/FENDL-Processed}. Therefore, interested users
can re-produce the application files with a single instruction if the Apptainer
application is installed on their system.
For completeness, a summary of the options used for the processing with NJOY2016
is given in~\cref{tbl:njoy2016options}. The Bodarenko $\sigma_0$ values
used in the calculation of self-shielded cross sections are provided in 
\cref{tbl:tablesig0}.

Regarding ENDF files in the \textbf{neutron transport sublibrary},
\cref{tbl:neutron-sublib1,tbl:neutron-sublib2} give an overview of its content.
As described in \cref{subsubsec:jaea-evaluations}, files were usually assembled
from two base evaluations, one being used for the low energy part up to 20 or 30\,MeV
and the other base library for the energy range above. The source for
the low energy range is indicated in the column \textit{Source \#1} and the one
for the high energy range in \textit{Source \#2}. The energy point where the
switch from one evaluation to the other occurs is given in column \textit{Ecut}.
In some cases, such as for TENDL-2019, one single evaluation covers the complete
energy range up to at least 150\,MeV. In this case, the \textit{Ecut} column is
empty and both \textit{Source \#1} and \textit{Source \#2} name the same library source.
As explained in~\cref{subsubsec:jaea-evaluations}, inconsistencies in the energy/angle
distributions led to non-physical KERMA factors in FENDL versions before 3.2.
These inconsistencies have been resolved by comparatively minor changes in FENDL-3.2.
Whether a file was affected by this issue and needed to be corrected is indicated
by a \textit{Y} in the column labeled \textit{H}.

The sources of the evaluations in the \textbf{proton transport sublibrary} are summarized
in~\cref{tbl:proton-sublib}. Also the upper energy limit of the evaluations is given
in column~\textit{Emax}. 

\end{minipage}

\begin{floatlist}[H]
    \caption{Summary of employed NJOY2016 options for the processing of the FENDL library} 
    \label{tbl:njoy2016options}
\begin{itemize}
    \item Cross-section reconstruction tolerance in RECONR: 0.1 percent.
    \item Temperature: 293.6K
    \item Thinning tolerance for Doppler broadening in BROADR: 0.1 percent.
    \item Maximum energy in BROADR: 20 MeV.
    \item Number of probability bins in PURR: 20.
    \item Number of resonance ladders in PURR: 100.
    \item Bondarenko $\sigma_0$ values: infinity(1.E+10), 100000, 10000, 1000, 300, 100, 30, 10, 3, 1, 0.3, 0.1, 0.001 barns, not more than ten values out of this list according to the material (\cref{tbl:tablesig0}).
    \item No thermal data.
    \item No thinning in ACER.
    \item Type 1 - ACE-formatted file.
    \item ZAID Suffix in ACER: .32
    \item New cumulative angle distributions in ACER.
    \item Detailed photon calculation in ACER.
    \item Neutron groups: 211 energy groups, 175 group Vitamin-J structure + 36 groups up to 55 MeV (1 MeV-width each)
    \item Gamma groups: 42 in Vitamin-J structure.
    \item Neutron weight function: VITAMIN-E (IWT=11 in NJOY).
    \item Gamma weight function: 1/E with roll-offs (IWT=3 in NJOY).
    \item Legendre order: P-6 for transport correction to P-5.
    \item Reactions included: all reactions contained in the evaluated file plus total kerma (MT = 301), partial kermas (MT=302, 304, 404), total kinematic kerma (MT = 443), total damage (MT = 444) and gas production (MT=203-207). For multi-group calculations, MT = 251 ($\mu$), MT = 252 ($\chi$), MT = 253 ($\gamma$) and MT = 259 (1/v) are also included.
\end{itemize}
\end{floatlist}

\begin{table*}
\caption{Bondarenko $\sigma_0$ values for processing neutron files}
\label{tbl:tablesig0}
\begin{tabular}{ | p{1cm} | p{6cm}| p{10.5cm} | }
\hline\hline
 Set & Bondarenko $\sigma_0$ values & Materials \\ [1ex]
 \hline
1 & (1.E+10) & H-1, H-3, He-3, He-4, Li-6, Li-7, B-10, B-11, N-15, F-19, Ne-20, Ne-21, Ne-22, Na-23, Mg-24, Mg-25, Mg-26, Al-27, P-31, Ta-180 \\[1ex]
 \hline
2 & (1.E+10, 1.E+02, 1.E+01, 1.E+00) & Si-28, Si-29, Si-30 \\ [1ex]
 \hline
3 & (1.E+10, 1.E+04, 3.E+03, 1.E+03) & U-234 \\ [1ex]
 \hline
4 & (1.E+10, 1.E+03, 1.E+02, 1.E+01, 1.E+00) & Pb-204, Pb-206, Pb-207, Pb-208 \\ [4ex]
 \hline
5 & (1.E+10, 1.E+04, 1.E+03, 1.E+02, 1.E+01) & Bi-209 \\ [4ex]
 \hline
6 & (1.E+10, 1.E+04, 1.E+03, 1.E+02, 1.E+01, 1.E+00) & H-2, Ta-181 \\ [4ex]
 \hline
7 & (1.E+10, 1.E+03, 3.E+02, 1.E+02, 3.E+01, 1.E+01) & S-32, S-33, S-34, S-36, Cl-35, Cl-37, Ar-36, Ar-38, Ar-40, K-39, K-40, K-41 \\ [4ex]
 \hline
8 & (1.E+10, 1.E+05, 1.E+03, 1.E+02, 1.E+01, 1.E+00) & Mn-55 \\ [4ex]
 \hline
9 & (1.E+10, 1.E+05, 1.E+04, 1.E+03, 1.E+02, 1.E+01) & Fe-54, Fe-57, Fe-58, Co-59 \\ [4ex]
 \hline
10 & (1.E+10, 1.E+03, 3.E+02, 1.E+02, 3.E+01, 1.E+01, 3.E+00, 1.E+00) & Ca-40, Ca-42, Ca-43, Ca-44, Ca-46, Ca-48, Cr-50, Cr-52, Cr-53, Cr-54, Ni-58, Ni-60, Ni-61, Ni-62, Ni-64 \\ [1ex]
 \hline
11 & (1.E+10, 1.E+04, 3.E+02, 1.E+02, 3.E+01, 1.E+01, 1.E+00, 1.E-01) & Cu-63, Cu-65, Zn-64, Zn-66, Zn-67, Zn-68, Zn-70 \\ [4ex]
 \hline
12 & (1.E+10, 1.E+04, 1.E+03, 3.E+02, 1.E+02, 3.E+01, 1.E+01, 1.E+00) & Pt-190, Pt-192, Pt-194, Pt-195, Pt-196, Pt-198, Au-197 \\ [4ex]
 \hline
13 & (1.E+10, 1.E+04, 1.E+03, 3.E+02, 1.E+02, 3.E+01, 1.E+01, 3.E+00, 1.E+00) & Sc-45, Ti-46, Ti-47, Ti-48, Ti-49, Ti-50 \\ [4ex]
 \hline
14 & (1.E+10, 1.E+04, 1.E+03, 3.E+02, 1.E+02, 3.E+01, 1.E+01, 1.E+00, 1.E-01, 1.E-03) & Be-9, C-12, C-13, N-14, O-16, O-17, O-18, V-50, V-51, Ga-69, Ga-71, Ge-70, Ge-72, Ge-73, Ge-74, Ge-76, Br-79, Br-81, Y-89, Zr-90, Zr-91, Zr-92, Zr-94, Zr-96, Nb-93, Mo-92, Mo-94, Mo-95, Mo-96, Mo-97, Mo-98, Mo-100, Rh-103, Ag-107, Ag-109, Cd-106, Cd-108, Cd-110, Cd-111, Cd-112, Cd-113, Cd-114, Cd-116, Sn-112, Sn-114, Sn-115, Sn-116, Sn-117, Sn-118, Sn-119, Sn-120, \newline Sn-122, Sn-124, Sb-121, Sb-123, I-127, Cs-133, Ba-130, Ba-132, Ba-134, \newline Ba-135, Ba-136, Ba-137, Ba-138, La-138, Ce-136, Ce-138, Ce-140, Ce-142, Sm-144, Sm-147, Sm-148, Sm-149, Sm-150, Sm-152, Sm-154, Gd-152, Gd-154, Gd-155, Gd-156, Gd-157, Gd-158, Gd-160, Er-162, Er-164, Er-166, Er-167, Er-168, Er-170, Lu-175, Lu-176, Hf-174, Hf-176, Hf-177, Hf-178, Hf-179, \newline Hf-180, W-180, W-182, W-183, W-184, W-186, Re-185, Re-187 \\ [1ex]
 \hline
15 & (1.E+10, 1.E+05, 1.E+04, 1.E+03, 1.E+02, 1.E+01, 3.E+00, 1.E+00, 3.E-01, 1.E-01) & Fe-56 \\ [4ex]
 \hline
16 & (1.E+10, 1.E+04, 1.E+03, 3.E+02, 1.E+02, 3.E+01, 1.E+01, 1.E+00, 0.3, 0.1) & La-139 \\[1ex]
 \hline
17 & (1.E+10, 1.E+04, 1.E+03, 1.E+02, 3.E+01, 1.E+01, 3.E+00, 1.E+00, 1.E-01, 1.E-03) & Th-232, U-238 \\ [4ex]
 \hline
18 & (1.E+10, 1.E+04, 3.E+03, 1.E+03, 3.E+02, 1.E+02, 3.E+01, 1.E+01, 3.E+00, 1.E+00) & U-235 \\ [4ex]
 \hline\hline
\end{tabular}
\end{table*}

\setsublib{neutron}

\registermaterial{H-1}
\setmatattr{H-1}{Emin}{1e-05}
\setmatattr{H-1}{Emax}{150}
\setmatattr{H-1}{lowsource}{ENDF/B-VII.1}
\setmatattr{H-1}{hisource}{JENDL-4.0-HE}
\setmatattr{H-1}{Ecut}{20}
\setmatattr{H-1}{konnocorrection}{}
\setmatattr{H-1}{hascov32}{}
\setmatattr{H-1}{hascov33}{A}
\setmatattr{H-1}{hascov34}{}
\setmatattr{H-1}{hascov35}{}
\setmatattr{H-1}{hascov36}{}

\registermaterial{H-2}
\setmatattr{H-2}{Emin}{1e-05}
\setmatattr{H-2}{Emax}{150}
\setmatattr{H-2}{lowsource}{ENDF/B-VII.0}
\setmatattr{H-2}{hisource}{ENDF/B-VII.0}
\setmatattr{H-2}{konnocorrection}{}
\setmatattr{H-2}{hascov32}{}
\setmatattr{H-2}{hascov33}{A}
\setmatattr{H-2}{hascov34}{}
\setmatattr{H-2}{hascov35}{}
\setmatattr{H-2}{hascov36}{}

\registermaterial{H-3}
\setmatattr{H-3}{Emin}{1e-05}
\setmatattr{H-3}{Emax}{60}
\setmatattr{H-3}{lowsource}{ENDF/B-VII.1}
\setmatattr{H-3}{hisource}{ENDF/B-VII.1}
\setmatattr{H-3}{comment}{differences found}
\setmatattr{H-3}{konnocorrection}{}
\setmatattr{H-3}{hascov32}{}
\setmatattr{H-3}{hascov33}{}
\setmatattr{H-3}{hascov34}{}
\setmatattr{H-3}{hascov35}{}
\setmatattr{H-3}{hascov36}{}

\registermaterial{He-3}
\setmatattr{He-3}{Emin}{1e-05}
\setmatattr{He-3}{Emax}{60}
\setmatattr{He-3}{lowsource}{JENDL-4}
\setmatattr{He-3}{hisource}{JENDL-4}
\setmatattr{He-3}{comment}{problem retrieving JENDL-4}
\setmatattr{He-3}{konnocorrection}{}
\setmatattr{He-3}{hascov32}{}
\setmatattr{He-3}{hascov33}{}
\setmatattr{He-3}{hascov34}{}
\setmatattr{He-3}{hascov35}{}
\setmatattr{He-3}{hascov36}{}

\registermaterial{He-4}
\setmatattr{He-4}{Emin}{1e-05}
\setmatattr{He-4}{Emax}{60}
\setmatattr{He-4}{lowsource}{ENDF/B-VII.1}
\setmatattr{He-4}{hisource}{ENDF/B-VII.1}
\setmatattr{He-4}{comment}{differences found}
\setmatattr{He-4}{konnocorrection}{}
\setmatattr{He-4}{hascov32}{}
\setmatattr{He-4}{hascov33}{A}
\setmatattr{He-4}{hascov34}{}
\setmatattr{He-4}{hascov35}{}
\setmatattr{He-4}{hascov36}{}

\registermaterial{Li-6}
\setmatattr{Li-6}{Emin}{1e-05}
\setmatattr{Li-6}{Emax}{200}
\setmatattr{Li-6}{lowsource}{ENDF-B-VII.1}
\setmatattr{Li-6}{hisource}{TENDL-2010}
\setmatattr{Li-6}{Ecut}{20}
\setmatattr{Li-6}{konnocorrection}{Y}
\setmatattr{Li-6}{hascov32}{}
\setmatattr{Li-6}{hascov33}{A}
\setmatattr{Li-6}{hascov34}{}
\setmatattr{Li-6}{hascov35}{}
\setmatattr{Li-6}{hascov36}{}

\registermaterial{Li-7}
\setmatattr{Li-7}{Emin}{1e-05}
\setmatattr{Li-7}{Emax}{200}
\setmatattr{Li-7}{lowsource}{ENDF-B-VII.1}
\setmatattr{Li-7}{hisource}{TENDL-2010}
\setmatattr{Li-7}{Ecut}{20}
\setmatattr{Li-7}{konnocorrection}{Y}
\setmatattr{Li-7}{hascov32}{}
\setmatattr{Li-7}{hascov33}{A}
\setmatattr{Li-7}{hascov34}{}
\setmatattr{Li-7}{hascov35}{}
\setmatattr{Li-7}{hascov36}{}

\registermaterial{Be-9}
\setmatattr{Be-9}{Emin}{1e-05}
\setmatattr{Be-9}{Emax}{200}
\setmatattr{Be-9}{lowsource}{ENDF-B-VII.1}
\setmatattr{Be-9}{hisource}{TENDL-2010}
\setmatattr{Be-9}{Ecut}{20}
\setmatattr{Be-9}{konnocorrection}{Y}
\setmatattr{Be-9}{hascov32}{}
\setmatattr{Be-9}{hascov33}{A}
\setmatattr{Be-9}{hascov34}{}
\setmatattr{Be-9}{hascov35}{}
\setmatattr{Be-9}{hascov36}{}

\registermaterial{B-10}
\setmatattr{B-10}{Emin}{1e-05}
\setmatattr{B-10}{Emax}{60}
\setmatattr{B-10}{lowsource}{FENDL-3.2}
\setmatattr{B-10}{hisource}{FENDL-3.2}
\setmatattr{B-10}{comment}{problem retrieving FENDL-3.2}
\setmatattr{B-10}{konnocorrection}{}
\setmatattr{B-10}{hascov32}{}
\setmatattr{B-10}{hascov33}{A}
\setmatattr{B-10}{hascov34}{}
\setmatattr{B-10}{hascov35}{}
\setmatattr{B-10}{hascov36}{}

\registermaterial{B-11}
\setmatattr{B-11}{Emin}{1e-05}
\setmatattr{B-11}{Emax}{60}
\setmatattr{B-11}{lowsource}{ENDF-B-VII.1}
\setmatattr{B-11}{hisource}{TENDL-2010}
\setmatattr{B-11}{Ecut}{20}
\setmatattr{B-11}{comment}{diff in low energy part / diff in high energy part}
\setmatattr{B-11}{konnocorrection}{}
\setmatattr{B-11}{hascov32}{}
\setmatattr{B-11}{hascov33}{A}
\setmatattr{B-11}{hascov34}{}
\setmatattr{B-11}{hascov35}{}
\setmatattr{B-11}{hascov36}{}

\registermaterial{C-12}
\setmatattr{C-12}{Emin}{1e-05}
\setmatattr{C-12}{Emax}{150}
\setmatattr{C-12}{lowsource}{JENDL-4.0}
\setmatattr{C-12}{hisource}{JENDL/HE-2007}
\setmatattr{C-12}{comment}{problem retrieving JENDL-4.0}
\setmatattr{C-12}{konnocorrection}{}
\setmatattr{C-12}{hascov32}{}
\setmatattr{C-12}{hascov33}{}
\setmatattr{C-12}{hascov34}{}
\setmatattr{C-12}{hascov35}{}
\setmatattr{C-12}{hascov36}{}

\registermaterial{C-13}
\setmatattr{C-13}{Emin}{1e-05}
\setmatattr{C-13}{Emax}{200}
\setmatattr{C-13}{lowsource}{TENDL-2014}
\setmatattr{C-13}{hisource}{TENDL-2014}
\setmatattr{C-13}{konnocorrection}{Y}
\setmatattr{C-13}{hascov32}{A}
\setmatattr{C-13}{hascov33}{A}
\setmatattr{C-13}{hascov34}{A}
\setmatattr{C-13}{hascov35}{}
\setmatattr{C-13}{hascov36}{}

\registermaterial{N-14}
\setmatattr{N-14}{Emin}{1e-05}
\setmatattr{N-14}{Emax}{150}
\setmatattr{N-14}{lowsource}{JENDL-4.0}
\setmatattr{N-14}{hisource}{JENDL-4.0-HE}
\setmatattr{N-14}{Ecut}{20}
\setmatattr{N-14}{konnocorrection}{}
\setmatattr{N-14}{hascov32}{}
\setmatattr{N-14}{hascov33}{A}
\setmatattr{N-14}{hascov34}{}
\setmatattr{N-14}{hascov35}{}
\setmatattr{N-14}{hascov36}{}

\registermaterial{N-15}
\setmatattr{N-15}{Emin}{1e-05}
\setmatattr{N-15}{Emax}{200}
\setmatattr{N-15}{lowsource}{RUSFOND-2010}
\setmatattr{N-15}{hisource}{TENDL-2010}
\setmatattr{N-15}{comment}{problem retrieving RUSFOND-2010}
\setmatattr{N-15}{konnocorrection}{Y}
\setmatattr{N-15}{hascov32}{}
\setmatattr{N-15}{hascov33}{}
\setmatattr{N-15}{hascov34}{}
\setmatattr{N-15}{hascov35}{}
\setmatattr{N-15}{hascov36}{}

\registermaterial{O-16}
\setmatattr{O-16}{Emin}{1e-05}
\setmatattr{O-16}{Emax}{150}
\setmatattr{O-16}{lowsource}{FENDL-3.2}
\setmatattr{O-16}{hisource}{FENDL-3.2}
\setmatattr{O-16}{comment}{problem retrieving FENDL-3.2}
\setmatattr{O-16}{konnocorrection}{}
\setmatattr{O-16}{hascov32}{}
\setmatattr{O-16}{hascov33}{A}
\setmatattr{O-16}{hascov34}{A}
\setmatattr{O-16}{hascov35}{}
\setmatattr{O-16}{hascov36}{}

\registermaterial{O-17}
\setmatattr{O-17}{Emin}{1e-05}
\setmatattr{O-17}{Emax}{200}
\setmatattr{O-17}{lowsource}{TENDL-2010}
\setmatattr{O-17}{hisource}{TENDL-2010}
\setmatattr{O-17}{konnocorrection}{Y}
\setmatattr{O-17}{hascov32}{A}
\setmatattr{O-17}{hascov33}{A}
\setmatattr{O-17}{hascov34}{A}
\setmatattr{O-17}{hascov35}{}
\setmatattr{O-17}{hascov36}{}

\registermaterial{O-18}
\setmatattr{O-18}{Emin}{1e-05}
\setmatattr{O-18}{Emax}{200}
\setmatattr{O-18}{lowsource}{FENDL-3.2}
\setmatattr{O-18}{hisource}{FENDL-3.2}
\setmatattr{O-18}{comment}{problem retrieving FENDL-3.2}
\setmatattr{O-18}{konnocorrection}{}
\setmatattr{O-18}{hascov32}{}
\setmatattr{O-18}{hascov33}{A}
\setmatattr{O-18}{hascov34}{}
\setmatattr{O-18}{hascov35}{}
\setmatattr{O-18}{hascov36}{}

\registermaterial{F-19}
\setmatattr{F-19}{Emin}{1e-05}
\setmatattr{F-19}{Emax}{150}
\setmatattr{F-19}{lowsource}{ENDF-B-VII.1}
\setmatattr{F-19}{hisource}{JENDL-4.0-HE}
\setmatattr{F-19}{Ecut}{20}
\setmatattr{F-19}{konnocorrection}{}
\setmatattr{F-19}{hascov32}{}
\setmatattr{F-19}{hascov33}{A}
\setmatattr{F-19}{hascov34}{}
\setmatattr{F-19}{hascov35}{}
\setmatattr{F-19}{hascov36}{}

\registermaterial{Ne-20}
\setmatattr{Ne-20}{Emin}{1e-05}
\setmatattr{Ne-20}{Emax}{200}
\setmatattr{Ne-20}{lowsource}{TENDL-2019}
\setmatattr{Ne-20}{hisource}{TENDL-2019}
\setmatattr{Ne-20}{konnocorrection}{}
\setmatattr{Ne-20}{hascov32}{}
\setmatattr{Ne-20}{hascov33}{A}
\setmatattr{Ne-20}{hascov34}{A}
\setmatattr{Ne-20}{hascov35}{}
\setmatattr{Ne-20}{hascov36}{}

\registermaterial{Ne-21}
\setmatattr{Ne-21}{Emin}{1e-05}
\setmatattr{Ne-21}{Emax}{200}
\setmatattr{Ne-21}{lowsource}{TENDL-2019}
\setmatattr{Ne-21}{hisource}{TENDL-2019}
\setmatattr{Ne-21}{konnocorrection}{}
\setmatattr{Ne-21}{hascov32}{}
\setmatattr{Ne-21}{hascov33}{A}
\setmatattr{Ne-21}{hascov34}{A}
\setmatattr{Ne-21}{hascov35}{}
\setmatattr{Ne-21}{hascov36}{}

\registermaterial{Ne-22}
\setmatattr{Ne-22}{Emin}{1e-05}
\setmatattr{Ne-22}{Emax}{200}
\setmatattr{Ne-22}{lowsource}{TENDL-2019}
\setmatattr{Ne-22}{hisource}{TENDL-2019}
\setmatattr{Ne-22}{konnocorrection}{}
\setmatattr{Ne-22}{hascov32}{}
\setmatattr{Ne-22}{hascov33}{A}
\setmatattr{Ne-22}{hascov34}{A}
\setmatattr{Ne-22}{hascov35}{}
\setmatattr{Ne-22}{hascov36}{}

\registermaterial{Na-23}
\setmatattr{Na-23}{Emin}{1e-05}
\setmatattr{Na-23}{Emax}{150}
\setmatattr{Na-23}{lowsource}{JENDL-4.0}
\setmatattr{Na-23}{hisource}{JENDL/HE-2007}
\setmatattr{Na-23}{Ecut}{20}
\setmatattr{Na-23}{konnocorrection}{}
\setmatattr{Na-23}{hascov32}{}
\setmatattr{Na-23}{hascov33}{}
\setmatattr{Na-23}{hascov34}{}
\setmatattr{Na-23}{hascov35}{}
\setmatattr{Na-23}{hascov36}{}

\registermaterial{Mg-24}
\setmatattr{Mg-24}{Emin}{1e-05}
\setmatattr{Mg-24}{Emax}{150}
\setmatattr{Mg-24}{lowsource}{JENDL-4.0}
\setmatattr{Mg-24}{hisource}{JENDL/HE-2007}
\setmatattr{Mg-24}{Ecut}{20}
\setmatattr{Mg-24}{konnocorrection}{}
\setmatattr{Mg-24}{hascov32}{}
\setmatattr{Mg-24}{hascov33}{}
\setmatattr{Mg-24}{hascov34}{}
\setmatattr{Mg-24}{hascov35}{}
\setmatattr{Mg-24}{hascov36}{}

\registermaterial{Mg-25}
\setmatattr{Mg-25}{Emin}{1e-05}
\setmatattr{Mg-25}{Emax}{150}
\setmatattr{Mg-25}{lowsource}{JENDL-4.0}
\setmatattr{Mg-25}{hisource}{JENDL/HE-2007}
\setmatattr{Mg-25}{Ecut}{20}
\setmatattr{Mg-25}{konnocorrection}{}
\setmatattr{Mg-25}{hascov32}{}
\setmatattr{Mg-25}{hascov33}{}
\setmatattr{Mg-25}{hascov34}{}
\setmatattr{Mg-25}{hascov35}{}
\setmatattr{Mg-25}{hascov36}{}

\registermaterial{Mg-26}
\setmatattr{Mg-26}{Emin}{1e-05}
\setmatattr{Mg-26}{Emax}{150}
\setmatattr{Mg-26}{lowsource}{JENDL-4.0}
\setmatattr{Mg-26}{hisource}{JENDL/HE-2007}
\setmatattr{Mg-26}{Ecut}{20}
\setmatattr{Mg-26}{konnocorrection}{}
\setmatattr{Mg-26}{hascov32}{}
\setmatattr{Mg-26}{hascov33}{}
\setmatattr{Mg-26}{hascov34}{}
\setmatattr{Mg-26}{hascov35}{}
\setmatattr{Mg-26}{hascov36}{}

\registermaterial{Al-27}
\setmatattr{Al-27}{Emin}{1e-05}
\setmatattr{Al-27}{Emax}{150}
\setmatattr{Al-27}{lowsource}{JEFF-311}
\setmatattr{Al-27}{hisource}{JEFF-311}
\setmatattr{Al-27}{konnocorrection}{}
\setmatattr{Al-27}{hascov32}{}
\setmatattr{Al-27}{hascov33}{}
\setmatattr{Al-27}{hascov34}{}
\setmatattr{Al-27}{hascov35}{}
\setmatattr{Al-27}{hascov36}{}

\registermaterial{Si-28}
\setmatattr{Si-28}{Emin}{1e-05}
\setmatattr{Si-28}{Emax}{150}
\setmatattr{Si-28}{lowsource}{ENDF/B-VII.0}
\setmatattr{Si-28}{hisource}{ENDF/B-VII.0}
\setmatattr{Si-28}{konnocorrection}{}
\setmatattr{Si-28}{hascov32}{}
\setmatattr{Si-28}{hascov33}{}
\setmatattr{Si-28}{hascov34}{}
\setmatattr{Si-28}{hascov35}{}
\setmatattr{Si-28}{hascov36}{}

\registermaterial{Si-29}
\setmatattr{Si-29}{Emin}{1e-05}
\setmatattr{Si-29}{Emax}{150}
\setmatattr{Si-29}{lowsource}{ENDF/B-VII.0}
\setmatattr{Si-29}{hisource}{ENDF/B-VII.0}
\setmatattr{Si-29}{konnocorrection}{}
\setmatattr{Si-29}{hascov32}{}
\setmatattr{Si-29}{hascov33}{}
\setmatattr{Si-29}{hascov34}{}
\setmatattr{Si-29}{hascov35}{}
\setmatattr{Si-29}{hascov36}{}

\registermaterial{Si-30}
\setmatattr{Si-30}{Emin}{1e-05}
\setmatattr{Si-30}{Emax}{150}
\setmatattr{Si-30}{lowsource}{ENDF/B-VII.0}
\setmatattr{Si-30}{hisource}{ENDF/B-VII.0}
\setmatattr{Si-30}{konnocorrection}{}
\setmatattr{Si-30}{hascov32}{}
\setmatattr{Si-30}{hascov33}{}
\setmatattr{Si-30}{hascov34}{}
\setmatattr{Si-30}{hascov35}{}
\setmatattr{Si-30}{hascov36}{}

\registermaterial{P-31}
\setmatattr{P-31}{Emin}{1e-05}
\setmatattr{P-31}{Emax}{200}
\setmatattr{P-31}{lowsource}{TENDL-2014}
\setmatattr{P-31}{hisource}{TENDL-2014}
\setmatattr{P-31}{konnocorrection}{Y}
\setmatattr{P-31}{hascov32}{A}
\setmatattr{P-31}{hascov33}{A}
\setmatattr{P-31}{hascov34}{A}
\setmatattr{P-31}{hascov35}{}
\setmatattr{P-31}{hascov36}{}

\registermaterial{S-32}
\setmatattr{S-32}{Emin}{1e-05}
\setmatattr{S-32}{Emax}{200}
\setmatattr{S-32}{lowsource}{TENDL-2010}
\setmatattr{S-32}{hisource}{TENDL-2010}
\setmatattr{S-32}{konnocorrection}{Y}
\setmatattr{S-32}{hascov32}{A}
\setmatattr{S-32}{hascov33}{A}
\setmatattr{S-32}{hascov34}{A}
\setmatattr{S-32}{hascov35}{}
\setmatattr{S-32}{hascov36}{}

\registermaterial{S-33}
\setmatattr{S-33}{Emin}{1e-05}
\setmatattr{S-33}{Emax}{200}
\setmatattr{S-33}{lowsource}{TENDL-2010}
\setmatattr{S-33}{hisource}{TENDL-2010}
\setmatattr{S-33}{konnocorrection}{Y}
\setmatattr{S-33}{hascov32}{A}
\setmatattr{S-33}{hascov33}{A}
\setmatattr{S-33}{hascov34}{A}
\setmatattr{S-33}{hascov35}{}
\setmatattr{S-33}{hascov36}{}

\registermaterial{S-34}
\setmatattr{S-34}{Emin}{1e-05}
\setmatattr{S-34}{Emax}{200}
\setmatattr{S-34}{lowsource}{TENDL-2014}
\setmatattr{S-34}{hisource}{TENDL-2014}
\setmatattr{S-34}{konnocorrection}{Y}
\setmatattr{S-34}{hascov32}{A}
\setmatattr{S-34}{hascov33}{A}
\setmatattr{S-34}{hascov34}{A}
\setmatattr{S-34}{hascov35}{}
\setmatattr{S-34}{hascov36}{}

\registermaterial{S-36}
\setmatattr{S-36}{Emin}{1e-05}
\setmatattr{S-36}{Emax}{200}
\setmatattr{S-36}{lowsource}{TENDL-2014}
\setmatattr{S-36}{hisource}{TENDL-2014}
\setmatattr{S-36}{konnocorrection}{Y}
\setmatattr{S-36}{hascov32}{A}
\setmatattr{S-36}{hascov33}{A}
\setmatattr{S-36}{hascov34}{A}
\setmatattr{S-36}{hascov35}{}
\setmatattr{S-36}{hascov36}{}

\registermaterial{Cl-35}
\setmatattr{Cl-35}{Emin}{1e-05}
\setmatattr{Cl-35}{Emax}{150}
\setmatattr{Cl-35}{lowsource}{ENDF/B-VII.0}
\setmatattr{Cl-35}{hisource}{JENDL-4.0-HE}
\setmatattr{Cl-35}{Ecut}{20}
\setmatattr{Cl-35}{konnocorrection}{}
\setmatattr{Cl-35}{hascov32}{}
\setmatattr{Cl-35}{hascov33}{}
\setmatattr{Cl-35}{hascov34}{}
\setmatattr{Cl-35}{hascov35}{}
\setmatattr{Cl-35}{hascov36}{}

\registermaterial{Cl-37}
\setmatattr{Cl-37}{Emin}{1e-05}
\setmatattr{Cl-37}{Emax}{150}
\setmatattr{Cl-37}{lowsource}{ENDF/B-VII.0}
\setmatattr{Cl-37}{hisource}{JENDL-4.0-HE}
\setmatattr{Cl-37}{Ecut}{20}
\setmatattr{Cl-37}{konnocorrection}{}
\setmatattr{Cl-37}{hascov32}{}
\setmatattr{Cl-37}{hascov33}{}
\setmatattr{Cl-37}{hascov34}{}
\setmatattr{Cl-37}{hascov35}{}
\setmatattr{Cl-37}{hascov36}{}

\registermaterial{Ar-36}
\setmatattr{Ar-36}{Emin}{1e-05}
\setmatattr{Ar-36}{Emax}{150}
\setmatattr{Ar-36}{lowsource}{JENDL/HE-2007}
\setmatattr{Ar-36}{hisource}{JENDL/HE-2007}
\setmatattr{Ar-36}{comment}{problem retrieving low and high energy library}
\setmatattr{Ar-36}{konnocorrection}{}
\setmatattr{Ar-36}{hascov32}{}
\setmatattr{Ar-36}{hascov33}{}
\setmatattr{Ar-36}{hascov34}{}
\setmatattr{Ar-36}{hascov35}{}
\setmatattr{Ar-36}{hascov36}{}

\registermaterial{Ar-38}
\setmatattr{Ar-38}{Emin}{1e-05}
\setmatattr{Ar-38}{Emax}{150}
\setmatattr{Ar-38}{lowsource}{JENDL-4.0}
\setmatattr{Ar-38}{hisource}{JENDL/HE-2007}
\setmatattr{Ar-38}{comment}{problem retrieving JENDL-4.0}
\setmatattr{Ar-38}{konnocorrection}{}
\setmatattr{Ar-38}{hascov32}{}
\setmatattr{Ar-38}{hascov33}{}
\setmatattr{Ar-38}{hascov34}{}
\setmatattr{Ar-38}{hascov35}{}
\setmatattr{Ar-38}{hascov36}{}

\registermaterial{Ar-40}
\setmatattr{Ar-40}{Emin}{1e-05}
\setmatattr{Ar-40}{Emax}{150}
\setmatattr{Ar-40}{lowsource}{JENDL-4.0}
\setmatattr{Ar-40}{hisource}{JENDL/HE-2007}
\setmatattr{Ar-40}{Ecut}{20}
\setmatattr{Ar-40}{konnocorrection}{}
\setmatattr{Ar-40}{hascov32}{}
\setmatattr{Ar-40}{hascov33}{}
\setmatattr{Ar-40}{hascov34}{}
\setmatattr{Ar-40}{hascov35}{}
\setmatattr{Ar-40}{hascov36}{}

\registermaterial{K-39}
\setmatattr{K-39}{Emin}{1e-05}
\setmatattr{K-39}{Emax}{200}
\setmatattr{K-39}{lowsource}{TENDL-2015}
\setmatattr{K-39}{hisource}{TENDL-2015}
\setmatattr{K-39}{comment}{differences found}
\setmatattr{K-39}{konnocorrection}{Y}
\setmatattr{K-39}{hascov32}{A}
\setmatattr{K-39}{hascov33}{A}
\setmatattr{K-39}{hascov34}{A}
\setmatattr{K-39}{hascov35}{}
\setmatattr{K-39}{hascov36}{}

\registermaterial{K-40}
\setmatattr{K-40}{Emin}{1e-05}
\setmatattr{K-40}{Emax}{200}
\setmatattr{K-40}{lowsource}{TENDL-2015}
\setmatattr{K-40}{hisource}{TENDL-2015}
\setmatattr{K-40}{konnocorrection}{Y}
\setmatattr{K-40}{hascov32}{A}
\setmatattr{K-40}{hascov33}{A}
\setmatattr{K-40}{hascov34}{A}
\setmatattr{K-40}{hascov35}{}
\setmatattr{K-40}{hascov36}{}

\registermaterial{K-41}
\setmatattr{K-41}{Emin}{1e-05}
\setmatattr{K-41}{Emax}{200}
\setmatattr{K-41}{lowsource}{TENDL-2015}
\setmatattr{K-41}{hisource}{TENDL-2015}
\setmatattr{K-41}{konnocorrection}{Y}
\setmatattr{K-41}{hascov32}{A}
\setmatattr{K-41}{hascov33}{A}
\setmatattr{K-41}{hascov34}{A}
\setmatattr{K-41}{hascov35}{}
\setmatattr{K-41}{hascov36}{}

\registermaterial{Ca-40}
\setmatattr{Ca-40}{Emin}{1e-05}
\setmatattr{Ca-40}{Emax}{150}
\setmatattr{Ca-40}{lowsource}{JENDL-4.0}
\setmatattr{Ca-40}{hisource}{JENDL/HE-2007}
\setmatattr{Ca-40}{Ecut}{20}
\setmatattr{Ca-40}{konnocorrection}{}
\setmatattr{Ca-40}{hascov32}{}
\setmatattr{Ca-40}{hascov33}{}
\setmatattr{Ca-40}{hascov34}{}
\setmatattr{Ca-40}{hascov35}{}
\setmatattr{Ca-40}{hascov36}{}

\registermaterial{Ca-42}
\setmatattr{Ca-42}{Emin}{1e-05}
\setmatattr{Ca-42}{Emax}{150}
\setmatattr{Ca-42}{lowsource}{JENDL-4.0}
\setmatattr{Ca-42}{hisource}{JENDL/HE-2007}
\setmatattr{Ca-42}{Ecut}{20}
\setmatattr{Ca-42}{konnocorrection}{}
\setmatattr{Ca-42}{hascov32}{}
\setmatattr{Ca-42}{hascov33}{}
\setmatattr{Ca-42}{hascov34}{}
\setmatattr{Ca-42}{hascov35}{}
\setmatattr{Ca-42}{hascov36}{}

\registermaterial{Ca-43}
\setmatattr{Ca-43}{Emin}{1e-05}
\setmatattr{Ca-43}{Emax}{150}
\setmatattr{Ca-43}{lowsource}{JENDL-4.0}
\setmatattr{Ca-43}{hisource}{JENDL/HE-2007}
\setmatattr{Ca-43}{Ecut}{20}
\setmatattr{Ca-43}{konnocorrection}{}
\setmatattr{Ca-43}{hascov32}{}
\setmatattr{Ca-43}{hascov33}{}
\setmatattr{Ca-43}{hascov34}{}
\setmatattr{Ca-43}{hascov35}{}
\setmatattr{Ca-43}{hascov36}{}

\registermaterial{Ca-44}
\setmatattr{Ca-44}{Emin}{1e-05}
\setmatattr{Ca-44}{Emax}{150}
\setmatattr{Ca-44}{lowsource}{JENDL-4.0}
\setmatattr{Ca-44}{hisource}{JENDL/HE-2007}
\setmatattr{Ca-44}{Ecut}{20}
\setmatattr{Ca-44}{konnocorrection}{}
\setmatattr{Ca-44}{hascov32}{}
\setmatattr{Ca-44}{hascov33}{}
\setmatattr{Ca-44}{hascov34}{}
\setmatattr{Ca-44}{hascov35}{}
\setmatattr{Ca-44}{hascov36}{}

\registermaterial{Ca-46}
\setmatattr{Ca-46}{Emin}{1e-05}
\setmatattr{Ca-46}{Emax}{150}
\setmatattr{Ca-46}{lowsource}{JENDL-4.0}
\setmatattr{Ca-46}{hisource}{JENDL/HE-2007}
\setmatattr{Ca-46}{Ecut}{20}
\setmatattr{Ca-46}{konnocorrection}{}
\setmatattr{Ca-46}{hascov32}{}
\setmatattr{Ca-46}{hascov33}{}
\setmatattr{Ca-46}{hascov34}{}
\setmatattr{Ca-46}{hascov35}{}
\setmatattr{Ca-46}{hascov36}{}

\registermaterial{Ca-48}
\setmatattr{Ca-48}{Emin}{1e-05}
\setmatattr{Ca-48}{Emax}{150}
\setmatattr{Ca-48}{lowsource}{JENDL-4.0}
\setmatattr{Ca-48}{hisource}{JENDL/HE-2007}
\setmatattr{Ca-48}{Ecut}{20}
\setmatattr{Ca-48}{konnocorrection}{}
\setmatattr{Ca-48}{hascov32}{}
\setmatattr{Ca-48}{hascov33}{}
\setmatattr{Ca-48}{hascov34}{}
\setmatattr{Ca-48}{hascov35}{}
\setmatattr{Ca-48}{hascov36}{}

\registermaterial{Sc-45}
\setmatattr{Sc-45}{Emin}{1e-05}
\setmatattr{Sc-45}{Emax}{200}
\setmatattr{Sc-45}{lowsource}{JEFF-311}
\setmatattr{Sc-45}{hisource}{JEFF-311}
\setmatattr{Sc-45}{konnocorrection}{}
\setmatattr{Sc-45}{hascov32}{}
\setmatattr{Sc-45}{hascov33}{}
\setmatattr{Sc-45}{hascov34}{}
\setmatattr{Sc-45}{hascov35}{}
\setmatattr{Sc-45}{hascov36}{}

\registermaterial{Ti-46}
\setmatattr{Ti-46}{Emin}{1e-05}
\setmatattr{Ti-46}{Emax}{150}
\setmatattr{Ti-46}{lowsource}{ENDF-B-VII.1}
\setmatattr{Ti-46}{hisource}{JENDL/HE-2007}
\setmatattr{Ti-46}{Ecut}{20}
\setmatattr{Ti-46}{konnocorrection}{}
\setmatattr{Ti-46}{hascov32}{A}
\setmatattr{Ti-46}{hascov33}{A}
\setmatattr{Ti-46}{hascov34}{}
\setmatattr{Ti-46}{hascov35}{}
\setmatattr{Ti-46}{hascov36}{}

\registermaterial{Ti-47}
\setmatattr{Ti-47}{Emin}{1e-05}
\setmatattr{Ti-47}{Emax}{150}
\setmatattr{Ti-47}{lowsource}{ENDF-B-VII.1}
\setmatattr{Ti-47}{hisource}{JENDL/HE-2007}
\setmatattr{Ti-47}{Ecut}{20}
\setmatattr{Ti-47}{konnocorrection}{}
\setmatattr{Ti-47}{hascov32}{A}
\setmatattr{Ti-47}{hascov33}{A}
\setmatattr{Ti-47}{hascov34}{}
\setmatattr{Ti-47}{hascov35}{}
\setmatattr{Ti-47}{hascov36}{}

\registermaterial{Ti-48}
\setmatattr{Ti-48}{Emin}{1e-05}
\setmatattr{Ti-48}{Emax}{150}
\setmatattr{Ti-48}{lowsource}{ENDF-B-VII.1}
\setmatattr{Ti-48}{hisource}{JENDL/HE-2007}
\setmatattr{Ti-48}{Ecut}{20}
\setmatattr{Ti-48}{konnocorrection}{}
\setmatattr{Ti-48}{hascov32}{A}
\setmatattr{Ti-48}{hascov33}{A}
\setmatattr{Ti-48}{hascov34}{}
\setmatattr{Ti-48}{hascov35}{}
\setmatattr{Ti-48}{hascov36}{}

\registermaterial{Ti-49}
\setmatattr{Ti-49}{Emin}{1e-05}
\setmatattr{Ti-49}{Emax}{150}
\setmatattr{Ti-49}{lowsource}{ENDF-B-VII.1}
\setmatattr{Ti-49}{hisource}{JENDL/HE-2007}
\setmatattr{Ti-49}{Ecut}{20}
\setmatattr{Ti-49}{konnocorrection}{}
\setmatattr{Ti-49}{hascov32}{A}
\setmatattr{Ti-49}{hascov33}{A}
\setmatattr{Ti-49}{hascov34}{}
\setmatattr{Ti-49}{hascov35}{}
\setmatattr{Ti-49}{hascov36}{}

\registermaterial{Ti-50}
\setmatattr{Ti-50}{Emin}{1e-05}
\setmatattr{Ti-50}{Emax}{150}
\setmatattr{Ti-50}{lowsource}{ENDF-B-VII.1}
\setmatattr{Ti-50}{hisource}{JENDL/HE-2007}
\setmatattr{Ti-50}{Ecut}{20}
\setmatattr{Ti-50}{konnocorrection}{}
\setmatattr{Ti-50}{hascov32}{A}
\setmatattr{Ti-50}{hascov33}{A}
\setmatattr{Ti-50}{hascov34}{}
\setmatattr{Ti-50}{hascov35}{}
\setmatattr{Ti-50}{hascov36}{}

\registermaterial{V-50}
\setmatattr{V-50}{Emin}{1e-05}
\setmatattr{V-50}{Emax}{200}
\setmatattr{V-50}{lowsource}{JENDL-4.0}
\setmatattr{V-50}{hisource}{TENDL-2010}
\setmatattr{V-50}{Ecut}{20}
\setmatattr{V-50}{konnocorrection}{Y}
\setmatattr{V-50}{hascov32}{}
\setmatattr{V-50}{hascov33}{}
\setmatattr{V-50}{hascov34}{}
\setmatattr{V-50}{hascov35}{}
\setmatattr{V-50}{hascov36}{}

\registermaterial{V-51}
\setmatattr{V-51}{Emin}{1e-05}
\setmatattr{V-51}{Emax}{150}
\setmatattr{V-51}{lowsource}{JENDL-4.0}
\setmatattr{V-51}{hisource}{JENDL/HE-2007}
\setmatattr{V-51}{Ecut}{20}
\setmatattr{V-51}{konnocorrection}{}
\setmatattr{V-51}{hascov32}{}
\setmatattr{V-51}{hascov33}{}
\setmatattr{V-51}{hascov34}{}
\setmatattr{V-51}{hascov35}{}
\setmatattr{V-51}{hascov36}{}

\registermaterial{Cr-50}
\setmatattr{Cr-50}{Emin}{1e-05}
\setmatattr{Cr-50}{Emax}{65}
\setmatattr{Cr-50}{lowsource}{INDEN-1.0}
\setmatattr{Cr-50}{hisource}{INDEN-1.0}
\setmatattr{Cr-50}{comment}{problem retrieving INDEN-1.0}
\setmatattr{Cr-50}{konnocorrection}{}
\setmatattr{Cr-50}{hascov32}{}
\setmatattr{Cr-50}{hascov33}{}
\setmatattr{Cr-50}{hascov34}{}
\setmatattr{Cr-50}{hascov35}{}
\setmatattr{Cr-50}{hascov36}{}

\registermaterial{Cr-52}
\setmatattr{Cr-52}{Emin}{1e-05}
\setmatattr{Cr-52}{Emax}{65}
\setmatattr{Cr-52}{lowsource}{INDEN-1.0}
\setmatattr{Cr-52}{hisource}{INDEN-1.0}
\setmatattr{Cr-52}{comment}{problem retrieving INDEN-1.0}
\setmatattr{Cr-52}{konnocorrection}{}
\setmatattr{Cr-52}{hascov32}{}
\setmatattr{Cr-52}{hascov33}{}
\setmatattr{Cr-52}{hascov34}{}
\setmatattr{Cr-52}{hascov35}{}
\setmatattr{Cr-52}{hascov36}{}

\registermaterial{Cr-53}
\setmatattr{Cr-53}{Emin}{1e-05}
\setmatattr{Cr-53}{Emax}{65}
\setmatattr{Cr-53}{lowsource}{INDEN-1.0}
\setmatattr{Cr-53}{hisource}{INDEN-1.0}
\setmatattr{Cr-53}{comment}{problem retrieving INDEN-1.0}
\setmatattr{Cr-53}{konnocorrection}{}
\setmatattr{Cr-53}{hascov32}{}
\setmatattr{Cr-53}{hascov33}{}
\setmatattr{Cr-53}{hascov34}{}
\setmatattr{Cr-53}{hascov35}{}
\setmatattr{Cr-53}{hascov36}{}

\registermaterial{Cr-54}
\setmatattr{Cr-54}{Emin}{1e-05}
\setmatattr{Cr-54}{Emax}{65}
\setmatattr{Cr-54}{lowsource}{INDEN-1.0}
\setmatattr{Cr-54}{hisource}{INDEN-1.0}
\setmatattr{Cr-54}{comment}{problem retrieving INDEN-1.0}
\setmatattr{Cr-54}{konnocorrection}{}
\setmatattr{Cr-54}{hascov32}{}
\setmatattr{Cr-54}{hascov33}{}
\setmatattr{Cr-54}{hascov34}{}
\setmatattr{Cr-54}{hascov35}{}
\setmatattr{Cr-54}{hascov36}{}

\registermaterial{Mn-55}
\setmatattr{Mn-55}{Emin}{1e-05}
\setmatattr{Mn-55}{Emax}{60}
\setmatattr{Mn-55}{lowsource}{ENDF/B-VII.1}
\setmatattr{Mn-55}{hisource}{ENDF/B-VII.1}
\setmatattr{Mn-55}{konnocorrection}{}
\setmatattr{Mn-55}{hascov32}{A}
\setmatattr{Mn-55}{hascov33}{A}
\setmatattr{Mn-55}{hascov34}{A}
\setmatattr{Mn-55}{hascov35}{}
\setmatattr{Mn-55}{hascov36}{}

\registermaterial{Fe-54}
\setmatattr{Fe-54}{Emin}{1e-05}
\setmatattr{Fe-54}{Emax}{150}
\setmatattr{Fe-54}{lowsource}{INDEN-1.0}
\setmatattr{Fe-54}{hisource}{INDEN-1.0}
\setmatattr{Fe-54}{comment}{problem retrieving INDEN-1.0}
\setmatattr{Fe-54}{konnocorrection}{}
\setmatattr{Fe-54}{hascov32}{}
\setmatattr{Fe-54}{hascov33}{A}
\setmatattr{Fe-54}{hascov34}{}
\setmatattr{Fe-54}{hascov35}{}
\setmatattr{Fe-54}{hascov36}{}

\registermaterial{Fe-56}
\setmatattr{Fe-56}{Emin}{1e-05}
\setmatattr{Fe-56}{Emax}{150}
\setmatattr{Fe-56}{lowsource}{INDEN-1.0}
\setmatattr{Fe-56}{hisource}{INDEN-1.0}
\setmatattr{Fe-56}{comment}{problem retrieving INDEN-1.0}
\setmatattr{Fe-56}{konnocorrection}{}
\setmatattr{Fe-56}{hascov32}{}
\setmatattr{Fe-56}{hascov33}{}
\setmatattr{Fe-56}{hascov34}{}
\setmatattr{Fe-56}{hascov35}{}
\setmatattr{Fe-56}{hascov36}{}

\registermaterial{Fe-57}
\setmatattr{Fe-57}{Emin}{1e-05}
\setmatattr{Fe-57}{Emax}{150}
\setmatattr{Fe-57}{lowsource}{INDEN-1.0}
\setmatattr{Fe-57}{hisource}{INDEN-1.0}
\setmatattr{Fe-57}{comment}{problem retrieving INDEN-1.0}
\setmatattr{Fe-57}{konnocorrection}{}
\setmatattr{Fe-57}{hascov32}{}
\setmatattr{Fe-57}{hascov33}{}
\setmatattr{Fe-57}{hascov34}{}
\setmatattr{Fe-57}{hascov35}{}
\setmatattr{Fe-57}{hascov36}{}

\registermaterial{Fe-58}
\setmatattr{Fe-58}{Emin}{1e-05}
\setmatattr{Fe-58}{Emax}{150}
\setmatattr{Fe-58}{lowsource}{ENDF/B-VIII.0}
\setmatattr{Fe-58}{hisource}{ENDF/B-VIII.0}
\setmatattr{Fe-58}{konnocorrection}{}
\setmatattr{Fe-58}{hascov32}{}
\setmatattr{Fe-58}{hascov33}{}
\setmatattr{Fe-58}{hascov34}{}
\setmatattr{Fe-58}{hascov35}{}
\setmatattr{Fe-58}{hascov36}{}

\registermaterial{Co-59}
\setmatattr{Co-59}{Emin}{1e-05}
\setmatattr{Co-59}{Emax}{150}
\setmatattr{Co-59}{lowsource}{ENDF/B-VII.0}
\setmatattr{Co-59}{hisource}{JENDL/HE-2007}
\setmatattr{Co-59}{comment}{problem parsing MAT 2725 in JENDL/HE-2007}
\setmatattr{Co-59}{konnocorrection}{}
\setmatattr{Co-59}{hascov32}{}
\setmatattr{Co-59}{hascov33}{A}
\setmatattr{Co-59}{hascov34}{}
\setmatattr{Co-59}{hascov35}{}
\setmatattr{Co-59}{hascov36}{}

\registermaterial{Ni-58}
\setmatattr{Ni-58}{Emin}{1e-05}
\setmatattr{Ni-58}{Emax}{150}
\setmatattr{Ni-58}{lowsource}{ENDF/B-VII.1}
\setmatattr{Ni-58}{hisource}{ENDF/B-VII.1}
\setmatattr{Ni-58}{konnocorrection}{}
\setmatattr{Ni-58}{hascov32}{A}
\setmatattr{Ni-58}{hascov33}{A}
\setmatattr{Ni-58}{hascov34}{}
\setmatattr{Ni-58}{hascov35}{}
\setmatattr{Ni-58}{hascov36}{}

\registermaterial{Ni-60}
\setmatattr{Ni-60}{Emin}{1e-05}
\setmatattr{Ni-60}{Emax}{150}
\setmatattr{Ni-60}{lowsource}{ENDF/B-VII.1}
\setmatattr{Ni-60}{hisource}{ENDF/B-VII.1}
\setmatattr{Ni-60}{konnocorrection}{}
\setmatattr{Ni-60}{hascov32}{A}
\setmatattr{Ni-60}{hascov33}{A}
\setmatattr{Ni-60}{hascov34}{}
\setmatattr{Ni-60}{hascov35}{}
\setmatattr{Ni-60}{hascov36}{}

\registermaterial{Ni-61}
\setmatattr{Ni-61}{Emin}{1e-05}
\setmatattr{Ni-61}{Emax}{150}
\setmatattr{Ni-61}{lowsource}{ENDF/B-VII.1}
\setmatattr{Ni-61}{hisource}{ENDF/B-VII.1}
\setmatattr{Ni-61}{konnocorrection}{}
\setmatattr{Ni-61}{hascov32}{}
\setmatattr{Ni-61}{hascov33}{}
\setmatattr{Ni-61}{hascov34}{}
\setmatattr{Ni-61}{hascov35}{}
\setmatattr{Ni-61}{hascov36}{}

\registermaterial{Ni-62}
\setmatattr{Ni-62}{Emin}{1e-05}
\setmatattr{Ni-62}{Emax}{150}
\setmatattr{Ni-62}{lowsource}{ENDF/B-VII.1}
\setmatattr{Ni-62}{hisource}{ENDF/B-VII.1}
\setmatattr{Ni-62}{konnocorrection}{}
\setmatattr{Ni-62}{hascov32}{}
\setmatattr{Ni-62}{hascov33}{}
\setmatattr{Ni-62}{hascov34}{}
\setmatattr{Ni-62}{hascov35}{}
\setmatattr{Ni-62}{hascov36}{}

\registermaterial{Ni-64}
\setmatattr{Ni-64}{Emin}{1e-05}
\setmatattr{Ni-64}{Emax}{150}
\setmatattr{Ni-64}{lowsource}{ENDF/B-VII.1}
\setmatattr{Ni-64}{hisource}{ENDF/B-VII.1}
\setmatattr{Ni-64}{konnocorrection}{}
\setmatattr{Ni-64}{hascov32}{}
\setmatattr{Ni-64}{hascov33}{}
\setmatattr{Ni-64}{hascov34}{}
\setmatattr{Ni-64}{hascov35}{}
\setmatattr{Ni-64}{hascov36}{}

\registermaterial{Cu-63}
\setmatattr{Cu-63}{Emin}{1e-05}
\setmatattr{Cu-63}{Emax}{150}
\setmatattr{Cu-63}{lowsource}{ENDF/B-VII}
\setmatattr{Cu-63}{hisource}{ENDF/B-VII}
\setmatattr{Cu-63}{comment}{problem retrieving ENDF/B-VII}
\setmatattr{Cu-63}{konnocorrection}{}
\setmatattr{Cu-63}{hascov32}{}
\setmatattr{Cu-63}{hascov33}{}
\setmatattr{Cu-63}{hascov34}{}
\setmatattr{Cu-63}{hascov35}{}
\setmatattr{Cu-63}{hascov36}{}

\registermaterial{Cu-65}
\setmatattr{Cu-65}{Emin}{1e-05}
\setmatattr{Cu-65}{Emax}{150}
\setmatattr{Cu-65}{lowsource}{ENDF/B-VII}
\setmatattr{Cu-65}{hisource}{ENDF/B-VII}
\setmatattr{Cu-65}{comment}{problem retrieving ENDF/B-VII}
\setmatattr{Cu-65}{konnocorrection}{}
\setmatattr{Cu-65}{hascov32}{}
\setmatattr{Cu-65}{hascov33}{}
\setmatattr{Cu-65}{hascov34}{}
\setmatattr{Cu-65}{hascov35}{}
\setmatattr{Cu-65}{hascov36}{}

\registermaterial{Zn-64}
\setmatattr{Zn-64}{Emin}{1e-05}
\setmatattr{Zn-64}{Emax}{150}
\setmatattr{Zn-64}{lowsource}{JENDL-4.0}
\setmatattr{Zn-64}{hisource}{JENDL/HE-2007}
\setmatattr{Zn-64}{Ecut}{20}
\setmatattr{Zn-64}{konnocorrection}{}
\setmatattr{Zn-64}{hascov32}{}
\setmatattr{Zn-64}{hascov33}{}
\setmatattr{Zn-64}{hascov34}{}
\setmatattr{Zn-64}{hascov35}{}
\setmatattr{Zn-64}{hascov36}{}

\registermaterial{Zn-66}
\setmatattr{Zn-66}{Emin}{1e-05}
\setmatattr{Zn-66}{Emax}{150}
\setmatattr{Zn-66}{lowsource}{JENDL-4.0}
\setmatattr{Zn-66}{hisource}{JENDL/HE-2007}
\setmatattr{Zn-66}{Ecut}{20}
\setmatattr{Zn-66}{konnocorrection}{}
\setmatattr{Zn-66}{hascov32}{}
\setmatattr{Zn-66}{hascov33}{}
\setmatattr{Zn-66}{hascov34}{}
\setmatattr{Zn-66}{hascov35}{}
\setmatattr{Zn-66}{hascov36}{}

\registermaterial{Zn-67}
\setmatattr{Zn-67}{Emin}{1e-05}
\setmatattr{Zn-67}{Emax}{150}
\setmatattr{Zn-67}{lowsource}{JENDL-4.0}
\setmatattr{Zn-67}{hisource}{JENDL/HE-2007}
\setmatattr{Zn-67}{Ecut}{20}
\setmatattr{Zn-67}{konnocorrection}{}
\setmatattr{Zn-67}{hascov32}{}
\setmatattr{Zn-67}{hascov33}{}
\setmatattr{Zn-67}{hascov34}{}
\setmatattr{Zn-67}{hascov35}{}
\setmatattr{Zn-67}{hascov36}{}

\registermaterial{Zn-68}
\setmatattr{Zn-68}{Emin}{1e-05}
\setmatattr{Zn-68}{Emax}{150}
\setmatattr{Zn-68}{lowsource}{JENDL-4.0}
\setmatattr{Zn-68}{hisource}{JENDL/HE-2007}
\setmatattr{Zn-68}{Ecut}{20}
\setmatattr{Zn-68}{konnocorrection}{}
\setmatattr{Zn-68}{hascov32}{}
\setmatattr{Zn-68}{hascov33}{}
\setmatattr{Zn-68}{hascov34}{}
\setmatattr{Zn-68}{hascov35}{}
\setmatattr{Zn-68}{hascov36}{}

\registermaterial{Zn-70}
\setmatattr{Zn-70}{Emin}{1e-05}
\setmatattr{Zn-70}{Emax}{150}
\setmatattr{Zn-70}{lowsource}{JENDL-4.0}
\setmatattr{Zn-70}{hisource}{JENDL/HE-2007}
\setmatattr{Zn-70}{Ecut}{20}
\setmatattr{Zn-70}{konnocorrection}{}
\setmatattr{Zn-70}{hascov32}{}
\setmatattr{Zn-70}{hascov33}{}
\setmatattr{Zn-70}{hascov34}{}
\setmatattr{Zn-70}{hascov35}{}
\setmatattr{Zn-70}{hascov36}{}

\registermaterial{Ga-69}
\setmatattr{Ga-69}{Emin}{1e-05}
\setmatattr{Ga-69}{Emax}{150}
\setmatattr{Ga-69}{lowsource}{JENDL-4.0}
\setmatattr{Ga-69}{hisource}{JENDL/HE-2007}
\setmatattr{Ga-69}{Ecut}{20}
\setmatattr{Ga-69}{konnocorrection}{}
\setmatattr{Ga-69}{hascov32}{}
\setmatattr{Ga-69}{hascov33}{}
\setmatattr{Ga-69}{hascov34}{}
\setmatattr{Ga-69}{hascov35}{}
\setmatattr{Ga-69}{hascov36}{}

\registermaterial{Ga-71}
\setmatattr{Ga-71}{Emin}{1e-05}
\setmatattr{Ga-71}{Emax}{150}
\setmatattr{Ga-71}{lowsource}{JENDL-4.0}
\setmatattr{Ga-71}{hisource}{JENDL/HE-2007}
\setmatattr{Ga-71}{Ecut}{20}
\setmatattr{Ga-71}{konnocorrection}{}
\setmatattr{Ga-71}{hascov32}{}
\setmatattr{Ga-71}{hascov33}{}
\setmatattr{Ga-71}{hascov34}{}
\setmatattr{Ga-71}{hascov35}{}
\setmatattr{Ga-71}{hascov36}{}

\registermaterial{Ge-70}
\setmatattr{Ge-70}{Emin}{1e-05}
\setmatattr{Ge-70}{Emax}{200}
\setmatattr{Ge-70}{lowsource}{JEFF-311}
\setmatattr{Ge-70}{hisource}{JEFF-311}
\setmatattr{Ge-70}{konnocorrection}{}
\setmatattr{Ge-70}{hascov32}{}
\setmatattr{Ge-70}{hascov33}{}
\setmatattr{Ge-70}{hascov34}{}
\setmatattr{Ge-70}{hascov35}{}
\setmatattr{Ge-70}{hascov36}{}

\registermaterial{Ge-72}
\setmatattr{Ge-72}{Emin}{1e-05}
\setmatattr{Ge-72}{Emax}{200}
\setmatattr{Ge-72}{lowsource}{JEFF-311}
\setmatattr{Ge-72}{hisource}{JEFF-311}
\setmatattr{Ge-72}{konnocorrection}{}
\setmatattr{Ge-72}{hascov32}{}
\setmatattr{Ge-72}{hascov33}{}
\setmatattr{Ge-72}{hascov34}{}
\setmatattr{Ge-72}{hascov35}{}
\setmatattr{Ge-72}{hascov36}{}

\registermaterial{Ge-73}
\setmatattr{Ge-73}{Emin}{1e-05}
\setmatattr{Ge-73}{Emax}{200}
\setmatattr{Ge-73}{lowsource}{JEFF-311}
\setmatattr{Ge-73}{hisource}{JEFF-311}
\setmatattr{Ge-73}{konnocorrection}{}
\setmatattr{Ge-73}{hascov32}{}
\setmatattr{Ge-73}{hascov33}{}
\setmatattr{Ge-73}{hascov34}{}
\setmatattr{Ge-73}{hascov35}{}
\setmatattr{Ge-73}{hascov36}{}

\registermaterial{Ge-74}
\setmatattr{Ge-74}{Emin}{1e-05}
\setmatattr{Ge-74}{Emax}{200}
\setmatattr{Ge-74}{lowsource}{JEFF-311}
\setmatattr{Ge-74}{hisource}{JEFF-311}
\setmatattr{Ge-74}{konnocorrection}{}
\setmatattr{Ge-74}{hascov32}{}
\setmatattr{Ge-74}{hascov33}{}
\setmatattr{Ge-74}{hascov34}{}
\setmatattr{Ge-74}{hascov35}{}
\setmatattr{Ge-74}{hascov36}{}

\registermaterial{Ge-76}
\setmatattr{Ge-76}{Emin}{1e-05}
\setmatattr{Ge-76}{Emax}{200}
\setmatattr{Ge-76}{lowsource}{JEFF-311}
\setmatattr{Ge-76}{hisource}{JEFF-311}
\setmatattr{Ge-76}{konnocorrection}{}
\setmatattr{Ge-76}{hascov32}{}
\setmatattr{Ge-76}{hascov33}{}
\setmatattr{Ge-76}{hascov34}{}
\setmatattr{Ge-76}{hascov35}{}
\setmatattr{Ge-76}{hascov36}{}

\registermaterial{Br-79}
\setmatattr{Br-79}{Emin}{1e-05}
\setmatattr{Br-79}{Emax}{200}
\setmatattr{Br-79}{lowsource}{JENDL-4.0}
\setmatattr{Br-79}{hisource}{TENDL-2010}
\setmatattr{Br-79}{Ecut}{20}
\setmatattr{Br-79}{konnocorrection}{Y}
\setmatattr{Br-79}{hascov32}{}
\setmatattr{Br-79}{hascov33}{}
\setmatattr{Br-79}{hascov34}{}
\setmatattr{Br-79}{hascov35}{}
\setmatattr{Br-79}{hascov36}{}

\registermaterial{Br-81}
\setmatattr{Br-81}{Emin}{1e-05}
\setmatattr{Br-81}{Emax}{200}
\setmatattr{Br-81}{lowsource}{JENDL-4.0}
\setmatattr{Br-81}{hisource}{TENDL-2010}
\setmatattr{Br-81}{Ecut}{20}
\setmatattr{Br-81}{konnocorrection}{Y}
\setmatattr{Br-81}{hascov32}{}
\setmatattr{Br-81}{hascov33}{}
\setmatattr{Br-81}{hascov34}{}
\setmatattr{Br-81}{hascov35}{}
\setmatattr{Br-81}{hascov36}{}

\registermaterial{Y-89}
\setmatattr{Y-89}{Emin}{1e-05}
\setmatattr{Y-89}{Emax}{200}
\setmatattr{Y-89}{lowsource}{ENDF/B-VII.1}
\setmatattr{Y-89}{hisource}{TENDL-2010}
\setmatattr{Y-89}{Ecut}{20}
\setmatattr{Y-89}{konnocorrection}{Y}
\setmatattr{Y-89}{hascov32}{}
\setmatattr{Y-89}{hascov33}{A}
\setmatattr{Y-89}{hascov34}{}
\setmatattr{Y-89}{hascov35}{}
\setmatattr{Y-89}{hascov36}{}

\registermaterial{Zr-90}
\setmatattr{Zr-90}{Emin}{1e-05}
\setmatattr{Zr-90}{Emax}{150}
\setmatattr{Zr-90}{lowsource}{JENDL-4.0}
\setmatattr{Zr-90}{hisource}{JENDL/HE-2007}
\setmatattr{Zr-90}{Ecut}{20}
\setmatattr{Zr-90}{konnocorrection}{}
\setmatattr{Zr-90}{hascov32}{}
\setmatattr{Zr-90}{hascov33}{A}
\setmatattr{Zr-90}{hascov34}{}
\setmatattr{Zr-90}{hascov35}{}
\setmatattr{Zr-90}{hascov36}{}

\registermaterial{Zr-91}
\setmatattr{Zr-91}{Emin}{1e-05}
\setmatattr{Zr-91}{Emax}{150}
\setmatattr{Zr-91}{lowsource}{JENDL-4.0}
\setmatattr{Zr-91}{hisource}{JENDL/HE-2007}
\setmatattr{Zr-91}{Ecut}{20}
\setmatattr{Zr-91}{konnocorrection}{}
\setmatattr{Zr-91}{hascov32}{}
\setmatattr{Zr-91}{hascov33}{}
\setmatattr{Zr-91}{hascov34}{}
\setmatattr{Zr-91}{hascov35}{}
\setmatattr{Zr-91}{hascov36}{}

\registermaterial{Zr-92}
\setmatattr{Zr-92}{Emin}{1e-05}
\setmatattr{Zr-92}{Emax}{150}
\setmatattr{Zr-92}{lowsource}{JENDL-4.0}
\setmatattr{Zr-92}{hisource}{JENDL/HE-2007}
\setmatattr{Zr-92}{Ecut}{20}
\setmatattr{Zr-92}{konnocorrection}{}
\setmatattr{Zr-92}{hascov32}{}
\setmatattr{Zr-92}{hascov33}{}
\setmatattr{Zr-92}{hascov34}{}
\setmatattr{Zr-92}{hascov35}{}
\setmatattr{Zr-92}{hascov36}{}

\registermaterial{Zr-94}
\setmatattr{Zr-94}{Emin}{1e-05}
\setmatattr{Zr-94}{Emax}{150}
\setmatattr{Zr-94}{lowsource}{JENDL-4.0}
\setmatattr{Zr-94}{hisource}{JENDL/HE-2007}
\setmatattr{Zr-94}{Ecut}{20}
\setmatattr{Zr-94}{konnocorrection}{}
\setmatattr{Zr-94}{hascov32}{}
\setmatattr{Zr-94}{hascov33}{}
\setmatattr{Zr-94}{hascov34}{}
\setmatattr{Zr-94}{hascov35}{}
\setmatattr{Zr-94}{hascov36}{}

\registermaterial{Zr-96}
\setmatattr{Zr-96}{Emin}{1e-05}
\setmatattr{Zr-96}{Emax}{150}
\setmatattr{Zr-96}{lowsource}{JENDL-4.0}
\setmatattr{Zr-96}{hisource}{JENDL/HE-2007}
\setmatattr{Zr-96}{Ecut}{20}
\setmatattr{Zr-96}{konnocorrection}{}
\setmatattr{Zr-96}{hascov32}{}
\setmatattr{Zr-96}{hascov33}{}
\setmatattr{Zr-96}{hascov34}{}
\setmatattr{Zr-96}{hascov35}{}
\setmatattr{Zr-96}{hascov36}{}

\registermaterial{Nb-93}
\setmatattr{Nb-93}{Emin}{1e-05}
\setmatattr{Nb-93}{Emax}{150}
\setmatattr{Nb-93}{lowsource}{JENDL-4.0}
\setmatattr{Nb-93}{hisource}{JENDL/HE-2007}
\setmatattr{Nb-93}{Ecut}{20}
\setmatattr{Nb-93}{konnocorrection}{}
\setmatattr{Nb-93}{hascov32}{}
\setmatattr{Nb-93}{hascov33}{}
\setmatattr{Nb-93}{hascov34}{}
\setmatattr{Nb-93}{hascov35}{}
\setmatattr{Nb-93}{hascov36}{}

\registermaterial{Mo-92}
\setmatattr{Mo-92}{Emin}{1e-05}
\setmatattr{Mo-92}{Emax}{150}
\setmatattr{Mo-92}{lowsource}{JENDL-4.0}
\setmatattr{Mo-92}{hisource}{JENDL/HE-2007}
\setmatattr{Mo-92}{Ecut}{20}
\setmatattr{Mo-92}{konnocorrection}{}
\setmatattr{Mo-92}{hascov32}{}
\setmatattr{Mo-92}{hascov33}{}
\setmatattr{Mo-92}{hascov34}{}
\setmatattr{Mo-92}{hascov35}{}
\setmatattr{Mo-92}{hascov36}{}

\registermaterial{Mo-94}
\setmatattr{Mo-94}{Emin}{1e-05}
\setmatattr{Mo-94}{Emax}{150}
\setmatattr{Mo-94}{lowsource}{JENDL-4.0}
\setmatattr{Mo-94}{hisource}{JENDL/HE-2007}
\setmatattr{Mo-94}{Ecut}{20}
\setmatattr{Mo-94}{konnocorrection}{}
\setmatattr{Mo-94}{hascov32}{}
\setmatattr{Mo-94}{hascov33}{}
\setmatattr{Mo-94}{hascov34}{}
\setmatattr{Mo-94}{hascov35}{}
\setmatattr{Mo-94}{hascov36}{}

\registermaterial{Mo-95}
\setmatattr{Mo-95}{Emin}{1e-05}
\setmatattr{Mo-95}{Emax}{150}
\setmatattr{Mo-95}{lowsource}{JENDL-4.0}
\setmatattr{Mo-95}{hisource}{JENDL/HE-2007}
\setmatattr{Mo-95}{Ecut}{20}
\setmatattr{Mo-95}{konnocorrection}{}
\setmatattr{Mo-95}{hascov32}{}
\setmatattr{Mo-95}{hascov33}{}
\setmatattr{Mo-95}{hascov34}{}
\setmatattr{Mo-95}{hascov35}{}
\setmatattr{Mo-95}{hascov36}{}

\registermaterial{Mo-96}
\setmatattr{Mo-96}{Emin}{1e-05}
\setmatattr{Mo-96}{Emax}{150}
\setmatattr{Mo-96}{lowsource}{JENDL-4.0}
\setmatattr{Mo-96}{hisource}{JENDL/HE-2007}
\setmatattr{Mo-96}{Ecut}{20}
\setmatattr{Mo-96}{konnocorrection}{}
\setmatattr{Mo-96}{hascov32}{}
\setmatattr{Mo-96}{hascov33}{}
\setmatattr{Mo-96}{hascov34}{}
\setmatattr{Mo-96}{hascov35}{}
\setmatattr{Mo-96}{hascov36}{}

\registermaterial{Mo-97}
\setmatattr{Mo-97}{Emin}{1e-05}
\setmatattr{Mo-97}{Emax}{150}
\setmatattr{Mo-97}{lowsource}{JENDL-4.0}
\setmatattr{Mo-97}{hisource}{JENDL/HE-2007}
\setmatattr{Mo-97}{Ecut}{20}
\setmatattr{Mo-97}{konnocorrection}{}
\setmatattr{Mo-97}{hascov32}{}
\setmatattr{Mo-97}{hascov33}{}
\setmatattr{Mo-97}{hascov34}{}
\setmatattr{Mo-97}{hascov35}{}
\setmatattr{Mo-97}{hascov36}{}

\registermaterial{Mo-98}
\setmatattr{Mo-98}{Emin}{1e-05}
\setmatattr{Mo-98}{Emax}{150}
\setmatattr{Mo-98}{lowsource}{JENDL-4.0}
\setmatattr{Mo-98}{hisource}{JENDL/HE-2007}
\setmatattr{Mo-98}{Ecut}{20}
\setmatattr{Mo-98}{konnocorrection}{}
\setmatattr{Mo-98}{hascov32}{}
\setmatattr{Mo-98}{hascov33}{}
\setmatattr{Mo-98}{hascov34}{}
\setmatattr{Mo-98}{hascov35}{}
\setmatattr{Mo-98}{hascov36}{}

\registermaterial{Mo-100}
\setmatattr{Mo-100}{Emin}{1e-05}
\setmatattr{Mo-100}{Emax}{150}
\setmatattr{Mo-100}{lowsource}{JENDL-4.0}
\setmatattr{Mo-100}{hisource}{JENDL/HE-2007}
\setmatattr{Mo-100}{Ecut}{20}
\setmatattr{Mo-100}{konnocorrection}{}
\setmatattr{Mo-100}{hascov32}{}
\setmatattr{Mo-100}{hascov33}{}
\setmatattr{Mo-100}{hascov34}{}
\setmatattr{Mo-100}{hascov35}{}
\setmatattr{Mo-100}{hascov36}{}

\registermaterial{Rh-103}
\setmatattr{Rh-103}{Emin}{1e-05}
\setmatattr{Rh-103}{Emax}{200}
\setmatattr{Rh-103}{lowsource}{JEFF-3.1.1}
\setmatattr{Rh-103}{hisource}{TENDL-2011}
\setmatattr{Rh-103}{Ecut}{30}
\setmatattr{Rh-103}{konnocorrection}{}
\setmatattr{Rh-103}{hascov32}{}
\setmatattr{Rh-103}{hascov33}{}
\setmatattr{Rh-103}{hascov34}{}
\setmatattr{Rh-103}{hascov35}{}
\setmatattr{Rh-103}{hascov36}{}

\registermaterial{Ag-107}
\setmatattr{Ag-107}{Emin}{1e-05}
\setmatattr{Ag-107}{Emax}{200}
\setmatattr{Ag-107}{lowsource}{ENDF-B-VII.1}
\setmatattr{Ag-107}{hisource}{TENDL-2010}
\setmatattr{Ag-107}{Ecut}{20}
\setmatattr{Ag-107}{konnocorrection}{Y}
\setmatattr{Ag-107}{hascov32}{}
\setmatattr{Ag-107}{hascov33}{}
\setmatattr{Ag-107}{hascov34}{}
\setmatattr{Ag-107}{hascov35}{}
\setmatattr{Ag-107}{hascov36}{}

\registermaterial{Ag-109}
\setmatattr{Ag-109}{Emin}{1e-05}
\setmatattr{Ag-109}{Emax}{200}
\setmatattr{Ag-109}{lowsource}{ENDF-B-VII.1}
\setmatattr{Ag-109}{hisource}{TENDL-2010}
\setmatattr{Ag-109}{Ecut}{20}
\setmatattr{Ag-109}{comment}{diff in low energy part}
\setmatattr{Ag-109}{konnocorrection}{Y}
\setmatattr{Ag-109}{hascov32}{}
\setmatattr{Ag-109}{hascov33}{}
\setmatattr{Ag-109}{hascov34}{}
\setmatattr{Ag-109}{hascov35}{}
\setmatattr{Ag-109}{hascov36}{}

\registermaterial{Cd-106}
\setmatattr{Cd-106}{Emin}{1e-05}
\setmatattr{Cd-106}{Emax}{200}
\setmatattr{Cd-106}{lowsource}{ENDF/B-VII.1}
\setmatattr{Cd-106}{hisource}{TENDL-2010}
\setmatattr{Cd-106}{Ecut}{20}
\setmatattr{Cd-106}{konnocorrection}{Y}
\setmatattr{Cd-106}{hascov32}{}
\setmatattr{Cd-106}{hascov33}{}
\setmatattr{Cd-106}{hascov34}{}
\setmatattr{Cd-106}{hascov35}{}
\setmatattr{Cd-106}{hascov36}{}

\registermaterial{Cd-108}
\setmatattr{Cd-108}{Emin}{1e-05}
\setmatattr{Cd-108}{Emax}{200}
\setmatattr{Cd-108}{lowsource}{ENDF/B-VII.1}
\setmatattr{Cd-108}{hisource}{TENDL-2010}
\setmatattr{Cd-108}{Ecut}{20}
\setmatattr{Cd-108}{konnocorrection}{Y}
\setmatattr{Cd-108}{hascov32}{}
\setmatattr{Cd-108}{hascov33}{}
\setmatattr{Cd-108}{hascov34}{}
\setmatattr{Cd-108}{hascov35}{}
\setmatattr{Cd-108}{hascov36}{}

\registermaterial{Cd-110}
\setmatattr{Cd-110}{Emin}{1e-05}
\setmatattr{Cd-110}{Emax}{200}
\setmatattr{Cd-110}{lowsource}{ENDF/B-VII.1}
\setmatattr{Cd-110}{hisource}{TENDL-2010}
\setmatattr{Cd-110}{Ecut}{20}
\setmatattr{Cd-110}{konnocorrection}{Y}
\setmatattr{Cd-110}{hascov32}{}
\setmatattr{Cd-110}{hascov33}{}
\setmatattr{Cd-110}{hascov34}{}
\setmatattr{Cd-110}{hascov35}{}
\setmatattr{Cd-110}{hascov36}{}

\registermaterial{Cd-111}
\setmatattr{Cd-111}{Emin}{1e-05}
\setmatattr{Cd-111}{Emax}{200}
\setmatattr{Cd-111}{lowsource}{ENDF/B-VII.1}
\setmatattr{Cd-111}{hisource}{TENDL-2010}
\setmatattr{Cd-111}{Ecut}{20}
\setmatattr{Cd-111}{konnocorrection}{Y}
\setmatattr{Cd-111}{hascov32}{}
\setmatattr{Cd-111}{hascov33}{}
\setmatattr{Cd-111}{hascov34}{}
\setmatattr{Cd-111}{hascov35}{}
\setmatattr{Cd-111}{hascov36}{}

\registermaterial{Cd-112}
\setmatattr{Cd-112}{Emin}{1e-05}
\setmatattr{Cd-112}{Emax}{200}
\setmatattr{Cd-112}{lowsource}{ENDF/B-VII.1}
\setmatattr{Cd-112}{hisource}{TENDL-2010}
\setmatattr{Cd-112}{Ecut}{20}
\setmatattr{Cd-112}{konnocorrection}{Y}
\setmatattr{Cd-112}{hascov32}{}
\setmatattr{Cd-112}{hascov33}{}
\setmatattr{Cd-112}{hascov34}{}
\setmatattr{Cd-112}{hascov35}{}
\setmatattr{Cd-112}{hascov36}{}

\registermaterial{Cd-113}
\setmatattr{Cd-113}{Emin}{1e-05}
\setmatattr{Cd-113}{Emax}{200}
\setmatattr{Cd-113}{lowsource}{ENDF/B-VII.1}
\setmatattr{Cd-113}{hisource}{TENDL-2010}
\setmatattr{Cd-113}{Ecut}{20}
\setmatattr{Cd-113}{konnocorrection}{Y}
\setmatattr{Cd-113}{hascov32}{}
\setmatattr{Cd-113}{hascov33}{}
\setmatattr{Cd-113}{hascov34}{}
\setmatattr{Cd-113}{hascov35}{}
\setmatattr{Cd-113}{hascov36}{}

\registermaterial{Cd-114}
\setmatattr{Cd-114}{Emin}{1e-05}
\setmatattr{Cd-114}{Emax}{200}
\setmatattr{Cd-114}{lowsource}{ENDF/B-VII.1}
\setmatattr{Cd-114}{hisource}{TENDL-2010}
\setmatattr{Cd-114}{Ecut}{20}
\setmatattr{Cd-114}{konnocorrection}{Y}
\setmatattr{Cd-114}{hascov32}{}
\setmatattr{Cd-114}{hascov33}{}
\setmatattr{Cd-114}{hascov34}{}
\setmatattr{Cd-114}{hascov35}{}
\setmatattr{Cd-114}{hascov36}{}

\registermaterial{Cd-116}
\setmatattr{Cd-116}{Emin}{1e-05}
\setmatattr{Cd-116}{Emax}{200}
\setmatattr{Cd-116}{lowsource}{ENDF/B-VII.1}
\setmatattr{Cd-116}{hisource}{TENDL-2010}
\setmatattr{Cd-116}{Ecut}{20}
\setmatattr{Cd-116}{konnocorrection}{Y}
\setmatattr{Cd-116}{hascov32}{}
\setmatattr{Cd-116}{hascov33}{}
\setmatattr{Cd-116}{hascov34}{}
\setmatattr{Cd-116}{hascov35}{}
\setmatattr{Cd-116}{hascov36}{}

\registermaterial{Sn-112}
\setmatattr{Sn-112}{Emin}{1e-05}
\setmatattr{Sn-112}{Emax}{200}
\setmatattr{Sn-112}{lowsource}{JENDL-4.0}
\setmatattr{Sn-112}{hisource}{TENDL-2010}
\setmatattr{Sn-112}{Ecut}{20}
\setmatattr{Sn-112}{konnocorrection}{Y}
\setmatattr{Sn-112}{hascov32}{}
\setmatattr{Sn-112}{hascov33}{}
\setmatattr{Sn-112}{hascov34}{}
\setmatattr{Sn-112}{hascov35}{}
\setmatattr{Sn-112}{hascov36}{}

\registermaterial{Sn-114}
\setmatattr{Sn-114}{Emin}{1e-05}
\setmatattr{Sn-114}{Emax}{200}
\setmatattr{Sn-114}{lowsource}{JENDL-4.0}
\setmatattr{Sn-114}{hisource}{TENDL-2010}
\setmatattr{Sn-114}{Ecut}{20}
\setmatattr{Sn-114}{konnocorrection}{Y}
\setmatattr{Sn-114}{hascov32}{}
\setmatattr{Sn-114}{hascov33}{}
\setmatattr{Sn-114}{hascov34}{}
\setmatattr{Sn-114}{hascov35}{}
\setmatattr{Sn-114}{hascov36}{}

\registermaterial{Sn-115}
\setmatattr{Sn-115}{Emin}{1e-05}
\setmatattr{Sn-115}{Emax}{200}
\setmatattr{Sn-115}{lowsource}{JENDL-4.0}
\setmatattr{Sn-115}{hisource}{TENDL-2010}
\setmatattr{Sn-115}{Ecut}{20}
\setmatattr{Sn-115}{konnocorrection}{Y}
\setmatattr{Sn-115}{hascov32}{}
\setmatattr{Sn-115}{hascov33}{}
\setmatattr{Sn-115}{hascov34}{}
\setmatattr{Sn-115}{hascov35}{}
\setmatattr{Sn-115}{hascov36}{}

\registermaterial{Sn-116}
\setmatattr{Sn-116}{Emin}{1e-05}
\setmatattr{Sn-116}{Emax}{200}
\setmatattr{Sn-116}{lowsource}{JENDL-4.0}
\setmatattr{Sn-116}{hisource}{TENDL-2010}
\setmatattr{Sn-116}{Ecut}{20}
\setmatattr{Sn-116}{konnocorrection}{Y}
\setmatattr{Sn-116}{hascov32}{}
\setmatattr{Sn-116}{hascov33}{}
\setmatattr{Sn-116}{hascov34}{}
\setmatattr{Sn-116}{hascov35}{}
\setmatattr{Sn-116}{hascov36}{}

\registermaterial{Sn-117}
\setmatattr{Sn-117}{Emin}{1e-05}
\setmatattr{Sn-117}{Emax}{200}
\setmatattr{Sn-117}{lowsource}{JENDL-4.0}
\setmatattr{Sn-117}{hisource}{TENDL-2010}
\setmatattr{Sn-117}{Ecut}{20}
\setmatattr{Sn-117}{konnocorrection}{Y}
\setmatattr{Sn-117}{hascov32}{}
\setmatattr{Sn-117}{hascov33}{}
\setmatattr{Sn-117}{hascov34}{}
\setmatattr{Sn-117}{hascov35}{}
\setmatattr{Sn-117}{hascov36}{}

\registermaterial{Sn-118}
\setmatattr{Sn-118}{Emin}{1e-05}
\setmatattr{Sn-118}{Emax}{200}
\setmatattr{Sn-118}{lowsource}{JENDL-4.0}
\setmatattr{Sn-118}{hisource}{TENDL-2010}
\setmatattr{Sn-118}{Ecut}{20}
\setmatattr{Sn-118}{konnocorrection}{Y}
\setmatattr{Sn-118}{hascov32}{}
\setmatattr{Sn-118}{hascov33}{}
\setmatattr{Sn-118}{hascov34}{}
\setmatattr{Sn-118}{hascov35}{}
\setmatattr{Sn-118}{hascov36}{}

\registermaterial{Sn-119}
\setmatattr{Sn-119}{Emin}{1e-05}
\setmatattr{Sn-119}{Emax}{200}
\setmatattr{Sn-119}{lowsource}{JENDL-4.0}
\setmatattr{Sn-119}{hisource}{TENDL-2010}
\setmatattr{Sn-119}{Ecut}{20}
\setmatattr{Sn-119}{konnocorrection}{Y}
\setmatattr{Sn-119}{hascov32}{}
\setmatattr{Sn-119}{hascov33}{}
\setmatattr{Sn-119}{hascov34}{}
\setmatattr{Sn-119}{hascov35}{}
\setmatattr{Sn-119}{hascov36}{}

\registermaterial{Sn-120}
\setmatattr{Sn-120}{Emin}{1e-05}
\setmatattr{Sn-120}{Emax}{200}
\setmatattr{Sn-120}{lowsource}{JENDL-4.0}
\setmatattr{Sn-120}{hisource}{TENDL-2010}
\setmatattr{Sn-120}{Ecut}{20}
\setmatattr{Sn-120}{konnocorrection}{Y}
\setmatattr{Sn-120}{hascov32}{}
\setmatattr{Sn-120}{hascov33}{}
\setmatattr{Sn-120}{hascov34}{}
\setmatattr{Sn-120}{hascov35}{}
\setmatattr{Sn-120}{hascov36}{}

\registermaterial{Sn-122}
\setmatattr{Sn-122}{Emin}{1e-05}
\setmatattr{Sn-122}{Emax}{200}
\setmatattr{Sn-122}{lowsource}{JENDL-4.0}
\setmatattr{Sn-122}{hisource}{TENDL-2010}
\setmatattr{Sn-122}{Ecut}{20}
\setmatattr{Sn-122}{konnocorrection}{Y}
\setmatattr{Sn-122}{hascov32}{}
\setmatattr{Sn-122}{hascov33}{}
\setmatattr{Sn-122}{hascov34}{}
\setmatattr{Sn-122}{hascov35}{}
\setmatattr{Sn-122}{hascov36}{}

\registermaterial{Sn-124}
\setmatattr{Sn-124}{Emin}{1e-05}
\setmatattr{Sn-124}{Emax}{200}
\setmatattr{Sn-124}{lowsource}{JENDL-4.0}
\setmatattr{Sn-124}{hisource}{TENDL-2010}
\setmatattr{Sn-124}{Ecut}{20}
\setmatattr{Sn-124}{konnocorrection}{Y}
\setmatattr{Sn-124}{hascov32}{}
\setmatattr{Sn-124}{hascov33}{}
\setmatattr{Sn-124}{hascov34}{}
\setmatattr{Sn-124}{hascov35}{}
\setmatattr{Sn-124}{hascov36}{}

\registermaterial{Sb-121}
\setmatattr{Sb-121}{Emin}{1e-05}
\setmatattr{Sb-121}{Emax}{200}
\setmatattr{Sb-121}{lowsource}{ENDF-B-VII.1}
\setmatattr{Sb-121}{hisource}{TENDL-2010}
\setmatattr{Sb-121}{Ecut}{20}
\setmatattr{Sb-121}{konnocorrection}{Y}
\setmatattr{Sb-121}{hascov32}{}
\setmatattr{Sb-121}{hascov33}{}
\setmatattr{Sb-121}{hascov34}{}
\setmatattr{Sb-121}{hascov35}{}
\setmatattr{Sb-121}{hascov36}{}

\registermaterial{Sb-123}
\setmatattr{Sb-123}{Emin}{1e-05}
\setmatattr{Sb-123}{Emax}{200}
\setmatattr{Sb-123}{lowsource}{ENDF-B-VII.1}
\setmatattr{Sb-123}{hisource}{TENDL-2010}
\setmatattr{Sb-123}{Ecut}{20}
\setmatattr{Sb-123}{konnocorrection}{Y}
\setmatattr{Sb-123}{hascov32}{}
\setmatattr{Sb-123}{hascov33}{}
\setmatattr{Sb-123}{hascov34}{}
\setmatattr{Sb-123}{hascov35}{}
\setmatattr{Sb-123}{hascov36}{}

\registermaterial{I-127}
\setmatattr{I-127}{Emin}{1e-05}
\setmatattr{I-127}{Emax}{200}
\setmatattr{I-127}{lowsource}{ENDF-B-VII.1}
\setmatattr{I-127}{hisource}{TENDL-2011}
\setmatattr{I-127}{Ecut}{30}
\setmatattr{I-127}{konnocorrection}{}
\setmatattr{I-127}{hascov32}{}
\setmatattr{I-127}{hascov33}{A}
\setmatattr{I-127}{hascov34}{}
\setmatattr{I-127}{hascov35}{}
\setmatattr{I-127}{hascov36}{}

\registermaterial{Cs-133}
\setmatattr{Cs-133}{Emin}{1e-05}
\setmatattr{Cs-133}{Emax}{200}
\setmatattr{Cs-133}{lowsource}{JENDL-4.0}
\setmatattr{Cs-133}{hisource}{TENDL-2010}
\setmatattr{Cs-133}{Ecut}{20}
\setmatattr{Cs-133}{konnocorrection}{Y}
\setmatattr{Cs-133}{hascov32}{}
\setmatattr{Cs-133}{hascov33}{}
\setmatattr{Cs-133}{hascov34}{}
\setmatattr{Cs-133}{hascov35}{}
\setmatattr{Cs-133}{hascov36}{}

\registermaterial{Ba-130}
\setmatattr{Ba-130}{Emin}{1e-05}
\setmatattr{Ba-130}{Emax}{200}
\setmatattr{Ba-130}{lowsource}{ENDF-B-VII.1}
\setmatattr{Ba-130}{hisource}{TENDL-2010}
\setmatattr{Ba-130}{Ecut}{20}
\setmatattr{Ba-130}{konnocorrection}{Y}
\setmatattr{Ba-130}{hascov32}{}
\setmatattr{Ba-130}{hascov33}{}
\setmatattr{Ba-130}{hascov34}{}
\setmatattr{Ba-130}{hascov35}{}
\setmatattr{Ba-130}{hascov36}{}

\registermaterial{Ba-132}
\setmatattr{Ba-132}{Emin}{1e-05}
\setmatattr{Ba-132}{Emax}{200}
\setmatattr{Ba-132}{lowsource}{ENDF-B-VII.1}
\setmatattr{Ba-132}{hisource}{TENDL-2010}
\setmatattr{Ba-132}{Ecut}{20}
\setmatattr{Ba-132}{konnocorrection}{Y}
\setmatattr{Ba-132}{hascov32}{}
\setmatattr{Ba-132}{hascov33}{}
\setmatattr{Ba-132}{hascov34}{}
\setmatattr{Ba-132}{hascov35}{}
\setmatattr{Ba-132}{hascov36}{}

\registermaterial{Ba-134}
\setmatattr{Ba-134}{Emin}{1e-05}
\setmatattr{Ba-134}{Emax}{200}
\setmatattr{Ba-134}{lowsource}{ENDF-B-VII.1}
\setmatattr{Ba-134}{hisource}{TENDL-2010}
\setmatattr{Ba-134}{comment}{problem retrieving TENDL-2010}
\setmatattr{Ba-134}{konnocorrection}{Y}
\setmatattr{Ba-134}{hascov32}{}
\setmatattr{Ba-134}{hascov33}{}
\setmatattr{Ba-134}{hascov34}{}
\setmatattr{Ba-134}{hascov35}{}
\setmatattr{Ba-134}{hascov36}{}

\registermaterial{Ba-135}
\setmatattr{Ba-135}{Emin}{1e-05}
\setmatattr{Ba-135}{Emax}{200}
\setmatattr{Ba-135}{lowsource}{ENDF-B-VII.1}
\setmatattr{Ba-135}{hisource}{TENDL-2010}
\setmatattr{Ba-135}{Ecut}{20}
\setmatattr{Ba-135}{konnocorrection}{Y}
\setmatattr{Ba-135}{hascov32}{}
\setmatattr{Ba-135}{hascov33}{}
\setmatattr{Ba-135}{hascov34}{}
\setmatattr{Ba-135}{hascov35}{}
\setmatattr{Ba-135}{hascov36}{}

\registermaterial{Ba-136}
\setmatattr{Ba-136}{Emin}{1e-05}
\setmatattr{Ba-136}{Emax}{200}
\setmatattr{Ba-136}{lowsource}{ENDF-B-VII.1}
\setmatattr{Ba-136}{hisource}{TENDL-2010}
\setmatattr{Ba-136}{Ecut}{20}
\setmatattr{Ba-136}{konnocorrection}{Y}
\setmatattr{Ba-136}{hascov32}{}
\setmatattr{Ba-136}{hascov33}{}
\setmatattr{Ba-136}{hascov34}{}
\setmatattr{Ba-136}{hascov35}{}
\setmatattr{Ba-136}{hascov36}{}

\registermaterial{Ba-137}
\setmatattr{Ba-137}{Emin}{1e-05}
\setmatattr{Ba-137}{Emax}{200}
\setmatattr{Ba-137}{lowsource}{ENDF-B-VII.1}
\setmatattr{Ba-137}{hisource}{TENDL-2010}
\setmatattr{Ba-137}{Ecut}{20}
\setmatattr{Ba-137}{konnocorrection}{Y}
\setmatattr{Ba-137}{hascov32}{}
\setmatattr{Ba-137}{hascov33}{}
\setmatattr{Ba-137}{hascov34}{}
\setmatattr{Ba-137}{hascov35}{}
\setmatattr{Ba-137}{hascov36}{}

\registermaterial{Ba-138}
\setmatattr{Ba-138}{Emin}{1e-05}
\setmatattr{Ba-138}{Emax}{200}
\setmatattr{Ba-138}{lowsource}{ENDF-B-VII.1}
\setmatattr{Ba-138}{hisource}{TENDL-2010}
\setmatattr{Ba-138}{Ecut}{20}
\setmatattr{Ba-138}{konnocorrection}{Y}
\setmatattr{Ba-138}{hascov32}{}
\setmatattr{Ba-138}{hascov33}{}
\setmatattr{Ba-138}{hascov34}{}
\setmatattr{Ba-138}{hascov35}{}
\setmatattr{Ba-138}{hascov36}{}

\registermaterial{La-138}
\setmatattr{La-138}{Emin}{1e-05}
\setmatattr{La-138}{Emax}{200}
\setmatattr{La-138}{lowsource}{TENDL-2014}
\setmatattr{La-138}{hisource}{TENDL-2014}
\setmatattr{La-138}{konnocorrection}{Y}
\setmatattr{La-138}{hascov32}{A}
\setmatattr{La-138}{hascov33}{A}
\setmatattr{La-138}{hascov34}{A}
\setmatattr{La-138}{hascov35}{}
\setmatattr{La-138}{hascov36}{}

\registermaterial{La-139}
\setmatattr{La-139}{Emin}{1e-05}
\setmatattr{La-139}{Emax}{200}
\setmatattr{La-139}{lowsource}{FENDL-3.2}
\setmatattr{La-139}{hisource}{FENDL-3.2}
\setmatattr{La-139}{comment}{problem retrieving FENDL-3.2}
\setmatattr{La-139}{konnocorrection}{}
\setmatattr{La-139}{hascov32}{A}
\setmatattr{La-139}{hascov33}{A}
\setmatattr{La-139}{hascov34}{A}
\setmatattr{La-139}{hascov35}{}
\setmatattr{La-139}{hascov36}{}

\registermaterial{Ce-136}
\setmatattr{Ce-136}{Emin}{1e-05}
\setmatattr{Ce-136}{Emax}{200}
\setmatattr{Ce-136}{lowsource}{ENDF-B-VII.1}
\setmatattr{Ce-136}{hisource}{TENDL-2010}
\setmatattr{Ce-136}{Ecut}{20}
\setmatattr{Ce-136}{konnocorrection}{Y}
\setmatattr{Ce-136}{hascov32}{}
\setmatattr{Ce-136}{hascov33}{}
\setmatattr{Ce-136}{hascov34}{}
\setmatattr{Ce-136}{hascov35}{}
\setmatattr{Ce-136}{hascov36}{}

\registermaterial{Ce-138}
\setmatattr{Ce-138}{Emin}{1e-05}
\setmatattr{Ce-138}{Emax}{200}
\setmatattr{Ce-138}{lowsource}{ENDF-B-VII.1}
\setmatattr{Ce-138}{hisource}{TENDL-2010}
\setmatattr{Ce-138}{Ecut}{20}
\setmatattr{Ce-138}{konnocorrection}{Y}
\setmatattr{Ce-138}{hascov32}{}
\setmatattr{Ce-138}{hascov33}{}
\setmatattr{Ce-138}{hascov34}{}
\setmatattr{Ce-138}{hascov35}{}
\setmatattr{Ce-138}{hascov36}{}

\registermaterial{Ce-140}
\setmatattr{Ce-140}{Emin}{1e-05}
\setmatattr{Ce-140}{Emax}{200}
\setmatattr{Ce-140}{lowsource}{ENDF-B-VII.1}
\setmatattr{Ce-140}{hisource}{TENDL-2010}
\setmatattr{Ce-140}{Ecut}{20}
\setmatattr{Ce-140}{konnocorrection}{Y}
\setmatattr{Ce-140}{hascov32}{}
\setmatattr{Ce-140}{hascov33}{}
\setmatattr{Ce-140}{hascov34}{}
\setmatattr{Ce-140}{hascov35}{}
\setmatattr{Ce-140}{hascov36}{}

\registermaterial{Ce-142}
\setmatattr{Ce-142}{Emin}{1e-05}
\setmatattr{Ce-142}{Emax}{200}
\setmatattr{Ce-142}{lowsource}{ENDF-B-VII.1}
\setmatattr{Ce-142}{hisource}{TENDL-2010}
\setmatattr{Ce-142}{Ecut}{20}
\setmatattr{Ce-142}{konnocorrection}{Y}
\setmatattr{Ce-142}{hascov32}{}
\setmatattr{Ce-142}{hascov33}{}
\setmatattr{Ce-142}{hascov34}{}
\setmatattr{Ce-142}{hascov35}{}
\setmatattr{Ce-142}{hascov36}{}

\registermaterial{Sm-144}
\setmatattr{Sm-144}{Emin}{1e-05}
\setmatattr{Sm-144}{Emax}{200}
\setmatattr{Sm-144}{lowsource}{TENDL-2019}
\setmatattr{Sm-144}{hisource}{TENDL-2019}
\setmatattr{Sm-144}{konnocorrection}{}
\setmatattr{Sm-144}{hascov32}{}
\setmatattr{Sm-144}{hascov33}{A}
\setmatattr{Sm-144}{hascov34}{A}
\setmatattr{Sm-144}{hascov35}{}
\setmatattr{Sm-144}{hascov36}{}

\registermaterial{Sm-147}
\setmatattr{Sm-147}{Emin}{1e-05}
\setmatattr{Sm-147}{Emax}{200}
\setmatattr{Sm-147}{lowsource}{TENDL-2019}
\setmatattr{Sm-147}{hisource}{TENDL-2019}
\setmatattr{Sm-147}{konnocorrection}{}
\setmatattr{Sm-147}{hascov32}{}
\setmatattr{Sm-147}{hascov33}{A}
\setmatattr{Sm-147}{hascov34}{A}
\setmatattr{Sm-147}{hascov35}{}
\setmatattr{Sm-147}{hascov36}{}

\registermaterial{Sm-148}
\setmatattr{Sm-148}{Emin}{1e-05}
\setmatattr{Sm-148}{Emax}{200}
\setmatattr{Sm-148}{lowsource}{TENDL-2019}
\setmatattr{Sm-148}{hisource}{TENDL-2019}
\setmatattr{Sm-148}{konnocorrection}{}
\setmatattr{Sm-148}{hascov32}{}
\setmatattr{Sm-148}{hascov33}{A}
\setmatattr{Sm-148}{hascov34}{A}
\setmatattr{Sm-148}{hascov35}{}
\setmatattr{Sm-148}{hascov36}{}

\registermaterial{Sm-149}
\setmatattr{Sm-149}{Emin}{1e-05}
\setmatattr{Sm-149}{Emax}{200}
\setmatattr{Sm-149}{lowsource}{TENDL-2019}
\setmatattr{Sm-149}{hisource}{TENDL-2019}
\setmatattr{Sm-149}{konnocorrection}{}
\setmatattr{Sm-149}{hascov32}{}
\setmatattr{Sm-149}{hascov33}{A}
\setmatattr{Sm-149}{hascov34}{A}
\setmatattr{Sm-149}{hascov35}{}
\setmatattr{Sm-149}{hascov36}{}

\registermaterial{Sm-150}
\setmatattr{Sm-150}{Emin}{1e-05}
\setmatattr{Sm-150}{Emax}{200}
\setmatattr{Sm-150}{lowsource}{TENDL-2019}
\setmatattr{Sm-150}{hisource}{TENDL-2019}
\setmatattr{Sm-150}{konnocorrection}{}
\setmatattr{Sm-150}{hascov32}{}
\setmatattr{Sm-150}{hascov33}{A}
\setmatattr{Sm-150}{hascov34}{A}
\setmatattr{Sm-150}{hascov35}{}
\setmatattr{Sm-150}{hascov36}{}

\registermaterial{Sm-152}
\setmatattr{Sm-152}{Emin}{1e-05}
\setmatattr{Sm-152}{Emax}{200}
\setmatattr{Sm-152}{lowsource}{TENDL-2019}
\setmatattr{Sm-152}{hisource}{TENDL-2019}
\setmatattr{Sm-152}{konnocorrection}{}
\setmatattr{Sm-152}{hascov32}{}
\setmatattr{Sm-152}{hascov33}{A}
\setmatattr{Sm-152}{hascov34}{A}
\setmatattr{Sm-152}{hascov35}{}
\setmatattr{Sm-152}{hascov36}{}

\registermaterial{Sm-154}
\setmatattr{Sm-154}{Emin}{1e-05}
\setmatattr{Sm-154}{Emax}{200}
\setmatattr{Sm-154}{lowsource}{TENDL-2019}
\setmatattr{Sm-154}{hisource}{TENDL-2019}
\setmatattr{Sm-154}{konnocorrection}{}
\setmatattr{Sm-154}{hascov32}{}
\setmatattr{Sm-154}{hascov33}{A}
\setmatattr{Sm-154}{hascov34}{A}
\setmatattr{Sm-154}{hascov35}{}
\setmatattr{Sm-154}{hascov36}{}

\registermaterial{Gd-152}
\setmatattr{Gd-152}{Emin}{1e-05}
\setmatattr{Gd-152}{Emax}{200}
\setmatattr{Gd-152}{lowsource}{JENDL-4.0}
\setmatattr{Gd-152}{hisource}{TENDL-2010}
\setmatattr{Gd-152}{Ecut}{20}
\setmatattr{Gd-152}{konnocorrection}{Y}
\setmatattr{Gd-152}{hascov32}{}
\setmatattr{Gd-152}{hascov33}{}
\setmatattr{Gd-152}{hascov34}{}
\setmatattr{Gd-152}{hascov35}{}
\setmatattr{Gd-152}{hascov36}{}

\registermaterial{Gd-154}
\setmatattr{Gd-154}{Emin}{1e-05}
\setmatattr{Gd-154}{Emax}{200}
\setmatattr{Gd-154}{lowsource}{JENDL-4.0}
\setmatattr{Gd-154}{hisource}{TENDL-2010}
\setmatattr{Gd-154}{Ecut}{20}
\setmatattr{Gd-154}{konnocorrection}{Y}
\setmatattr{Gd-154}{hascov32}{}
\setmatattr{Gd-154}{hascov33}{}
\setmatattr{Gd-154}{hascov34}{}
\setmatattr{Gd-154}{hascov35}{}
\setmatattr{Gd-154}{hascov36}{}

\registermaterial{Gd-155}
\setmatattr{Gd-155}{Emin}{1e-05}
\setmatattr{Gd-155}{Emax}{200}
\setmatattr{Gd-155}{lowsource}{JENDL-4.0}
\setmatattr{Gd-155}{hisource}{TENDL-2010}
\setmatattr{Gd-155}{Ecut}{20}
\setmatattr{Gd-155}{konnocorrection}{Y}
\setmatattr{Gd-155}{hascov32}{}
\setmatattr{Gd-155}{hascov33}{}
\setmatattr{Gd-155}{hascov34}{}
\setmatattr{Gd-155}{hascov35}{}
\setmatattr{Gd-155}{hascov36}{}

\registermaterial{Gd-156}
\setmatattr{Gd-156}{Emin}{1e-05}
\setmatattr{Gd-156}{Emax}{200}
\setmatattr{Gd-156}{lowsource}{JENDL-4.0}
\setmatattr{Gd-156}{hisource}{TENDL-2010}
\setmatattr{Gd-156}{Ecut}{20}
\setmatattr{Gd-156}{konnocorrection}{Y}
\setmatattr{Gd-156}{hascov32}{}
\setmatattr{Gd-156}{hascov33}{}
\setmatattr{Gd-156}{hascov34}{}
\setmatattr{Gd-156}{hascov35}{}
\setmatattr{Gd-156}{hascov36}{}

\registermaterial{Gd-157}
\setmatattr{Gd-157}{Emin}{1e-05}
\setmatattr{Gd-157}{Emax}{200}
\setmatattr{Gd-157}{lowsource}{JENDL-4.0}
\setmatattr{Gd-157}{hisource}{TENDL-2010}
\setmatattr{Gd-157}{Ecut}{20}
\setmatattr{Gd-157}{konnocorrection}{Y}
\setmatattr{Gd-157}{hascov32}{}
\setmatattr{Gd-157}{hascov33}{}
\setmatattr{Gd-157}{hascov34}{}
\setmatattr{Gd-157}{hascov35}{}
\setmatattr{Gd-157}{hascov36}{}

\registermaterial{Gd-158}
\setmatattr{Gd-158}{Emin}{1e-05}
\setmatattr{Gd-158}{Emax}{200}
\setmatattr{Gd-158}{lowsource}{JENDL-4.0}
\setmatattr{Gd-158}{hisource}{TENDL-2010}
\setmatattr{Gd-158}{Ecut}{20}
\setmatattr{Gd-158}{konnocorrection}{Y}
\setmatattr{Gd-158}{hascov32}{}
\setmatattr{Gd-158}{hascov33}{}
\setmatattr{Gd-158}{hascov34}{}
\setmatattr{Gd-158}{hascov35}{}
\setmatattr{Gd-158}{hascov36}{}

\registermaterial{Gd-160}
\setmatattr{Gd-160}{Emin}{1e-05}
\setmatattr{Gd-160}{Emax}{200}
\setmatattr{Gd-160}{lowsource}{JENDL-4.0}
\setmatattr{Gd-160}{hisource}{TENDL-2010}
\setmatattr{Gd-160}{Ecut}{20}
\setmatattr{Gd-160}{konnocorrection}{Y}
\setmatattr{Gd-160}{hascov32}{}
\setmatattr{Gd-160}{hascov33}{}
\setmatattr{Gd-160}{hascov34}{}
\setmatattr{Gd-160}{hascov35}{}
\setmatattr{Gd-160}{hascov36}{}

\registermaterial{Er-162}
\setmatattr{Er-162}{Emin}{1e-05}
\setmatattr{Er-162}{Emax}{200}
\setmatattr{Er-162}{lowsource}{ENDF-B-VII.1}
\setmatattr{Er-162}{hisource}{TENDL-2010}
\setmatattr{Er-162}{Ecut}{20}
\setmatattr{Er-162}{konnocorrection}{Y}
\setmatattr{Er-162}{hascov32}{}
\setmatattr{Er-162}{hascov33}{}
\setmatattr{Er-162}{hascov34}{}
\setmatattr{Er-162}{hascov35}{}
\setmatattr{Er-162}{hascov36}{}

\registermaterial{Er-164}
\setmatattr{Er-164}{Emin}{1e-05}
\setmatattr{Er-164}{Emax}{200}
\setmatattr{Er-164}{lowsource}{ENDF-B-VII.1}
\setmatattr{Er-164}{hisource}{TENDL-2010}
\setmatattr{Er-164}{Ecut}{20}
\setmatattr{Er-164}{konnocorrection}{Y}
\setmatattr{Er-164}{hascov32}{}
\setmatattr{Er-164}{hascov33}{}
\setmatattr{Er-164}{hascov34}{}
\setmatattr{Er-164}{hascov35}{}
\setmatattr{Er-164}{hascov36}{}

\registermaterial{Er-166}
\setmatattr{Er-166}{Emin}{1e-05}
\setmatattr{Er-166}{Emax}{200}
\setmatattr{Er-166}{lowsource}{ENDF-B-VII.1}
\setmatattr{Er-166}{hisource}{TENDL-2010}
\setmatattr{Er-166}{Ecut}{20}
\setmatattr{Er-166}{konnocorrection}{Y}
\setmatattr{Er-166}{hascov32}{}
\setmatattr{Er-166}{hascov33}{}
\setmatattr{Er-166}{hascov34}{}
\setmatattr{Er-166}{hascov35}{}
\setmatattr{Er-166}{hascov36}{}

\registermaterial{Er-167}
\setmatattr{Er-167}{Emin}{1e-05}
\setmatattr{Er-167}{Emax}{200}
\setmatattr{Er-167}{lowsource}{ENDF-B-VII.1}
\setmatattr{Er-167}{hisource}{TENDL-2010}
\setmatattr{Er-167}{Ecut}{20}
\setmatattr{Er-167}{konnocorrection}{Y}
\setmatattr{Er-167}{hascov32}{}
\setmatattr{Er-167}{hascov33}{}
\setmatattr{Er-167}{hascov34}{}
\setmatattr{Er-167}{hascov35}{}
\setmatattr{Er-167}{hascov36}{}

\registermaterial{Er-168}
\setmatattr{Er-168}{Emin}{1e-05}
\setmatattr{Er-168}{Emax}{200}
\setmatattr{Er-168}{lowsource}{ENDF-B-VII.1}
\setmatattr{Er-168}{hisource}{TENDL-2010}
\setmatattr{Er-168}{Ecut}{20}
\setmatattr{Er-168}{konnocorrection}{Y}
\setmatattr{Er-168}{hascov32}{}
\setmatattr{Er-168}{hascov33}{}
\setmatattr{Er-168}{hascov34}{}
\setmatattr{Er-168}{hascov35}{}
\setmatattr{Er-168}{hascov36}{}

\registermaterial{Er-170}
\setmatattr{Er-170}{Emin}{1e-05}
\setmatattr{Er-170}{Emax}{200}
\setmatattr{Er-170}{lowsource}{ENDF-B-VII.1}
\setmatattr{Er-170}{hisource}{TENDL-2010}
\setmatattr{Er-170}{Ecut}{20}
\setmatattr{Er-170}{konnocorrection}{Y}
\setmatattr{Er-170}{hascov32}{}
\setmatattr{Er-170}{hascov33}{}
\setmatattr{Er-170}{hascov34}{}
\setmatattr{Er-170}{hascov35}{}
\setmatattr{Er-170}{hascov36}{}

\registermaterial{Lu-175}
\setmatattr{Lu-175}{Emin}{1e-05}
\setmatattr{Lu-175}{Emax}{200}
\setmatattr{Lu-175}{lowsource}{TENDL-2010}
\setmatattr{Lu-175}{hisource}{TENDL-2010}
\setmatattr{Lu-175}{konnocorrection}{Y}
\setmatattr{Lu-175}{hascov32}{A}
\setmatattr{Lu-175}{hascov33}{A}
\setmatattr{Lu-175}{hascov34}{A}
\setmatattr{Lu-175}{hascov35}{}
\setmatattr{Lu-175}{hascov36}{}

\registermaterial{Lu-176}
\setmatattr{Lu-176}{Emin}{1e-05}
\setmatattr{Lu-176}{Emax}{200}
\setmatattr{Lu-176}{lowsource}{TENDL-2010}
\setmatattr{Lu-176}{hisource}{TENDL-2010}
\setmatattr{Lu-176}{konnocorrection}{Y}
\setmatattr{Lu-176}{hascov32}{A}
\setmatattr{Lu-176}{hascov33}{A}
\setmatattr{Lu-176}{hascov34}{A}
\setmatattr{Lu-176}{hascov35}{}
\setmatattr{Lu-176}{hascov36}{}

\registermaterial{Hf-174}
\setmatattr{Hf-174}{Emin}{1e-05}
\setmatattr{Hf-174}{Emax}{200}
\setmatattr{Hf-174}{lowsource}{JENDL-4.0}
\setmatattr{Hf-174}{hisource}{TENDL-2010}
\setmatattr{Hf-174}{Ecut}{20}
\setmatattr{Hf-174}{konnocorrection}{Y}
\setmatattr{Hf-174}{hascov32}{}
\setmatattr{Hf-174}{hascov33}{}
\setmatattr{Hf-174}{hascov34}{}
\setmatattr{Hf-174}{hascov35}{}
\setmatattr{Hf-174}{hascov36}{}

\registermaterial{Hf-176}
\setmatattr{Hf-176}{Emin}{1e-05}
\setmatattr{Hf-176}{Emax}{200}
\setmatattr{Hf-176}{lowsource}{JENDL-4.0}
\setmatattr{Hf-176}{hisource}{TENDL-2010}
\setmatattr{Hf-176}{Ecut}{20}
\setmatattr{Hf-176}{konnocorrection}{Y}
\setmatattr{Hf-176}{hascov32}{}
\setmatattr{Hf-176}{hascov33}{}
\setmatattr{Hf-176}{hascov34}{}
\setmatattr{Hf-176}{hascov35}{}
\setmatattr{Hf-176}{hascov36}{}

\registermaterial{Hf-177}
\setmatattr{Hf-177}{Emin}{1e-05}
\setmatattr{Hf-177}{Emax}{200}
\setmatattr{Hf-177}{lowsource}{JENDL-4.0}
\setmatattr{Hf-177}{hisource}{TENDL-2010}
\setmatattr{Hf-177}{Ecut}{20}
\setmatattr{Hf-177}{konnocorrection}{Y}
\setmatattr{Hf-177}{hascov32}{}
\setmatattr{Hf-177}{hascov33}{}
\setmatattr{Hf-177}{hascov34}{}
\setmatattr{Hf-177}{hascov35}{}
\setmatattr{Hf-177}{hascov36}{}

\registermaterial{Hf-178}
\setmatattr{Hf-178}{Emin}{1e-05}
\setmatattr{Hf-178}{Emax}{200}
\setmatattr{Hf-178}{lowsource}{JENDL-4.0}
\setmatattr{Hf-178}{hisource}{TENDL-2010}
\setmatattr{Hf-178}{Ecut}{20}
\setmatattr{Hf-178}{konnocorrection}{Y}
\setmatattr{Hf-178}{hascov32}{}
\setmatattr{Hf-178}{hascov33}{}
\setmatattr{Hf-178}{hascov34}{}
\setmatattr{Hf-178}{hascov35}{}
\setmatattr{Hf-178}{hascov36}{}

\registermaterial{Hf-179}
\setmatattr{Hf-179}{Emin}{1e-05}
\setmatattr{Hf-179}{Emax}{200}
\setmatattr{Hf-179}{lowsource}{JENDL-4.0}
\setmatattr{Hf-179}{hisource}{TENDL-2010}
\setmatattr{Hf-179}{Ecut}{20}
\setmatattr{Hf-179}{konnocorrection}{Y}
\setmatattr{Hf-179}{hascov32}{}
\setmatattr{Hf-179}{hascov33}{}
\setmatattr{Hf-179}{hascov34}{}
\setmatattr{Hf-179}{hascov35}{}
\setmatattr{Hf-179}{hascov36}{}

\registermaterial{Hf-180}
\setmatattr{Hf-180}{Emin}{1e-05}
\setmatattr{Hf-180}{Emax}{200}
\setmatattr{Hf-180}{lowsource}{JENDL-4.0}
\setmatattr{Hf-180}{hisource}{TENDL-2010}
\setmatattr{Hf-180}{Ecut}{20}
\setmatattr{Hf-180}{konnocorrection}{Y}
\setmatattr{Hf-180}{hascov32}{}
\setmatattr{Hf-180}{hascov33}{}
\setmatattr{Hf-180}{hascov34}{}
\setmatattr{Hf-180}{hascov35}{}
\setmatattr{Hf-180}{hascov36}{}

\registermaterial{Ta-180m}
\setmatattr{Ta-180m}{Emin}{1e-05}
\setmatattr{Ta-180m}{Emax}{200}
\setmatattr{Ta-180m}{lowsource}{TENDL-2019}
\setmatattr{Ta-180m}{hisource}{TENDL-2019}
\setmatattr{Ta-180m}{konnocorrection}{}

\registermaterial{Ta-181}
\setmatattr{Ta-181}{Emin}{1e-05}
\setmatattr{Ta-181}{Emax}{150}
\setmatattr{Ta-181}{lowsource}{JENDL-4.0}
\setmatattr{Ta-181}{hisource}{JENDL/HE-2007}
\setmatattr{Ta-181}{Ecut}{20}
\setmatattr{Ta-181}{konnocorrection}{}
\setmatattr{Ta-181}{hascov32}{}
\setmatattr{Ta-181}{hascov33}{}
\setmatattr{Ta-181}{hascov34}{}
\setmatattr{Ta-181}{hascov35}{}
\setmatattr{Ta-181}{hascov36}{}

\registermaterial{W-180}
\setmatattr{W-180}{Emin}{1e-05}
\setmatattr{W-180}{Emax}{150}
\setmatattr{W-180}{lowsource}{ENDF/B-VII.1}
\setmatattr{W-180}{hisource}{ENDF/B-VII.1}
\setmatattr{W-180}{konnocorrection}{}
\setmatattr{W-180}{hascov32}{}
\setmatattr{W-180}{hascov33}{A}
\setmatattr{W-180}{hascov34}{A}
\setmatattr{W-180}{hascov35}{}
\setmatattr{W-180}{hascov36}{}

\registermaterial{W-182}
\setmatattr{W-182}{Emin}{1e-05}
\setmatattr{W-182}{Emax}{150}
\setmatattr{W-182}{lowsource}{ENDF/B-VII.1}
\setmatattr{W-182}{hisource}{ENDF/B-VII.1}
\setmatattr{W-182}{konnocorrection}{}
\setmatattr{W-182}{hascov32}{A}
\setmatattr{W-182}{hascov33}{A}
\setmatattr{W-182}{hascov34}{A}
\setmatattr{W-182}{hascov35}{}
\setmatattr{W-182}{hascov36}{}

\registermaterial{W-183}
\setmatattr{W-183}{Emin}{1e-05}
\setmatattr{W-183}{Emax}{150}
\setmatattr{W-183}{lowsource}{ENDF/B-VII.1}
\setmatattr{W-183}{hisource}{ENDF/B-VII.1}
\setmatattr{W-183}{konnocorrection}{}
\setmatattr{W-183}{hascov32}{A}
\setmatattr{W-183}{hascov33}{A}
\setmatattr{W-183}{hascov34}{A}
\setmatattr{W-183}{hascov35}{}
\setmatattr{W-183}{hascov36}{}

\registermaterial{W-184}
\setmatattr{W-184}{Emin}{1e-05}
\setmatattr{W-184}{Emax}{150}
\setmatattr{W-184}{lowsource}{ENDF/B-VII.1}
\setmatattr{W-184}{hisource}{ENDF/B-VII.1}
\setmatattr{W-184}{konnocorrection}{}
\setmatattr{W-184}{hascov32}{A}
\setmatattr{W-184}{hascov33}{A}
\setmatattr{W-184}{hascov34}{A}
\setmatattr{W-184}{hascov35}{}
\setmatattr{W-184}{hascov36}{}

\registermaterial{W-186}
\setmatattr{W-186}{Emin}{1e-05}
\setmatattr{W-186}{Emax}{150}
\setmatattr{W-186}{lowsource}{ENDF/B-VII.1}
\setmatattr{W-186}{hisource}{ENDF/B-VII.1}
\setmatattr{W-186}{konnocorrection}{}
\setmatattr{W-186}{hascov32}{A}
\setmatattr{W-186}{hascov33}{A}
\setmatattr{W-186}{hascov34}{A}
\setmatattr{W-186}{hascov35}{}
\setmatattr{W-186}{hascov36}{}

\registermaterial{Re-185}
\setmatattr{Re-185}{Emin}{1e-05}
\setmatattr{Re-185}{Emax}{200}
\setmatattr{Re-185}{lowsource}{TENDL-2010}
\setmatattr{Re-185}{hisource}{TENDL-2010}
\setmatattr{Re-185}{konnocorrection}{Y}
\setmatattr{Re-185}{hascov32}{A}
\setmatattr{Re-185}{hascov33}{A}
\setmatattr{Re-185}{hascov34}{A}
\setmatattr{Re-185}{hascov35}{}
\setmatattr{Re-185}{hascov36}{}

\registermaterial{Re-187}
\setmatattr{Re-187}{Emin}{1e-05}
\setmatattr{Re-187}{Emax}{200}
\setmatattr{Re-187}{lowsource}{TENDL-2010}
\setmatattr{Re-187}{hisource}{TENDL-2010}
\setmatattr{Re-187}{konnocorrection}{Y}
\setmatattr{Re-187}{hascov32}{A}
\setmatattr{Re-187}{hascov33}{A}
\setmatattr{Re-187}{hascov34}{A}
\setmatattr{Re-187}{hascov35}{}
\setmatattr{Re-187}{hascov36}{}

\registermaterial{Pt-190}
\setmatattr{Pt-190}{Emin}{1e-05}
\setmatattr{Pt-190}{Emax}{200}
\setmatattr{Pt-190}{lowsource}{TENDL-2010}
\setmatattr{Pt-190}{hisource}{TENDL-2010}
\setmatattr{Pt-190}{konnocorrection}{Y}
\setmatattr{Pt-190}{hascov32}{A}
\setmatattr{Pt-190}{hascov33}{A}
\setmatattr{Pt-190}{hascov34}{A}
\setmatattr{Pt-190}{hascov35}{}
\setmatattr{Pt-190}{hascov36}{}

\registermaterial{Pt-192}
\setmatattr{Pt-192}{Emin}{1e-05}
\setmatattr{Pt-192}{Emax}{200}
\setmatattr{Pt-192}{lowsource}{TENDL-2010}
\setmatattr{Pt-192}{hisource}{TENDL-2010}
\setmatattr{Pt-192}{konnocorrection}{Y}
\setmatattr{Pt-192}{hascov32}{A}
\setmatattr{Pt-192}{hascov33}{A}
\setmatattr{Pt-192}{hascov34}{A}
\setmatattr{Pt-192}{hascov35}{}
\setmatattr{Pt-192}{hascov36}{}

\registermaterial{Pt-194}
\setmatattr{Pt-194}{Emin}{1e-05}
\setmatattr{Pt-194}{Emax}{200}
\setmatattr{Pt-194}{lowsource}{TENDL-2010}
\setmatattr{Pt-194}{hisource}{TENDL-2010}
\setmatattr{Pt-194}{konnocorrection}{Y}
\setmatattr{Pt-194}{hascov32}{A}
\setmatattr{Pt-194}{hascov33}{A}
\setmatattr{Pt-194}{hascov34}{A}
\setmatattr{Pt-194}{hascov35}{}
\setmatattr{Pt-194}{hascov36}{}

\registermaterial{Pt-195}
\setmatattr{Pt-195}{Emin}{1e-05}
\setmatattr{Pt-195}{Emax}{200}
\setmatattr{Pt-195}{lowsource}{TENDL-2010}
\setmatattr{Pt-195}{hisource}{TENDL-2010}
\setmatattr{Pt-195}{konnocorrection}{Y}
\setmatattr{Pt-195}{hascov32}{A}
\setmatattr{Pt-195}{hascov33}{A}
\setmatattr{Pt-195}{hascov34}{A}
\setmatattr{Pt-195}{hascov35}{}
\setmatattr{Pt-195}{hascov36}{}

\registermaterial{Pt-196}
\setmatattr{Pt-196}{Emin}{1e-05}
\setmatattr{Pt-196}{Emax}{200}
\setmatattr{Pt-196}{lowsource}{TENDL-2010}
\setmatattr{Pt-196}{hisource}{TENDL-2010}
\setmatattr{Pt-196}{konnocorrection}{Y}
\setmatattr{Pt-196}{hascov32}{A}
\setmatattr{Pt-196}{hascov33}{A}
\setmatattr{Pt-196}{hascov34}{A}
\setmatattr{Pt-196}{hascov35}{}
\setmatattr{Pt-196}{hascov36}{}

\registermaterial{Pt-198}
\setmatattr{Pt-198}{Emin}{1e-05}
\setmatattr{Pt-198}{Emax}{200}
\setmatattr{Pt-198}{lowsource}{TENDL-2010}
\setmatattr{Pt-198}{hisource}{TENDL-2010}
\setmatattr{Pt-198}{konnocorrection}{Y}
\setmatattr{Pt-198}{hascov32}{A}
\setmatattr{Pt-198}{hascov33}{A}
\setmatattr{Pt-198}{hascov34}{A}
\setmatattr{Pt-198}{hascov35}{}
\setmatattr{Pt-198}{hascov36}{}

\registermaterial{Au-197}
\setmatattr{Au-197}{Emin}{1e-05}
\setmatattr{Au-197}{Emax}{150}
\setmatattr{Au-197}{lowsource}{ENDF/B-VII.0}
\setmatattr{Au-197}{hisource}{JENDL/HE-2007}
\setmatattr{Au-197}{Ecut}{30}
\setmatattr{Au-197}{konnocorrection}{}
\setmatattr{Au-197}{hascov32}{}
\setmatattr{Au-197}{hascov33}{A}
\setmatattr{Au-197}{hascov34}{}
\setmatattr{Au-197}{hascov35}{}
\setmatattr{Au-197}{hascov36}{}

\registermaterial{Pb-204}
\setmatattr{Pb-204}{Emin}{1e-05}
\setmatattr{Pb-204}{Emax}{200}
\setmatattr{Pb-204}{lowsource}{JEFF-311}
\setmatattr{Pb-204}{hisource}{JEFF-311}
\setmatattr{Pb-204}{konnocorrection}{}
\setmatattr{Pb-204}{hascov32}{}
\setmatattr{Pb-204}{hascov33}{}
\setmatattr{Pb-204}{hascov34}{}
\setmatattr{Pb-204}{hascov35}{}
\setmatattr{Pb-204}{hascov36}{}

\registermaterial{Pb-206}
\setmatattr{Pb-206}{Emin}{1e-05}
\setmatattr{Pb-206}{Emax}{200}
\setmatattr{Pb-206}{lowsource}{JEFF-311}
\setmatattr{Pb-206}{hisource}{JEFF-311}
\setmatattr{Pb-206}{konnocorrection}{}
\setmatattr{Pb-206}{hascov32}{}
\setmatattr{Pb-206}{hascov33}{}
\setmatattr{Pb-206}{hascov34}{}
\setmatattr{Pb-206}{hascov35}{}
\setmatattr{Pb-206}{hascov36}{}

\registermaterial{Pb-207}
\setmatattr{Pb-207}{Emin}{1e-05}
\setmatattr{Pb-207}{Emax}{200}
\setmatattr{Pb-207}{lowsource}{JEFF-311}
\setmatattr{Pb-207}{hisource}{JEFF-311}
\setmatattr{Pb-207}{konnocorrection}{}
\setmatattr{Pb-207}{hascov32}{}
\setmatattr{Pb-207}{hascov33}{}
\setmatattr{Pb-207}{hascov34}{}
\setmatattr{Pb-207}{hascov35}{}
\setmatattr{Pb-207}{hascov36}{}

\registermaterial{Pb-208}
\setmatattr{Pb-208}{Emin}{1e-05}
\setmatattr{Pb-208}{Emax}{200}
\setmatattr{Pb-208}{lowsource}{JEFF-311}
\setmatattr{Pb-208}{hisource}{JEFF-311}
\setmatattr{Pb-208}{konnocorrection}{}
\setmatattr{Pb-208}{hascov32}{}
\setmatattr{Pb-208}{hascov33}{}
\setmatattr{Pb-208}{hascov34}{}
\setmatattr{Pb-208}{hascov35}{}
\setmatattr{Pb-208}{hascov36}{}

\registermaterial{Bi-209}
\setmatattr{Bi-209}{Emin}{1e-05}
\setmatattr{Bi-209}{Emax}{200}
\setmatattr{Bi-209}{lowsource}{JEFF-311}
\setmatattr{Bi-209}{hisource}{JEFF-311}
\setmatattr{Bi-209}{konnocorrection}{}
\setmatattr{Bi-209}{hascov32}{}
\setmatattr{Bi-209}{hascov33}{}
\setmatattr{Bi-209}{hascov34}{}
\setmatattr{Bi-209}{hascov35}{}
\setmatattr{Bi-209}{hascov36}{}

\registermaterial{Th-232}
\setmatattr{Th-232}{Emin}{1e-05}
\setmatattr{Th-232}{Emax}{60}
\setmatattr{Th-232}{lowsource}{ENDF/B-VII}
\setmatattr{Th-232}{hisource}{ENDF/B-VII}
\setmatattr{Th-232}{comment}{problem retrieving ENDF/B-VII}
\setmatattr{Th-232}{konnocorrection}{}
\setmatattr{Th-232}{hascov32}{A}
\setmatattr{Th-232}{hascov33}{A}
\setmatattr{Th-232}{hascov34}{A}
\setmatattr{Th-232}{hascov35}{}
\setmatattr{Th-232}{hascov36}{}

\registermaterial{U-234}
\setmatattr{U-234}{Emin}{1e-05}
\setmatattr{U-234}{Emax}{200}
\setmatattr{U-234}{lowsource}{TENDL-2019}
\setmatattr{U-234}{hisource}{TENDL-2019}
\setmatattr{U-234}{comment}{differences found}
\setmatattr{U-234}{konnocorrection}{}
\setmatattr{U-234}{hascov32}{}
\setmatattr{U-234}{hascov33}{A}
\setmatattr{U-234}{hascov34}{A}
\setmatattr{U-234}{hascov35}{A}
\setmatattr{U-234}{hascov36}{}

\registermaterial{U-235}
\setmatattr{U-235}{Emin}{1e-05}
\setmatattr{U-235}{Emax}{150}
\setmatattr{U-235}{lowsource}{ENDF/B-VII.1}
\setmatattr{U-235}{hisource}{JENDL/HE-2007}
\setmatattr{U-235}{Ecut}{20}
\setmatattr{U-235}{konnocorrection}{}
\setmatattr{U-235}{hascov32}{}
\setmatattr{U-235}{hascov33}{A}
\setmatattr{U-235}{hascov34}{}
\setmatattr{U-235}{hascov35}{A}
\setmatattr{U-235}{hascov36}{}

\registermaterial{U-238}
\setmatattr{U-238}{Emin}{1e-05}
\setmatattr{U-238}{Emax}{150}
\setmatattr{U-238}{lowsource}{ENDF/B-VII.1}
\setmatattr{U-238}{hisource}{JENDL/HE-2007}
\setmatattr{U-238}{Ecut}{30}
\setmatattr{U-238}{konnocorrection}{}
\setmatattr{U-238}{hascov32}{}
\setmatattr{U-238}{hascov33}{A}
\setmatattr{U-238}{hascov34}{}
\setmatattr{U-238}{hascov35}{A}
\setmatattr{U-238}{hascov36}{}

\setsublib{atom}


\registermaterial{1-H}
\setmatattr{1-H}{source}{ENDF/B-VI}
\setmatattr{1-H}{Emax}{0}

\registermaterial{2-HE}
\setmatattr{2-HE}{source}{ENDF/B-VI}
\setmatattr{2-HE}{Emax}{0}

\registermaterial{3-LI}
\setmatattr{3-LI}{source}{ENDF/B-VI}
\setmatattr{3-LI}{Emax}{0}

\registermaterial{4-BE}
\setmatattr{4-BE}{source}{ENDF/B-VI}
\setmatattr{4-BE}{Emax}{0}

\registermaterial{5-B}
\setmatattr{5-B}{source}{ENDF/B-VI}
\setmatattr{5-B}{Emax}{0}

\registermaterial{6-C}
\setmatattr{6-C}{source}{ENDF/B-VI}
\setmatattr{6-C}{Emax}{0}

\registermaterial{7-N}
\setmatattr{7-N}{source}{ENDF/B-VI}
\setmatattr{7-N}{Emax}{0}

\registermaterial{8-O}
\setmatattr{8-O}{source}{ENDF/B-VI}
\setmatattr{8-O}{Emax}{0}

\registermaterial{9-F}
\setmatattr{9-F}{source}{ENDF/B-VI}
\setmatattr{9-F}{Emax}{0}

\registermaterial{10-Ne-0}
\setmatattr{10-Ne-0}{source}{ENDF/B-VI}
\setmatattr{10-Ne-0}{Emax}{100000}

\registermaterial{11-NA}
\setmatattr{11-NA}{source}{ENDF/B-VI}
\setmatattr{11-NA}{Emax}{0}

\registermaterial{12-MG}
\setmatattr{12-MG}{source}{ENDF/B-VI}
\setmatattr{12-MG}{Emax}{0}

\registermaterial{13-AL}
\setmatattr{13-AL}{source}{ENDF/B-VI}
\setmatattr{13-AL}{Emax}{0}

\registermaterial{14-SI}
\setmatattr{14-SI}{source}{ENDF/B-VI}
\setmatattr{14-SI}{Emax}{0}

\registermaterial{15-P}
\setmatattr{15-P}{source}{ENDF/B-VI}
\setmatattr{15-P}{Emax}{0}

\registermaterial{16-S}
\setmatattr{16-S}{source}{ENDF/B-VI}
\setmatattr{16-S}{Emax}{0}

\registermaterial{17-CL}
\setmatattr{17-CL}{source}{ENDF/B-VI}
\setmatattr{17-CL}{Emax}{0}

\registermaterial{18-AR}
\setmatattr{18-AR}{source}{ENDF/B-VI}
\setmatattr{18-AR}{Emax}{0}

\registermaterial{19-K}
\setmatattr{19-K}{source}{ENDF/B-VI}
\setmatattr{19-K}{Emax}{0}

\registermaterial{20-CA}
\setmatattr{20-CA}{source}{ENDF/B-VI}
\setmatattr{20-CA}{Emax}{0}

\registermaterial{21-SC}
\setmatattr{21-SC}{source}{ENDF/B-VI}
\setmatattr{21-SC}{Emax}{0}

\registermaterial{22-TI}
\setmatattr{22-TI}{source}{ENDF/B-VI}
\setmatattr{22-TI}{Emax}{0}

\registermaterial{23-V}
\setmatattr{23-V}{source}{ENDF/B-VI}
\setmatattr{23-V}{Emax}{0}

\registermaterial{24-CR}
\setmatattr{24-CR}{source}{ENDF/B-VI}
\setmatattr{24-CR}{Emax}{0}

\registermaterial{25-MN}
\setmatattr{25-MN}{source}{ENDF/B-VI}
\setmatattr{25-MN}{Emax}{0}

\registermaterial{26-FE}
\setmatattr{26-FE}{source}{ENDF/B-VI}
\setmatattr{26-FE}{Emax}{0}

\registermaterial{27-CO}
\setmatattr{27-CO}{source}{ENDF/B-VI}
\setmatattr{27-CO}{Emax}{0}

\registermaterial{28-NI}
\setmatattr{28-NI}{source}{ENDF/B-VI}
\setmatattr{28-NI}{Emax}{0}

\registermaterial{29-CU}
\setmatattr{29-CU}{source}{ENDF/B-VI}
\setmatattr{29-CU}{Emax}{0}

\registermaterial{30-ZN}
\setmatattr{30-ZN}{source}{ENDF/B-VI}
\setmatattr{30-ZN}{Emax}{0}

\registermaterial{31-GA}
\setmatattr{31-GA}{source}{ENDF/B-VI}
\setmatattr{31-GA}{Emax}{0}

\registermaterial{32-GE}
\setmatattr{32-GE}{source}{ENDF/B-VI}
\setmatattr{32-GE}{Emax}{0}

\registermaterial{35-BR}
\setmatattr{35-BR}{source}{ENDF/B-VI}
\setmatattr{35-BR}{Emax}{0}

\registermaterial{39-Y}
\setmatattr{39-Y}{source}{ENDF/B-VI}
\setmatattr{39-Y}{Emax}{0}

\registermaterial{40-ZR}
\setmatattr{40-ZR}{source}{ENDF/B-VI}
\setmatattr{40-ZR}{Emax}{0}

\registermaterial{41-NB}
\setmatattr{41-NB}{source}{ENDF/B-VI}
\setmatattr{41-NB}{Emax}{0}

\registermaterial{42-MO}
\setmatattr{42-MO}{source}{ENDF/B-VI}
\setmatattr{42-MO}{Emax}{0}

\registermaterial{45-RH}
\setmatattr{45-RH}{source}{ENDF/B-VI}
\setmatattr{45-RH}{Emax}{0}

\registermaterial{47-AG}
\setmatattr{47-AG}{source}{ENDF/B-VI}
\setmatattr{47-AG}{Emax}{0}

\registermaterial{48-CD}
\setmatattr{48-CD}{source}{ENDF/B-VI}
\setmatattr{48-CD}{Emax}{0}

\registermaterial{50-SN}
\setmatattr{50-SN}{source}{ENDF/B-VI}
\setmatattr{50-SN}{Emax}{0}

\registermaterial{51-SB}
\setmatattr{51-SB}{source}{ENDF/B-VI}
\setmatattr{51-SB}{Emax}{0}

\registermaterial{53-I}
\setmatattr{53-I}{source}{ENDF/B-VI}
\setmatattr{53-I}{Emax}{0}

\registermaterial{55-CS}
\setmatattr{55-CS}{source}{ENDF/B-VI}
\setmatattr{55-CS}{Emax}{0}

\registermaterial{56-BA}
\setmatattr{56-BA}{source}{ENDF/B-VI}
\setmatattr{56-BA}{Emax}{0}

\registermaterial{57-LA}
\setmatattr{57-LA}{source}{ENDF/B-VI}
\setmatattr{57-LA}{Emax}{0}

\registermaterial{58-CE}
\setmatattr{58-CE}{source}{ENDF/B-VI}
\setmatattr{58-CE}{Emax}{0}

\registermaterial{62-Sm-0}
\setmatattr{62-Sm-0}{source}{ENDF/B-VI}
\setmatattr{62-Sm-0}{Emax}{100000}

\registermaterial{64-GD}
\setmatattr{64-GD}{source}{ENDF/B-VI}
\setmatattr{64-GD}{Emax}{0}

\registermaterial{68-ER}
\setmatattr{68-ER}{source}{ENDF/B-VI}
\setmatattr{68-ER}{Emax}{0}

\registermaterial{71-LU}
\setmatattr{71-LU}{source}{ENDF/B-VI}
\setmatattr{71-LU}{Emax}{0}

\registermaterial{72-HF}
\setmatattr{72-HF}{source}{ENDF/B-VI}
\setmatattr{72-HF}{Emax}{0}

\registermaterial{73-TA}
\setmatattr{73-TA}{source}{ENDF/B-VI}
\setmatattr{73-TA}{Emax}{0}

\registermaterial{74-W}
\setmatattr{74-W}{source}{ENDF/B-VI}
\setmatattr{74-W}{Emax}{0}

\registermaterial{75-RE}
\setmatattr{75-RE}{source}{ENDF/B-VI}
\setmatattr{75-RE}{Emax}{0}

\registermaterial{78-PT}
\setmatattr{78-PT}{source}{ENDF/B-VI}
\setmatattr{78-PT}{Emax}{0}

\registermaterial{79-AU}
\setmatattr{79-AU}{source}{ENDF/B-VI}
\setmatattr{79-AU}{Emax}{0}

\registermaterial{82-PB}
\setmatattr{82-PB}{source}{ENDF/B-VI}
\setmatattr{82-PB}{Emax}{0}

\registermaterial{83-BI}
\setmatattr{83-BI}{source}{ENDF/B-VI}
\setmatattr{83-BI}{Emax}{0}

\registermaterial{90-TH}
\setmatattr{90-TH}{source}{ENDF/B-VI}
\setmatattr{90-TH}{Emax}{0}

\registermaterial{92-U}
\setmatattr{92-U}{source}{ENDF/B-VI}
\setmatattr{92-U}{Emax}{0}

\setsublib{deuteron}


\registermaterial{3-Li-6}
\setmatattr{3-Li-6}{source}{JENDL/DEU-2020}
\setmatattr{3-Li-6}{Emax}{200}

\registermaterial{3-Li-7}
\setmatattr{3-Li-7}{source}{JENDL/DEU-2020}
\setmatattr{3-Li-7}{Emax}{200}

\registermaterial{4-Be-9}
\setmatattr{4-Be-9}{source}{JENDL/DEU-2020}
\setmatattr{4-Be-9}{Emax}{200}

\registermaterial{5-B-10}
\setmatattr{5-B-10}{source}{TENDL-2011}
\setmatattr{5-B-10}{Emax}{200}

\registermaterial{5-B-11}
\setmatattr{5-B-11}{source}{TENDL-2011}
\setmatattr{5-B-11}{Emax}{200}

\registermaterial{6-C-12}
\setmatattr{6-C-12}{source}{JENDL/DEU-2020}
\setmatattr{6-C-12}{Emax}{200}

\registermaterial{6-C-13}
\setmatattr{6-C-13}{source}{JENDL/DEU-2020}
\setmatattr{6-C-13}{Emax}{200}

\registermaterial{7-N-14}
\setmatattr{7-N-14}{source}{TENDL-2011}
\setmatattr{7-N-14}{Emax}{200}

\registermaterial{7-N-15}
\setmatattr{7-N-15}{source}{TENDL-2011}
\setmatattr{7-N-15}{Emax}{200}

\registermaterial{8-O-16}
\setmatattr{8-O-16}{source}{TENDL-2011}
\setmatattr{8-O-16}{Emax}{200}

\registermaterial{8-O-17}
\setmatattr{8-O-17}{source}{TENDL-2011}
\setmatattr{8-O-17}{Emax}{200}

\registermaterial{8-O-18}
\setmatattr{8-O-18}{source}{TENDL-2011}
\setmatattr{8-O-18}{Emax}{200}

\registermaterial{9-F-19}
\setmatattr{9-F-19}{source}{TENDL-2011}
\setmatattr{9-F-19}{Emax}{200}

\registermaterial{11-Na-23}
\setmatattr{11-Na-23}{source}{TENDL-2011}
\setmatattr{11-Na-23}{Emax}{200}

\registermaterial{12-Mg-24}
\setmatattr{12-Mg-24}{source}{TENDL-2011}
\setmatattr{12-Mg-24}{Emax}{200}

\registermaterial{12-Mg-25}
\setmatattr{12-Mg-25}{source}{TENDL-2011}
\setmatattr{12-Mg-25}{Emax}{200}

\registermaterial{12-Mg-26}
\setmatattr{12-Mg-26}{source}{TENDL-2011}
\setmatattr{12-Mg-26}{Emax}{200}

\registermaterial{13-Al-27}
\setmatattr{13-Al-27}{source}{TENDL-2011}
\setmatattr{13-Al-27}{Emax}{200}

\registermaterial{14-Si-28}
\setmatattr{14-Si-28}{source}{TENDL-2011}
\setmatattr{14-Si-28}{Emax}{200}

\registermaterial{14-Si-29}
\setmatattr{14-Si-29}{source}{TENDL-2011}
\setmatattr{14-Si-29}{Emax}{200}

\registermaterial{14-Si-30}
\setmatattr{14-Si-30}{source}{TENDL-2011}
\setmatattr{14-Si-30}{Emax}{200}

\registermaterial{15-P-31}
\setmatattr{15-P-31}{source}{TENDL-2011}
\setmatattr{15-P-31}{Emax}{200}

\registermaterial{16-S-32}
\setmatattr{16-S-32}{source}{TENDL-2011}
\setmatattr{16-S-32}{Emax}{200}

\registermaterial{16-S-33}
\setmatattr{16-S-33}{source}{TENDL-2011}
\setmatattr{16-S-33}{Emax}{200}

\registermaterial{16-S-34}
\setmatattr{16-S-34}{source}{TENDL-2011}
\setmatattr{16-S-34}{Emax}{200}

\registermaterial{16-S-36}
\setmatattr{16-S-36}{source}{TENDL-2011}
\setmatattr{16-S-36}{Emax}{200}

\registermaterial{17-Cl-35}
\setmatattr{17-Cl-35}{source}{TENDL-2011}
\setmatattr{17-Cl-35}{Emax}{200}

\registermaterial{17-Cl-37}
\setmatattr{17-Cl-37}{source}{TENDL-2011}
\setmatattr{17-Cl-37}{Emax}{200}

\registermaterial{18-Ar-36}
\setmatattr{18-Ar-36}{source}{TENDL-2011}
\setmatattr{18-Ar-36}{Emax}{200}

\registermaterial{18-Ar-38}
\setmatattr{18-Ar-38}{source}{TENDL-2011}
\setmatattr{18-Ar-38}{Emax}{200}

\registermaterial{18-Ar-40}
\setmatattr{18-Ar-40}{source}{TENDL-2011}
\setmatattr{18-Ar-40}{Emax}{200}

\registermaterial{19-K-39}
\setmatattr{19-K-39}{source}{TENDL-2011}
\setmatattr{19-K-39}{Emax}{200}

\registermaterial{19-K-40}
\setmatattr{19-K-40}{source}{TENDL-2011}
\setmatattr{19-K-40}{Emax}{200}

\registermaterial{19-K-41}
\setmatattr{19-K-41}{source}{TENDL-2011}
\setmatattr{19-K-41}{Emax}{200}

\registermaterial{20-Ca-40}
\setmatattr{20-Ca-40}{source}{TENDL-2011}
\setmatattr{20-Ca-40}{Emax}{200}

\registermaterial{20-Ca-42}
\setmatattr{20-Ca-42}{source}{TENDL-2011}
\setmatattr{20-Ca-42}{Emax}{200}

\registermaterial{20-Ca-43}
\setmatattr{20-Ca-43}{source}{TENDL-2011}
\setmatattr{20-Ca-43}{Emax}{200}

\registermaterial{20-Ca-44}
\setmatattr{20-Ca-44}{source}{TENDL-2011}
\setmatattr{20-Ca-44}{Emax}{200}

\registermaterial{20-Ca-46}
\setmatattr{20-Ca-46}{source}{TENDL-2011}
\setmatattr{20-Ca-46}{Emax}{200}

\registermaterial{20-Ca-48}
\setmatattr{20-Ca-48}{source}{TENDL-2011}
\setmatattr{20-Ca-48}{Emax}{200}

\registermaterial{21-Sc-45}
\setmatattr{21-Sc-45}{source}{TENDL-2011}
\setmatattr{21-Sc-45}{Emax}{200}

\registermaterial{22-Ti-46}
\setmatattr{22-Ti-46}{source}{TENDL-2011}
\setmatattr{22-Ti-46}{Emax}{200}

\registermaterial{22-Ti-47}
\setmatattr{22-Ti-47}{source}{TENDL-2011}
\setmatattr{22-Ti-47}{Emax}{200}

\registermaterial{22-Ti-48}
\setmatattr{22-Ti-48}{source}{TENDL-2011}
\setmatattr{22-Ti-48}{Emax}{200}

\registermaterial{22-Ti-49}
\setmatattr{22-Ti-49}{source}{TENDL-2011}
\setmatattr{22-Ti-49}{Emax}{200}

\registermaterial{22-Ti-50}
\setmatattr{22-Ti-50}{source}{TENDL-2011}
\setmatattr{22-Ti-50}{Emax}{200}

\registermaterial{23-V-50}
\setmatattr{23-V-50}{source}{TENDL-2011}
\setmatattr{23-V-50}{Emax}{200}

\registermaterial{23-V-51}
\setmatattr{23-V-51}{source}{TENDL-2011}
\setmatattr{23-V-51}{Emax}{200}

\registermaterial{24-Cr-50}
\setmatattr{24-Cr-50}{source}{TENDL-2011}
\setmatattr{24-Cr-50}{Emax}{200}

\registermaterial{24-Cr-52}
\setmatattr{24-Cr-52}{source}{TENDL-2011}
\setmatattr{24-Cr-52}{Emax}{200}

\registermaterial{24-Cr-53}
\setmatattr{24-Cr-53}{source}{TENDL-2011}
\setmatattr{24-Cr-53}{Emax}{200}

\registermaterial{24-Cr-54}
\setmatattr{24-Cr-54}{source}{TENDL-2011}
\setmatattr{24-Cr-54}{Emax}{200}

\registermaterial{25-Mn-55}
\setmatattr{25-Mn-55}{source}{TENDL-2011}
\setmatattr{25-Mn-55}{Emax}{200}

\registermaterial{26-Fe-54}
\setmatattr{26-Fe-54}{source}{TENDL-2011}
\setmatattr{26-Fe-54}{Emax}{200}

\registermaterial{26-Fe-56}
\setmatattr{26-Fe-56}{source}{TENDL-2011}
\setmatattr{26-Fe-56}{Emax}{200}

\registermaterial{26-Fe-57}
\setmatattr{26-Fe-57}{source}{TENDL-2011}
\setmatattr{26-Fe-57}{Emax}{200}

\registermaterial{26-Fe-58}
\setmatattr{26-Fe-58}{source}{TENDL-2011}
\setmatattr{26-Fe-58}{Emax}{200}

\registermaterial{27-Co-59}
\setmatattr{27-Co-59}{source}{TENDL-2011}
\setmatattr{27-Co-59}{Emax}{200}

\registermaterial{28-Ni-58}
\setmatattr{28-Ni-58}{source}{TENDL-2011}
\setmatattr{28-Ni-58}{Emax}{200}

\registermaterial{28-Ni-60}
\setmatattr{28-Ni-60}{source}{TENDL-2011}
\setmatattr{28-Ni-60}{Emax}{200}

\registermaterial{28-Ni-61}
\setmatattr{28-Ni-61}{source}{TENDL-2011}
\setmatattr{28-Ni-61}{Emax}{200}

\registermaterial{28-Ni-62}
\setmatattr{28-Ni-62}{source}{TENDL-2011}
\setmatattr{28-Ni-62}{Emax}{200}

\registermaterial{28-Ni-64}
\setmatattr{28-Ni-64}{source}{TENDL-2011}
\setmatattr{28-Ni-64}{Emax}{200}

\registermaterial{29-Cu-63}
\setmatattr{29-Cu-63}{source}{TENDL-2011}
\setmatattr{29-Cu-63}{Emax}{200}

\registermaterial{29-Cu-65}
\setmatattr{29-Cu-65}{source}{TENDL-2011}
\setmatattr{29-Cu-65}{Emax}{200}

\registermaterial{30-Zn-64}
\setmatattr{30-Zn-64}{source}{TENDL-2011}
\setmatattr{30-Zn-64}{Emax}{200}

\registermaterial{30-Zn-66}
\setmatattr{30-Zn-66}{source}{TENDL-2011}
\setmatattr{30-Zn-66}{Emax}{200}

\registermaterial{30-Zn-67}
\setmatattr{30-Zn-67}{source}{TENDL-2011}
\setmatattr{30-Zn-67}{Emax}{200}

\registermaterial{30-Zn-68}
\setmatattr{30-Zn-68}{source}{TENDL-2011}
\setmatattr{30-Zn-68}{Emax}{200}

\registermaterial{30-Zn-70}
\setmatattr{30-Zn-70}{source}{TENDL-2011}
\setmatattr{30-Zn-70}{Emax}{200}

\registermaterial{31-Ga-69}
\setmatattr{31-Ga-69}{source}{TENDL-2011}
\setmatattr{31-Ga-69}{Emax}{200}

\registermaterial{31-Ga-71}
\setmatattr{31-Ga-71}{source}{TENDL-2011}
\setmatattr{31-Ga-71}{Emax}{200}

\registermaterial{32-Ge-70}
\setmatattr{32-Ge-70}{source}{TENDL-2011}
\setmatattr{32-Ge-70}{Emax}{200}

\registermaterial{32-Ge-72}
\setmatattr{32-Ge-72}{source}{TENDL-2011}
\setmatattr{32-Ge-72}{Emax}{200}

\registermaterial{32-Ge-73}
\setmatattr{32-Ge-73}{source}{TENDL-2011}
\setmatattr{32-Ge-73}{Emax}{200}

\registermaterial{32-Ge-74}
\setmatattr{32-Ge-74}{source}{TENDL-2011}
\setmatattr{32-Ge-74}{Emax}{200}

\registermaterial{32-Ge-76}
\setmatattr{32-Ge-76}{source}{TENDL-2011}
\setmatattr{32-Ge-76}{Emax}{200}

\registermaterial{35-Br-79}
\setmatattr{35-Br-79}{source}{TENDL-2011}
\setmatattr{35-Br-79}{Emax}{200}

\registermaterial{35-Br-81}
\setmatattr{35-Br-81}{source}{TENDL-2011}
\setmatattr{35-Br-81}{Emax}{200}

\registermaterial{39-Y-89}
\setmatattr{39-Y-89}{source}{TENDL-2011}
\setmatattr{39-Y-89}{Emax}{200}

\registermaterial{40-Zr-90}
\setmatattr{40-Zr-90}{source}{TENDL-2011}
\setmatattr{40-Zr-90}{Emax}{200}

\registermaterial{40-Zr-91}
\setmatattr{40-Zr-91}{source}{TENDL-2011}
\setmatattr{40-Zr-91}{Emax}{200}

\registermaterial{40-Zr-92}
\setmatattr{40-Zr-92}{source}{TENDL-2011}
\setmatattr{40-Zr-92}{Emax}{200}

\registermaterial{40-Zr-94}
\setmatattr{40-Zr-94}{source}{TENDL-2011}
\setmatattr{40-Zr-94}{Emax}{200}

\registermaterial{40-Zr-96}
\setmatattr{40-Zr-96}{source}{TENDL-2011}
\setmatattr{40-Zr-96}{Emax}{200}

\registermaterial{41-Nb-93}
\setmatattr{41-Nb-93}{source}{TENDL-2011}
\setmatattr{41-Nb-93}{Emax}{200}

\registermaterial{42-Mo-92}
\setmatattr{42-Mo-92}{source}{TENDL-2011}
\setmatattr{42-Mo-92}{Emax}{200}

\registermaterial{42-Mo-94}
\setmatattr{42-Mo-94}{source}{TENDL-2011}
\setmatattr{42-Mo-94}{Emax}{200}

\registermaterial{42-Mo-95}
\setmatattr{42-Mo-95}{source}{TENDL-2011}
\setmatattr{42-Mo-95}{Emax}{200}

\registermaterial{42-Mo-96}
\setmatattr{42-Mo-96}{source}{TENDL-2011}
\setmatattr{42-Mo-96}{Emax}{200}

\registermaterial{42-Mo-97}
\setmatattr{42-Mo-97}{source}{TENDL-2011}
\setmatattr{42-Mo-97}{Emax}{200}

\registermaterial{42-Mo-98}
\setmatattr{42-Mo-98}{source}{TENDL-2011}
\setmatattr{42-Mo-98}{Emax}{200}

\registermaterial{42-Mo-100}
\setmatattr{42-Mo-100}{source}{TENDL-2011}
\setmatattr{42-Mo-100}{Emax}{200}

\registermaterial{45-Rh-103}
\setmatattr{45-Rh-103}{source}{TENDL-2011}
\setmatattr{45-Rh-103}{Emax}{200}

\registermaterial{47-Ag-107}
\setmatattr{47-Ag-107}{source}{TENDL-2011}
\setmatattr{47-Ag-107}{Emax}{200}

\registermaterial{47-Ag-109}
\setmatattr{47-Ag-109}{source}{TENDL-2011}
\setmatattr{47-Ag-109}{Emax}{200}

\registermaterial{48-Cd-106}
\setmatattr{48-Cd-106}{source}{TENDL-2011}
\setmatattr{48-Cd-106}{Emax}{200}

\registermaterial{48-Cd-108}
\setmatattr{48-Cd-108}{source}{TENDL-2011}
\setmatattr{48-Cd-108}{Emax}{200}

\registermaterial{48-Cd-110}
\setmatattr{48-Cd-110}{source}{TENDL-2011}
\setmatattr{48-Cd-110}{Emax}{200}

\registermaterial{48-Cd-111}
\setmatattr{48-Cd-111}{source}{TENDL-2011}
\setmatattr{48-Cd-111}{Emax}{200}

\registermaterial{48-Cd-112}
\setmatattr{48-Cd-112}{source}{TENDL-2011}
\setmatattr{48-Cd-112}{Emax}{200}

\registermaterial{48-Cd-113}
\setmatattr{48-Cd-113}{source}{TENDL-2011}
\setmatattr{48-Cd-113}{Emax}{200}

\registermaterial{48-Cd-114}
\setmatattr{48-Cd-114}{source}{TENDL-2011}
\setmatattr{48-Cd-114}{Emax}{200}

\registermaterial{48-Cd-116}
\setmatattr{48-Cd-116}{source}{TENDL-2011}
\setmatattr{48-Cd-116}{Emax}{200}

\registermaterial{50-Sn-112}
\setmatattr{50-Sn-112}{source}{TENDL-2011}
\setmatattr{50-Sn-112}{Emax}{200}

\registermaterial{50-Sn-114}
\setmatattr{50-Sn-114}{source}{TENDL-2011}
\setmatattr{50-Sn-114}{Emax}{200}

\registermaterial{50-Sn-115}
\setmatattr{50-Sn-115}{source}{TENDL-2011}
\setmatattr{50-Sn-115}{Emax}{200}

\registermaterial{50-Sn-116}
\setmatattr{50-Sn-116}{source}{TENDL-2011}
\setmatattr{50-Sn-116}{Emax}{200}

\registermaterial{50-Sn-117}
\setmatattr{50-Sn-117}{source}{TENDL-2011}
\setmatattr{50-Sn-117}{Emax}{200}

\registermaterial{50-Sn-118}
\setmatattr{50-Sn-118}{source}{TENDL-2011}
\setmatattr{50-Sn-118}{Emax}{200}

\registermaterial{50-Sn-119}
\setmatattr{50-Sn-119}{source}{TENDL-2011}
\setmatattr{50-Sn-119}{Emax}{200}

\registermaterial{50-Sn-120}
\setmatattr{50-Sn-120}{source}{TENDL-2011}
\setmatattr{50-Sn-120}{Emax}{200}

\registermaterial{50-Sn-122}
\setmatattr{50-Sn-122}{source}{TENDL-2011}
\setmatattr{50-Sn-122}{Emax}{200}

\registermaterial{50-Sn-124}
\setmatattr{50-Sn-124}{source}{TENDL-2011}
\setmatattr{50-Sn-124}{Emax}{200}

\registermaterial{51-Sb-121}
\setmatattr{51-Sb-121}{source}{TENDL-2011}
\setmatattr{51-Sb-121}{Emax}{200}

\registermaterial{51-Sb-123}
\setmatattr{51-Sb-123}{source}{TENDL-2011}
\setmatattr{51-Sb-123}{Emax}{200}

\registermaterial{53-I-127}
\setmatattr{53-I-127}{source}{TENDL-2011}
\setmatattr{53-I-127}{Emax}{200}

\registermaterial{55-Cs-133}
\setmatattr{55-Cs-133}{source}{TENDL-2011}
\setmatattr{55-Cs-133}{Emax}{200}

\registermaterial{56-Ba-130}
\setmatattr{56-Ba-130}{source}{TENDL-2011}
\setmatattr{56-Ba-130}{Emax}{200}

\registermaterial{56-Ba-132}
\setmatattr{56-Ba-132}{source}{TENDL-2011}
\setmatattr{56-Ba-132}{Emax}{200}

\registermaterial{56-Ba-134}
\setmatattr{56-Ba-134}{source}{TENDL-2011}
\setmatattr{56-Ba-134}{Emax}{200}

\registermaterial{56-Ba-135}
\setmatattr{56-Ba-135}{source}{TENDL-2011}
\setmatattr{56-Ba-135}{Emax}{200}

\registermaterial{56-Ba-136}
\setmatattr{56-Ba-136}{source}{TENDL-2011}
\setmatattr{56-Ba-136}{Emax}{200}

\registermaterial{56-Ba-137}
\setmatattr{56-Ba-137}{source}{TENDL-2011}
\setmatattr{56-Ba-137}{Emax}{200}

\registermaterial{56-Ba-138}
\setmatattr{56-Ba-138}{source}{TENDL-2011}
\setmatattr{56-Ba-138}{Emax}{200}

\registermaterial{57-La-138}
\setmatattr{57-La-138}{source}{TENDL-2011}
\setmatattr{57-La-138}{Emax}{200}

\registermaterial{57-La-139}
\setmatattr{57-La-139}{source}{TENDL-2011}
\setmatattr{57-La-139}{Emax}{200}

\registermaterial{58-Ce-136}
\setmatattr{58-Ce-136}{source}{TENDL-2011}
\setmatattr{58-Ce-136}{Emax}{200}

\registermaterial{58-Ce-138}
\setmatattr{58-Ce-138}{source}{TENDL-2011}
\setmatattr{58-Ce-138}{Emax}{200}

\registermaterial{58-Ce-140}
\setmatattr{58-Ce-140}{source}{TENDL-2011}
\setmatattr{58-Ce-140}{Emax}{200}

\registermaterial{58-Ce-142}
\setmatattr{58-Ce-142}{source}{TENDL-2011}
\setmatattr{58-Ce-142}{Emax}{200}

\registermaterial{64-Gd-152}
\setmatattr{64-Gd-152}{source}{TENDL-2011}
\setmatattr{64-Gd-152}{Emax}{200}

\registermaterial{64-Gd-154}
\setmatattr{64-Gd-154}{source}{TENDL-2011}
\setmatattr{64-Gd-154}{Emax}{200}

\registermaterial{64-Gd-155}
\setmatattr{64-Gd-155}{source}{TENDL-2011}
\setmatattr{64-Gd-155}{Emax}{200}

\registermaterial{64-Gd-156}
\setmatattr{64-Gd-156}{source}{TENDL-2011}
\setmatattr{64-Gd-156}{Emax}{200}

\registermaterial{64-Gd-157}
\setmatattr{64-Gd-157}{source}{TENDL-2011}
\setmatattr{64-Gd-157}{Emax}{200}

\registermaterial{64-Gd-158}
\setmatattr{64-Gd-158}{source}{TENDL-2011}
\setmatattr{64-Gd-158}{Emax}{200}

\registermaterial{64-Gd-160}
\setmatattr{64-Gd-160}{source}{TENDL-2011}
\setmatattr{64-Gd-160}{Emax}{200}

\registermaterial{68-Er-162}
\setmatattr{68-Er-162}{source}{TENDL-2011}
\setmatattr{68-Er-162}{Emax}{200}

\registermaterial{68-Er-164}
\setmatattr{68-Er-164}{source}{TENDL-2011}
\setmatattr{68-Er-164}{Emax}{200}

\registermaterial{68-Er-166}
\setmatattr{68-Er-166}{source}{TENDL-2011}
\setmatattr{68-Er-166}{Emax}{200}

\registermaterial{68-Er-167}
\setmatattr{68-Er-167}{source}{TENDL-2011}
\setmatattr{68-Er-167}{Emax}{200}

\registermaterial{68-Er-168}
\setmatattr{68-Er-168}{source}{TENDL-2011}
\setmatattr{68-Er-168}{Emax}{200}

\registermaterial{68-Er-170}
\setmatattr{68-Er-170}{source}{TENDL-2011}
\setmatattr{68-Er-170}{Emax}{200}

\registermaterial{71-Lu-175}
\setmatattr{71-Lu-175}{source}{TENDL-2011}
\setmatattr{71-Lu-175}{Emax}{200}

\registermaterial{71-Lu-176}
\setmatattr{71-Lu-176}{source}{TENDL-2011}
\setmatattr{71-Lu-176}{Emax}{200}

\registermaterial{72-Hf-174}
\setmatattr{72-Hf-174}{source}{TENDL-2011}
\setmatattr{72-Hf-174}{Emax}{200}

\registermaterial{72-Hf-176}
\setmatattr{72-Hf-176}{source}{TENDL-2011}
\setmatattr{72-Hf-176}{Emax}{200}

\registermaterial{72-Hf-177}
\setmatattr{72-Hf-177}{source}{TENDL-2011}
\setmatattr{72-Hf-177}{Emax}{200}

\registermaterial{72-Hf-178}
\setmatattr{72-Hf-178}{source}{TENDL-2011}
\setmatattr{72-Hf-178}{Emax}{200}

\registermaterial{72-Hf-179}
\setmatattr{72-Hf-179}{source}{TENDL-2011}
\setmatattr{72-Hf-179}{Emax}{200}

\registermaterial{72-Hf-180}
\setmatattr{72-Hf-180}{source}{TENDL-2011}
\setmatattr{72-Hf-180}{Emax}{200}

\registermaterial{73-Ta-181}
\setmatattr{73-Ta-181}{source}{TENDL-2011}
\setmatattr{73-Ta-181}{Emax}{200}

\registermaterial{74-W-180}
\setmatattr{74-W-180}{source}{TENDL-2011}
\setmatattr{74-W-180}{Emax}{200}

\registermaterial{74-W-182}
\setmatattr{74-W-182}{source}{TENDL-2011}
\setmatattr{74-W-182}{Emax}{200}

\registermaterial{74-W-183}
\setmatattr{74-W-183}{source}{TENDL-2011}
\setmatattr{74-W-183}{Emax}{200}

\registermaterial{74-W-184}
\setmatattr{74-W-184}{source}{TENDL-2011}
\setmatattr{74-W-184}{Emax}{200}

\registermaterial{74-W-186}
\setmatattr{74-W-186}{source}{TENDL-2011}
\setmatattr{74-W-186}{Emax}{200}

\registermaterial{75-Re-185}
\setmatattr{75-Re-185}{source}{TENDL-2011}
\setmatattr{75-Re-185}{Emax}{200}

\registermaterial{75-Re-187}
\setmatattr{75-Re-187}{source}{TENDL-2011}
\setmatattr{75-Re-187}{Emax}{200}

\registermaterial{78-Pt-190}
\setmatattr{78-Pt-190}{source}{TENDL-2011}
\setmatattr{78-Pt-190}{Emax}{200}

\registermaterial{78-Pt-192}
\setmatattr{78-Pt-192}{source}{TENDL-2011}
\setmatattr{78-Pt-192}{Emax}{200}

\registermaterial{78-Pt-194}
\setmatattr{78-Pt-194}{source}{TENDL-2011}
\setmatattr{78-Pt-194}{Emax}{200}

\registermaterial{78-Pt-195}
\setmatattr{78-Pt-195}{source}{TENDL-2011}
\setmatattr{78-Pt-195}{Emax}{200}

\registermaterial{78-Pt-196}
\setmatattr{78-Pt-196}{source}{TENDL-2011}
\setmatattr{78-Pt-196}{Emax}{200}

\registermaterial{78-Pt-198}
\setmatattr{78-Pt-198}{source}{TENDL-2011}
\setmatattr{78-Pt-198}{Emax}{200}

\registermaterial{79-Au-197}
\setmatattr{79-Au-197}{source}{TENDL-2011}
\setmatattr{79-Au-197}{Emax}{200}

\registermaterial{82-Pb-204}
\setmatattr{82-Pb-204}{source}{TENDL-2011}
\setmatattr{82-Pb-204}{Emax}{200}

\registermaterial{82-Pb-206}
\setmatattr{82-Pb-206}{source}{TENDL-2011}
\setmatattr{82-Pb-206}{Emax}{200}

\registermaterial{82-Pb-207}
\setmatattr{82-Pb-207}{source}{TENDL-2011}
\setmatattr{82-Pb-207}{Emax}{200}

\registermaterial{82-Pb-208}
\setmatattr{82-Pb-208}{source}{TENDL-2011}
\setmatattr{82-Pb-208}{Emax}{200}

\registermaterial{83-Bi-209}
\setmatattr{83-Bi-209}{source}{TENDL-2011}
\setmatattr{83-Bi-209}{Emax}{200}

\registermaterial{90-Th-232}
\setmatattr{90-Th-232}{source}{TENDL-2011}
\setmatattr{90-Th-232}{Emax}{200}

\registermaterial{92-U-235}
\setmatattr{92-U-235}{source}{TENDL-2011}
\setmatattr{92-U-235}{Emax}{200}

\registermaterial{92-U-238}
\setmatattr{92-U-238}{source}{TENDL-2011}
\setmatattr{92-U-238}{Emax}{200}

\setsublib{proton}


\registermaterial{1-H-1}
\setmatattr{1-H-1}{source}{JENDL/HE-2007}
\setmatattr{1-H-1}{Emax}{3000}

\registermaterial{1-H-2}
\setmatattr{1-H-2}{source}{ENDF/B-VII}
\setmatattr{1-H-2}{Emax}{150}

\registermaterial{1-H-3}
\setmatattr{1-H-3}{source}{ENDF/B-VII}
\setmatattr{1-H-3}{Emax}{20}
\setmatattr{1-H-3}{comment}{Example description. What can we see about the evaluation of tritium?}

\registermaterial{2-He-3}
\setmatattr{2-He-3}{source}{ENDF/B-VII}
\setmatattr{2-He-3}{Emax}{20}

\registermaterial{3-Li-6}
\setmatattr{3-Li-6}{source}{JENDL-4.0/HE}
\setmatattr{3-Li-6}{Emax}{200}

\registermaterial{3-Li-7}
\setmatattr{3-Li-7}{source}{JENDL-4.0/HE}
\setmatattr{3-Li-7}{Emax}{200}

\registermaterial{4-Be-9}
\setmatattr{4-Be-9}{source}{ENDF/B-VII}
\setmatattr{4-Be-9}{Emax}{113}

\registermaterial{5-B-10}
\setmatattr{5-B-10}{source}{ENDF/B-VII}
\setmatattr{5-B-10}{Emax}{3}

\registermaterial{5-B-11}
\setmatattr{5-B-11}{source}{TENDL-2011}
\setmatattr{5-B-11}{Emax}{200}

\registermaterial{6-C-12}
\setmatattr{6-C-12}{source}{JENDL/HE-2007}
\setmatattr{6-C-12}{Emax}{3000}

\registermaterial{6-C-13}
\setmatattr{6-C-13}{source}{JENDL/HE-2007}
\setmatattr{6-C-13}{Emax}{3000}

\registermaterial{7-N-14}
\setmatattr{7-N-14}{source}{JENDL/HE-2007}
\setmatattr{7-N-14}{Emax}{3000}

\registermaterial{7-N-15}
\setmatattr{7-N-15}{source}{TENDL-2011}
\setmatattr{7-N-15}{Emax}{200}

\registermaterial{8-O-16}
\setmatattr{8-O-16}{source}{JENDL/HE-2007}
\setmatattr{8-O-16}{Emax}{3000}

\registermaterial{8-O-17}
\setmatattr{8-O-17}{source}{TENDL-2011}
\setmatattr{8-O-17}{Emax}{200}

\registermaterial{8-O-18}
\setmatattr{8-O-18}{source}{TENDL-2011}
\setmatattr{8-O-18}{Emax}{200}

\registermaterial{9-F-19}
\setmatattr{9-F-19}{source}{JENDL/HE-2007}
\setmatattr{9-F-19}{Emax}{3000}

\registermaterial{11-Na-23}
\setmatattr{11-Na-23}{source}{JENDL/HE-2007}
\setmatattr{11-Na-23}{Emax}{3000}

\registermaterial{12-Mg-24}
\setmatattr{12-Mg-24}{source}{JENDL/HE-2007}
\setmatattr{12-Mg-24}{Emax}{3000}

\registermaterial{12-Mg-25}
\setmatattr{12-Mg-25}{source}{JENDL/HE-2007}
\setmatattr{12-Mg-25}{Emax}{3000}

\registermaterial{12-Mg-26}
\setmatattr{12-Mg-26}{source}{JENDL/HE-2007}
\setmatattr{12-Mg-26}{Emax}{3000}

\registermaterial{13-Al-27}
\setmatattr{13-Al-27}{source}{JENDL/HE-2007}
\setmatattr{13-Al-27}{Emax}{3000}

\registermaterial{14-Si-28}
\setmatattr{14-Si-28}{source}{JENDL/HE-2007}
\setmatattr{14-Si-28}{Emax}{3000}

\registermaterial{14-Si-29}
\setmatattr{14-Si-29}{source}{JENDL/HE-2007}
\setmatattr{14-Si-29}{Emax}{3000}

\registermaterial{14-Si-30}
\setmatattr{14-Si-30}{source}{JENDL/HE-2007}
\setmatattr{14-Si-30}{Emax}{3000}

\registermaterial{15-P-31}
\setmatattr{15-P-31}{source}{ENDF/B-VII}
\setmatattr{15-P-31}{Emax}{150}

\registermaterial{16-S-32}
\setmatattr{16-S-32}{source}{TENDL-2011}
\setmatattr{16-S-32}{Emax}{200}

\registermaterial{16-S-33}
\setmatattr{16-S-33}{source}{TENDL-2011}
\setmatattr{16-S-33}{Emax}{200}

\registermaterial{16-S-34}
\setmatattr{16-S-34}{source}{TENDL-2011}
\setmatattr{16-S-34}{Emax}{200}

\registermaterial{16-S-36}
\setmatattr{16-S-36}{source}{TENDL-2011}
\setmatattr{16-S-36}{Emax}{200}

\registermaterial{17-Cl-35}
\setmatattr{17-Cl-35}{source}{JENDL/HE-2007}
\setmatattr{17-Cl-35}{Emax}{3000}

\registermaterial{17-Cl-37}
\setmatattr{17-Cl-37}{source}{JENDL/HE-2007}
\setmatattr{17-Cl-37}{Emax}{3000}

\registermaterial{18-Ar-36}
\setmatattr{18-Ar-36}{source}{JENDL/HE-2007}
\setmatattr{18-Ar-36}{Emax}{3000}

\registermaterial{18-Ar-38}
\setmatattr{18-Ar-38}{source}{JENDL/HE-2007}
\setmatattr{18-Ar-38}{Emax}{3000}

\registermaterial{18-Ar-40}
\setmatattr{18-Ar-40}{source}{JENDL/HE-2007}
\setmatattr{18-Ar-40}{Emax}{3000}

\registermaterial{19-K-39}
\setmatattr{19-K-39}{source}{JENDL/HE-2007}
\setmatattr{19-K-39}{Emax}{3000}

\registermaterial{19-K-40}
\setmatattr{19-K-40}{source}{TENDL-2011}
\setmatattr{19-K-40}{Emax}{200}

\registermaterial{19-K-41}
\setmatattr{19-K-41}{source}{JENDL/HE-2007}
\setmatattr{19-K-41}{Emax}{3000}

\registermaterial{20-Ca-40}
\setmatattr{20-Ca-40}{source}{JENDL/HE-2007}
\setmatattr{20-Ca-40}{Emax}{3000}

\registermaterial{20-Ca-42}
\setmatattr{20-Ca-42}{source}{JENDL/HE-2007}
\setmatattr{20-Ca-42}{Emax}{3000}

\registermaterial{20-Ca-43}
\setmatattr{20-Ca-43}{source}{JENDL/HE-2007}
\setmatattr{20-Ca-43}{Emax}{3000}

\registermaterial{20-Ca-44}
\setmatattr{20-Ca-44}{source}{JENDL/HE-2007}
\setmatattr{20-Ca-44}{Emax}{3000}

\registermaterial{20-Ca-46}
\setmatattr{20-Ca-46}{source}{JENDL/HE-2007}
\setmatattr{20-Ca-46}{Emax}{3000}

\registermaterial{20-Ca-48}
\setmatattr{20-Ca-48}{source}{JENDL/HE-2007}
\setmatattr{20-Ca-48}{Emax}{3000}

\registermaterial{21-Sc-45}
\setmatattr{21-Sc-45}{source}{TENDL-2011}
\setmatattr{21-Sc-45}{Emax}{200}

\registermaterial{22-Ti-46}
\setmatattr{22-Ti-46}{source}{JENDL/HE-2007}
\setmatattr{22-Ti-46}{Emax}{3000}

\registermaterial{22-Ti-47}
\setmatattr{22-Ti-47}{source}{JENDL/HE-2007}
\setmatattr{22-Ti-47}{Emax}{3000}

\registermaterial{22-Ti-48}
\setmatattr{22-Ti-48}{source}{JENDL/HE-2007}
\setmatattr{22-Ti-48}{Emax}{3000}

\registermaterial{22-Ti-49}
\setmatattr{22-Ti-49}{source}{JENDL/HE-2007}
\setmatattr{22-Ti-49}{Emax}{3000}

\registermaterial{22-Ti-50}
\setmatattr{22-Ti-50}{source}{JENDL/HE-2007}
\setmatattr{22-Ti-50}{Emax}{3000}

\registermaterial{23-V-50}
\setmatattr{23-V-50}{source}{TENDL-2011}
\setmatattr{23-V-50}{Emax}{200}
\setmatattr{23-V-50}{comment}{Here is another example comment for vanadium 50}

\registermaterial{23-V-51}
\setmatattr{23-V-51}{source}{JENDL/HE-2007}
\setmatattr{23-V-51}{Emax}{3000}

\registermaterial{24-Cr-50}
\setmatattr{24-Cr-50}{source}{JENDL/HE-2007}
\setmatattr{24-Cr-50}{Emax}{3000}

\registermaterial{24-Cr-52}
\setmatattr{24-Cr-52}{source}{JENDL/HE-2007}
\setmatattr{24-Cr-52}{Emax}{3000}

\registermaterial{24-Cr-53}
\setmatattr{24-Cr-53}{source}{JENDL/HE-2007}
\setmatattr{24-Cr-53}{Emax}{3000}

\registermaterial{24-Cr-54}
\setmatattr{24-Cr-54}{source}{JENDL/HE-2007}
\setmatattr{24-Cr-54}{Emax}{3000}

\registermaterial{25-Mn-55}
\setmatattr{25-Mn-55}{source}{JENDL/HE-2007}
\setmatattr{25-Mn-55}{Emax}{3000}

\registermaterial{26-Fe-54}
\setmatattr{26-Fe-54}{source}{JENDL/HE-2007}
\setmatattr{26-Fe-54}{Emax}{3000}

\registermaterial{26-Fe-56}
\setmatattr{26-Fe-56}{source}{JENDL/HE-2007}
\setmatattr{26-Fe-56}{Emax}{3000}

\registermaterial{26-Fe-57}
\setmatattr{26-Fe-57}{source}{JENDL/HE-2007}
\setmatattr{26-Fe-57}{Emax}{3000}

\registermaterial{26-Fe-58}
\setmatattr{26-Fe-58}{source}{JENDL/HE-2007}
\setmatattr{26-Fe-58}{Emax}{3000}

\registermaterial{27-Co-59}
\setmatattr{27-Co-59}{source}{JENDL/HE-2007}
\setmatattr{27-Co-59}{Emax}{3000}

\registermaterial{28-Ni-58}
\setmatattr{28-Ni-58}{source}{JENDL/HE-2007}
\setmatattr{28-Ni-58}{Emax}{3000}

\registermaterial{28-Ni-60}
\setmatattr{28-Ni-60}{source}{JENDL/HE-2007}
\setmatattr{28-Ni-60}{Emax}{3000}

\registermaterial{28-Ni-61}
\setmatattr{28-Ni-61}{source}{JENDL/HE-2007}
\setmatattr{28-Ni-61}{Emax}{3000}

\registermaterial{28-Ni-62}
\setmatattr{28-Ni-62}{source}{JENDL/HE-2007}
\setmatattr{28-Ni-62}{Emax}{3000}

\registermaterial{28-Ni-64}
\setmatattr{28-Ni-64}{source}{JENDL/HE-2007}
\setmatattr{28-Ni-64}{Emax}{3000}

\registermaterial{29-Cu-63}
\setmatattr{29-Cu-63}{source}{JENDL/HE-2007}
\setmatattr{29-Cu-63}{Emax}{3000}

\registermaterial{29-Cu-65}
\setmatattr{29-Cu-65}{source}{JENDL/HE-2007}
\setmatattr{29-Cu-65}{Emax}{3000}

\registermaterial{30-Zn-64}
\setmatattr{30-Zn-64}{source}{JENDL/HE-2007}
\setmatattr{30-Zn-64}{Emax}{3000}

\registermaterial{30-Zn-66}
\setmatattr{30-Zn-66}{source}{JENDL/HE-2007}
\setmatattr{30-Zn-66}{Emax}{3000}

\registermaterial{30-Zn-67}
\setmatattr{30-Zn-67}{source}{JENDL/HE-2007}
\setmatattr{30-Zn-67}{Emax}{3000}

\registermaterial{30-Zn-68}
\setmatattr{30-Zn-68}{source}{JENDL/HE-2007}
\setmatattr{30-Zn-68}{Emax}{3000}

\registermaterial{30-Zn-70}
\setmatattr{30-Zn-70}{source}{JENDL/HE-2007}
\setmatattr{30-Zn-70}{Emax}{3000}

\registermaterial{31-Ga-69}
\setmatattr{31-Ga-69}{source}{JENDL/HE-2007}
\setmatattr{31-Ga-69}{Emax}{3000}

\registermaterial{31-Ga-71}
\setmatattr{31-Ga-71}{source}{JENDL/HE-2007}
\setmatattr{31-Ga-71}{Emax}{3000}

\registermaterial{32-Ge-70}
\setmatattr{32-Ge-70}{source}{JENDL/HE-2007}
\setmatattr{32-Ge-70}{Emax}{3000}

\registermaterial{32-Ge-72}
\setmatattr{32-Ge-72}{source}{JENDL/HE-2007}
\setmatattr{32-Ge-72}{Emax}{3000}

\registermaterial{32-Ge-73}
\setmatattr{32-Ge-73}{source}{JENDL/HE-2007}
\setmatattr{32-Ge-73}{Emax}{3000}

\registermaterial{32-Ge-74}
\setmatattr{32-Ge-74}{source}{JENDL/HE-2007}
\setmatattr{32-Ge-74}{Emax}{3000}

\registermaterial{32-Ge-76}
\setmatattr{32-Ge-76}{source}{JENDL/HE-2007}
\setmatattr{32-Ge-76}{Emax}{3000}

\registermaterial{35-Br-79}
\setmatattr{35-Br-79}{source}{TENDL-2011}
\setmatattr{35-Br-79}{Emax}{200}

\registermaterial{35-Br-81}
\setmatattr{35-Br-81}{source}{TENDL-2011}
\setmatattr{35-Br-81}{Emax}{200}

\registermaterial{39-Y-89}
\setmatattr{39-Y-89}{source}{TENDL-2011}
\setmatattr{39-Y-89}{Emax}{200}

\registermaterial{40-Zr-90}
\setmatattr{40-Zr-90}{source}{JENDL/HE-2007}
\setmatattr{40-Zr-90}{Emax}{3000}

\registermaterial{40-Zr-91}
\setmatattr{40-Zr-91}{source}{JENDL/HE-2007}
\setmatattr{40-Zr-91}{Emax}{3000}

\registermaterial{40-Zr-92}
\setmatattr{40-Zr-92}{source}{JENDL/HE-2007}
\setmatattr{40-Zr-92}{Emax}{3000}

\registermaterial{40-Zr-94}
\setmatattr{40-Zr-94}{source}{JENDL/HE-2007}
\setmatattr{40-Zr-94}{Emax}{3000}

\registermaterial{40-Zr-96}
\setmatattr{40-Zr-96}{source}{JENDL/HE-2007}
\setmatattr{40-Zr-96}{Emax}{3000}

\registermaterial{41-Nb-93}
\setmatattr{41-Nb-93}{source}{JENDL/HE-2007}
\setmatattr{41-Nb-93}{Emax}{3000}

\registermaterial{42-Mo-92}
\setmatattr{42-Mo-92}{source}{JENDL/HE-2007}
\setmatattr{42-Mo-92}{Emax}{3000}

\registermaterial{42-Mo-94}
\setmatattr{42-Mo-94}{source}{JENDL/HE-2007}
\setmatattr{42-Mo-94}{Emax}{3000}

\registermaterial{42-Mo-95}
\setmatattr{42-Mo-95}{source}{JENDL/HE-2007}
\setmatattr{42-Mo-95}{Emax}{3000}

\registermaterial{42-Mo-96}
\setmatattr{42-Mo-96}{source}{JENDL/HE-2007}
\setmatattr{42-Mo-96}{Emax}{3000}

\registermaterial{42-Mo-97}
\setmatattr{42-Mo-97}{source}{JENDL/HE-2007}
\setmatattr{42-Mo-97}{Emax}{3000}

\registermaterial{42-Mo-98}
\setmatattr{42-Mo-98}{source}{JENDL/HE-2007}
\setmatattr{42-Mo-98}{Emax}{3000}

\registermaterial{42-Mo-100}
\setmatattr{42-Mo-100}{source}{JENDL/HE-2007}
\setmatattr{42-Mo-100}{Emax}{3000}

\registermaterial{45-Rh-103}
\setmatattr{45-Rh-103}{source}{TENDL-2011}
\setmatattr{45-Rh-103}{Emax}{200}

\registermaterial{47-Ag-107}
\setmatattr{47-Ag-107}{source}{TENDL-2011}
\setmatattr{47-Ag-107}{Emax}{200}

\registermaterial{47-Ag-109}
\setmatattr{47-Ag-109}{source}{TENDL-2011}
\setmatattr{47-Ag-109}{Emax}{200}

\registermaterial{48-Cd-106}
\setmatattr{48-Cd-106}{source}{TENDL-2011}
\setmatattr{48-Cd-106}{Emax}{200}

\registermaterial{48-Cd-108}
\setmatattr{48-Cd-108}{source}{TENDL-2011}
\setmatattr{48-Cd-108}{Emax}{200}

\registermaterial{48-Cd-110}
\setmatattr{48-Cd-110}{source}{TENDL-2011}
\setmatattr{48-Cd-110}{Emax}{200}

\registermaterial{48-Cd-111}
\setmatattr{48-Cd-111}{source}{TENDL-2011}
\setmatattr{48-Cd-111}{Emax}{200}

\registermaterial{48-Cd-112}
\setmatattr{48-Cd-112}{source}{TENDL-2011}
\setmatattr{48-Cd-112}{Emax}{200}

\registermaterial{48-Cd-113}
\setmatattr{48-Cd-113}{source}{TENDL-2011}
\setmatattr{48-Cd-113}{Emax}{200}

\registermaterial{48-Cd-114}
\setmatattr{48-Cd-114}{source}{TENDL-2011}
\setmatattr{48-Cd-114}{Emax}{200}

\registermaterial{48-Cd-116}
\setmatattr{48-Cd-116}{source}{TENDL-2011}
\setmatattr{48-Cd-116}{Emax}{200}

\registermaterial{50-Sn-112}
\setmatattr{50-Sn-112}{source}{TENDL-2011}
\setmatattr{50-Sn-112}{Emax}{200}

\registermaterial{50-Sn-114}
\setmatattr{50-Sn-114}{source}{TENDL-2011}
\setmatattr{50-Sn-114}{Emax}{200}

\registermaterial{50-Sn-115}
\setmatattr{50-Sn-115}{source}{TENDL-2011}
\setmatattr{50-Sn-115}{Emax}{200}

\registermaterial{50-Sn-116}
\setmatattr{50-Sn-116}{source}{TENDL-2011}
\setmatattr{50-Sn-116}{Emax}{200}

\registermaterial{50-Sn-117}
\setmatattr{50-Sn-117}{source}{TENDL-2011}
\setmatattr{50-Sn-117}{Emax}{200}

\registermaterial{50-Sn-118}
\setmatattr{50-Sn-118}{source}{TENDL-2011}
\setmatattr{50-Sn-118}{Emax}{200}

\registermaterial{50-Sn-119}
\setmatattr{50-Sn-119}{source}{TENDL-2011}
\setmatattr{50-Sn-119}{Emax}{200}

\registermaterial{50-Sn-120}
\setmatattr{50-Sn-120}{source}{TENDL-2011}
\setmatattr{50-Sn-120}{Emax}{200}

\registermaterial{50-Sn-122}
\setmatattr{50-Sn-122}{source}{TENDL-2011}
\setmatattr{50-Sn-122}{Emax}{200}

\registermaterial{50-Sn-124}
\setmatattr{50-Sn-124}{source}{TENDL-2011}
\setmatattr{50-Sn-124}{Emax}{200}

\registermaterial{51-Sb-121}
\setmatattr{51-Sb-121}{source}{TENDL-2011}
\setmatattr{51-Sb-121}{Emax}{200}

\registermaterial{51-Sb-123}
\setmatattr{51-Sb-123}{source}{TENDL-2011}
\setmatattr{51-Sb-123}{Emax}{200}

\registermaterial{53-I-127}
\setmatattr{53-I-127}{source}{TENDL-2011}
\setmatattr{53-I-127}{Emax}{200}

\registermaterial{55-Cs-133}
\setmatattr{55-Cs-133}{source}{TENDL-2011}
\setmatattr{55-Cs-133}{Emax}{200}

\registermaterial{56-Ba-130}
\setmatattr{56-Ba-130}{source}{TENDL-2011}
\setmatattr{56-Ba-130}{Emax}{200}

\registermaterial{56-Ba-132}
\setmatattr{56-Ba-132}{source}{TENDL-2011}
\setmatattr{56-Ba-132}{Emax}{200}

\registermaterial{56-Ba-134}
\setmatattr{56-Ba-134}{source}{TENDL-2011}
\setmatattr{56-Ba-134}{Emax}{200}

\registermaterial{56-Ba-135}
\setmatattr{56-Ba-135}{source}{TENDL-2011}
\setmatattr{56-Ba-135}{Emax}{200}

\registermaterial{56-Ba-136}
\setmatattr{56-Ba-136}{source}{TENDL-2011}
\setmatattr{56-Ba-136}{Emax}{200}

\registermaterial{56-Ba-137}
\setmatattr{56-Ba-137}{source}{TENDL-2011}
\setmatattr{56-Ba-137}{Emax}{200}

\registermaterial{56-Ba-138}
\setmatattr{56-Ba-138}{source}{TENDL-2011}
\setmatattr{56-Ba-138}{Emax}{200}

\registermaterial{57-La-138}
\setmatattr{57-La-138}{source}{TENDL-2011}
\setmatattr{57-La-138}{Emax}{200}

\registermaterial{57-La-139}
\setmatattr{57-La-139}{source}{TENDL-2011}
\setmatattr{57-La-139}{Emax}{200}

\registermaterial{58-Ce-136}
\setmatattr{58-Ce-136}{source}{TENDL-2011}
\setmatattr{58-Ce-136}{Emax}{200}

\registermaterial{58-Ce-138}
\setmatattr{58-Ce-138}{source}{TENDL-2011}
\setmatattr{58-Ce-138}{Emax}{200}

\registermaterial{58-Ce-140}
\setmatattr{58-Ce-140}{source}{TENDL-2011}
\setmatattr{58-Ce-140}{Emax}{200}

\registermaterial{58-Ce-142}
\setmatattr{58-Ce-142}{source}{TENDL-2011}
\setmatattr{58-Ce-142}{Emax}{200}

\registermaterial{64-Gd-152}
\setmatattr{64-Gd-152}{source}{TENDL-2011}
\setmatattr{64-Gd-152}{Emax}{200}

\registermaterial{64-Gd-154}
\setmatattr{64-Gd-154}{source}{TENDL-2011}
\setmatattr{64-Gd-154}{Emax}{200}

\registermaterial{64-Gd-155}
\setmatattr{64-Gd-155}{source}{TENDL-2011}
\setmatattr{64-Gd-155}{Emax}{200}

\registermaterial{64-Gd-156}
\setmatattr{64-Gd-156}{source}{TENDL-2011}
\setmatattr{64-Gd-156}{Emax}{200}

\registermaterial{64-Gd-157}
\setmatattr{64-Gd-157}{source}{TENDL-2011}
\setmatattr{64-Gd-157}{Emax}{200}

\registermaterial{64-Gd-158}
\setmatattr{64-Gd-158}{source}{TENDL-2011}
\setmatattr{64-Gd-158}{Emax}{200}

\registermaterial{64-Gd-160}
\setmatattr{64-Gd-160}{source}{TENDL-2011}
\setmatattr{64-Gd-160}{Emax}{200}

\registermaterial{68-Er-162}
\setmatattr{68-Er-162}{source}{TENDL-2011}
\setmatattr{68-Er-162}{Emax}{200}

\registermaterial{68-Er-164}
\setmatattr{68-Er-164}{source}{TENDL-2011}
\setmatattr{68-Er-164}{Emax}{200}

\registermaterial{68-Er-166}
\setmatattr{68-Er-166}{source}{TENDL-2011}
\setmatattr{68-Er-166}{Emax}{200}

\registermaterial{68-Er-167}
\setmatattr{68-Er-167}{source}{TENDL-2011}
\setmatattr{68-Er-167}{Emax}{200}

\registermaterial{68-Er-168}
\setmatattr{68-Er-168}{source}{TENDL-2011}
\setmatattr{68-Er-168}{Emax}{200}

\registermaterial{68-Er-170}
\setmatattr{68-Er-170}{source}{TENDL-2011}
\setmatattr{68-Er-170}{Emax}{200}

\registermaterial{71-Lu-175}
\setmatattr{71-Lu-175}{source}{TENDL-2011}
\setmatattr{71-Lu-175}{Emax}{200}

\registermaterial{71-Lu-176}
\setmatattr{71-Lu-176}{source}{TENDL-2011}
\setmatattr{71-Lu-176}{Emax}{200}

\registermaterial{72-Hf-174}
\setmatattr{72-Hf-174}{source}{TENDL-2011}
\setmatattr{72-Hf-174}{Emax}{200}

\registermaterial{72-Hf-176}
\setmatattr{72-Hf-176}{source}{TENDL-2011}
\setmatattr{72-Hf-176}{Emax}{200}

\registermaterial{72-Hf-177}
\setmatattr{72-Hf-177}{source}{TENDL-2011}
\setmatattr{72-Hf-177}{Emax}{200}

\registermaterial{72-Hf-178}
\setmatattr{72-Hf-178}{source}{TENDL-2011}
\setmatattr{72-Hf-178}{Emax}{200}

\registermaterial{72-Hf-179}
\setmatattr{72-Hf-179}{source}{TENDL-2011}
\setmatattr{72-Hf-179}{Emax}{200}

\registermaterial{72-Hf-180}
\setmatattr{72-Hf-180}{source}{TENDL-2011}
\setmatattr{72-Hf-180}{Emax}{200}

\registermaterial{73-Ta-181}
\setmatattr{73-Ta-181}{source}{JENDL/HE-2007}
\setmatattr{73-Ta-181}{Emax}{3000}

\registermaterial{74-W-180}
\setmatattr{74-W-180}{source}{JENDL/HE-2007}
\setmatattr{74-W-180}{Emax}{3000}

\registermaterial{74-W-182}
\setmatattr{74-W-182}{source}{JENDL/HE-2007}
\setmatattr{74-W-182}{Emax}{3000}

\registermaterial{74-W-183}
\setmatattr{74-W-183}{source}{JENDL/HE-2007}
\setmatattr{74-W-183}{Emax}{3000}

\registermaterial{74-W-184}
\setmatattr{74-W-184}{source}{JENDL/HE-2007}
\setmatattr{74-W-184}{Emax}{3000}

\registermaterial{74-W-186}
\setmatattr{74-W-186}{source}{JENDL/HE-2007}
\setmatattr{74-W-186}{Emax}{3000}

\registermaterial{75-Re-185}
\setmatattr{75-Re-185}{source}{TENDL-2011}
\setmatattr{75-Re-185}{Emax}{200}

\registermaterial{75-Re-187}
\setmatattr{75-Re-187}{source}{TENDL-2011}
\setmatattr{75-Re-187}{Emax}{200}

\registermaterial{78-Pt-190}
\setmatattr{78-Pt-190}{source}{TENDL-2011}
\setmatattr{78-Pt-190}{Emax}{200}

\registermaterial{78-Pt-192}
\setmatattr{78-Pt-192}{source}{TENDL-2011}
\setmatattr{78-Pt-192}{Emax}{200}

\registermaterial{78-Pt-194}
\setmatattr{78-Pt-194}{source}{TENDL-2011}
\setmatattr{78-Pt-194}{Emax}{200}

\registermaterial{78-Pt-195}
\setmatattr{78-Pt-195}{source}{TENDL-2011}
\setmatattr{78-Pt-195}{Emax}{200}

\registermaterial{78-Pt-196}
\setmatattr{78-Pt-196}{source}{TENDL-2011}
\setmatattr{78-Pt-196}{Emax}{200}

\registermaterial{78-Pt-198}
\setmatattr{78-Pt-198}{source}{TENDL-2011}
\setmatattr{78-Pt-198}{Emax}{200}

\registermaterial{79-Au-197}
\setmatattr{79-Au-197}{source}{JENDL/HE-2007}
\setmatattr{79-Au-197}{Emax}{3000}

\registermaterial{82-Pb-204}
\setmatattr{82-Pb-204}{source}{JENDL/HE-2007}
\setmatattr{82-Pb-204}{Emax}{3000}

\registermaterial{82-Pb-206}
\setmatattr{82-Pb-206}{source}{JENDL/HE-2007}
\setmatattr{82-Pb-206}{Emax}{3000}

\registermaterial{82-Pb-207}
\setmatattr{82-Pb-207}{source}{JENDL/HE-2007}
\setmatattr{82-Pb-207}{Emax}{3000}

\registermaterial{82-Pb-208}
\setmatattr{82-Pb-208}{source}{JENDL/HE-2007}
\setmatattr{82-Pb-208}{Emax}{3000}

\registermaterial{83-Bi-209}
\setmatattr{83-Bi-209}{source}{JENDL/HE-2007}
\setmatattr{83-Bi-209}{Emax}{3000}

\registermaterial{90-Th-232}
\setmatattr{90-Th-232}{source}{TENDL-2011}
\setmatattr{90-Th-232}{Emax}{200}

\registermaterial{92-U-235}
\setmatattr{92-U-235}{source}{JENDL/HE-2007}
\setmatattr{92-U-235}{Emax}{3000}

\registermaterial{92-U-238}
\setmatattr{92-U-238}{source}{JENDL/HE-2007}
\setmatattr{92-U-238}{Emax}{3000}


\setsublib{neutron}

\def\sublibtableheader{
    \T\# & Material & Source \#1 & Ecut  & Source \#2 & Emax  & HDC \\
         &          &            & [MeV] &            & [MeV] &     \\
\hline
}

\def\sublibtableheaderb{
    \T\# & Mat. & Source \#1 & Ecut  & Source \#2 & Emax  & H & \T\# & Mat. & Source \#1 & Ecut  & Source \#2 & Emax  & H   \\
         &          &            & [MeV] &            & [MeV] &     &      &          &            & [MeV] &            & [MeV] &   \\
\hline
}

\begin{table*}
\caption{neutron transport sublibrary}
\label{tbl:neutron-sublib1}
\def\tand{&}
    \begin{tabular}{r|llllll|r|llllll}
    \hline \hline
\sublibtableheaderb
  \ifcounter{ita}{}{\newcounter{ita}}
  \ifcounter{itb}{}{\newcounter{itb}}
  \setcounter{ita}{1}
  \setcounter{itb}{51}
  \whiledo{\theita<51}{
    \theita \tand
	\getmaterial{\theita} \tand
	\getmatattr{\getmaterial{\theita}}{lowsource} \tand
	\getmatattr{\getmaterial{\theita}}{Ecut} \tand
	\getmatattr{\getmaterial{\theita}}{hisource} \tand
	\getmatattr{\getmaterial{\theita}}{Emax} \tand
	\getmatattr{\getmaterial{\theita}}{konnocorrection} \tand 
    \theitb \tand
	\getmaterial{\theitb} \tand
	\getmatattr{\getmaterial{\theitb}}{lowsource} \tand
	\getmatattr{\getmaterial{\theitb}}{Ecut} \tand
	\getmatattr{\getmaterial{\theitb}}{hisource} \tand
	\getmatattr{\getmaterial{\theitb}}{Emax} \tand
	\getmatattr{\getmaterial{\theitb}}{konnocorrection} \\
    \ifnum\value{ita}=50\hline\hline\end{tabular}\fi
  \stepcounter{ita}
  \stepcounter{itb}
  }%
\end{table*}

\begin{table*}
\caption{neutron transport sublibrary}
\label{tbl:neutron-sublib2}
\def\tand{&}
    \begin{tabular}{r|llllll|r|llllll}
    \hline\hline
\sublibtableheaderb
  \ifcounter{ita}{}{\newcounter{ita}}
  \ifcounter{itb}{}{\newcounter{itb}}
  \setcounter{ita}{101}
  \setcounter{itb}{147}
  \whiledo{\theita<147}{
    \theita \tand
	\getmaterial{\theita} \tand
	\getmatattr{\getmaterial{\theita}}{lowsource} \tand
	\getmatattr{\getmaterial{\theita}}{Ecut} \tand
	\getmatattr{\getmaterial{\theita}}{hisource} \tand
	\getmatattr{\getmaterial{\theita}}{Emax} \tand
	\getmatattr{\getmaterial{\theita}}{konnocorrection} 
    \tand
    \theitb \tand
    \getmaterial{\theitb} \tand
    \getmatattr{\getmaterial{\theitb}}{lowsource} \tand
    \getmatattr{\getmaterial{\theitb}}{Ecut} \tand
    \getmatattr{\getmaterial{\theitb}}{hisource} \tand
    \getmatattr{\getmaterial{\theitb}}{Emax} \tand
    \getmatattr{\getmaterial{\theitb}}{konnocorrection}
    \\
    \ifnum\value{ita}=146\hline\hline\end{tabular}\fi
  \stepcounter{ita}
  \stepcounter{itb}
  }%
\end{table*}


\setsublib{proton}

\def\sublibtableheader{
\T\# & Material & Source & Emax  & \T\# & Material & Source & Emax  & \T\# & Material & Source & Emax \\
     &          &        & [MeV] &      &          &        & [MeV] &      &          &        & [MeV]\\
\hline
}

\begin{table*}
\caption{proton transport sublibrary}
\label{tbl:proton-sublib}
\def\tand{&}
\begin{tabular}{r|lll|r|lll|r|lll}
\hline\hline
\sublibtableheader
  \ifcounter{ita}{}{\newcounter{ita}}
  \ifcounter{itb}{}{\newcounter{itb}}
  \ifcounter{itc}{}{\newcounter{itc}}
  \setcounter{ita}{1}
  \setcounter{itb}{61}
  \setcounter{itc}{121}
  \whiledo{\theita<61}{
    \theita \tand
	\getmaterial{\theita} \tand
	\getmatattr{\getmaterial{\theita}}{source} \tand
	\getmatattr{\getmaterial{\theita}}{Emax} \tand
    \theitb \tand
	\getmaterial{\theitb} \tand
	\getmatattr{\getmaterial{\theitb}}{source} \tand
	\getmatattr{\getmaterial{\theitb}}{Emax}
    \ifnum\value{itc}<180
        \tand
		\theitc \tand
		\getmaterial{\theitc} \tand
		\getmatattr{\getmaterial{\theitc}}{source} \tand
		\getmatattr{\getmaterial{\theitc}}{Emax}
    \else
        \tand \tand \tand
    \fi
    \\
    \ifnum\value{ita}=60\hline\hline\end{tabular}\fi
  \stepcounter{ita}
  \stepcounter{itb}
  \stepcounter{itc}
  }%
\end{table*}

\clearpage

\section{Abbreviations used in the paper}

\label{apx:acronyms}
  \begin{tabular}{ll}
    \hline
    Abbreviation & Explanation \\
    \hline
      ACE & A Compact ENDF (file format) \\
      ADVANTG & Automated Variance Reduction Generator \\
      AISI-321 & Stainless Steel (SS321) \\
      ASPIS & Designation for a group of shielding benchmarks \\
      BB & Breeding Blanket \\
      BM & Blanket Module \\
      BROND & Bibliographic Referential, Evaluations, and Numerical Data Library (Russia) \\
      BZ & Breeding Zone \\
      CAD & Computer-Aided Design \\
      CCONE & A computer code system for nuclear data evaluation  \\
      CENDL & Chinese Evaluated Nuclear Data Library \\
      CIAE & Chinese Institute of Atomic Energy \\ 
      CIELO & Collaborative International Evaluated Library Organization \\
      CoNDERC & Compilation of Nuclear Experiments for Radiation Characterisation \\
      CPNG & Chinese Pulse Neutron Generator \\
      CRP & Coordinated Research Project \\
      CSEWG & Cross Section Evaluation Working Group (U.S.A and Canada) \\
      DENISMET-180 & A tungsten-nickel-iron alloy \\
      DEURACS & Code system dedicated for deuteron-induced reactions \\
      DPA & Displacements Per Atom \\
      EAF & European Activation File \\
      EFF & European Fusion File \\
      ELBE & Electron Linac for Beams with high Brilliance and low Emittance \\
      EMPIRE & A nuclear reaction code system for data evaluation\\
      ENDF-6 & Version 6 of the Evaluated Nuclear Data File Format \\
      ENEA & Italian National Agency for New Technologies, Energy and Sustainable Economic Development \\
      ENDF/B & Evaluated Nuclear Data Library B (U.S.A) \\
      EU DEMO & European demonstration power plant  \\
      EUROFER97 & A European reduced activation steel for plasma-facing components in nuclear fusion technology \\
      EXFOR & Experimental nuclear reaction database maintained by NRDC \\
      FENDL & Fusion Evaluated Nuclear Data Library \\
      FESS & Fusion Energy Systems Study \\
      FISPACT & Multiphysics inventory and source-term code system \\
      FNG & Frascati Neutron Generator \\
      FNS & Fusion Neutronics Source \\
      FNSF & Fusion Nuclear Science Facility \\
      FS & Ferritic Steel \\
      FW & First Wall \\
      FW-CADIS & Forward-Weighted Consistent Adjoint Driven Importance Sampling \\
      F4E & Fusion for Energy \\
      GNASH & Nuclear model code for calculation of cross sections and emission spectra \\
      HCPB-TBM & Helium Cooled Pebble Bed Test Blanket Module \\
      HPGe & High Purity Germanium \\
      IAEA & International Atomic Energy Agency \\
      IAEA/NDS & Nuclear Data Section of the IAEA \\
      IB & Inboard (region) \\
      IBCC & Inboard Region Magnet Coil Case \\
      IBVV & Inboard Region Vacuum Vessel \\
      ICSBEP & International Criticality Safety Benchmark Evaluation Project \\
      IFJ-PAN & The Henryk Niewodniczanski Institute of Nuclear Physics, Polish Academy of Sciences \\
      IFMIF & International Fusion Materials Irradiation Facility \\
      INCONEL & A nickel-chromium-based superalloy for extreme conditions \\
      INDEN & International Nuclear Data Evaluation Network \\
      IRDFF & International Reactor Dosimetry and Fusion File \\
      ITER & International Thermonuclear Experimental Reactor \\
    \hline 
  \end{tabular}

\clearpage

  \begin{tabular}{ll}
    \hline
    Abbreviation & Explanation \\
    \hline
      JADE & Python-based open-source software for verification and validation of nuclear data \\
      JAEA & Japanese Atomic Energy Agency \\
      JAERI & Japan Atomic Energy Research Institute \\
      JEFF & Joint Evaluation Fission and Fusion File (maintained at NEA Data Bank) \\
      JENDL & Japanese Evaluated Nuclear Data Library \\
      JENDL/HE & JENDL High Energy File \\
      JENDL/DEU & JENDL Deuteron Reaction Data File \\
      JET & Joint European Torus \\
      JRC Geel & Joint Research Centre Geel \\
      KERMA & Kinetic Energy Released per unit Mass \\
      LANL & Los Alamos National Laboratory \\
      LLNL & Lawrence Livermore National Laboratory \\
      LTIS & Long-term Irradiation Station \\
      LTsh & Low Temperature Shield \\
      LAW & Variable in ENDF-6 file to specify representation of energy-distribution of reaction product  \\
      MATXS & File format to store cross-section data used by TRANSX \\
      McDeLicious & An extension to the MCNP Monte Carlo code to simulate neutron generation from D-Li interaction \\
      MCNP & Monte Carlo N-Particle code developed at LANL \\
      MF/MT & Index numbers pointing to subdivisions in an ENDF-6 formatted file \\
      MF82H & Low activation modified ferritic steel (90\,Fe, 7.5\,Cr, 2.0\,W, 0.2\,V, 0.02\,Ta, 0.1\,C (in wt.\%)) \\
      NCSRD & National Centre for Scientific Research `Demokritos' \\
      nELBE & Neutrons at ELBE \\
      NJOY & Nuclear data processing code developed at LANL \\
      NPA & Net Peak Area \\
      NRDC & Nuclear Reaction Data Centres Network \\ 
      OB & Outboard (region) \\
      OBLTsh & Outboard Low Temperature Shield \\
      OECD & Organisation for Economic Co-operation and Development \\ 
      OECD/NEA & OECD Nuclear Energy Agency \\
      OKTAVIAN & Experimental facility located at the Osaka University \\
      PbLi & Alloy of Pb and Li investigated as breeding blanket material  \\
      PF & Poloidal Field \\ 
      PFC & Plasma Facing Component \\
      QST & National Institutes for Quantum Science and Technology (Japan) \\
      RAFM & Reduced-Activation Ferritic-Martensitic (steel) \\ 
      RRR & Resolved Resonance Region \\
      RUSFOND & Russian National Library of Evaluated Neutron Data \\
      SB & Shield Block \\
      SINBAD & Shielding Integral Benchmark Archive and Database \\
      SR & Structural Ring \\
      SS316 & Stainless Steel 316 \\
      TALYS & A nuclear models code \\
      TBM & Test Blanket Module \\
      TBR & Tritium Breeding Ratio \\
      TENDL & TALYS Evaluated Nuclear Data Library  \\
      TF & Toroidal Field \\
      THshield & Thermal shield \\
      TIARA & Takasaki Ion Accelerators for Advanced Radiation Application \\
      TOF & Time Of Flight \\
      TRANSP & A free-boundary equilibrium and transport solver \\
      TRANSX & Computer code to produce input files for transport codes from MATXS files \\
      UKAEA & United Kingdom Atomic Energy Authority \\
      VV & Vacuum Vessel \\
      V\&V & Verification and Validation \\
      WCLL-TBM & Water Cooled Lithium Lead Test Blanket Module \\
      WPEC & Working Party on International Nuclear Data Evaluation Co-operation \\
    \hline 
  \end{tabular}

\clearpage

\end{document}